%% file: diss_main.tex
\documentclass[12pt,titlepage]{report} 

\usepackage{epsfig}
\usepackage{lscape}
\usepackage{afterpage}

\input{mytitlepage}
\input{mythesismacros}

\RequirePackage{thesisla}	

\def\Journal#1#2#3#4{{#1} {\bf #2}, #3 (#4)}

\def\NPA{{\rm Nucl. Phys.} A}
\def\NPB{{\rm Nucl. Phys.} B}
\def\PLB{{\rm Phys. Lett.}  B}
\def\PRL{\rm Phys. Rev. Lett.}
\def\PRD{{\rm Phys. Rev.} D}
\def\PRC{{\rm Phys. Rev.} C}
\def\ZPC{{\rm Z. Phys.} C}
\def\PAN{\rm Phys. At. Nuclei }
\def\la{\langle}
\def\ra{\rangle}

\def\lam{\lambda}
\def\gam{\gamma}

\def\vep{\varepsilon}
\def\ra{\rangle}
\def\la{\langle}

\def\al{\alpha}

\def\be{\begin{equation}}
\def\ee{\end{equation}}
\def\bea{\begin{eqnarray}}
\def\eea{\end{eqnarray}}

\begin{document}

\title{ Light-Front Quark Model Analysis of
  Electroweak Decays of Pseudoscalar
  and Vector Mesons}

\author{Ho-Meoyng Choi}

\date{1999}

\pagenumbering{empty}
\thispagestyle{empty}
\baselineskip 1.95em 

\input{abstract}

\pagenumbering{roman}		
\thispagestyle{empty}
\baselineskip 1em

\maketitle

\baselineskip 2em
\setcounter{page}{2}		




\newpage
\thispagestyle{myheadings}
\baselineskip 1em

\tableofcontents		
\listoftables			
\listoffigures			

\newpage
\pagenumbering{arabic}		
\baselineskip 2em 

\input{psfig} 
\input{Intro}
\input{NPA}

\input{Relations}

\input{Mixing}

\input{anal_final}
\input{zm}

\input{Semi}

\input{Conclusion}
\newpage
\thispagestyle{myheadings}
\baselineskip 2em

\input{references}
\newpage
\appendix
\input{appendix1}
\input{appendix2}
\input{appendix3}
\input{appendix4}
\input{My_resume}

\end{document}

%% file: mytitlepage.tex
\makeatletter          

\def\maketitle{\begin{titlepage}%
 \let\footnotesize\small
 \let\footnoterule\relax
 \setcounter{page}{1}%
 \begin{center}%
   {\large \bf \@title \\}%
   \vskip 2.2cm
   {by \par}%
   \vskip 1em 
   {\bf
     \begin{tabular}[t]{c}\@author
     \end{tabular}\\}%
   \vskip 2.2cm
   {A \thesisname\ submitted to the Graduate Faculty of \\}
   {North Carolina State University \\}
   {in partial fulfillment of the \\}
   {requirements for the Degree of \\}
   {Doctor of Philosophy \\}
   \vskip 1.6em
   {\bf Department of Physics \\}
   \vskip 1.8cm
   {Raleigh \\}
   \vskip 1.6em
   {\@date \\}%
   \vskip 1.8cm
   {\bf APPROVED BY: \\}
   \vskip 1.8cm
   \rule{6.8cm}{0.16mm} \hspace{1.2cm} \rule{6.8cm}{0.16mm}\\
   \vskip 1.5cm
   \rule{6.8cm}{0.16mm} \hspace{1.2cm} \rule{6.8cm}{0.16mm}\\
   \hspace{8.1cm}
   \makebox[6.5cm]{Chair of Advisory Committee}
 \end{center}\par
\@thanks
\vfil
\end{titlepage}%
\setcounter{footnote}{0}%
\let\thanks\relax
\gdef\@thanks{}\gdef\@author{}\gdef\@title{}\let\maketitle\relax}

\def\thesisname{dissertation}

\def\abstract{\titlepage
\null
\begin{center}
 \bf \abstractname
\end{center}}

\def\prepage{\@restonecolfalse\if@twocolumn\@restonecoltrue\onecolumn
     \else \newpage \fi}

\def\endprepage{\if@restonecol\twocolumn \else \newpage \fi}

\makeatother           

%% file: mythesismacros.tex
\makeatletter                         

\def\newitheorem#1{\@ifnextchar[{\@iothm{#1}}{\@inthm{#1}}}
 
\def\@inthm#1#2{%
\@ifnextchar[{\@ixnthm{#1}{#2}}{\@iynthm{#1}{#2}}}
 
\def\@ixnthm#1#2[#3]{\expandafter\@ifdefinable\csname #1\endcsname
{\@definecounter{#1}\@addtoreset{#1}{#3}%
\expandafter\xdef\csname the#1\endcsname{\expandafter\noexpand
  \csname the#3\endcsname \@thmcountersep \@thmcounter{#1}}%
\global\@namedef{#1}{\@thm{#1}{#2}}\global\@namedef{end#1}{\@enditheorem}}}
 
\def\@iynthm#1#2{\expandafter\@ifdefinable\csname #1\endcsname
{\@definecounter{#1}%
\expandafter\xdef\csname the#1\endcsname{\@thmcounter{#1}}%
\global\@namedef{#1}{\@ithm{#1}{#2}}\global\@namedef{end#1}{\@enditheorem}}}

\def\@iothm#1[#2]#3{%
  \@ifundefined{c@#2}{\@latexerr{No theorem environment `#2' defined}\@eha}%
  {\expandafter\@ifdefinable\csname #1\endcsname
  {\global\@namedef{the#1}{\@nameuse{the#2}}%
\global\@namedef{#1}{\@thm{#2}{#3}}%
\global\@namedef{end#1}{\@enditheorem}}}}
 
\def\@ithm#1#2{\refstepcounter
    {#1}\@ifnextchar[{\@iythm{#1}{#2}}{\@ixthm{#1}{#2}}}
 
\def\@ixthm#1#2{\@beginitheorem{#2}{\csname the#1\endcsname}\ignorespaces}
\def\@iythm#1#2[#3]{\@opargbeginitheorem{#2}{\csname
       the#1\endcsname}{#3}\ignorespaces}

\def\@beginitheorem#1#2{\trivlist \item[\hskip \labelsep{\it #1\ #2.}]}
\def\@opargbeginitheorem#1#2#3{\trivlist
      \item[\hskip \labelsep{\it #1\ #2\ (#3).}]}
\def\@enditheorem{\endtrivlist}

\def\ps@cornerplain{\let\@mkboth\@gobbletwo
     \let\@oddhead\@empty\def\@oddfoot{\reset@font\rm\hfil\thepage}%
     \let\@evenhead\@empty\let\@evenfoot\@oddfoot}

\def\ps@centermyheadings{\let\@mkboth\@gobbletwo
 \def\@oddhead{{\sl\rightmark}\hfil \rm\thepage \hfil}%
 \def\@oddfoot{}\def\@evenhead{\hfil \rm \thepage\hfil\sl\leftmark}%
 \def\@evenfoot{}\def\chaptermark##1{}\def\sectionmark##1{}%
 \def\subsectionmark##1{}}

\makeatother                          

\newcommand{\off}[1]{} 

\newitheorem{definition}{Definition}[chapter]
\newitheorem{property}{Property}[chapter]

\newcounter{ralgocounter}[chapter]
\renewcommand{\theralgocounter}{\thechapter.\arabic{ralgocounter}}
{\\ \rule{6.0in}{1mm} \end{tabbing}  
\end{trivlist} } 

{\hspace*{\fill}\ {\bf (Q.E.D)} \end{trivlist}}

%% file: abstract.tex
\begin{center}
{\Large\bf ABSTRACT}
\end{center}
Choi, Ho-Meoyng. Light-Front Quark Model Analysis of
  Electroweak Decays of Pseudoscalar and Vector Mesons.
(Under the direction of Chueng-Ryong Ji.)

We investigate the electroweak form factors and semileptonic 
decay rates of pseudoscalar and vector mesons using 
the light-front constituent quark model.
Exploring various detailed schemes of model building, we attempt to
fill the gap between the model wave function and the QCD-motivated
Hamiltonian which includes not only
the Coulomb plus confining potential but also the hyperfine
interaction to obtain the correct $\rho$-$\pi$ splitting.
The variational principle to this Hamiltonian 
significantly constrains our model.
For the confining potential, we use both (1) harmonic oscillator 
potential and (2) linear potential to compare the numerical 
results for these two cases. Furthermore, 
our method of analytic continuation in the Drell-Yan-West 
($q^+$=$q^0+q^3$=0) frame to obtain the 
weak form factors avoids the difficulty associated with the 
contributions from the nonvalence quark-antiquark pair creation. 
This method seems to provide a promising step forward to handle the exclusive
processes in the timelike region.
In the exactly solvable model of scalar field theory interacting with 
gauge field, we explicitly show that our method of analytic continuation 
automatically yields the effect of complicated nonvalence contributions. 
Our calculation reveals that the zero mode 
contribution is crucial to obtaining the 
correct values of the light-front current $J^{-}$ in the $q^+$=0 frame. 
We quantify the zero mode contributions and observe that the zero mode
effects are very large for the light meson form factors even though
they may be reduced for the heavy meson cases.
The mass spectra of ground state pseudoscalar and vector mesons and 
mixing angles of $\omega$-$\phi$ and $\eta$-$\eta'$ are predicted 
and various physical observables such as decay constants, charge radii, 
radiative and semileptonic decay rates, etc., are calculated. 
Our numerical results are overall in a very good agreement with the available 
exprimental data.


%% file: Intro.tex
\newpage 
\setcounter{chapter}{0}
\setcounter{equation}{0}
\setcounter{figure}{0}
\renewcommand{\theequation}{\mbox{1.\arabic{equation}}}
\chapter{Introduction}
Quantum Chromodynamics (QCD)~\cite{QCD}, which describes the strong 
interactions of quarks and glouns, has been a very active field
of research in nuclear and particle physics. 
To calculate the hadronic properties from QCD, however, is such
a formidable task that progress has been slow and many 
problems are still not solved today. The equations to be solved
are highly non-linear and one is dealing with formulas for infinitely
many relativistic particles, i.e., a quantum
field theory requiring careful renormalization. Thus QCD presents
one of the most challenging problems of theoretical physics. 

Presently, however, this situation is changing. The continuous slow
progress in computational and analytic techniques has now reached 
a point where real QCD calculations have become feasible for a 
rapidly increasing number of problems in hadron physics. Together
with the much improved experimental situation this should trigger
an enormous boost for hadron physics, e.g., various static and
non-static hadron properties such as form factors, structure functions
and cross sections etc., in the years ahead. 
These results are thought to be very important as they will help us
in understanding the role of QCD or the quark substructure and their
interactions inside the nuclear matter. For best comparison with the
experiment, reliable and versatile theoretical calculations have to
be at hand. 
To avoid the extreme complexity of real QCD, many phenomenological
``QCD-inspired" models have been developed, often with astonishing
success. 

The main purpose of this thesis is to develope ``QCD-inspired" model 
describing the mass spectrum and hadron properties of low-lying
hadrons at small momentum transfer region(nonperturbative QCD). 
Even though more fundamental nonperturbative QCD methods such as QCD
sum-rule techniques~\cite{chernyak,narison,rad} and lattice
QCD calculations~\cite{gott,loft,gupta} are available,
there is still growing interest in using simple
relativistic quark models~\cite
{teren,jaus1,jaus,chung2,IM,Mix,Bag,azn,Kar,card,tao,sch,dziem1,ji,spin}
to describe hadron properties.
Our relativistic quark model~\cite{IM,Mix,ji,spin}
is based upon the Fock state decomposition of hadronic state
which arises naturally in the ``light-front quantization" of 
QCD\footnote{The interested readers are referred to extensive 
review by Brodsky, Pauli and Pinsky~\cite{BPP}}.

\section{Light-Front Degrees of Freedom}

The foundations of light-front (LF) quantization date back to 
Dirac~\cite{Dirac} who showed that there are remarkable advantages in 
quantizing relativistic field theories at a particular value of LF 
time $\tau=t + z/c$ rather than at a particular ordinary time $t$.  
In this formalism, a hadron is characterized by a set of Fock state wave
functions, the probability amplitudes for finding different combinations of
bare quarks and gluons in the hadron at a given LF time $\tau$.
For example, a meson can be described by 
\be\label{fock} 
|M\ra = \sum_{q\bar{q}}\psi_{q\bar{q}/M} + 
\sum_{q\bar{q}g}\psi_{q\bar{q}g}|q\bar{q}g\ra\psi_{q\bar{q}g/M}
+ \cdots.
\ee 
These wave functions provide the essential link between hadronic
phenomena at short distances(perturbative) and at long
distances(non-perturbative)~\cite{Lepage}.
The quantum field theory at equal-$\tau$ is quite different
from the one at equal-$t$. The distinguished features in the LF  
approach are the dynamical property of rotation operators
and the simplicity of the vacuum except the 
zero modes~\cite{zm,zero,BH,Fred}. 
Let us begin by illustrating these two distinguished features.

The boost operators of the ten Poincar\'{e} generators 
contain interactions (i.e. dynamical operators) changing particle 
numbers in equal $t$. Thus, solving the relativistic scattering and bound 
state problems in the cannonical quantization at the equal $t$, 
depends on the reference frame due to the boost non-invariance 
when the Fock-space is truncated for practical calculations.  
The problem of boost operators changing
particle numbers at equal $t$ can be cured by the LF quantization  
since the quantization surface $\tau$ = 0 is invariant under
both longitudinal and transverse boosts (i.e. kinematical operators)
defined at equal $\tau$.
In return, however, the rotational invariance is violated
at equal $\tau$ when the Fock space is truncated for practical
calculations~\cite{JS}. The transverse angular momentum whose 
direction is perpendicular to the direction of the quantization axis $z$ 
in equal $\tau$ involves interactions (i.e. dynamical operators)
and thus it is not easy to specify the
total angular momentum of a particular hadronic state.  
Also $\tau$ is not invariant under parity~\cite{Sop}.
The number of problems seems to conserve
even though the boost problem at equal $t$ is now replaced by the
rotation problem at equal $\tau$.
However, we point out that the rotation problem is
much easier to deal with compared to the boost problem
because the rotation is compact, i.e., closed and periodic~\cite{JCon}.
Moreover, the problem of assigning quantum numbers $J^{PC}$ to hadrons
can be circumvented by the Melosh transformation~\cite{Mel}. 
This ought to be regarded as an advantage of the LF quantization
over the ordinary equal $t$ quantization.
The restoration of the rotational symmetry in the
LF quantization has also been discussed~\cite{JKM}.
  
The characteristics of vacuum at equal $\tau$ has also
a dramatic difference compared to the vacuum properties at equal $t$.
Suppose that a particle has mass $m$ and four-momentum
$k = (k^{0},k^{1},k^{2},k^{3})$, then the relativistic energy-momentum
relation of the particle at equal-$\tau$ is given by
\be \label{dis}
k^{-} = \frac{{\bf k}_{\bot}^{2} + m^{2}}{k^{+}},
\ee
where the LF energy conjugate to $\tau$ is given by
$k^{-} = k^{0} - k^{3}$ and the LF momenta $k^{+} =
k^{0} + k^{3}$ and ${\bf k}_{\bot} = (k^{1},k^{2})$ are orthogonal
to $k^{-}$ and form the LF three-momentum $\underline{k} =
(k^{+},{\bf k}_{\bot})$.
The rational relation given by Eq. (\ref{dis}) is
drastically different from the irrational energy-momentum
relation at equal-$t$ given by
\be \label{dis2}
k^{0} = \sqrt{{\bf k}^{2} + m^{2}},
\ee
where the energy $k^{0}$ is conjugate to $t$ and the three-momentum
vector ${\bf k}$ is given by ${\bf k} = (k^{1},k^{2},k^{3})$.
The important point here is that the signs of $k^{+}$ and $k^{-}$ are
correlated and thus the momentum $k^{+}$ is always positive because
only the positive energy $k^{-}$ makes the system evolve to the future
direction (i.e. positive $\tau$), while at equal-$t$ the signs of
$k^{0}$ and ${\bf k}$ are not correlated and thus the momentum $k^{3}$
corresponding to $k^{+}$ of equal-$\tau$ can be either positive or negative.
This provides a remarkable feature to the LF vacuum; i.e.,
the trivial vacuum of the free LF theory is an
eigenstate of the full Hamiltonian, viz., the true vacuum~\cite{JCon}.
This can be proved by showing that the full LF Hamiltonian
annihilates the trivial perturbative vacuum~\cite{BMPP}.
For example, in QED, the application of the interaction
$H^{I}_{LF} = \int {{d^{3}{\underline x}} {\overline \psi} \gamma^{\mu}
\psi A_{\mu}}$ to the perturbative vacuum $|0 \rangle$
results in a sum of terms $b^{\dag}({\underline k}_{1})
a^{\dag}({\underline k}_{2})d^{\dag}({\underline k}_{3})|0 \rangle$.
While the conservation of the LF momentum requires
$\sum_{i=1}^{3} k_{i}^{+} = 0$, the massive fermions with finite
$k_{i}^{-}$ cannot have $k_{i}^{+} = 0$ due to Eq.(\ref{dis}).
Thus, $H^{I}_{LF}$ annihilates the trivial vacuum $|0 \rangle$
and so does the full Hamiltonian $H_{LF} = H^{I}_{LF} + H^{0}_{LC}$
since $|0 \rangle$ is annihilated by the free Hamiltonian $H^{0}_{LF}$
by definition. This feature is drastically different from the
equal-$t$ quantization where the state $H |0 \rangle$ is a
highly complex composite of pair fluctuations.
This greatly complicates the interpretation of the hadronic wave 
functions at equal-$t$ quantization.

This apparent simplicity of the LF vacuum may yield
a problem to understand the novel phenomena such as the spontaneous
symmetry breaking, Higgs mechanism, chiral symmetry breaking, axial
anomaly, $\theta$-vacuua, etc., because these were known as the direct
consequencies of the nontrivial vacuum structures of various
field theories. Thus, the question of how one can realize
these nontrivial vacuum phenomena from the trivial LF vacuum 
arises~\cite{Horn}. In fact, some efforts have been made to show that 
nontrivial vacuum phenomena 
can still be realized in the LF quantization approach if one takes into 
account the nontrivial zero-mode ($k^{+}=0$) contributions~\cite{zero,Rey}.
As an example, it was shown~\cite{Rey} that the axial anomaly in the 
Schwinger model can be obtained in the LF quantization approach by 
carefully analyzing the contributions from zero modes.

\section{Construction of Light-Front Quark Model}   
The distinguished features in the LF approach can now be summarized 
as (1) the dynamical property of the rotation operators and 
(2) the simplicity of the vacuum except the zero modes. 
Thus, in the LF quantization approach, one can take advantage
of the rational energy-momentum dispersion relation and build a 
clean Fock state expansion of hadronic wave functions
based on a simple vacuum by decoupling the complicated nontrivial zero modes.
The decoupling of zero modes can be achieved in the light-front quark 
model (LFQM) since the constituent quark and antiquark acquire 
appreciable constituent masses.
Furthermore, recent lattice QCD results~\cite{Ku} indicated that the mass
difference between $\eta'$ and pseudoscalar octet mesons due to the
complicated nontrivial vacuum effect increases (or decreases) as
the quark mass $m_{q}$ decreases (or increases); i.e.,
the effect of the topological charge contribution should be small
as $m_{q}$ increases. This supports us in building the constituent 
quark model in the LF quantization approach because
the complicated nontrivial vacuum effect in QCD
can be traded off by the rather large constituent quark masses.

One can also provide a well-established formulation of various
form factor calculations in the LF quantization method using the 
well-known Drell-Yan-West ($q^{+}=q^{0} + q^{3}=0$) frame~\cite{Drell,LB}, 
which provided an effective formulation for the calculation of
various form factors in the space-like momentum transfer region
$q^{2}=-Q^{2}<0$. In $q^{+}=0$ frame, only parton-number-conserving Fock
state (valence) contribution is needed when the ``good" components of the
current, $J^{+}$ and ${\bf J}_{\perp}=(J_{x},J_{y})$, are used~\cite{Kaon}.
For example, only the valence diagram shown in 
Fig.~1.1(a)  is used in the
LFQM analysis of spacelike meson form factors.
\begin{figure}\label{toy1}
\centerline{\psfig{figure=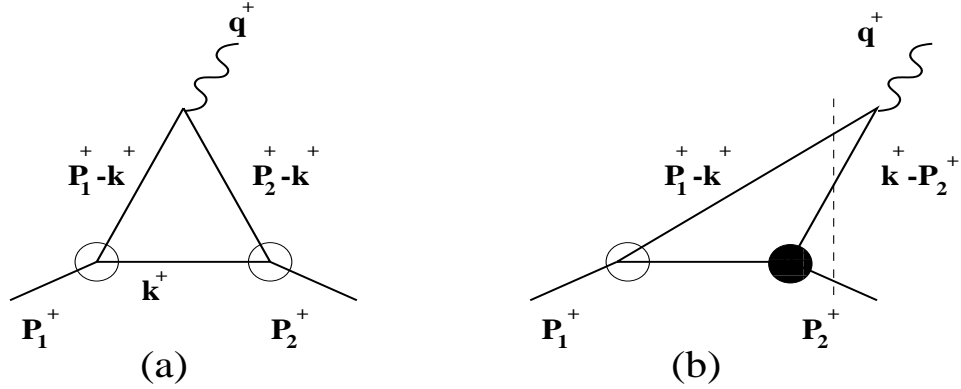,width=5in,height=2.0in}}
\caption{The LFQM description of a electroweak
meson form factor:
(a) the usual LF valence diagram and (b) the
nonvalence(pair-creation) diagram. The vertical dashed line in (b)
indicates the energy-denominator for the nonvalence contributions.
While the white blob represents the usual LF
valence wave function, the modeling of black blob has not yet been made.}
\end{figure}

The key approximation of the LFQM is the mock-hadron 
approximation~\cite{HI} to truncate the expansion by retaining only the 
lowest Fock state and treat the lowest Fock state as a
free state as far as the spin-orbit part is concerned while the radial
part is given by the ground state of the harmonic oscillator 
wave function.  
Then, the assignment of the quantum numbers such as angular
momentum, parity and charge conjugation can be given to the LF 
wave functions by the Melosh transformation~\cite{Mel}. 
For example, the meson bound state $|M\rangle$ is represented by
\be\label{eq.121} 
|M\rangle = \Psi^{M}_{Q\bar{Q}}|Q\bar{Q}\rangle,
\ee
where $Q$ and $\bar{Q}$ are the effective dressed quark and
antiquark. The model wave function in momentum space is given 
by\footnote{The wave function in Eq.(\ref{eq.122}) is represented by the 
Lorentz-invariant variables
$x_{i} = p^{+}_{i}/P^{+}$, ${{\bf k}_{\perp i}} = {{\bf p}_{\perp i}}
- x_{i}{\bf P_{\perp}}$ and $\lam_{i}$, where
$P^{\mu} = (P^{+},P^{-},{\bf P_{\perp}})= ( P^{0}+P^{3},
(M_{0}^{2}+{\bf P}^2_\perp)/P^{+},{\bf P_{\perp}})$
is the momentum of the meson $M$, and $p^{\mu}_{i}$ and $\lam_{i}$
are the momentum and the helicity of constituent quarks, respectively.}
\be\label{eq.122}
\Psi^{M}_{Q\bar{Q}}=
\Psi^{JJ_{3}}_{\lam_{q},\lam_{\bar{q}}}(x,{\bf k}_{\perp})=
\sqrt{\frac{\partial k_{n}}{\partial x}}
\phi(x,{\bf k}_{\perp})
{\cal R}^{JJ_{3}}_{\lam_{q},\lam_{\bar{q}}}(x,{\bf k}_{\perp}),
\ee
where $\phi(x,{\bf k}_{\perp})$ is the radial wave function and 
$\partial k_{n}/\partial x$ is a Jacobi factor, and
${\cal R}^{JJ_{3}}_{\lam_{q},\lam_{\bar{q}}}(x,\\{\bf k}_{\perp})$ 
is the spin-orbit wave function obtained by the interaction-independent
Melosh transformation from the ordinary equal-time
static spin-orbit wave function assigned by the quantum numbers $J^{PC}$.
When the longitudinal
component $k_{n}$ is defined by $k_{n}= (x-1/2)M_{0} +
(m^{2}_{\bar{q}}-m^{2}_{q})/2M_{0}$, the Jacobian of the variable
transformation $\{x,{\bf k}_{\perp}\}\rightarrow {\bf k}=
(k_{n},{\bf k}_{\perp})$ is given by
\be\label{jacob}
\frac{\partial k_{n}}{\partial x}
= \frac{M_{0}}{4x(1-x)}\biggl\{ 1 - \biggl[\frac{(m^{2}_{q}
- m^{2}_{\bar{q}})}{M^{2}_{0}}\biggr]^{2}\biggr\}.
\ee
The explicit form of the spin-orbit wave function is given by
\bea\label{eq.123} 
{\cal R}^{JJ_{3}}_{\lam_{q}\lam_{\bar{q}}}(x,{\bf k}_{\perp})&=&
\sum_{\lam'_{q},\lam'_{\bar{q}}}
\langle\lam_{q}|{\cal R}^{\dagger}_{{\rm M}}(x,{\bf k}_{\perp},m_{q})
|\lam'_{q}\rangle \nonumber\\
&\times&
\langle\lam_{\bar{q}}|{\cal R}^{\dagger}_{{\rm M}}(1-x,-{\bf k}_{\perp},
m_{\bar{q}})|\lam'_{\bar{q}}\rangle
\langle\frac{1}{2}\lam'_{q}\frac{1}{2}\lam'_{\bar{q}}
|JJ_{3}\rangle,
\eea 
where 
\be\label{eq.124}
{\cal R}_{{\rm M}}(x,{\bf k}_{\perp},m)=
\frac{m + xM_{0} - i\sigma\cdot(\hat{\bf n}
\times \hat{\bf k})}{\sqrt{(m+xM_{0})^{2} + {\bf k}^{2}_{\perp}}},
\ee
is the Melosh transformation operator with $\hat{\bf n}$=(0,0,1) 
being a unit vector in the $z$ direction.

Although the spin-orbit wave function is in principle uniquely determined
by the Melosh transformation given by Eq. (\ref{eq.124}), a couple of
different schemes for handling the meson mass $M_{0}$ in 
Eq. (\ref{eq.124}) have appeared in the literature~
\cite{teren,jaus1,jaus,chung2,IM,Mix,Bag,azn,card,tao,sch,dziem1,ji,spin}.
While in the invariant meson mass (IM) 
scheme~\cite{teren,jaus1,jaus,chung2,IM,Mix,Bag,azn,card,tao,sch},
the meson mass square $M^{2}_{0}$ is given by
\be\label{inv} 
M^{2}_{0}= \frac{{\bf k}_{\perp}^{2} + m^{2}_{q}}{x}
+ \frac{{\bf k}_{\perp}^{2} + m^{2}_{\bar{q}}}{1-x},
\ee
in the spin-averaged meson mass (SM) scheme~\cite{dziem1,ji,spin}, 
$M_{0}$ was taken as the average of physical masses with appropriate 
weighting factors from the spin degrees of freedom.

This thesis is divided mainly into two parts, i.e., the analyses
of the spacelike ($q^{2}<0$) and the timelike ($q^{2}>0$) form factors.
In Chapter 2 of this thesis~\cite{spin}, we first study the radiative 
decays of pseudoscalar ($\pi, K,\eta,\eta'$), 
vector ($\rho, K^{*}, \omega, \phi$) and axial vector ($A_{1}$) mesons 
using the SM scheme, which was originally argued by Dziembowsky 
and Mankiewicz~\cite{dziem1} and later developed by 
Ji and Cotanch~\cite{ji}. The purpose of this work is to investigate more 
observables with the same model as Refs.~\cite{dziem1,ji}. 
Our overall predictions of pseudoscalar, vector and axial-vector 
meson radiative decay processes are in remarkably good agreement with the 
experimental data.

In Chapter 3~\cite{IM}, we discuss the relations among the 
LF quark models~\cite{jaus,chung2,tao,sch} which
are based on the IM scheme. We noticed that there are a couple of
differences within the same IM sheme, i.e., (a) the presence-absence of
the Jacobi factor $\partial k_n/\partial x$ 
in choosing harmonic oscillator (HO) wave function and (b) 
the difference in the choice of radial wave function, e.g., 
HO versus power-law wave functions. 
We found that the difference from the Jacobi factor is substantially
reduced in the numerical predictions for the physical observables
once the best fit parameters are chosen. The difference in the 
choice of radial wave function is, however, still appreciable even 
though the best fit parameters are used.
Comparing the two different meson mass schemes, i.e., SM and IM schemes,
we also found that once the best fit parameters were used both schemes 
provided the predictions that were not only pretty similar with each other 
but also remarkably good compared to the available experimental data for 
form factors, decay constants, charge radii, etc., of various light
pseudoscalar and vector mesons as well as their radiative decay widths.
Through the analyses in Chapters 2 and 3, we model the radial wave function
rather than a potential.

In Chapter 4~\cite{Mix}, we develop LFQM attempting to fill the gap 
between the model wave function and the QCD-motivated
potential, which includes not only the Coulomb plus confining potential
but also the hyperfine interaction, to obtain the correct $\rho$-$\pi$
splitting. In our LFQM~\cite{Mix}, we analyzed both the mass spectra and 
the wave functions of the light pseudoscalar and vector
mesons and predicted mixing angles of $\omega$-$\phi$ and $\eta$-$\eta'$ 
and various physical observables such as decay constants, charge radii,
radiative decay rates, etc.  

The timelike ($q^{2}>0$) form factor analysis in the LFQM 
is more subtle than the analysis of the spacelike form factors since
the $q^{+}=0$ frame is defined only in the spacelike region ($q^{2}<0$).
While the $q^{+}\neq0$ frame can be used in principle to compute the 
timelike form factors, it is inevitable (if $q^{+}\neq 0$) to encounter 
the nonvalence diagram arising from the quark-antiquark pair creation 
(so called ``Z-graph"). For example, the nonvalence diagram in the case 
of semileptonic meson decays is shown in Fig.~1.1(b).
The main source of the difficulty, however, in calculating the nonvalence
diagram is the lack of information on the black blob which
should contrast with the white blob representing the usual LF 
valence wave function. In fact, we recently found~\cite{Kaon} in the
analysis of kaon electorweak ($k_{\ell3}$) decays that
the omission of nonvalence contribution
leads to a large deviation from the full results.
The timelike form factors associated with the hadron pair productions in
$e^{+}e^{-}$ annihilations also involve the nonvalence contributions.
Therefore, it would be very useful to avoid encountering the nonvalence
diagram and still be able to generate the results of timelike form factors.
This can be done by the analytic continuation from the spacelike 
form factor calculated in the Drell-Yan-West ($q^{+}=0$) frame 
to the timelike region. 

Before applying our LFQM~\cite{Mix} to the electroweak form factors and
semileptonic decay rates of pseudoscalar and vector mesons,
we start from the heuristic model calculations
to compare the results of the timelike form factors obtained from
both $q^{+}\neq 0$ (i.e. valence + nonvalence contributions) and
$q^{+}=0$ (i.e. valence contribution only) and to show the analytic
continuation method is correct. The choice of the components of the 
LF current $J^\mu$ is especially important
when one needs to calculate timelike processes in $q^{+}=0$ frame such 
as the semileptonic decays of pseudoscalar mesons. We also come to this
point later in this thesis.   

In Chapter 5 of this thesis~\cite{anal}, 
we investigate the form factors of $q\bar{Q}$ bound states
both in spacelike and timelike region using an exactly solvable model
of $(3+1)$ dimensional scalar field theory interacting with gauge fields.
Based on the LF quantization, the Drell-Yan-West ($q^{+}$=0) frame
as well as the purely longitudinal momentum ($q^{+}\neq0$ and 
${\bf q}_{\perp}$=0) frame were used for the calculations of the
$M\to\gamma^{*}+M$ form factors.
We then analytically continue the form factors in the spacelike region to
the timelike region and compare those with the direct results of the
timelike $\gamma^{*}\to M+\bar{M}$ form factors.
Analytically continued results coincide exactly
with the direct results and it verifies
that the method of analytic continuation is capable of yielding
the effect of complicate nonvalence contributions.
The meson peaks analogous to the vector meson dominance(VMD) phenomena
are also generated at the usual VMD positions.

In Chapter 6~\cite{zm}, we discuss the zero mode($q^{+}=0$ mode of a 
continuum theory) contribution, which is crucial to obtain the
correct values of the LF current $J^{-}$ in the
Drell-Yan-West($q^{+}=0$) frame. In the exactly solvable model of
(1+1)-dimensional scalar field theory interacting with gauge fields,
we quantify the zero mode contribution
and observe that the zero mode effects are very large for the
light meson form factors even though they are substantially reduced
for the heavy meson cases.

Finally, we present in Chapter 7~\cite{Kaon,Semi} the analysis of 
exclusive $0^{-}$$\to$$0^{-}(1^{-})$ semileptonic
meson decays using our LFQM~\cite{Mix}.  
Our method of analytic continuation to obtain the weak form factors avoids
the difficulty associated with the contribution from the nonvalence
quark-antiquark pair creation. Our numerical results for the 
decay rates are in a good agreement with the
available experimental data and the lattice QCD results.
In addition, our model predicts the two unmeasured mass spectra of
$^{1}S_{0}(b\bar{b})$ and $^{3}S_{1}(b\bar{s})$ systems as
$M_{b\bar{b}}$= 9295 (9657) MeV and $M_{b\bar{s}}$= 5471 (5424) 
MeV, for the HO (liner) confining potential, respectively.
Conclusions and discussions of this thesis follow in Chapter 8.
In Appendix A, the spin-orbit wave functions 
${\cal R}^{JJ_{3}}_{\lam_{q},\lam_{\bar{q}}}(x,{\bf k}_{\perp})$
of pseudoscalar and vector mesons are presented for the IM scheme.
In Appendix B, the details of how to obtain the spin-averaged masses
of $\eta,\eta',\omega$ and $\phi$ are presented.
In Appendix C, we present the derivation of the
formulas used for the electromagnetic decay widths.
In Appendices D and E, we show how to fix the model parameters 
and determine the mixing angles of $\omega$-$\phi$ and 
$\eta$-$\eta'$, respectively, from our QCD-motivated quark potential model 
discussed in Chapter 4. The derivations of the matrix element 
of the form factors for $0^{-}\to 0^{-}$ semileptonic decays
in both $q^{+}$=0 and $q^+$$\neq$0 frames follow in Appendix F.
 

%% file: NPA.tex
\setcounter{equation}{0}
\setcounter{figure}{0}
\renewcommand{\theequation}{\mbox{2.\arabic{equation}}}
\chapter{LFQM Predictions for Radiative Meson Decays: SM 
Scheme}
As we discussed briefly in the Introduction, the main idea of the 
spin-averaged meson mass (SM) 
scheme~\cite{dziem1,ji,spin}, which was originally argued by 
Dziembowski and Mankiewicz~\cite{dziem1}, is
to take $M_0$ in the spin-orbit wave function
${\cal R}^{JJ_3}_{\lam\bar\lam}(x,{\bf k}_{\perp})$ given by 
Eq. (\ref{eq.123}) as the average of physical masses with
appropriate weighting factors from the spin degrees of freedom.
For distinction from the invariant meson mass (IM)
scheme (see Appendix A), we use the following covariant form of the 
spin-orbit wave function in the SM scheme\footnote{See the Appendix of 
Ref.~\cite{ji} for more detailed derivations, where the
Lepage-Brodsky convention~\cite{BPP,LB} was used to obtain the Dirac
spinors $u$ and $v$ given by Eq. (\ref{sm_mel}).}:
\begin{equation}\label{sm_mel}
\chi^{JJ_3}_{\lam_1\lam_2}(x,{{\bf k}_{\perp}})
= \bar{u}(p_1,\lambda_{1})\Gamma_{M,\mu}v(p_2,\lambda_{2}),
\end{equation}
where the operators $\Gamma_{M,\mu}$ are given by
\begin{eqnarray}\label{jpc}
J^{PC} &=& 0^{-+},\;\;\;
\Gamma_{p} = (m_{p} + \not\!P)\gamma_{5},\\
&& 1^{--},\;\;\;
\Gamma_{v,\mu} = m_{v}\not\!\vep(\mu)
+ \frac{[\not\!P, \not\!\vep(\mu)]}{2},\\
&& 1^{++},\;\;\;
\Gamma_{a,\mu} = (m_{a} + \not\!P)
\biggl[\frac{k\cdot P}{m_{a}}\not\!\vep(\mu)
+ \frac{[\not\!\vep(\mu),\not\!k]}{2}\biggr]\;
\gamma_{5}.
\end{eqnarray}
Here $m_{p(v),a}$ is the spin-averaged meson mass of the pseudoscalar
(vector), axial vector mesons and the space components $\vec{\vep}(\mu)$ 
of the polarization four-vectors $\vep(\mu)$ in the rest frame have 
the components $\vec{\vep}(\pm) =\mp(1,\pm i,0)/\sqrt{2}$,
$\vec{\vep}(0)=(0,0,1)$.

In addition to the above treatment of the dynamical character of the 
angular momentum operator in LF dynamics, we assume the radial wave 
function of the ground state mesons are described by the 
harmonic-oscillator (HO) wave functions~\cite{Lepage}:
\be\label{BHL}
\phi(x,{{\bf k}_{\perp}})
= A\exp\biggl[- \sum_{i=1}^{2}\frac{{\bf k}^{2}_{\perp i}+
m_{i}^2}{x_{i}}/8\beta^{2}\biggr]\;,
\ee
This Brodsky-Huang-Lepage (BHL) oscillator 
prescription~\cite{Lepage,BHL1} 
of Eq. (\ref{BHL}) can be connected to the HO wave function
in equal-$t$ formulation by equating the off-shell propagator
$E=M^2 - (\sum^{n}_{i=1} k^2_i)$ in the two frames~\cite{tao}.
As explained in Ref.~\cite{nisgur}, the equal-$t$ HO wave functions are 
known to give a reasonable first-approximation description of the 
static properties in a scheme without short-range hyperfine 
interactions. Therefore, it is reasonable to assume the spin-averaged 
mass as the meson mass in Eqs.~(2.2)-(2.4). 
For example, the $\pi$ and
$\rho$ masses are equal and both approximated by the spin-averaged 
value $m_{M}=(\frac{1}{4}m_\pi + \frac{3}{4}m_\rho)_{\rm exp.}$

As the SM scheme in LFQM~\cite{dziem1,ji}
provided a remarkably good description of the static
properties for the pion and $K$ mesons and reproduced the basic
features
of the lattice QCD and the QCD sum-rule results for $\pi, K, \rho$,
and $A_{1}$ mesons, it is our intention to investigate
more observables with the same model.
In this work~\cite{spin}, we present a comprehensive study of the
radiative decays of pseudoscalar($\pi,K, \eta,\eta'$),
vector($\rho,K^{*},\omega,\phi$) and axial vector($A_{1}$) mesons.
We shall also contrast our results with the results based on a different
treatment of meson mass such as the invariant mass (IM)
scheme~\cite{Bag,card} in the calculation of the magnetic and quadrupole 
moments of the $\rho$ meson, and the transition form factors of 
$\rho\to\pi\gamma^{*}$ and $A_{1}\to\pi\gamma^{*}$.
The spin-averaged masses of $\pi, K, \rho$, and $A_{1}$ mesons are
given by $m_{\pi}= m_{\rho}=0.612$ GeV, 
$m_{K} = m_{K^{*}}$ = 0.793 GeV and $m_{A_{1}}$ = 1.120
GeV~\cite{ji}.
Since we now consider $\eta,\eta',\omega$ and $\phi$ mesons also
in this work, we present the details of how to obtain the spin-averaged
values of $\eta,\eta',\omega$ and $\phi$ mesons in Appendix B.
The flavor assignment of $\eta$ and $\eta'$ mesons
in the quark and antiquark basis are as follows:
\begin{eqnarray}\label{eta}
&&\eta = X_{\eta}\frac{(u\bar{u} + d\bar{d})}{\sqrt{2}}
- Y_{\eta}s\bar{s},\\
&&\eta' = X_{\eta'}\frac{(u\bar{u} + d\bar{d})}{\sqrt{2}} +
Y_{\eta'}s\bar{s},
\end{eqnarray}
where $X_{\eta} = Y_{\eta'} = -\sin\delta_{P}$ and
$Y_{\eta} = X_{\eta'} = \cos\delta_{P}$ with $\delta_{P}
= \theta_{SU(3)} - \theta_{\rm ideal}\approx \theta_{SU(3)} - 35^\circ$.
Of particular interest are the values of mixing angle
$\theta_{SU(3)} = -10^\circ$(so called 
``perfect mixing")~\cite{isgur1,isgur2}
and $-23^\circ$ that are used in our analysis.
The spin-averaged masses of $\eta$ and $\eta'$ for each scheme are
given in Table~\ref{T1}. For $\omega$ and $\phi$ mesons, we use the scheme 
of ideal mixing~\cite{data} and obtain the spin-averaged masses 
$m_{\omega}$ = 928 MeV and $m_{\phi}$ = 799 MeV.
Once the spin-averaged masses are fixed, then besides the well-known
constituent quark masses of $(u,d,s)$ quarks, i.e.,  $m_u = m_d = 330$
MeV and $m_s = 450$ MeV, the only papameter in this
model is the Gaussian parameter $\beta$ which
determines the broadness (or sharpness) of radial
wave function. We will present our numerical results for a typical
$\beta$ value of $\beta = 360$ MeV throughout this section, unless
stated otherwise.
\begin{table}
\centering
\caption{Three different mixing schemes for $\eta$ and $\eta'$ and the
corresponding spin-averaged masses.}\label{T1}
\begin{tabular}{|c|c|c|c|c|c|c|}\hline
$\theta_{SU(3)}$ & $X_{\eta}$ & $Y_{\eta}$ & $X_{\eta'}$ & $Y_{\eta'}$ &
$m_{\eta}$[MeV] & $m_{\eta'}$[MeV]\\ \hline
0$^\circ$ & $\sqrt{\frac{1}{3}}$ & $\sqrt{\frac{2}{3}}$ &
$\sqrt{\frac{2}{3}}$ & $\sqrt{\frac{1}{3}}$ & 842 & 885\\ \hline
$-10^\circ$ & $\sqrt{\frac{1}{2}}$ & $\sqrt{\frac{1}{2}}$ &
$\sqrt{\frac{1}{2}}$ & $\sqrt{\frac{1}{2}}$ & 843 & 884\\ \hline
$-23^\circ$ & 0.85 & 0.53 & 0.53 & 0.85 & 834 & 873 \\  \hline
\end{tabular}
\end{table}

This Chapter is organized as follows: 
In Section 2.1, we calculate
the form factors, charge radii, magnetic and quadrupole moments
of the $\rho$ and $A_{1}$ mesons and compare with the results of
QCD sum rules~\cite{smilga}.
In Section 2.2, the transition form factors and the decay widths of
the transitions $V\to P\gamma^{*}$, $P\to V\gamma^{*}$
($V = \rho, K^{*}, \omega, \phi$ and $P = \pi, K, \eta, \eta'$)
and $A_{1}\to\pi\gamma^{*}$ are presented including the
comparison with other theoretical results as well as the experimental 
data.  In Section 2.3, we present the calculation of the transition 
form factors of the $\pi^{0}\to\gamma^{*}\gamma$, 
$\eta\to\gamma^{*}\gamma$ and $\eta'\to\gamma^{*}\gamma$ transitions 
and compare our results with the
recent experimental data~\cite{cello1,cello2,tpc}.
Summary and discussions of our major results follow in Section 2.4.
\section{The Form Factors of the $\rho$ and $A1$ Mesons}
Our analysis will be carried out using the Drell-Yan-West  
($q^{+}=0$) frame~\cite{Drell}:
\begin{eqnarray}\label{LF}
&&P = (P^{+},P^{-},{\bf P}_{\perp})
= ( P^{+},\frac{M^{2}}{P^{+}},{\bf 0}_{\perp}),\nonumber\\
&&q = (q^{+},q^{-},{\bf q}_{\perp})
= (0,\frac{Q^{2}}{P^{+}},{\bf q}_{\perp}),
\end{eqnarray}
where $M$ is the meson mass and the photon momentum is transverse to the
direction of the incident spin-one system,
with $q_{\perp}^{2} = Q^{2} = -q^{2}$.

The Lorentz invariant electromagnetic form factors $F_{i}(i=1,2,3)$
for a spin 1 particle are defined~\cite{arnold} by the matrix elements
of the current operator $J^{\mu}$ between the initial $|P,\lambda\ra$ and
the final $|P',\lambda'\ra$ eigenstate of momentum $P$ and helicity
$\lambda$ as follows:
\begin{eqnarray}\label{spin1}
\la P',\lambda'|J^{\mu}|P,\lambda\ra
&=& \vep_{\alpha}^{*'}\vep_{\beta}
\biggr[-{\rm g}^{\alpha\beta}(P + P')^{\mu}F_{1}(Q^{2})
+ ({\rm g}^{\mu\beta}q^{\alpha}
- {\rm g}^{\mu\alpha}q^{\beta})F_{2}(Q^{2})\nonumber\\
&+& q^{\alpha}q^{\beta}(P + P')^{\mu}F_{3}(Q^{2})/(2M^{2})\biggr]\;,
\end{eqnarray}
where $q = P'- P$ and  the polarization vectors of the
initial and final mesons $\vep\equiv\vep_{\lambda}$ and
$\vep'\equiv\vep_{\lambda'}$, respectively, are defined by
\begin{eqnarray}\label{pol}
&&\vep(0) = \frac{1}{M}( P^{+}, -\frac{M^{2}}{P^{+}},
{\bf 0}_{\perp})\;,\;\;\; \vep'(0) = \frac{1}{M}( P^{+},
\frac{-M^{2} + Q^{2}}{P^{+}}, {\bf q}_{\perp})\;,\nonumber\\
&&\vep(\pm 1)= \frac{\mp 1}{\sqrt{2}}(0,0,1,\pm i)\;,
\;\;\;
\vep'(\pm 1) =
\frac{\mp 1}{\sqrt{2}}(0,\frac{2{\bf q}_{\perp}}{P^{+}}, 1,\pm i)\;.
\end{eqnarray}
Also, the Lorentz invariant form factors $F_{i}(Q^{2})$ are related to
the charge, magnetic and quadrupole form factors of
a meson~\cite{arnold} as follows:
\begin{eqnarray}\label{cmq}
&&F_{C} = F_{1} + \frac{2}{3}\kappa F_{Q}\;,\nonumber\\
&&F_{M} = F_{2}\;,\\
&&F_{Q} = F_{1} - F_{2} + (1 + \kappa)F_{3}\;,\nonumber
\end{eqnarray}
where $\kappa = Q^{2}/4M^{2}$ is a kinematic factor.
At zero momentum transfer,these form factors are proportional to the
usual static quantities of charge $e$, magnetic moment $\mu_{1}$, and
quadrupole moment $Q_{1}$:
\begin{eqnarray}\label{cmq0}
F_{C}(0) = 1,\hspace{.1in} eF_{M}(0)= 2M\mu_{1},
\hspace{.1in}eF_{Q}(0) = M^{2}Q_{1}.
\end{eqnarray}
In the LFQM, the matrix element can be calculated by the convolution 
of initial and final LF wave functions of a meson:
\begin{eqnarray}
\la P',\lambda'|J^{\mu}|P,\lambda\ra &=& e_{1}\int_{0}^{1}dx\int
\frac{d^{2}{\bf k}_{\perp}}{16\pi^{3}}\phi(x,{\bf k}_{\perp} +
(1-x){\bf q}_{\perp})\phi(x,{\bf k}_{\perp})\nonumber\\
&\times&\sum_{\lambda,\lambda'}\chi^{\dagger}_{\lambda'_{1},\lambda'_{2}}
(x,{\bf k}_{\perp} + (1-x){\bf q}_{\perp})\Gamma^{\mu}
\chi_{\lambda_{1},\lambda_{2}}(x,{\bf k}_{\perp}) \nonumber\\
&+& e_{2}(1\leftrightarrow 2 \hspace{.1in}\mbox{of the first term}),
\end{eqnarray}
where the spin-covariant functions $\chi(x,{\bf k}_{\perp})$
are given by Eqs.~(2.2)-(2.4) and the vertex $\Gamma^{\mu}$ is obtained
from the expression $[\bar{u}(p')/(p'^{+})^{1/2}]\gamma^{\mu}
[u(p)/(p^{+})^{1/2}]$~\cite{LB}. 
Note that the jacobi factor $\partial k_n /\partial x$ given by
Eq. (\ref{eq.122}) is absorbed into the normalization constant
$A$ in Eq. (\ref{BHL}) throughout the Chapter 2.  

The relationship in Eq. (\ref{spin1}) between the covariant form factors 
and current matrix elements can be applied~\cite{chung,gross}, in
principle, to any choice of Lorentz frame.
As discussed by Brodsky and Hiller~\cite{hiller}, in
the standard LF frame~\cite{Drell}, i.e., $q^{+}=0$,
$q_{y}=0$, and $q_{x}=Q$, the three form factors can be obtained from
the `$+$' component of three helicity matrix elements:
\begin{eqnarray}\label{hel}
&&F_{C} =\frac{1}{2P^{+}(2\kappa + 1)}
\biggr[\frac{16}{3}\kappa\frac{F^{+}_{+0}}{\sqrt{2\kappa}} -
\frac{2\kappa - 3}{3}F^{+}_{00}
+ \frac{2}{3}(2\kappa - 1)F^{+}_{+-}\biggr]\;,\nonumber\\
&&F_{M} =\frac{2}{2P^{+}(2\kappa + 1)}
\biggr[(2\kappa - 1)\frac{F^{+}_{+0}}{\sqrt{2\kappa}}
+ F^{+}_{00} - F^{+}_{+-}\biggr]\;,\\
&&F_{Q} = \frac{1}{2P^{+}(2\kappa + 1)}
\biggr[2\frac{F^{+}_{+0}}{\sqrt{2\kappa}} - F^{+}_{00}
- \frac{\kappa + 1}{\kappa}F^{+}_{+-}\biggr]\;.\nonumber
\end{eqnarray}
where we defined $\la P',\lambda'|J^{\mu}|P,\lambda\ra \equiv
F^{\mu}_{\lambda'\lambda}$.
After a tedious but straightforward calculation,
we find the following expressions for the helicity form factors of
the $\rho$ meson:
\begin{eqnarray}
F^{+}_{00}&=&N_{\rho}\int_{0}^{1}\frac{dx}{x(1-x)}
\exp\biggl[-\frac{\xi^{2} + \tilde{m}^{2}_{1}
+ x(\tilde{m}^{2}_{2}-\tilde{m}^{2}_{1})}{x(1-x)}\biggr]\nonumber\\
&&\; \times\biggl[ 2x^{2}(1-x)^{2} + x(1-x)( a_{1}^{2} +
a_{2}^{2} - 2\xi^2) + (a_{1}a_{2})^{2}\nonumber\\
&&\; - (a_{1}^{2} + a_{2}^{2} - 4a_{1}a_{2})
\xi^{2} + \xi^{4}\biggr]\;,
\\
F^{+}_{+0}&=&\frac{Q}{2\sqrt{2}\beta}N_{\rho}\int_{0}^{1}
\frac{dx}{x}\exp\biggl [-\frac{\xi^{2} +
\tilde{m}^{2}_{1} + x(\tilde{m}^{2}_{2}-
\tilde{m}^{2}_{1})}{x(1-x)}\biggr ](a_{1}- a_{2})\nonumber\\
&&\; \times [a_{1}a_{2} + \xi^{2} - x(1-x)],
\end{eqnarray}
\begin{equation}
F^{+}_{+-}=-\frac{Q^{2}}{4\beta^{2}}N_{\rho}
\int_{0}^{1}\frac{(1-x)dx}{x}a_{1}a_{2}
\exp\biggl [-\frac{\xi^{2} + \tilde{m}^{2}_{1}
+ x(\tilde{m}^{2}_{2}-\tilde{m}^{2}_{1})}{x(1-x)}\biggr],
\end{equation}
where
\begin{eqnarray}
& &\xi^{2}= \frac{(1-x)^{2}Q^{2}}{16\beta^{2}},\hspace{.2in}
\tilde{m_{i}} = \frac{m_{i}}{2\beta},\nonumber\\
& &a_{1} = \frac{(xm_{\rho} + m_{1})}{2\beta},\hspace{.5cm}
a_{2} = \frac{((1-x)m_{\rho} + m_{2})}{2\beta},
\end{eqnarray}
and $m_{1}$, $m_{2}$ are the constituent masses of the quark and
antiquark. The normalization constant $N_{\rho}=(4A_{\rho}\beta^{3}/
\pi P^{+})^{2}$ is fixed by the definition of charge, $F_{C}(0)=1$.
For the systems of spin 1 or greater, in addition to the parity and
time-reversal invariance of the current operator
$J^{+}(0)$~\cite{arnold,hiller}, an additional constraint on the
current operator comes from the rotational covariance
requirement~\cite{chung,grach,keister}. 
The angular condition for the spin-1 system
is given by~\cite{card,grach,keister}:
\begin{eqnarray}\label{angl}
\Delta(Q^{2})= (1 + 2\kappa)F^{+}_{++} + F^{+}_{+-}
- \sqrt{8\kappa}F^{+}_{+0} - F^{+}_{00}= 0.
\end{eqnarray}
As a matter of fact, the expressions of the right-hand side in 
Eq. (\ref{hel}) are not unique because of the angular condition in 
Eq. (\ref{angl}).  As pointed out in Refs.~\cite{card,keister},
unless the exact Poincar\'{e} covariant current operator beyond
one-body sector is used, the angular condition is in general
violated (i.e., $\Delta(Q^{2})\neq 0$) and the
calculation of the form factors $F_{i}$ is dependent on the
expressions on the r.h.s. of Eq. (\ref{hel}).  Examples of
different choices of current combinations can be found in
Ref.~\cite{card} for the calculation of $\rho$ meson form factors.
As shown in Fig. 2.1(a), our result obtained for $\Delta(Q^{2})$ 
is comparable with other choices given in Ref.~\cite{card}.
\begin{figure}\label{NPA1a}
\centerline{\psfig{figure=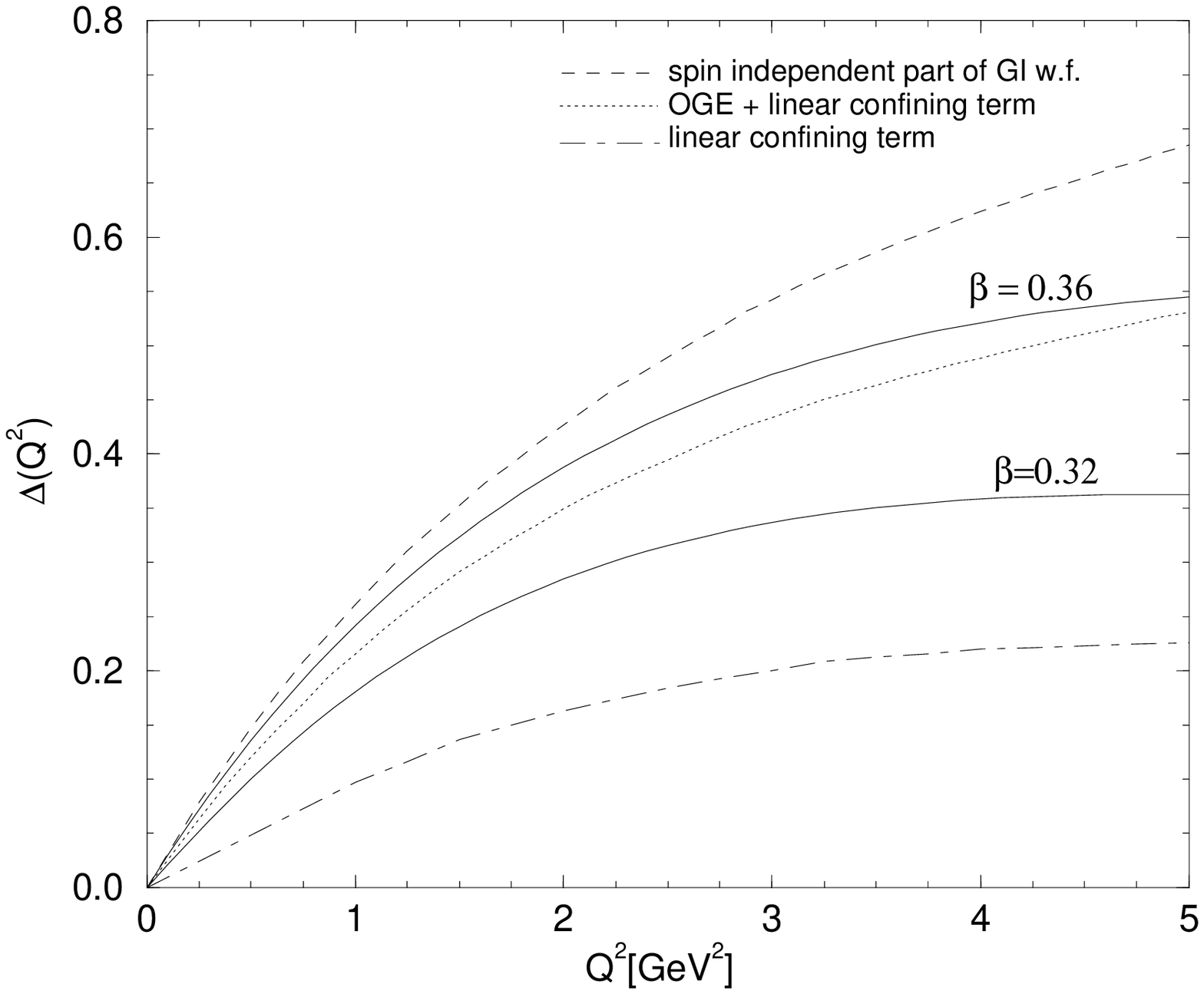,width=3.5in,height=3.3in}} 
\centerline{\psfig{figure=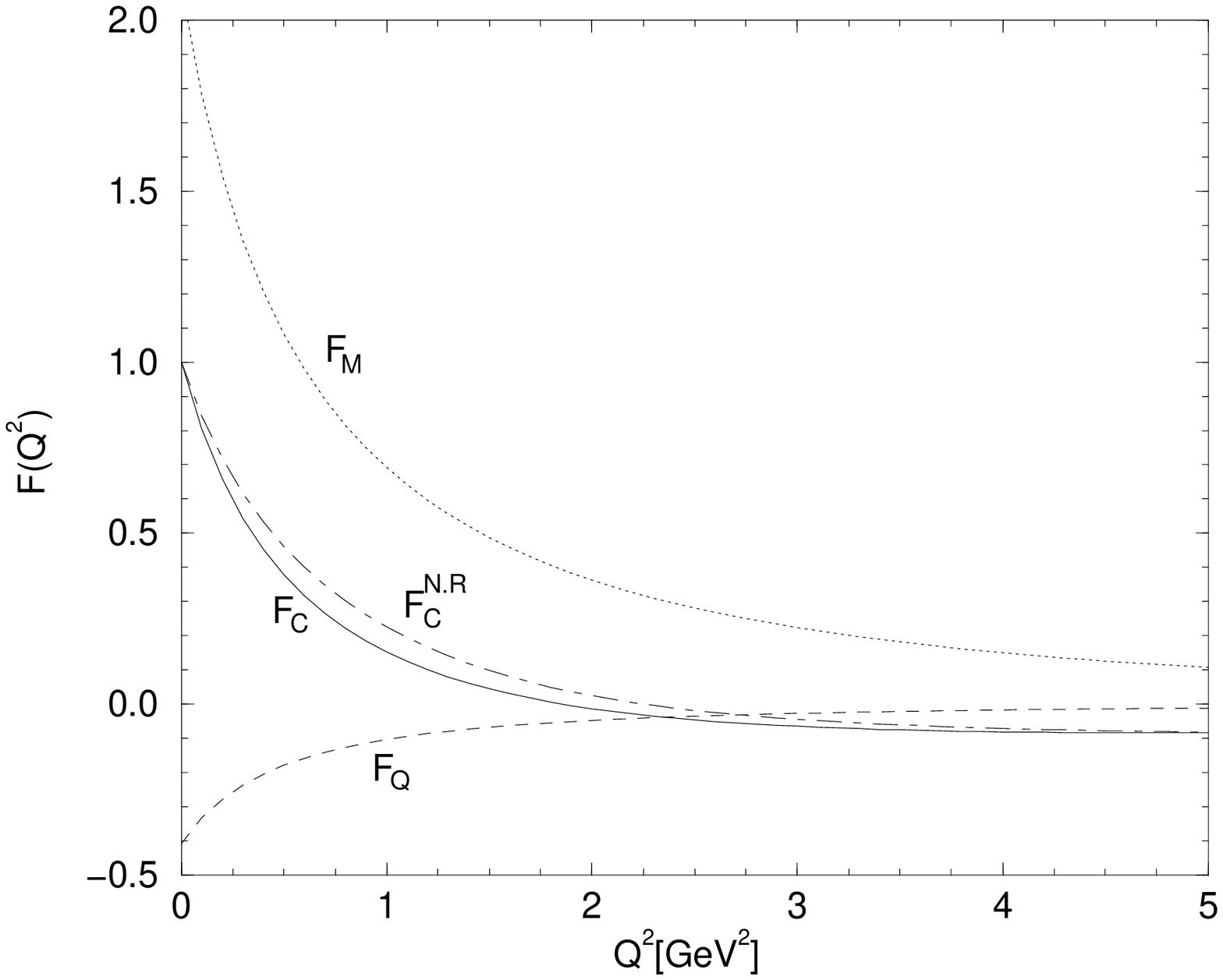,width=3.5in,height=3.3in}}
\caption{(a) $\Delta(Q^{2})$ testing the angular condition
is shown as a function of $Q^{2}$.
The solid lines are our results with $\beta = 0.32$ and $0.36$ GeV.
For comparison, we include various choices
of the wave function $w^{\rho}$ introduced by Godfrey and Isgur(GI)
in Ref.~\protect\cite{card}. The dashed, dotted and
dashed-dotted lines correspond to $w_{si}$(spin-independent part),
$w_{GI}$ (OGE + linear confining term) and 
$w_{conf}$ (linear confinging term), respectively.
(b) The form factors of $\rho$ meson with
parameter $\beta = 0.36$ GeV.  The solid, dotted, and dashed lines
correspond to $F_{C}(Q^{2})$,$F_{M}(Q^{2})$ and $F_{Q}(Q^{2})$,
respectively. The dashed-dotted line is the result of non-relativistic
limit of $F_{C}(Q^{2})$ by turning off the Melosh transformation.}
\end{figure}

The magnetic (in unit of $e/2M$) and quadrupole moments (in unit
of $e/M^{2}$) of the $\rho$ meson are obtained by Eq. (\ref{cmq0}),
\begin{eqnarray}
\mu_{1} = 2.1,\hspace{.2in} Q_{1} = 0.41.
\end{eqnarray}
These values are not much different from the values of other model
predictions based on the IM scheme presented in 
Ref.~\cite{Bag} ($\mu_{1}= 2.3$, $Q_{1}= 0.45$) and 
Ref.~\cite{card} ($\mu_{1}= 2.26$, $Q_{1}= 0.37$).
We also calculated the electromagnetic radii associated with the
form factors, $F^{\rho}_{C}$, $F^{\rho}_{M}$ and $F^{\rho}_{Q}$ as
$\la r^{2}_{F_{C}}\ra$= 14 GeV$^{-2}$,
$\la r^{2}_{F_{M}}\ra$= 22 GeV$^{-2}$ and
$\la r^{2}_{F_{Q}}\ra$= $-5$ GeV$^{-2}$, respectively.
The results of $F_{C}(Q^{2}), F_{M}(Q^{2})$ and $F_{Q}(Q^{2})$ for
$0\leq Q^{2}\leq 5 \hspace{.1in}\mbox{GeV}^{2}$ are shown in 
Fig. 2.1(b).
To see the effect of the Melosh transformation (the measure of
relativistic effects), we calculated the charge form factor by
turning the Melosh rotations off and included the result in 
Fig. 2.1(b).
As one can see in this figure, the breakdown of rotational covariance
occurs around $Q^{2}= 1-2$ GeV$^{2}$. Therefore, the model calculations
of the form factor may not be reliable beyond $1-2$ GeV$^{2}$ range.
However, for the comparison with other model calculations, we have
displayed our results beyond this region. The charge radius from
this nonrelativistic form fator is $\la r^{2}\ra_{\mbox{non-rel}} =
10.6 \hspace{.1in}\mbox{GeV}^{-2}$ which is about 30$\%$ smaller than 
that of relativistic charge radius. In Fig. 2.1(c), we also
compared our results with the previous calculations of these form
factors which were made in the framework of QCD sum rules by Ioffe
and Smilga~\cite{smilga}.\footnote{ The definition of the form
factors($G_{i}$) by QCD sum-rule~\cite{smilga} and
the definition in this paper are related as follows:
$G_{1} = F_{1}$, $G_{2}= F_{2} - F_{1}$ and $G_{3}=F_{3}/2$.}
\setcounter{figure}{0}
\begin{figure}
\centerline{\psfig{figure=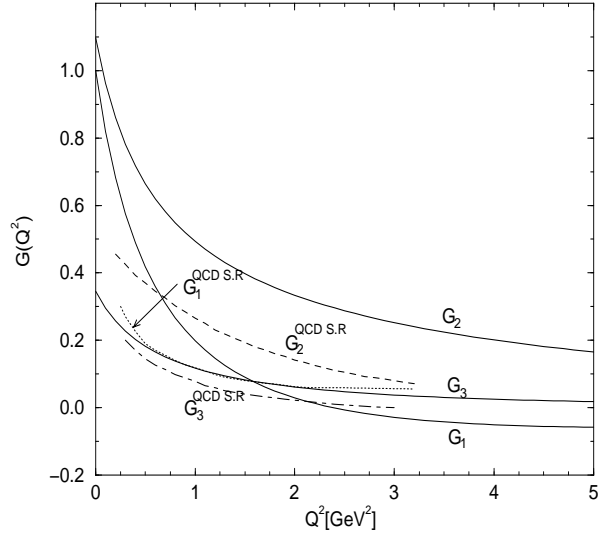,width=3.5in,height=3.3in}}
\caption{(c) The form factors, $G_{1}= F_{1}$, $G_{2}= F_{2} - F_{1}$
and $G_{3} = F_{3}/2$, of $\rho$ meson
are compared with the results of QCD sum rules~\protect\cite{smilga}.
The dotted, dashed and dashed-dotted lines correspond to $G_{1}$,
$G_{2}$, and $G_{3}$ of Ref.~\protect\cite{smilga}.}
\end{figure}

The electromagnetic form factors of the $A1$ meson are defined by
Eq. (\ref{spin1}) as in the case of the $\rho$ meson.
Using a similar method taken in the $\rho$ meson case, we find
the following helicitiy component of the $A1$ form factors analogous
to Eqs. (2.15)-(2.17):
\begin{eqnarray}
F^{+}_{00}&=& N_{A1}\int^{1}_{0}\frac{dx}{x(1-x)}
\exp\biggl [-\frac{\xi^{2} + \tilde{m}^{2}_{1}
+ x(\tilde{m}^{2}_{2}-\tilde{m}^{2}_{1})}{x(1-x)}\biggr ]
\nonumber\\
&&\;\times \biggl[6x^{3}(1-x)^{3} + 2x^{2}(1-x)^{2}
[a_{1}^{2} + a_{2}^{2} - \xi^{2}]
+ x(1-x)[(a_{1}a_{2})^{2} + \xi^{4}]\nonumber\\
&&\;- \xi^{2}[(a_{1}a_{2})^{2} -(a_{1}^{2} +a_{2}^{2} +
4a_{1}a_{2})\xi^{2} + \xi^{4}]\biggr]\;,\\
F^{+}_{+0}&=&\frac{Q}{2\beta}N_{A1}\int^{1}_{0}\frac{dx}{x}
\exp\biggl [-\frac{\xi^{2} + \tilde{m}^{2}_{1}
+ x(\tilde{m}^{2}_{2}-\tilde{m}^{2}_{1})}{x(1-x)}\biggr ]
(a_{2} - a_{1})\nonumber\\
&&\;\times \biggl[ 2x^{2}(1-x)^{2} + x(1-x)(4\xi^{2} - a_{1}a_{2})
+ \xi^{4} - a_{1}a_{2}\xi^{2}\biggr]\;,\\
F^{+}_{+-}&=&\frac{Q^{2}}{8\beta^{2}}N_{A1}\int^{1}_{0}\frac{(1-x)dx}{x}
\exp\biggl [-\frac{\xi^{2} + \tilde{m}^{2}_{1}
+ x(\tilde{m}^{2}_{2}-\tilde{m}^{2}_{1})}{x(1-x)}\biggl ]
\nonumber\\
&&\;\times(a_{2} - a_{1})^{2}[x(1-x) + \xi^{2}]\;.
\end{eqnarray}
As one can see in Fig.~2.2(a), the behavior of each form factor looks
similar to the $\rho$ meson case. In Fig.~2.2(a), we also included
$|\Delta(Q^{2})|$ for $A_{1}$. The amount of deviation from the
rotational covariance is almost the same as that of the $\rho$ meson case.
\begin{figure}
\centerline{\psfig{figure=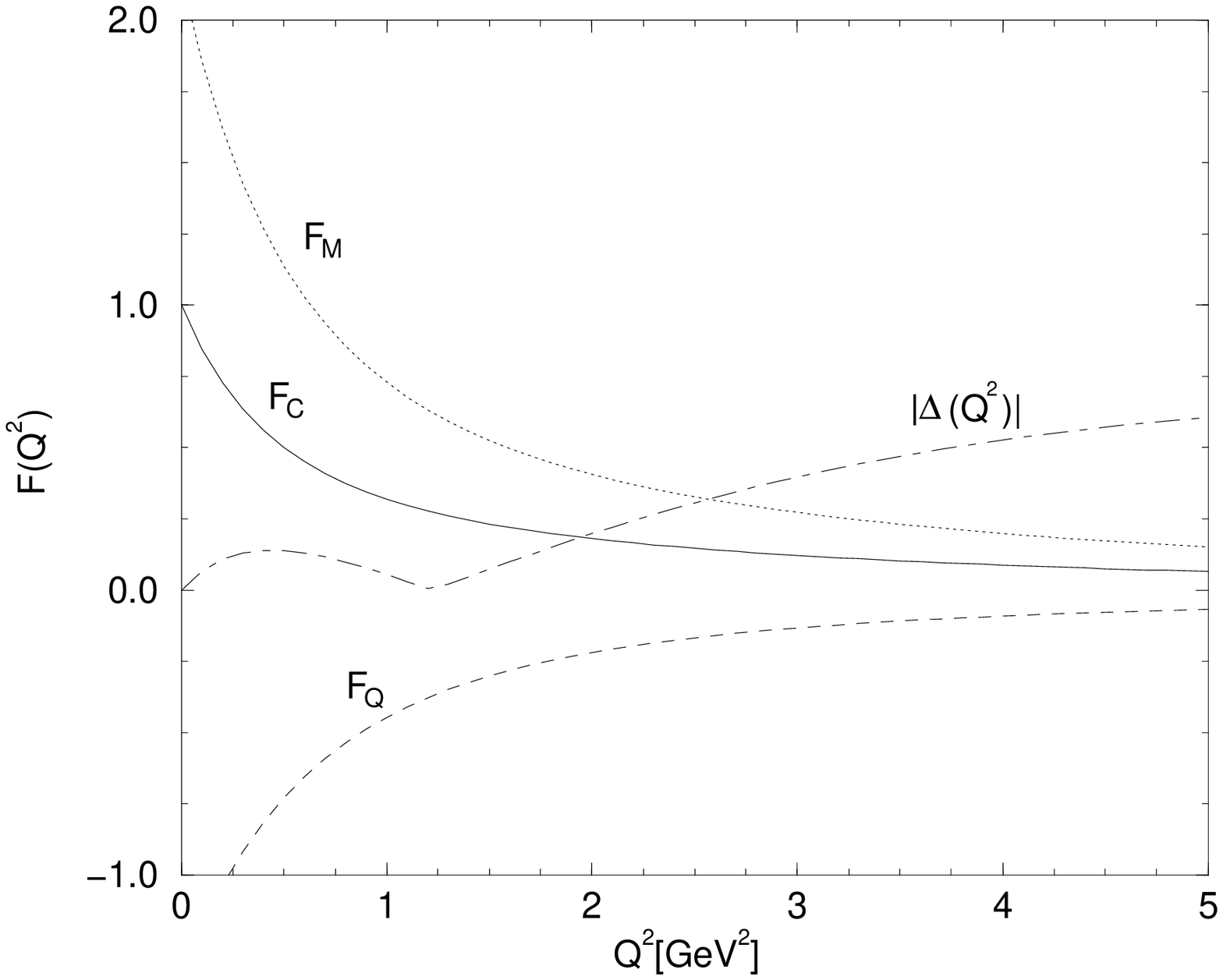,width=3.5in,height=3.3in}}
\centerline{\psfig{figure=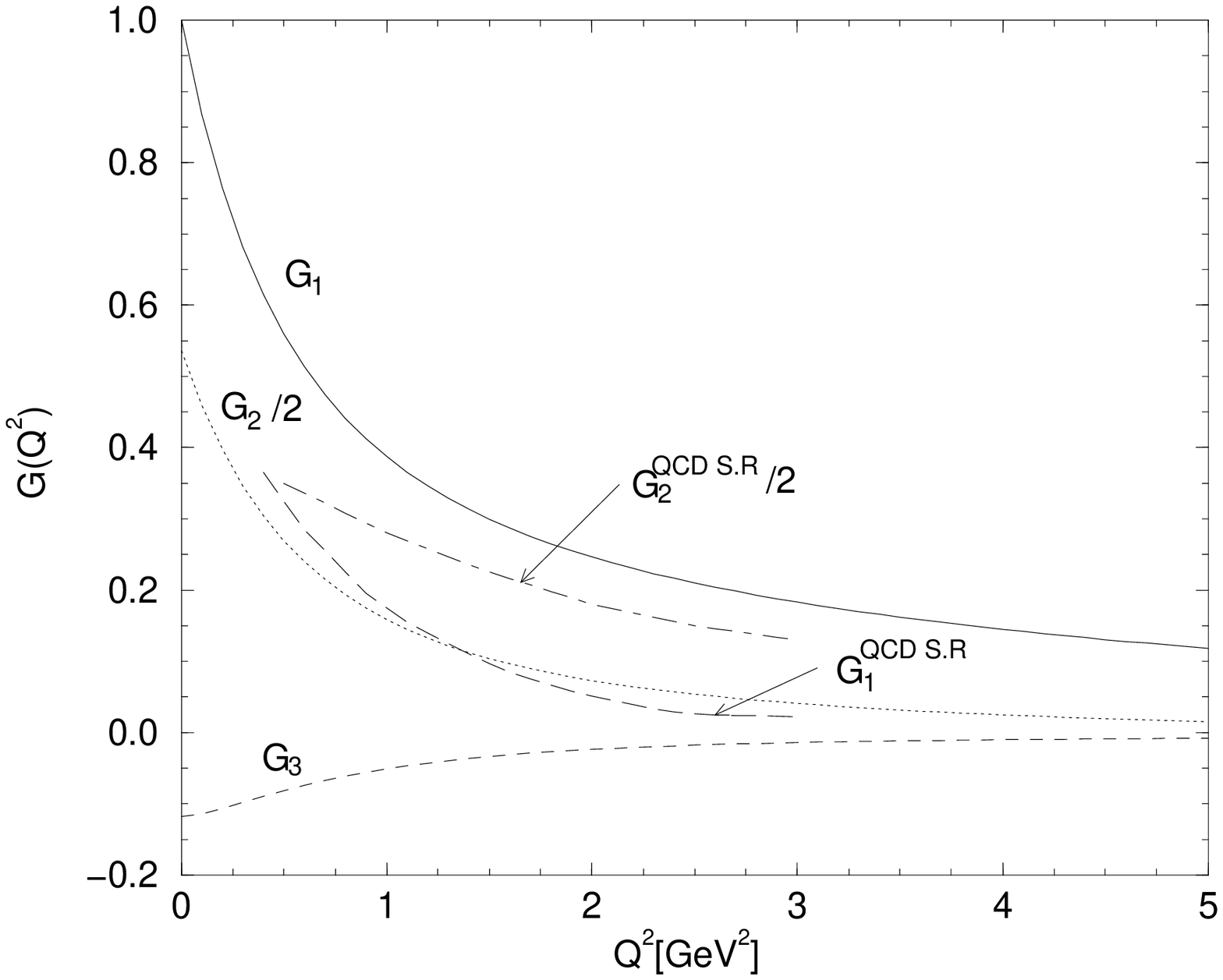,width=3.5in,height=3.3in}}
\caption{ (a) The form factors of $A_{1}$ meson with the
parameter $\beta = 0.36$ GeV. The solid, dotted, and dashed lines
correspond to $F_{C}$, $F_{M}$, and $F_{Q}$, respectively. The
quantity $|\Delta(Q^{2})|$ is shown as dashed-dotted line.
(b) The form factors, $G_{1}= F_{1}$, $G_{2}= F_{2} - F_{1}$
and $G_{3} = F_{3}/2$, of $A_{1}$ meson are compared
with the result of QCD sum rules~\protect\cite{smilga}. 
The solid, dotted, and dashed
lines are our predictions of $G_{1}$, $G_{2}/2$, and $G_{3}$, 
respectively.  The QCD sum-rule results are $G_{1}(Q^{2})$(long 
dashed line) and $G_{2}(Q^{2})/2$(dashed-dotted line)}
\end{figure}

The electromagnetic radii associated with
$F^{A_{1}}_{C}$, $F^{A_{1}}_{M}$ and $F^{A_{1}}_{Q}$ are obtained as
$\la r^{2}_{F_{C}}\ra$= 10 GeV$^{-2}$,
$\la r^{2}_{F_{M}}\ra$= 20 GeV$^{-2}$ and
$\la r^{2}_{F_{Q}}\ra$= $-10$ GeV$^{-2}$,
respectively.
The magnetic and quadrupole moments of the $A_{1}$ meson are
also obtained as
\begin{eqnarray}
\mu_{1} = 2.16,\hspace{.2in} Q_{1} = 1.34.
\end{eqnarray}
While the magnetic moment of $A_{1}$ does not differ from that
of the $\rho$ meson, the quadrupole moment of $A_{1}$ is about 4
times greater than that of the $\rho$ meson in accordance with the fact
that the $A_{1}$ meson is a bound state
with nonzero orbital angular momentum $l=1$\cite{Bag}.
In Fig.~2.2(b), we compare our results with those of QCD sum
rules\cite{smilga}.

\section{The Form Factors for $V(P)\to P(V)\gamma^{*}$
and $A_{1}\to\pi\gamma^{*}$ Transitions}
The transition form factors of $A\to B\gamma^{*}$
$((A,B) = (\rho,\pi),(\rho,\eta),
(\omega,\pi),(\omega,\eta),(K^{*},K),\\(\eta',\rho),(\eta',\omega),
(\phi,\eta),(\phi,\eta'))$ and $A_{1}\to\pi\gamma^{*}$
are defined by
\begin{eqnarray}\label{PV_form}
\la B(P')|J^{\mu}|A(P,\lambda)\ra= eG_{AB}(Q^{2})
\epsilon^{\mu\nu\alpha\beta}\vep_{\nu}(P,\lambda)P'_{\alpha}
P_{\beta},\\
\la\pi(P')|J^{\mu}|A_{1}(P,\lambda)\ra= \frac{e}{m_{A_{1}}}\biggl[
({\cal P}\cdot q{\rm g^{\mu\nu}} - {\cal P}^{\mu}q^{\nu})G_{1}(Q^{2})
\nonumber\\
+\frac{1}{m_{A_{1}}^{2}}({\cal P}\cdot qq^{\mu} - q^{2}{\cal P}^{\mu})
 q^{\nu}G_{2}(Q^{2})\biggr]\vep_{\nu}(P,\lambda),
\end{eqnarray}
where $\vep(P,\lambda)$ denotes the polarization vector of the
initial particles and ${\cal P} = P + P'$. We used the same 
$q^{+}=0$ frame for the calculation of transition form
factors as defined in Eq. (\ref{LF}).
As shown in Appendix C, the width of the decay $A\to B\gamma$ is
given by
\begin{eqnarray}\label{PV_rate}
\Gamma(A\to B\gamma)= \frac{\alpha}{2S_{A} + 1}|G_{AB}(0)|^{2}
\biggl( \frac{M_{A}^{2} - M_{B}^{2}}{2M_{A}}\biggr)^{3},
\end{eqnarray}
where $\alpha$ is the fine-structure constant, $S_{A}$ is the spin of
the initial particle and $M_{A(B)}$ is the mass\footnote{This must be
the physical mass rather than the spin-averaged
mass because the phase factor is nothing to do with the model.
Spin-averaged masses are used only for the calculation of form
factors.} of the meson A (B).
In the case of $A_{1}\to \pi\gamma$ transition, the decay width is
expressed in terms of $G_{1}(0)$(see Appendix C),
\begin{eqnarray}\label{A1_rate}
\Gamma(A_{1}\to\pi\gamma) = \frac{4\alpha}{3}
\biggl |\frac{G_{1}(0)}{M_{A_{1}}}\biggr |^{2}
\biggl(\frac{M_{A_{1}}^{2} - M_{\pi}^{2}}{2M_{A_{1}}}\biggr)^{3}.
\end{eqnarray}
In the calculations of the transition form factors $V\to
P\gamma^{*}$, we used the `+' component of the matrix elements of the 
current operator $J^{\mu}$. Since both sides of Eq. (\ref{PV_form}) are 
vanishing for the longitudinal polarization $\lambda_{\rho(\omega)}=0$ 
for any $Q^{2}$ value, we considered $\lambda_{\rho(\omega)}= 1$ to 
calculate the transition form factor. Even though the matrix element 
$J^{+}$ of this component($\lambda_{\rho(\omega)}= 1$) also vanishes for
real-photon($Q^{2}=0$) limit, this factor of $Q^{2}=0$ is nothing to
do with the form factor itself because the r.h.s. of
Eq. (\ref{PV_form}) also vanishes in the limit of $Q^{2}=0$. 
In other words, both sides of Eq. (\ref{PV_form}) in the case of 
$\lambda_{\rho(\omega)}= 1$ have the same kinematic factor which does 
not vanish for any $Q^{2}$ except $Q^{2}=0$.  Therefore, one can extract 
the transition form factor.
We thus find the following expressions for the form factors
$G_{\rho\pi}(Q^{2})$ and $G_{\omega\pi}(Q^{2})$:
\begin{eqnarray}
&&G_{\rho\pi}(Q^{2})= (e_{u} + e_{d})I_{PV\gam^*}(m_{M},m_{q},\beta),
\nonumber\\
&&G_{\omega\pi}(Q^{2})= (e_{u} - e_{d})I_{PV\gam^*}(m_{M},m_{q},\beta),
\end{eqnarray}
\begin{eqnarray}\label{PV_com}
I_{PV\gam^*}(m_{M},m_{q},\beta) &=&
\frac{1}{2\beta}\sqrt{\frac{N_{M}N_{\pi}}{2}}
\int_{0}^{1}\frac{dx}{x}
\exp[-\frac{\xi^{2} + \tilde{m}^{2}_{1} + x(\tilde{m}^{2}_{2}
- \tilde{m}^{2}_{1})}{x(1-x)}]\nonumber\\
&&\times\biggl[a^{\rm i}_{1} + a^{\rm i}_{2} +
a^{\rm f}_{1} + a^{\rm f}_{2}][x(1-x) - \xi^{2}]
+ a^{\rm i}_{1}a^{\rm i}_{2}(a^{\rm f}_{1} + a^{\rm
f}_{2})\nonumber\\
&&+ a^{\rm f}_{1}a^{\rm f}_{2}(a^{\rm i}_{1} + a^{\rm i}_{2})\biggr],
\end{eqnarray}
and $a^{\rm i(f)} = (xm_{M_{\rm i(f)}} + m_{q})/2\beta$ with
${\rm i}$ and ${\rm f}$ meaning incident and outgoing mesons.
The normalization constants of the $\rho$ and $\pi$-mesons are given
by $N_{\pi} =\frac{1}{2}N_{\rho} = 2(2A_{\pi(\rho)}\beta^{3}/\pi
P^{+})^{2}$.

We obtained the following prediction of decay widths of the
$\rho^{\pm}(770)\to\pi^{\pm}\gamma$  and
$\omega(782)\to\pi\gamma$ transitions:
\begin{eqnarray}
&&\Gamma(\rho^{\pm}\to\pi^{\pm}\gamma) = 69 \hspace{.1in}\mbox{keV}
\hspace{.2in}
(\Gamma^{\exp}_{\rho^{\pm}\to\pi^{\pm}\gamma} = 68\pm 8
\hspace{.1in} \mbox{keV}),\nonumber\\
&&\Gamma(\omega\to\pi\gamma) = 708 \hspace{.1in}\mbox{keV}
\hspace{.2in}
(\Gamma^{\exp}_{\omega\to\pi\gamma} = 717\pm 51 \hspace{.1in}
\mbox{keV}).
\end{eqnarray}
Our results for the decay widths $\Gamma_{\rho\pi}$ and
$\Gamma_{\omega\pi}$ are in a very good agreement with the
experimental data.  
The electromagnetic radii of these form factors are obtained as
$\la r^{2}_{G_{\rho\pi}}\ra$= 7 GeV$^{-2}$ and
$\la r^{2}_{G_{\omega\pi}}\ra$= 20 GeV$^{-2}$,
respectively. In Fig.~2.3(a), we present the transition form factor of
$\rho\to\pi\gamma^{*}$
for $0\leq Q^{2}\leq 8\hspace{.1in}\mbox{GeV}^{2}$ and compare with
other model predictions of Refs.~\cite{card,Ito}. In Fig.~2.3(b),
our prediction(solid line) of $\omega\to\pi\gamma^{*}$ transition
form factor in the spacelike region is compared with the VDM (dashed line)
with $F_{\omega\pi} = 1/(1 + Q^{2}/M_{\rho}^{2})$ and the pole 
fit (dotted line) of the experimental data in the timelike 
region~\cite{dzh,munz}.
\begin{figure}
\centerline{\psfig{figure=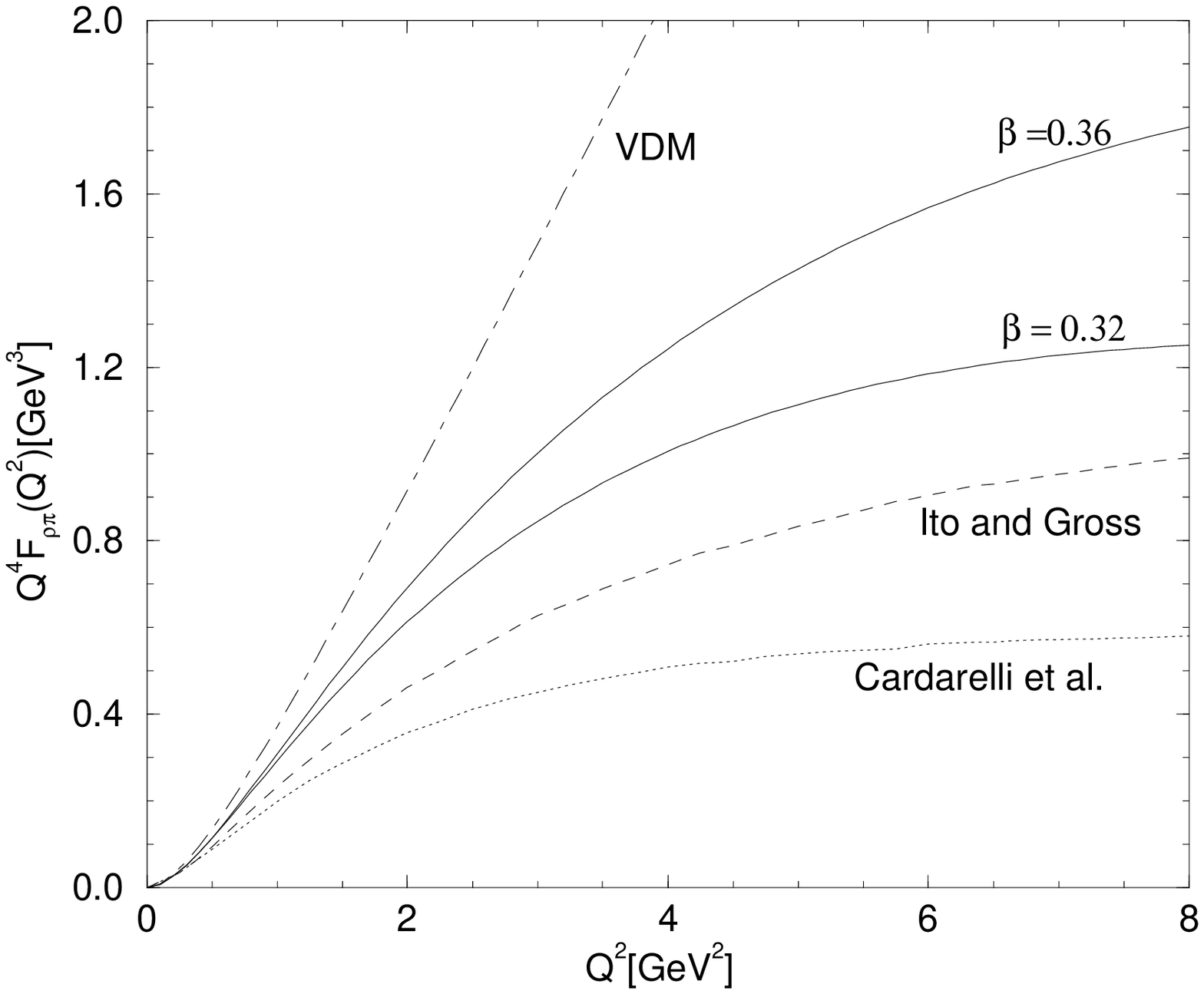,width=3.5in,height=3.3in}}
\centerline{\psfig{figure=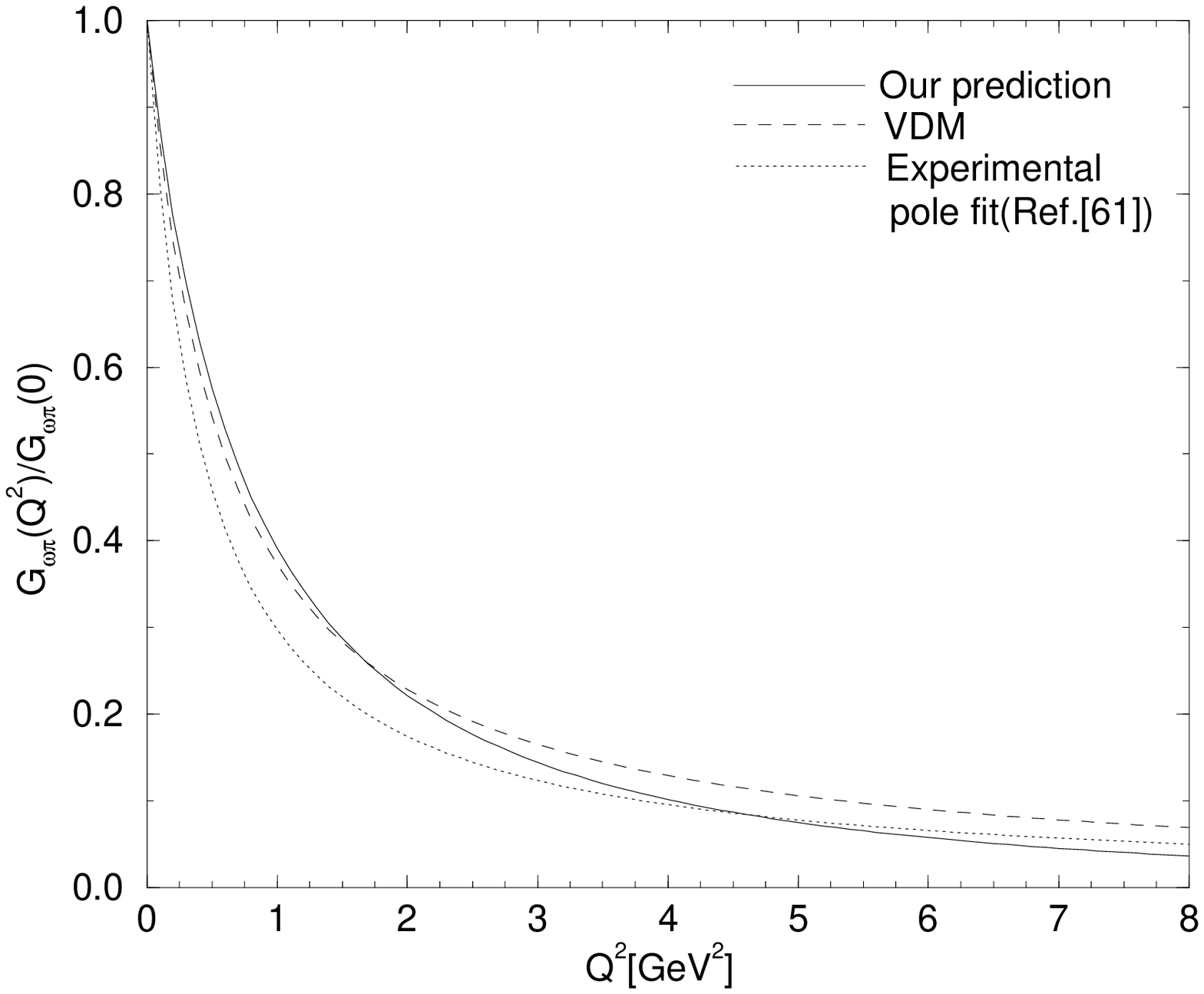,width=3.5in,height=3.3in}}
\caption{ (a) The transition form factor of
$\rho^{+}\rightarrow\pi^{+}\gamma^{*}$
multiplied by $Q^{4}$ is shown as a function of $Q^{2}$.
The solid lines are our results with $\beta = 0.32$ and $0.36$ GeV.
The dotted line, dashed, and dashed-dotted lines correspond to the
results obtained using $w_{GI}$ in Ref.~\protect\cite{card}, 
the BS approach of Ref.~\protect\cite{Ito} 
and the predictions of VDM model with $G_{\rho\pi}
= 1/(1 + Q^{2}/M_{\rho}^{2})$. (b) The normalized transition form
factor of $\omega\to\pi\gamma^{*}$. 
Our result with $\beta = 0.36$ GeV (solid line)
is compared with VDM (dashed line) with 
$G_{\omega\pi}= 1/(1 + Q^{2}/M_{\rho}^{2})$
and the pole fit of the experimental data (dotted line)~\protect\cite{dzh}.}
\end{figure}

Similarly, we calculated all other radiative decay processes between
vector ($1^{--}$) and pseudoscalar mesons($0^{-+}$) using the one loop 
integral formula $I_{PV\gam^*}(m_{M},m_{q},\beta)$ given by 
Eq. (\ref{PV_com}).
We needed to change only the spin-averaged meson masses $m_{M}$
and constituent quark masses $m_{i}(i=1,2)$, accordingly.
Thus, for the $\rho\to\eta\gamma^{*}$ and $\omega\to\eta\gamma^{*}$
decays, we obtain the following transition form factors:
\begin{eqnarray}
&&G_{\rho\eta}(Q^{2})= X_{\eta}(e_{u} -
e_{d})I_{PV\gam^*}(m_{M},m_{q},\beta),
\nonumber\\
&&G_{\omega\eta}(Q^{2}) = X_{\eta}(e_{u} +
e_{d})I_{PV\gam^*}(m_{M},m_{q},\beta).
\end{eqnarray}
Using the two different mixing schemes, i.e., $\theta_{SU(3)}= -10^\circ$
and $-23^\circ$, we obtain the decay widths of the transitions
$\rho\to\eta\gamma$ and $\omega\to\eta\gamma$ as
$\Gamma^{-10^\circ}_{\rho\to\eta\gamma} = 56$ keV,
$\Gamma^{-10^\circ}_{\omega\to\eta\gamma} = 6.4$ keV
and $\Gamma^{-23^\circ}_{\rho\to\eta\gamma} = 65$ keV,
$\Gamma^{-23^\circ}_{\omega\to\eta\gamma} = 7.4$ keV, respectively.
Both schemes are in excellent agreement with the experimental data of
$\Gamma^{\rm exp}_{\rho\to\eta\gamma} = 58\pm 10$ keV and
$\Gamma^{\rm exp}_{\omega\to\eta\gamma} = 7.0\pm 1.8$ keV.
The electromagnetic charge radii of $\rho\to\eta\gamma^{*}$ transition
are also predicted as $\la r^{2}_{G_{\rho\eta}}\ra = 16$ ${\rm GeV}^{-2}$
for $-10^\circ$ and 18 ${\rm GeV}^{-2}$ for $-23^\circ$ mixing angle.
The charge radii for $\omega\to\eta\gamma^{*}$ transition are
$\la r^{2}_{G_{\omega\eta}}\ra = 5$ ${\rm GeV}^{-2}$ for
$-10^\circ$ and 6 ${\rm GeV}^{-2}$ for $-23^\circ$, respectively.

The form factors of the $\eta'\to\rho\gamma^{*}$ and
$\eta'\to\omega\gamma^{*}$ are given by
\begin{eqnarray}
&&G_{\eta'\rho}(Q^{2})= X_{\eta'}(e_{u} 
- e_{d})I_{PV\gam^*}(m_{M},m_{q},\beta),
\nonumber\\
&&G_{\eta'\omega}(Q^{2})=X_{\eta'}(e_{u} 
+ e_{d})I_{PV\gam^*}(m_{M},m_{q},\beta).
\end{eqnarray}
Our predictions of the decay widths are given by
$\Gamma^{-10^\circ}_{\eta'\to\rho\gamma} = 117$ keV,
$\Gamma^{-10^\circ}_{\eta'\to\omega\gamma} = 9.7$ keV
and $\Gamma^{-23^\circ}_{\eta'\to\rho\gamma} = 72$ keV,
$\Gamma^{-23^\circ}_{\eta'\to\omega\gamma} = 6.0$ keV.
The experimental data of $\Gamma^{\rm exp}_{\eta'\to\rho\gamma}
= 61\pm 8$ keV and $\Gamma^{\rm exp}_{\eta'\to\omega\gamma} = 5.9\pm 0.9$ keV.
It is interesting to note that in case of transitions involving $\eta'$,
the result of $-23^\circ$ mixing scheme is much better than that of
$-10^\circ$ mixing scheme.  The electromagnetic charge radii
for $-23^\circ$ mixing angle are
predicted as $\la r^{2}_{G_{\eta'\rho}}\ra$= 12 ${\rm GeV}^{-2}$ and
$\la r^{2}_{G_{\eta'\omega}}\ra$= 4 ${\rm GeV}^{-2}$, respectively.

The transitions of the $K^{*\pm}\to K^{\pm}\gamma^{*}$ and
$K^{0}\to K^{0}\gamma^{*}$ in which the constituent quarks have unequal
masses are also interesting processes to test our model predictions.
As shown in Ref.~\cite{ji}, the predictions for the
kaon charge radius $\la r^{2}_{K}\ra^{1/2}$, the kaon form factor $F_{K}$,
and the decay constant $f_{K}$ are consistent with available
experimental data.
The transition form factors of the charged and netural $K^{*}$ decays
are given by
\begin{eqnarray}
&&G_{K^{*\pm}K^{\pm}}(Q^{2})=\pm(e_{u} +
e_{s})I_{PV\gam^*}(m_{M},m_{q},\beta),\nonumber\\
&&G_{K^{*0}K^{0}}(Q^{2})= (e_{d} 
+ e_{s})I_{PV\gam^*}(m_{M},m_{q},\beta).
\end{eqnarray}
\begin{figure}
\centerline{\psfig{figure=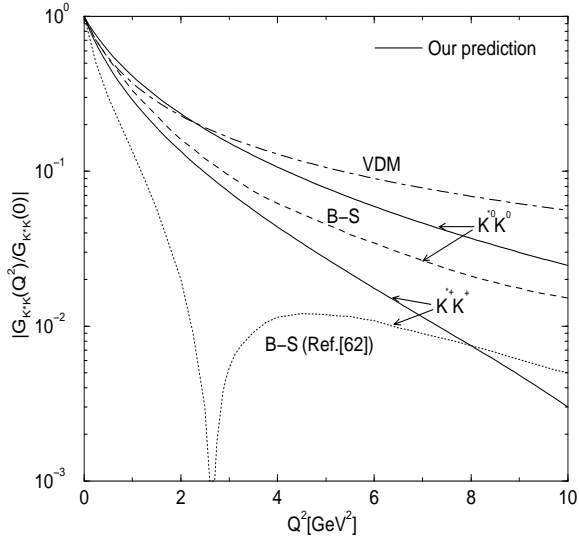,width=3.4in,height=3.25in}}
\caption{ The normalized form factors of
$K^{*+}\to K^{+}\gamma^{*}$(solid line) and
$K^{*0}\to K^{0}\gamma^{*}$(dotted line) transitions with the parameter
$\beta$= 0.36 GeV. The dotted and dashed lines correspond to charged and
neutral vector kaon transition form factors by Ref.~\protect\cite{munz}.
The dashed-dotted line corresponds to the VDM prediction. }
\end{figure}
Using the spin-averaged masses of
$m_{K^{*\pm}}$= 792.1 MeV and $m_{K^{*0}}$= 796.5 MeV
and the constituent quark mass $m_{s}$= 0.45 GeV~\cite{ji},
we obtain the decay widths for these charged and neutral vector kaon
decay processes as $\Gamma_{K^{*\pm}\to K^{\pm}\gamma}$= 53 keV,
$\Gamma_{K^{*0}\to K^{0}\gamma}$= 122 keV, respectively.
The experimental data are
$\Gamma^{\rm exp}_{K^{*\pm}\to K^{\pm}\gamma}$= 50 $\pm$ 5 keV,
$\Gamma^{\rm exp}_{K^{*0}\to K^{0}\gamma}$= 117 $\pm$ 10 keV. The
electromagnetic transition charge radii are predicted as
$\la r^{2}_{G_{K^{*\pm}K^{\pm}}}\ra = 11$ ${\rm GeV}^{-2}$
and $\la r^{2}_{G_{K^{*0}K^{0}}}\ra = 10$ ${\rm GeV}^{-2}$.
In Fig.~2.4, we report the $Q^{2}$-dependence of transition form factor of
charged and neutral vector kaon, i.e.,
$G_{K^{*+}K^{+}}(Q^{2})$ and $G_{K^{*0}K^{0}}(Q^{2})$, respectively,
for $0\leq Q^{2}\leq 10$ ${\rm GeV}^{2}$.
Even though we showed the result of $\beta = 0.36$ GeV only, we note
that our results become much closer to those of Bethe-Salpeter(BS) quark
model prediction~\cite{munz} with smaller values of $\beta$, i.e.,
$\beta\sim 0.3$ GeV.

For the decay processes of $\phi\to\eta(\eta')\gamma^{*}$,
the transition form factors of
$\phi\to\eta\gamma^{*}$ and $\phi\to\eta'\gamma^{*}$ are given by
\begin{eqnarray}
&&G_{\phi\eta}(Q^{2})=
Y_{\eta}2e_{s}I_{PV\gam^*}(m_{M},m_{q},\beta)\nonumber\\
&&G_{\phi\eta'}(Q^{2})= Y_{\eta'}2e_{s}I_{PV\gam^*}(m_{M},m_{q},\beta).
\end{eqnarray}
As shown in Appendix B, we obtain the spin-averaged mass of
$\phi$-meson as $m_{\phi} = 0.799$ GeV. Our predictions of
the decay widths of the transitions $\phi\to\eta\gamma$
and $\phi\to\eta'\gamma$ with $\theta_{SU(3)} = -10^\circ$ and
$-23^\circ$ are given by $\Gamma^{-10^\circ}_{\phi\to\eta\gamma} = 61$ keV,
$\Gamma^{-10^\circ}_{\phi\to\eta'\gamma} = 0.28$ keV and
$\Gamma^{-23^\circ}_{\phi\to\eta\gamma} = 45$ keV,
$\Gamma^{-23^\circ}_{\phi\to\eta'\gamma} = 0.45$ keV, respectively.
The current experimental data are $\Gamma^{\rm exp}_{\phi\to\eta\gamma}
= 56.9\pm 2.9 $ keV and $\Gamma^{\rm exp}_{\phi\to\eta'\gamma} < 1.8$ keV.
It will be very interesting to compare our results, especially for
$\phi\to\eta'\gamma$, with the more precise measurements envisioned at
TJNAF. The electromagnetic charge radii for $-10^\circ$ mixing angle
are also predicted as
$\la r^{2}_{G_{\phi\eta}}\ra = 5$ ${\rm GeV}^{-2}$ and
$\la r^{2}_{G_{\phi\eta'}}\ra = 5.5$ ${\rm GeV}^{-2}$.

In the case of transition $A_{1}\to\pi\gamma^{*}$,
the kinematical factors in front of the form factor
$G_{2}$ in Eq. (2.26) yields zero for a certain $Q^2$ in
the standard Drell-Yan frame(${\bf P}_{\perp} = 0, q^{+} =0)$ as
discussed in Ref.~\cite{Bag}.
This leads to a technical difficulty in extracting the form factors
numerically from Eq. (2.26). Thus, as pointed out in Ref.~\cite{Bag}, we
use the following symmetric coordinate frame for the calculation of
of $A_{1}\to\pi\gamma^{*}$ transition form factors:
\begin{eqnarray}
&&P = \biggl(P^{+},\frac{m_{A_{1}}^{2} +\frac{1}{4}Q^2}{P^{+}}, 
-\frac{1}{2}{\bf q}_{\perp}\biggr),\hspace{.2in}
P' = \biggl(P^{+},\frac{m_{\pi}^{2} +\frac{1}{4}Q^2}{P^{+}},
\frac{1}{2}{\bf q}_{\perp}\biggr),\nonumber\\
&& q = \biggl(0, \frac{m_{\pi}^{2} - m_{A_{1}}^{2}}{P^{+}},
{\bf q}_{\perp}\biggr).
\end{eqnarray}
Then, we find the following expressions for the $A_{1}$ form factors:
\begin{eqnarray}
&&G_{1}(Q^{2}) = \frac{m_{A_{1}}^{2}}{P^{+}(
m_{\pi}^{2} - m_{A_{1}}^{2})}
\biggl[ G^{+}_{0} + \frac{(m_{\pi}^{2} - m_{A_{1}}^{2}+ Q^{2})}
{\sqrt{2}m_{A_{1}}}\frac{G^{+}_{+}}{Q}\biggr],\\
&&G_{2}(Q^{2})= \frac{2m_{A_{1}}^{4}}
{(m_{A_{1}}^{2} - m_{\pi}^{2})^{2}Q}\biggl[ G^{x}_{0}
- \frac{(m_{\pi}^{2} - m_{A_{1}}^{2})}{2\sqrt{2}m_{A_{1}}P^{+}}
G^{+}_{+}\biggr],
\end{eqnarray}
where $G^{\mu}_{\lambda}\equiv\la\pi(P')|J^{\mu}|A_{1}(P,\lambda)\ra$
and its matrix elements are given by
\begin{eqnarray}
&&G^{+}_{+}=\frac{Q}{2\beta}\sqrt{\frac{N_{A_{1}}N_{\pi}}{2}}
\int_{0}^{1}\frac{dx}{x}\exp\biggl[-\frac{\xi^{2}
+ \tilde{m}^{2}_{1} + x(\tilde{m}^{2}_{2}
- \tilde{m}^{2}_{1})}{x(1-x)}\biggr ](a^{\rm f}_{1} + a^{\rm f}_{2})\;\;
\nonumber\\
&&\hspace{1cm}\times(a^{\rm i}_{1} - a^{\rm i}_{2})[x(1-x) + \xi^{2}],
\\ &&G^{+}_{0}=\sqrt{2N_{A_{1}}N_{\pi}}
\int_{0}^{1}\frac{dx}{x(1-x)}\exp\biggl [-\frac{\xi^{2}
+ \tilde{m}^{2}_{1} + x(\tilde{m}^{2}_{2}
- \tilde{m}^{2}_{1})}{x(1-x)}\biggr] {\cal G}_{0},\;\;\\
&&G^{x}_{0}=\frac{Q}{P^{+}}\sqrt{\frac{N_{A_{1}}N_{\pi}}{2}}
\int_{0}^{1}\frac{dx}{x^{2}}\exp\biggl[-\frac{\xi^{2}
+ \tilde{m}^{2}_{1} + x(\tilde{m}^{2}_{2} -
\tilde{m}^{2}_{1})}{x(1-x)}\biggr]{\cal G}_{0},\;\;
\end{eqnarray}
with a common facfor ${\cal G}_{0}$ in the longitudinal
component of $G$:
\begin{eqnarray}
&&{\cal G}_{0}= 2x^{2}(1-x)^{2}[(a^{\rm f}_{1} +
a^{\rm f}_{2}) -(a^{\rm i}_{1} + a^{\rm i}_{2})] \nonumber\\
&&\hspace{0.5cm}+ x(1-x)[a^{\rm f}_{1}a^{f}_{2}(a^{\rm i}_{1} +
a^{\rm i}_{2}) - a^{\rm i}_{1}a^{\rm i}_{2}(a^{\rm f}_{1}
+ a^{\rm f}_{2})] - \xi^{4}[(a^{\rm i}_{1} + a^{\rm i}_{2})
+ (a^{\rm f}_{1} + a^{\rm f}_{2})] \nonumber\\
&&\hspace{0.5cm}+ \xi^{2}[a^{\rm i}_{1}a^{\rm i}_{2}(a^{\rm f}_{1}
+ a^{\rm f}_{2}) + a^{\rm f}_{1}a^{\rm f}_{2}( a^{\rm i}_{1}
+ a^{\rm i}_{2})].
\end{eqnarray}
In Fig.~2.5, we show the result of the $A_{1}$ transition form 
factors, $G_{1}(Q^{2})$ (solid line) and $G_{2}(Q^{2})$ (dashed line).
While the form factor $G_{2}(Q^{2})$ agrees with the predictions of
the QCD sum-rules(dot-dashed line)~\cite{smilga} in the region
$1\leq Q^{2}\leq 3\hspace{.05in}\mbox{GeV}^{2}$,
the form factor $G_{1}(Q^{2})$ seems to be quite
different from the QCD sum-rule result(dotted line)~\cite{smilga}.
However, there are other QCD sum rule calculations of $G_{1}$
and $G_{2}$~\cite{braun,bel} and the results are rather different
from each other.
The authors of QCD sum-rules~\cite{smilga} also pointed out that
their predictions for the transition $A_{1}\to\pi\gamma^{*}$ are of
semiqualitative results. The electromagnetic radii corresponding to
these form factors are evaluated as
$\la r^{2}_{G_{1}}\ra = 15$ ${\rm GeV}^{-2}$ and
$\la r^{2}_{G_{2}}\ra = -30$ ${\rm GeV}^{-2}$, respectively.

\begin{figure}[t]
\centerline{\psfig{figure=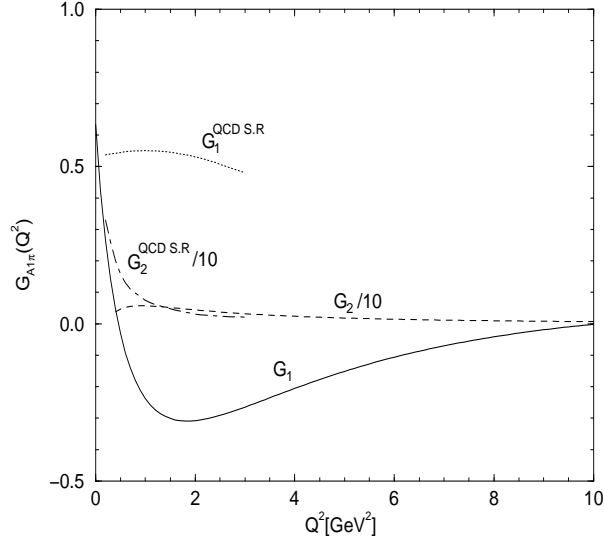,width=3.5in,height=3.3in}}
\caption{The form factors of $A^{+}_{1}\rightarrow\pi^{+}\gamma^{*}$
transition with the parameter $\beta =0.36$ GeV.
$G_{1}(Q^{2})$ and $G_{2}(Q^{2})/10$ of our results are represented by
the solid and dashed lines, respectively. For comparison, we show also
the QCD sum-rule results~\protect\cite{smilga}:
$G_{1}(Q^{2})$ (dotted-line) and
$G_{2}(Q^{2})/10$ (dotted-dashed line)}.
\end{figure}
The decay width of $A_{1}^{+}(1260)\to\pi^{+}\gamma$
is obtained from Eq. (\ref{A1_rate}) as
\be
\Gamma(A_{1}\rightarrow\pi\gamma)= 705\hspace{.1in} \mbox{keV}.
\ee
The prediction of VDM~\cite{rose1,rose2} is given by $\Gamma^{VDM}_{A_{1}
\to\pi\gamma} = 1000 - 1500$ keV and the experimental value has
been reported~\cite{ziel1,ziel2} as $\Gamma(A_{1}^{+}\to\pi^{+}
\gamma) = 640 \pm 246$ keV  using the measurment of Primakoff production
of the $A_{1}$ resonance. Thus, our predicted width
is quite consistent with the corresponding experimental data.
In Refs.~\cite{Bag,azn}, the same decay width was calculated using the
invariant mass of $A_{1}$ meson instead of the spin-averaged mass.
Their results $\Gamma(A_{1}^{+}\to\pi^{+}\gamma) = 319$ keV~\cite{Bag}
and $250$ keV~\cite{azn} seem to have rather large discrepancy from the
experimental data.
\section{The Form Factors for $(\pi^{0},\eta,\eta')\to\gamma^{*}\gamma$
Transitions}
The transition form factor of
$P\to\gamma^{*}\gamma$($P = \pi^{0},\eta$, and $\eta'$)
is defined from the matrix element of electromagnetic current
$\Gamma_{\mu}= \la\gamma(P+q)|J_{\mu}|P(P)\ra$ as follows:
\begin{eqnarray}\label{Gam}
\Gamma_{\mu}
= ie^{2}G_{P\gamma}(Q^{2})\epsilon_{\mu\nu\rho\sigma}P^{\nu}
\vep^{\rho}q^{\sigma},
\end{eqnarray}
where $P$ and $q$ are the momenta of the incident pseudoscalar meson
and virtual photon, respectively, and $\vep$ is the transverse
polarization vector of the final (on-shell) photon satisfying
${\vec{\vep}}_{\perp}\cdot{\bf q}_{\perp}=0$. We again used the
standard $q^{+}=0$ frame for the calculation of the decays
involving two photons, 
\begin{eqnarray}
P= (P^{+},P^{-},{\bf P}_{\perp})
= (1,m^{2}_{M},{\bf 0}_{\perp}), \hspace{0.3cm}
q= (0, Q^{2} - m^{2}_{M},{\bf q}_{\perp}),
\end{eqnarray}
where $P^{+}$ is arbitrary but for simplicity we choose $P^{+}=1$ and
thus we have $q^{2}= -q^{2}_{\perp}= -Q^{2}$.
Each component of the final-state real photon is uniquely determined
in this reference frame by the conservation of momentum and the
on-mass shell condition of the real photon in the final state.

The $Q^{2}$-dependent decay
rate for $P\to\gamma^{*}\gamma$ is given by~\cite{Lepage}
\be\label{2gam_rate}
\Gamma_{P\to\gamma^{*}\gamma}(Q^{2}) =
\frac{\pi}{4}\alpha^{2}M_{P}^{3}G^{2}_{P\gamma}(Q^{2}),
\ee 
where $M_{P}$ is the physical mass of $P=\pi^{0},\eta$, and $\eta'$.
Here, the decay width is given by $\Gamma_{P\to\gamma\gamma}$ at
$Q^{2}=0$. If we choose the `$+$' component of the current, the vertex factor
in Eq. (\ref{Gam}) is given by
\begin{eqnarray}
\Gamma^{+} &=&\sqrt{n_{c}}\sum_{q}e^{2}_{q}
\sum_{\lambda,\lambda'}\int^{1}_{0}dx
\int\frac{d^{2}{\bf{k_{\perp}}}}{16\pi^{3}}
\Psi_{P}(x,{\bf{k_{\perp}}})\nonumber\\
&\times&
\biggl[\frac{\bar{v}_{\lambda'}(x_{2},{\bf{k_{\perp}}})}{\sqrt{x_{2}}}
\not\!\vep\frac{u_{\lambda}(x_{1},{\bf{k_{\perp}}} +
{\bf{q_{\perp}}})}{\sqrt{x_{1}}}
\frac{\bar{u}_{\lambda}(x_{1},{\bf{k_{\perp}}}
+ {\bf{q_{\perp}}})}{\sqrt{x_{1}}}\gamma^{+}
\frac{u_{\lambda}(x_{1},{\bf{k_{\perp}}})}{\sqrt{x_{1}}}\nonumber\\
&\times& \frac{1}{{\bf q}^2_{\perp}- [({\bf{k_{\perp}}} +
{\bf{q_{\perp}}})^{2} + m^{2}]/x_{1} -
({\bf k}^2_{\perp} + m^{2})/x_{2}}+(1\leftrightarrow 2)\biggr].
\end{eqnarray}
Here, only anti-parallel helicities of constituents contribute to the
integrations and our model wave function
$\Psi_{P}(x,{\bf{k_{\perp}}})$
for anti-parallel helicities is given by
\begin{eqnarray}
\Psi_{P}(x,{\bf{k_{\perp}}}) = \frac{\pi}{2\beta^{3}}
(\frac{N_{P}}{2})^{1/2}
\frac{(a_{1}a_{2} - {\bf k}^2_{\perp})} {x(1-x)}
\exp\biggl(-\frac{{\bf k}^2_{\perp} 
+ m^{2}}{8x(1-x)\beta^{2}}\biggr).
\end{eqnarray}
A straightforward calculation for the transition
$P\to\gamma^{*}\gamma$ gives the following result:
\begin{eqnarray}
&&G_{\pi\gamma} =(e_{u}^{2} -
e_{d}^{2})I_{P\gamma^{*}\gamma}(m_{\pi},m_{u(d)},\beta),
\nonumber\\
&&G_{\eta\gamma}= X_{\eta}(e_{u}^{2} + e_{d}^{2})
I_{P\gamma^{*}\gamma}(m_{\eta},m_{u(d)},\beta) -
Y_{\eta}e_{s}^{2}\sqrt{2}I_{P\gamma^{*}\gamma}(m_{\eta},m_{s},\beta), 
\nonumber\\
&&G_{\eta'\gamma}= X_{\eta'}(e_{u}^{2} + e_{d}^{2})
I_{P\gamma^{*}\gamma}(m_{\eta'},m_{u(d)},\beta) +
Y_{\eta'}e_{s}^{2}\sqrt{2}I_{P\gamma^{*}\gamma}(m_{\eta'},m_{s},\beta),
\nonumber\\
\end{eqnarray}
where
\begin{eqnarray}\label{I2}
I_{P\gamma^{*}\gamma}(m_{M},m_{q},\beta)&=&
-\sqrt{n_c}\frac{\sqrt{N_{P}}}{\pi\beta}\int^{1}_{0}
\frac{dx}{x}\exp\biggl(-\frac{\tilde{m}^{2}}{2x(1-x)}\biggr)
\nonumber\\
&\times& \biggl[ 2x(1-x) -
\exp\biggl(\frac{4\xi^{2} + \tilde{m}^{2}}{2x(1-x)}\biggr)
(a_{1}a_{2} + \tilde{m}^{2} + 4\xi^{2})\nonumber\\
&\times& \int^{1}_{\infty}dt\exp\biggl(-\frac{4\xi^{2}
+ \tilde{m}^{2}}{2x(1-x)}t\biggr)/t\biggr].
\end{eqnarray}
Using Eqs. (\ref{2gam_rate})-(\ref{I2}), 
the decay widths for $P\to\gamma\gamma$ are obtained as
\begin{eqnarray}
&&\Gamma(\pi^{0}\to\gamma\gamma)= 6.50\hspace{.05in}\mbox{eV},
\hspace{0.3cm}
\Gamma(\eta\to\gamma\gamma)= 0.47 (0.65)\hspace{.05in}\mbox{keV},\nonumber\\
&&\Gamma(\eta'\to\gamma\gamma)= 7.9 (5.6)\hspace{.05in}\mbox{keV},
\end{eqnarray}
where the values for $\eta\to\gamma\gamma$ and $\eta'\to\gamma\gamma$
are obtained for the $-10^\circ$ ($-23^\circ$) mixing scheme.
The experimental data\cite{data} are given by
$\Gamma^{\rm exp}_{\pi\gamma} = 7.8\pm 0.5$ eV,
$\Gamma^{\rm exp}_{\eta\gamma} = 0.47\pm 0.05$ keV and
$\Gamma^{\rm exp}_{\eta'\gamma} = 4.3\pm 0.6$ keV.
The agreement of our results with the experimental data is not
unreasonable. In Figs. 2.6-2.8,
the $Q^{2}$-dependence of the decay rate $\Gamma_{P\gamma}(Q^{2})$
for $P = \pi^{0},\eta,$ and $\eta'$ are shown and compared
with the recent experimental data~\cite{cello1,cello2,tpc}.
Our predictions for all of these processes are overall in a good
agreement with the experimental data up to a rather large $Q^{2}$.
\begin{figure}
\centerline{\psfig{figure=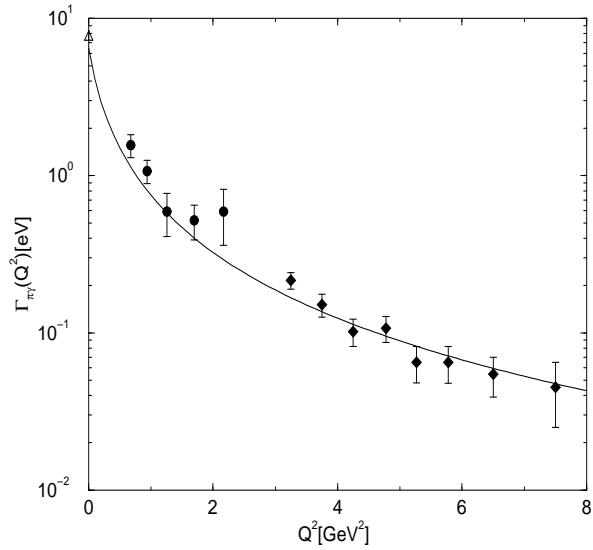,width=3.5in,height=3.3in}}
\caption{The decay rate for $\pi^{0}\to\gamma^{*}\gamma$ transition
with the parameter $\beta =0.36$ GeV.
Data are taken from Refs.~\protect\cite{cello1,cello2}.}
\end{figure}
\begin{figure}
\centerline{\psfig{figure=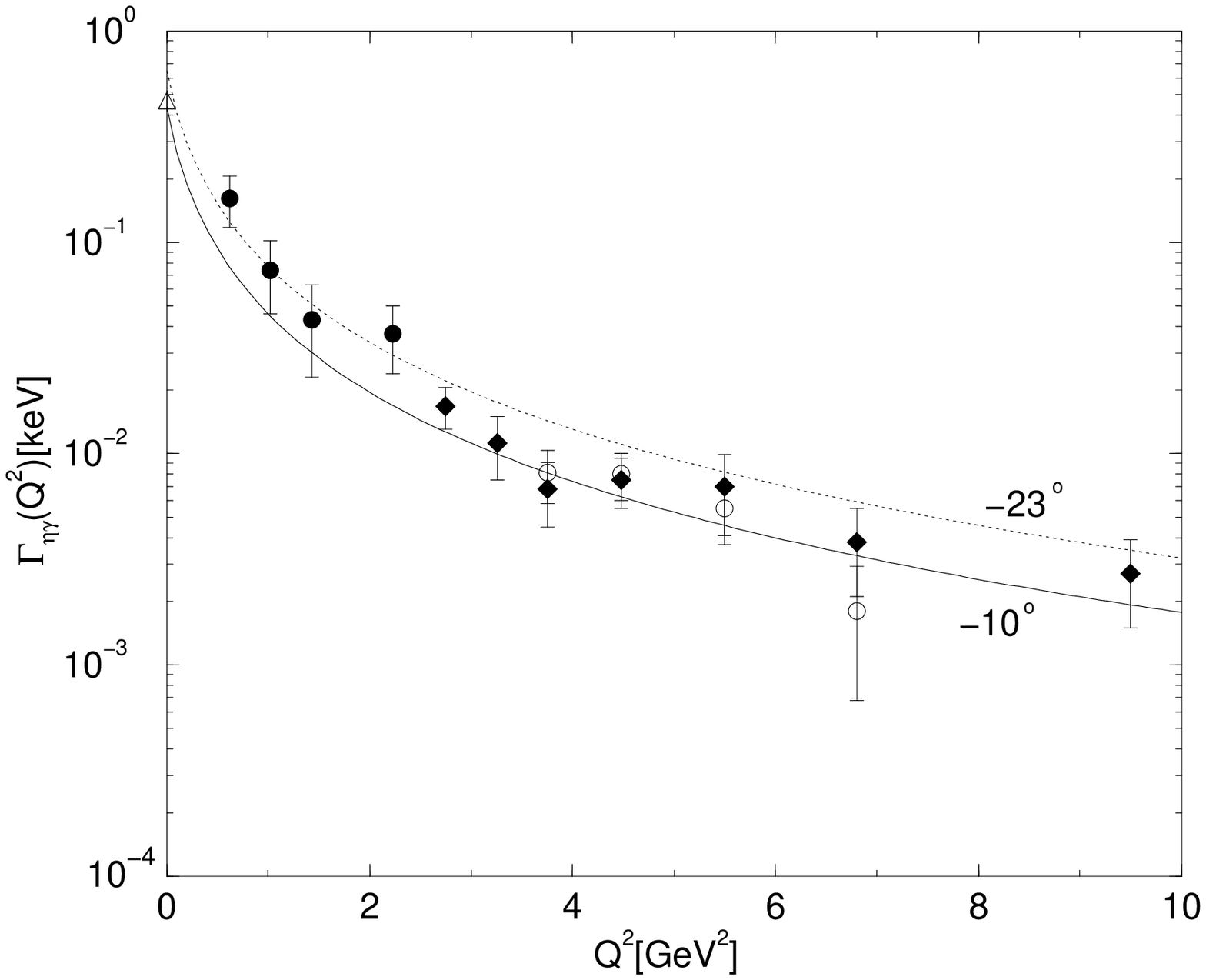,width=3.5in,height=3.3in}}
\caption{The decay rate for $\eta\to\gamma^{*}\gamma$ transition
with the parameter $\beta =0.36$ GeV. The solid and dotted lines
correspond to the $\theta_{SU(3)} = -10^\circ$ and $-23^\circ$ mixing
schemes, respectively. Data are taken from
Refs.~\protect\cite{cello1,cello2,tpc}.}
\end{figure}
\begin{figure}
\centerline{\psfig{figure=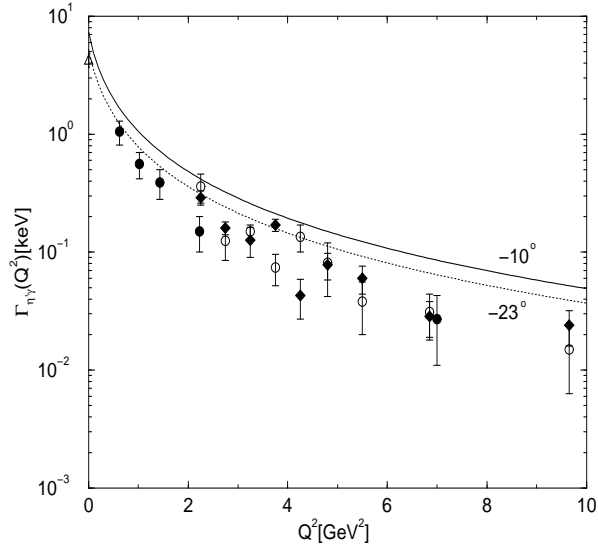,width=3.5in,height=3.3in}}
\caption{The decay rate for $\eta'\to\gamma^{*}\gamma$ transition
with the parameter $\beta =0.36$ GeV. The solid and dotted lines correspond
to the $\theta_{SU(3)} = -10^\circ$ and $-23^\circ$ mixing schemes,
respectively. Data are taken from
Refs.~\protect\cite{cello1,cello2,tpc}.}
\end{figure}
\section{Summary and Discussion}
In this work, we have investigated the radiative decays of
pseudoscalar($\pi,K, \eta,\eta'$), vector($\rho,K^{*},\omega,\phi$) and
axial vector($A_{1}$) mesons as well as the form factors of $\rho$ and
$A_{1}$ mesons using a simple relativistic constituent quark model.
We summarized all of our predictions on the meson decay widths for
various values of $\beta$ ($\beta = 0.32, 0.34, 0.36, 0.38$ GeV) in
Table~\ref{T22}.
Remarkably, most of our predictions with the parameter $\beta$ in this
region are within the experimental errors.
We have also investigated the sensitivity of our  results by varying
quark masses. For $\pm 10\%$ variation of the non-strange quark masses,
the decay width $\Gamma$ and the transition charge radius
$\la r^{2}\ra^{1/2}$ of the radiative meson decays change by
3-4$\%$ and 2-4$\%$, respectively.
Changing strange quark mass by $\pm 10 \%$ for the process of
$K^{*\pm}\to K^{\pm}\gamma$ yields 13-15$\%$ and 3-4$\%$ difference
in the decay width and the transition charge radius, respectively.
\begin{table}
\centering
\caption{Radiative decay widths for the $V(P)\to P(V)\gamma$,
$A_{1}\to\pi\gamma$ and $P\to 2\gamma$ for various model parameters
$\beta$ (in unit of GeV) with the $\eta$-$\eta'$ mixing angle
$\theta_{SU(3)}=-10^\circ[-23^\circ]$. The experiment is the result from
~\protect\cite{data}, 
unless otherwise noted. The unit of decay width is [keV], unless
otherwise noted. }\label{T22}
{\small
\begin{tabular}{|l|c|c|c|c|c|c|} \hline
Widths ($\Gamma$)&$\beta=0.32$ & 0.34 & 0.36 & 0.38 & Expt.\\ \hline
$\rho^{\pm}\to\pi^{\pm}\gamma$ & 78 & 73 & 69 & 64& $60\pm 8$ \\
\hline
$\omega\to\pi\gamma$ & 775& 742 & 708 & 674 & $717\pm 51$ \\
\hline
$K^{*\pm}\to K^{\pm}\gamma$ & 60 & 57 & 53 & 50 & $50\pm 5$ \\
\hline
$K^{*0}\to K^{0}\gamma$ & 134 & 128 & 122 & 116 & $117\pm 10$ \\
\hline
$\rho\to\eta\gamma$ & 66 [77] & 60 [70] & 56 [65] & 51 [60] 
& $58\pm 10$ \\ 
\hline
$\omega\to\eta\gamma$ & 7.4 [8.5] & 6.9 [7.9] & 6.4 [7.4] & 6.0 [6.8] &
$7.0\pm 1.8$ \\
\hline
$\eta'\to\rho\gamma$ & 137 [89] & 126 [80] & 117 [72] &108 [66] 
& 61 $\pm8 $ \\ 
\hline
$\eta'\to\omega\gamma$ & 11.2 [7.3] & 10.4 [6.6] & 9.7 [6.0]
& 9.1 [5.6] & $5.9\pm 0.9$ \\
\hline
$\phi\to\eta\gamma$ & 54 [40] & 58 [42] & 61 [45]& 65 [47] 
& $56.9\pm 2.9$\\
\hline
$\phi\to\eta'\gamma$ & 0.26 [0.43] & 0.27 [0.44] & 0.28 [0.45]
& 0.29 [0.46] & $< 1.8$\\
\hline
$A_{1}\to\pi\gamma$ & 620 & 664 & 705 & 742
& $640\pm 246^{\cite{ziel1,ziel2}}$ \\
\hline
$\pi^{0}\to 2\gamma$ & 7.58 & 7.06 & 6.50 & 5.91& $7.8\pm 0.5$ [eV]\\
\hline
$\eta\to 2\gamma$ & 0.61 [0.78] & 0.53 [0.71] &0.47 [0.65] &
0.42 [0.58] & $0.47\pm 0.05$\\
\hline
$\eta'\to 2\gamma$ & 8.8 [6.5] & 8.3 [6.1] & 7.9 [5.6]& 7.3 [5.1]
& $4.3\pm 0.6$\\ \hline
\end{tabular} }
\end{table}

In our point of view, the success of this
model hinges upon the simplicity of the light-front vacuum.
The recent lattice QCD results~\cite{Ku} indicate that the mass
difference between $\eta'$ and pseudoscalar octet mesons due to the
complicated nontrivial vacuum effect increases(or decreases) as
the quark mass $m_{q}$ decreases(or increases), i.e.,
the effect of the topological charge contribution should be small
as $m_{q}$ increases.
This makes us believe that the complicated nontrivial vacuum effect
can be traded off by the constituent quark masses.
This may mean that there is a suppression of complicate zero-mode
contribution~\cite{Rey} from the light-front vacuum in our model due
to the rather large constituent quark masses.

Our approach in this work was to model the wave fuction rather
than to model the potential. However, we have attempted to compare our
results with various other available theoretical results 
including the potential models and the QCD sum rules. At the very least,
our results seem to be quite comparable with the results from modelling the
potential.
The results on the angular condition are also not drastically different
from the result of the potential models (See Fig. 2.1(a)). Furthermore, the
agreement with the QCD sum rule results is not unreasonable. Our model
has a predictive power and more experimentally measurable quantities 
should be calculated and compared with data.
A particularly interesting prediction from our model is the branching
ratio of $\phi\to\eta'\gamma$ estimated as 10$^{-4}$ for the $\eta$-$\eta'$
mixing angle of $-23^\circ$. For the mixing angle $-10^\circ$, the estimation
for this branching ratio is reduced by 60$\%$ from that of $-23^\circ$.
Thus, it will be very interesting to compare our results with the precise
measurements envisioned at TJNAF.


%% file: Relations.tex
\newpage
\setcounter{equation}{0}
\renewcommand{\theequation}{\mbox{3.\arabic{equation}}}
\chapter{Relations Among the Light-Front Quark Models
and the Model Prediction of $\eta-\eta'$ Mixing Angle: IM Scheme}
One of the popular quark models in the LF formalism is
the invariant meson mass (IM)
scheme~\cite{jaus,chung2,card,tao,sch}
in which the IM square $M^{2}_{0}$ is given by
\be\label{in_mass3}
M_{0}^{2}=\sum_{i}^{2}\frac{ {\bf k}^{2}_{i\perp} +
m_{i}^{2}}{x_{i}}.
\ee
The corresponding spin-orbit wave functions
${\cal R}^{JJ_{3}}_{\lam_{1},\lam_{2}}(x,{\bf k}_{\perp})$
of pseudoscalar and vector
mesons are obtained by the Melosh transformation and their
expressions are given in the Appendix A.

While the expressions of the spin-orbit wave functions for
pseudoscalar and vector mesons are unique, the models differ in 
choosing the radial wave function.
For examples, Huang, Ma, and Shen~\cite{tao} adopted the
Brodsky-Huang-Lepage (BHL) oscillator 
prescription~\cite{Lepage,BHL1}
(model $H$):
\be\label{h3}
\phi^{H}(k^{2})
= A\exp\biggl[- \sum_{i=1}^{2}\frac{{\bf k}^{2}_{\perp i}+
m_{i}^2}{x_{i}}/8\beta^{2}\biggr]\;.
\ee
On the other hand, the authors of Ref.~\cite{chung2} (model $C$) and
Ref.~\cite{jaus} (model $J$) used the following radial wave functions:
\be\label{c3}
\phi^{C(J)}(k^{2}) = N_{C(J)}\exp(-{\bf k}^{2}/2\beta^{2}),
\ee
where ${\bf k}=(k_{n},{\bf k}_{\perp})$ is the three momentum and
the normalization constants $N_{C(J)}$ are
$N_{C}= (4/\sqrt{\pi}\beta^{3})^{1/2}$
and $N_{J}=\pi\sqrt{2/3}N_{C}$, respectively.
In addition to the harmonic oscillator (HO) wave functions,
Schlumpf~\cite{sch} used a power-law (PL) wave function
given by
\be\label{sch3}
\phi^{PL}(k^{2})= N_{\rm power}(1 + {\bf k}^{2}/\beta^{2})^{-s},
\ee
where the power $s$ may be chosen typically as $s$=2 (model PL).
Furthermore, more realistic model wave function from the
Godfrey-Isgur potential~\cite{isgur2} was investigated in 
Ref.~\cite{card}.

Regarding the differences in choosing HO radial wave functions,
Huang, Ma, and Shen~\cite{tao} pointed out the ambiguity of
introducing
a factor, such as $\sqrt{M_{0}/4x(1-x)}$ in Ref.~\cite{chung2} or
$\sqrt{1/2x(1-x)}$ in Ref.~\cite{coester}, to the BHL wave function
as a consequence of the jacobian
relating the instant-form momentum to the LF momentum.
This ambiguity which we will discuss in detail in the following
section distinguishes the model $H$ from the models $C$ and $J$.

The purpose of this work is to (1) show explicitly the difference of
the model $H$ from the other two HO models ($C$ and $J$),
(2) point out nevertheless that the numerical predictions
from all of these HO models ($H$,$J$ and $C$) are almost equivalent
once the best fit parameters(i.e., constituent quark masses and Gaussian
parameters $\beta$) are chosen, (3) apply these models to the two
photon decay processes and the radiative decays between pseudoscalar
($P$) and vector ($V$) mesons, and (4) obtain the best value of
$\eta$-$\eta'$ mixing angle. In addition, we change the form of the radial 
wave function from HO to PL (model PL) and investigate the 
presence-absence effect of the Jacobi factor to further 
identify the differences in the model wave functions.

This Chapter is organized as follows:
In Secion 3.1, we analyze the relations among the three HO model
wave functions and show that they are basically equivalent with
each other modulo a Jacobi factor.  In Section 3.2, we list several
constraints to fix the free parameters of the relativistic quark model 
in comparison with the relavant experimental data for the decay constants.
In our analysis we consider both HO and PL
wave functions and restrict ourselves on the light-meson sector
($u$-, $d$- and $s$-quarks) with equal quark and anti-quark
masses($m_{q}=m_{\bar{q}}$).
In Section 3.3, we analyze the two-photon decays of $\pi,\eta,$ and
$\eta'$ and
the transitions of $V(P)\to P(V)\gamma$ searching for the
best value of $\eta$-$\eta'$ mixing angles.
Summary and discussion are followed in Section 3.4.

\section{Relations among the Light-Front Quark Models}
Since the spin-orbit wave functions
${\cal R}^{JJ_{3}}_{\lam_{1},\lam_{2}}(x,{\bf k}_{\perp})$
obtained from the Melosh rotation
are independent from the radial wave functions as far as we consider
the same IM scheme, the relations among the three HO
models ($H,J$ and $C$) are obtained by comparing only the radial wave
functions in Eqs. (\ref{h3}) and (\ref{c3}).

The relation between $\phi^{C}(k^{2})$~\cite{chung2} and
$\phi^{J}(k^{2})$~\cite{jaus}
can be obtained by the definition of the normalization of the wave
functions
\be\label{norm}
\frac{1}{4\pi}\int d^{3}k |\phi^{C}(k^{2})|^{2}
= \frac{n_{c}}{(2\pi)^{3}}\int d^{3}k |\phi^{J}(k^{2})|^{2}=1,
\ee
which leads to
\be\label{eqiv}
\phi^{J}(k^{2})= \pi\sqrt{\frac{2}{3}}\phi^{C}(k^{2}),
\ee
where $n_{c}(=3)$ in Eq. (\ref{norm}) is the color factor.
Eq. (\ref{eqiv}) implies that the two models $C$ and $J$ are equivalent 
to each other.
Comparing $\phi^{H}(k^{2})$ and $\phi^{C(J)}(k^{2})$, we use the
jacobian ${\cal J}={\partial k_{n}}/{\partial x}= M_{0}/4x(1-x)$ of the
variable transformation $\{x,{\bf k}_{\perp}\}\to {\bf k}=(k_{n},{\bf
k}_{\perp})$ in which the longitudinal component $k_{n}$ is defined as 
a function of $x$ and 
${\bf k}_{\perp}$ by $k_{n}= (x-\frac{1}{2})M_{0}$~\cite{chung2}.
Then the phase space [$d^{3}k$] in Eq. (\ref{norm}) is transformed 
into $(x,{\bf k}_{\perp})$ variables as follows:
\begin{eqnarray}
d^{3}k= dxd^{2}{\bf k}_{\perp}\frac{M_{0}}{4x(1-x)}.
\end{eqnarray}
The wave function $\phi^{H}(k^{2})$ can then be compared directly
with $\phi^{C}(k^{2})$ (or $\phi^{J}(k^{2})$) in 
$(x,{\bf k}_{\perp})$-space by the definition of the normalization 
of the wave functions
\begin{eqnarray}\label{norm2}
&&\int^{1}_{0}dx\int\frac{d^{2}{\bf k}_{\perp}}
{16\pi^{3}}|\phi^{H}(k^{2})|^{2}\nonumber\\
&&\hspace{0.5cm}= \frac{1}{4\pi}\int^{1}_{0}dx\int d^{2}{\bf
k}_{\perp}
\frac{M_{0}}{4 x(1-x)}|\phi^{C}(k^{2})|^{2}=1.
\end{eqnarray}
From Eqs. (\ref{eqiv}) and (\ref{norm2}), we obtain
\begin{eqnarray}\label{eq.39}
&&\phi^{H}(k^{2})=\pi\sqrt{\frac{M_{0}}{x(1-x)}}\phi^{C}(k^{2})
\nonumber\\
&&\hspace{1.4cm}= 
\sqrt{n_{c}}\sqrt{\frac{M_{0}}{2x(1-x)}}\phi^{J}(k^{2}).
\end{eqnarray}
Here, we should note that in order to satisfy the above equivalence
relation the normalization constant $A$ in Eq. (\ref{h3}) is no longer 
a constant but a function of the LF variables $(x,{\bf k}_{\perp})$:
\be\label{const}
A(x,{\bf k}_{\perp})= \pi\sqrt{\frac{M_{0}}{x(1-x)}}
\biggl(\frac{4}{\sqrt{\pi}\beta^{3}}\biggr)^{1/2}\exp(m^2/2\beta^2),
\ee
for $\phi^{H}(k^{2})$. However, in the model $H$~\cite{tao}, $A$ is
not a function of $(x,{\bf k}_{\perp})$ given by Eq. (\ref{const}) but 
a constant.
Therefore, the predictions of the model $H$ are different from those
of the models $C$ and $J$.

With the analysis of the wave functions of each model, we have now
shown that the two models $C$ and $J$ are completely equivalent to 
each other and the model $H$ is different from $C$ and $J$ 
unless $A$ satisfies Eq. (\ref{const}).
If we were to use $A(x,{\bf k}_{\perp})$ given by Eq. (\ref{const}),
then we can explicitly show the equivalence of all the observables
among the three models.
Similarly, one can consider the presence-absence effect of the Jacobi
factor for the PL model wave function. In the next section, we will
present the constraints to fix the free parameters in both HO and PL 
wave functions.

\section{Constraints on the Wave Functions of 
$0^{-+}$ and $1^{--}$ Mesons}
In order to fix the free parameters($m_{q}$ and $\beta_{q\bar{q}}$)
of the models $H$ and PL, we use the method adopted by
Ref.~\cite{jaus} (model $J$).  Considering the mesons with equal quark 
and anti-quark masses, we need four parameters in the $u-,d-$ and 
$s$-quark sector, i.e., $m_{u(d)}, m_{s}$, and 
$\beta_{q\bar{q}}$($q=u(d)$,$s$).
We fix these four parameters by fitting the pion decay constant
$f_{\pi}$ and the vector meson decay constants $f_{V}$($V=\rho,\omega$ 
and $\phi$).
The decay constant $f_{P}$ of a pseudoscalar meson $P(q_1{\bar q}_2)$
defined by 
$\la 0|{\bar q}_2\gam^\mu\gam_5q_1|P\ra=i\sqrt{2}f_{P}P^{\mu}$ can
be evaluated as follows\footnote{For convenience, we use the 
wave function of the model $C$~\cite{chung2}, i.e.,
$\phi^C(k^2)=(4/\sqrt{\pi}\beta^3)^{1/2}\exp(-k^2/2\beta^2)$.}
\be
f_{P}\label{P_decay}
= \frac{\sqrt{6}}{(2\pi)^{3/2}}\int_{0}^{1}dx\int d^{2}{\bf k}_{\perp}
\sqrt{\frac{\partial k_3}{\partial x}}
\frac{\phi^{C}(k^{2})}{\sqrt{4\pi}}
\frac{{\cal A}}{\sqrt{{\cal A}^2 + {\bf k}^2_\perp}},
\ee 
where ${\cal A}= (1-x)m_1 + xm_2$. Likewise, the decay constant
$f_V$ of a vector meson is defined by 
$\la 0|{\bar q}_2\gam^\mu q_1|V\ra=M_{V}f_{V}\vep^{\mu}$.
For the decay of $\rho^0=(u\bar{u}-d\bar{d})/\sqrt{2}$, we obtain
\be
f_\rho = I_V(m_1,m_2,\beta)(e_u-e_d)/\sqrt{2},
\ee
where\footnote{We have given the general result for unequal quark
and antiquark masses since we shall use it to calculate, e.g., 
the decay constants $f_{K^*},f_{D^*}$, and $f_{B^*}$ etc. in the
following chapters.}
\be\label{V_decay}
I_V(m_1,m_2,\beta)
=\frac{\sqrt{6}}{4\pi^2}\int_{0}^{1}dx\int d^{2}{\bf k}_{\perp}
\sqrt{\frac{\partial k_3}{\partial x}}
\frac{\phi^{C}(k^{2})}{\sqrt{{\cal A}^2 + {\bf k}^2_\perp}}
\biggl[{\cal A} + \frac{2{\bf k}^2_\perp}{M_0 + m_1 +m_2}\biggr].
\ee    
While the $\pi$ and $\rho$ decay constants($f_{\pi},f_{\rho}$) are
used to fix the parameters $m=m_{u}=m_{d}$ and
$\beta=\beta_{u\bar{u}}=\beta_{d\bar{d}}$,
the $\omega$ and $\phi$ decay constants($f_{\omega},f_{\phi}$) are
used to fix $m_{s}$ and $\beta_{s\bar{s}}$.
Since the flavor mixing of $\omega$ and $\phi$ in the quark basis
is given by~\cite{jaus}
\begin{eqnarray}\label{angle}
&&\phi=-\sin\delta_{V}(u\bar{u}+d\bar{d})/\sqrt{2}-
\cos\delta_{V}s\bar{s},
\nonumber\\
&&\omega=
\cos\delta_{V}(u\bar{u}+d\bar{d})/\sqrt{2}-\sin\delta_{V}s\bar{s},
\end{eqnarray}
where $\delta_{V}=\theta_{SU(3)}-35.26^{\circ}$, we use the vector mixing 
angle as $\delta_{V}= -3.3^{\circ}$ for the direct comparison with
the results\footnote{The $f_\phi$ and $f_\omega$ are
obtained as $f_\phi$=$-\sin\delta_{V}I_V(m_1,m_2,\beta)(e_u+e_d)/\sqrt{2}
-\cos\delta_{V}I_V(m_1,m_2,\beta)e_s$ and
$f_\omega$= $\cos\delta_{V}I_V(m_1,m_2,\beta)(e_u+e_d)/\sqrt{2}
-\sin\delta_{V}I_V(m_1,m_2,\beta)e_s$, respectively.} 
of the decay constants $f_{\omega}$ and $f_{\phi}$ in
Ref.~\cite{jaus}.
\begin{table}
\centering
\caption{Best fit quark masses and the model parameters $\beta$.
$PLJ$ and $PLH$ are the power law(PL) models for $s$=2 in Eq. (3.4)
with and without Jacobi factor, respectively. $q$=$u$ and $d$.}
\begin{tabular}{|c|c|c|c|c|}\hline
 & $H$ & $J$ & $PLH$ & $PLJ$\\
\hline
$m_{q}$[GeV] & 0.25 & 0.25 & 0.28 & 0.28\\
\hline
$m_{s}$[GeV]& 0.37 & 0.37 & 0.37 & 0.37\\
\hline
$\beta_{q\bar{q}}$[GeV]& 0.36 & 0.3194&0.40&0.307\\
\hline
$\beta_{s\bar{s}}$[GeV]& 0.38 & 0.3478 &0.415&0.328\\
\hline
\end{tabular}
\end{table}

The constituent quark masses and the model parameters
$\beta_{u\bar{u}} =\beta_{d\bar{d}}$ and $\beta_{s\bar{s}}$ 
fixed by the available experimental
data~\cite{data} for the decay constants $f_{\pi},f_{\rho},f_{\omega}$
and $f_{\phi}$ are summarized in Table 3.1 for both HO and PL models.
The PL models with and without Jacobi factor are denoted by $PLJ$ and
$PLH$, respectively. In Table 3.2, we summarize the results of the model
$H(PLH)$ for the decay constants and compare these results with those 
of the model $J(PLJ)$ and the experimental data. In order to see the 
effect of the presence-absence of the Jacobi factor given by 
Eq. (\ref{eq.39}), we present the results of the model $H(PLH)$ using 
the same parameters of the
model $J(PLJ)$, which we call $H'(PLH')$, as well as the results
of the model $H(PLH)$ using the best fit parameters.
\begin{table}
\centering
\caption{Decay constants for $\pi\to\mu\nu$ and $V\to e^{+}e^{-}$,
where the mixing angle $\delta_{V}=-3.3^{\circ}$ is used.
$H'(PLH')$ uses the parameters of $J(PLJ)$ to see the effect of the
presence-absence of the Jacobi factor.
The experimental data are taken from Ref.~\protect\cite{data}.}
\begin{tabular}{|c|c|c|c|c|c|c|c|}\hline
 $f_{P(V)}$& $H'$& $H$& $J$& $PLH'$& $PLH$ & $PLJ$& Expt.[MeV]\\
\hline
$f_{\pi}$& 86.5& 92.1 & 92.4&79.2&92.8 & 92.5& 92.4$\pm$0.25\\
\hline
$f_{\rho}$& 138.5& 156.1 & 151.9&120.1&155.5& 154.7&  152.8$\pm$
3.6\\
\hline
$f_{\omega}$& 42.0& 47.5 & 46.1&43.4&47.3& 47.1&  45.9$\pm$ 0.7\\
\hline
$f_{\phi}$ & 73.0& 80.2 & 79.7&62.3&79.4& 79.6& 79.1$\pm$ 1.3\\
\hline
\end{tabular}
\end{table}
As one can see from the comparison between $H'(PLH')$ and $J(PLJ)$ in
Table 3.2, the effect of the Jacobi factor is not negligible.
However, once the best fit parameters for the model $H(PLH)$ are chosen
as shown in Table 3.1, the numerical results of the decay constants are
almost equivalent between the models $H(PLH)$ and $J(PLJ)$
regardless of the Jacobi factor.

The quark distribution amplitude defined by
$\phi_{M,\lam}(x_i,Q)$=$\int^Q[d^2 k_{\perp}]
\Psi_{M,\lam}(x_i,{\bf k}_{i\perp},\lam_i)$,
i.e., the probability amplitude for 
finding quarks in the $L_{z}$=0 ($s$-wave) projection of the wave function
collinear up to the scale $Q$~\cite{ji,LB}, and the form factor of the pion
from the models $H'(PLH')$, $H(PLH)$ and $J(PLJ)$ are plotted
in Figs. 3.1(a)[3.1(b)] and 3.2(a)[3.2(b)], respectively.
As shown in Fig. 3.1(a)[3.1(b)], a rather substantial difference between
$\phi^{H'}_{\pi}(\phi^{PLH'}_{\pi})$ and
$\phi^{J}_{\pi}(\phi^{PLJ}_{\pi})$\footnote{Because of the presence
of the damping factor in Eqs.~(\ref{h3})-(\ref{sch3}), we extended
the integral limit to infinity without loss of accuracy.}
is shown due to the presence-absence of Jacobi factor.
Because the quark distribution
amplitude depicts just functional dependence of the model wave
function, the deviation in the quark distribution amplitude due to the
presence-absence of Jacobi factor does not diminish substantially 
even though the best fit parameters are used as one can see from the 
comparison between $\phi^{H}_{\pi}(\phi^{PLH}_{\pi})$ and
$\phi^{J}_{\pi}(\phi^{PLJ}_{\pi})$ in Fig. 3.1(a)[3.1(b)]. 
However, the deviation between model $H(PLH)$ and model $J(PLJ)$
is reduced more substantially in the physical observable such as the
form factor of pion in Fig. 3.2(a)[3.2(b)] once the best fit parameters
are used.  This observation is more pronounced in the calculation of 
decay widths as we present in the next section.

\begin{figure}
\centerline{\psfig{figure=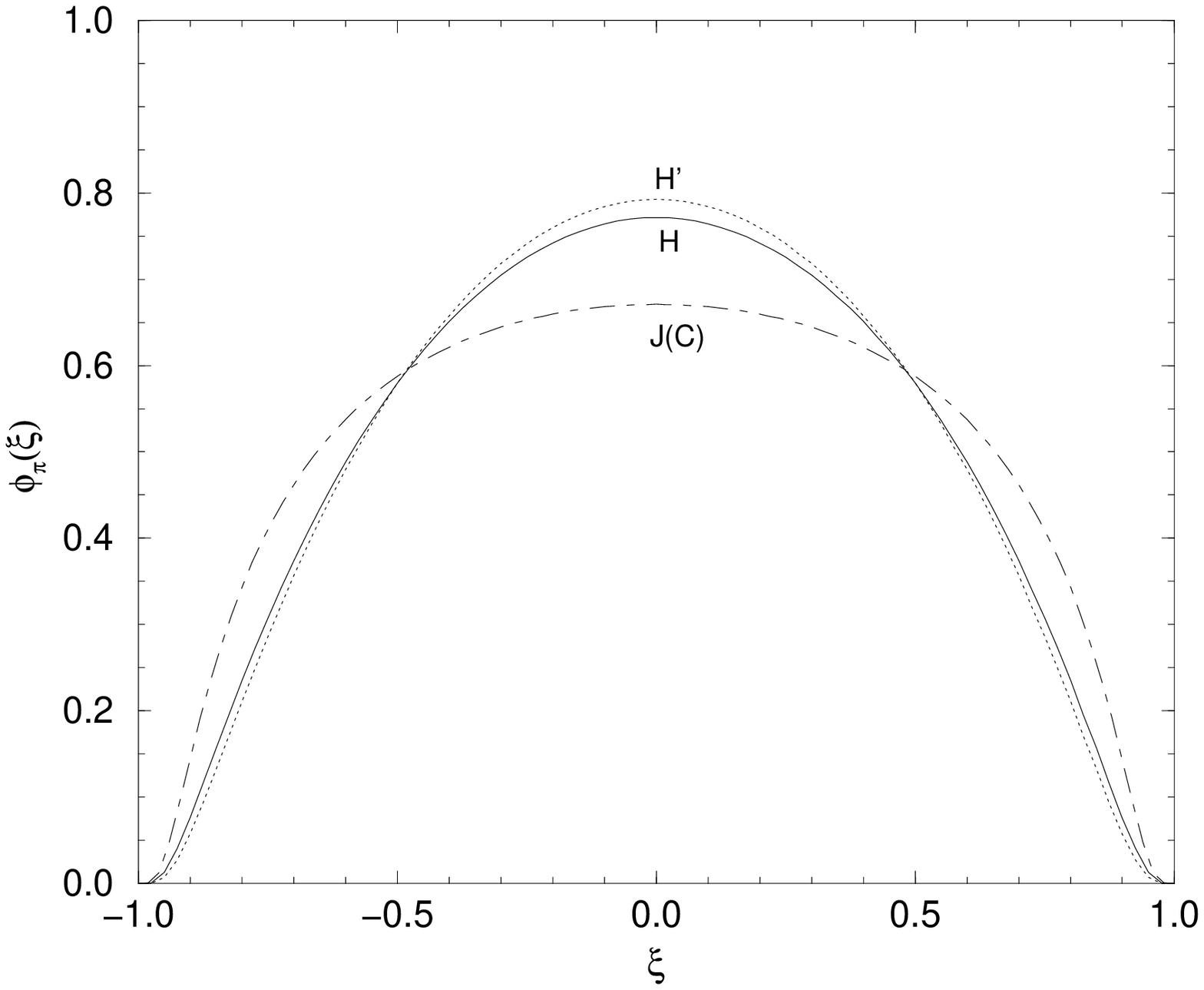,width=3.5in,height=3.3in}}
\centerline{\psfig{figure=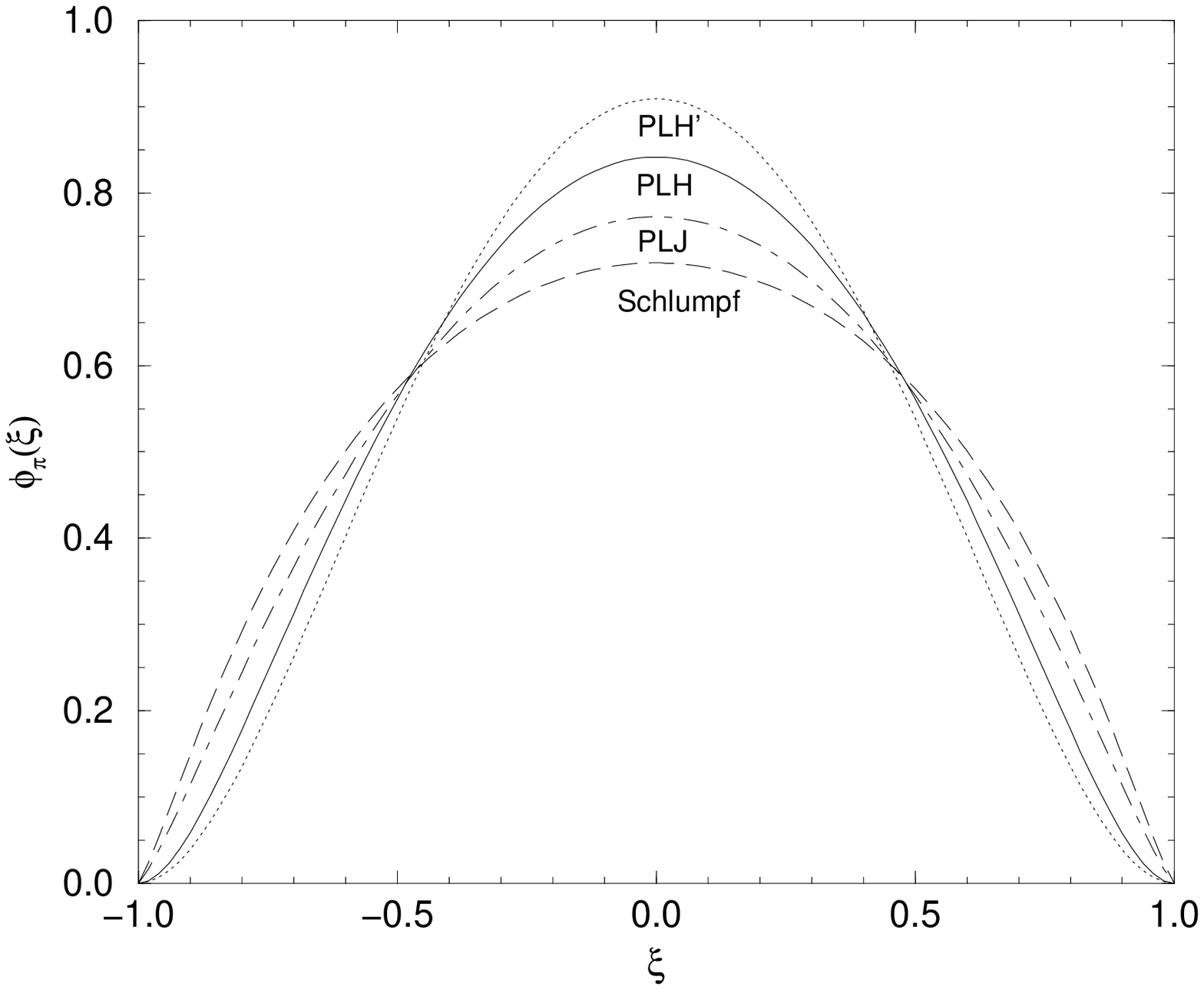,width=3.5in,height=3.3in}}
\caption{(a) The normalized quark distribution amplitude of the pion in 
HO model. The solid, dotted, dot-dashed lines represent the results of the
models $H$, $H'$, and $J$, respectively.
The normalization is fixed by the zeroth moment(the area underneath
each curve) $\la\xi^{0}\ra$=$\int^{+1}_{-1}d\xi\xi^0\phi_\pi(\xi)$=1 
with $\xi$=$x_{1}-x_{2}$. (b) The normalized
quark distribution amplitude of the pion in PL model.
The solid, dotted, dot-dashed lines represent the results of the
models $PLH$, $PLH'$, and $PLJ$, respectively.
The long-dashed line represents the Schlumpf's result~\protect\cite{sch}.}
\end{figure}
\begin{figure}
\centerline{\psfig{figure=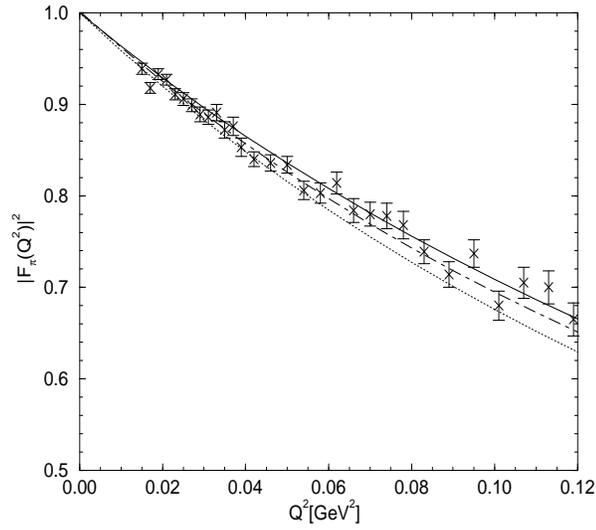,width=3.5in,height=3.2in}}
\centerline{\psfig{figure=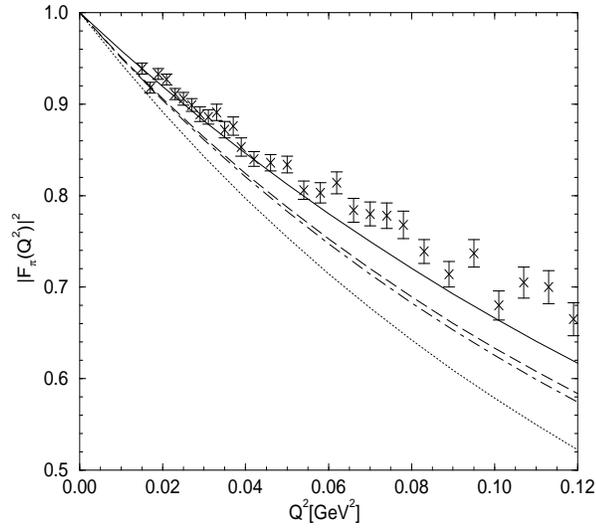,width=3.5in,height=3.2in}}
\caption{(a) The square of the pion charge form factor in HO model for
low values of $Q^{2}$ compared with the experimental
data~\protect\cite{amendolia}. 
The same line code as in Fig. 3.1(a) is used. (b)
The square of the pion charge form factor in PL model.
The same line code as in Fig. 3.1(b) is used.}
\end{figure} 

\section{ Decay Widths for $P\to\gamma\gamma$ and
$V(P)\to P(V)\gamma$ Transitions}
Applying these models to calculate the decay widths for
$P\to\gamma\gamma$($P=\pi,\eta,\eta'$) and
$V(P)\to P(V)\gamma$($V=\rho,\omega,\phi$) transitions, the decay width for
$P\to\gamma\gamma$ is given by
\begin{eqnarray}
\Gamma_{P\to\gamma\gamma}=
\frac{\pi}{4}\alpha^{2}g_{P\gamma\gamma}^{2}m^{3}_{P},
\end{eqnarray}
where the coupling constants $g_{P\gamma\gamma}$ are expressed in terms of
three pseudoscalar decay constants obtained by the axial-vector anomaly plus
PCAC (partial conservation of axial-vector
current)~\cite{jaus,adler,gasser,qft,john}:
\begin{eqnarray}\label{pcac}
&&g_{\pi\gamma\gamma}= \frac{1}{4\pi^{2}f_{\pi}},\nonumber\\
&&g_{\eta\gamma\gamma}= \frac{1}{4\pi^{2}\sqrt{3}}\biggr[
\frac{1}{f_{8}}\cos\theta_{SU(3)} -
\frac{2\sqrt{2}}{f_{0}}\sin\theta_{SU(3)}\biggl],
\\
&&g_{\eta'\gamma\gamma}= \frac{1}{4\pi^{2}\sqrt{3}}\biggr[
\frac{1}{f_{8}}\sin\theta_{SU(3)}
+ \frac{2\sqrt{2}}{f_{0}}\cos\theta_{SU(3)}\biggl].
\nonumber
\end{eqnarray}
Here, the predictions of the decay constants $f_{8}$ and $f_{0}$ of
the model $H(PLH)$ are given by
\be\label{f08}
f_{8}/f_{\pi}= 1.148(1.096),\hspace{.2cm}
f_{0}/f_{\pi}= 1.074(1.048).
\ee
These results are very close to the predictions of the model
$J(PLJ)$~\cite{jaus} which are given by $f_{8}/f_{\pi}= 1.156(1.094)$
and $f_{0}/f_{\pi}=1.078(1.047)$.
The predictions of the chiral perturbation theory in Ref.~\cite{john} has
been reported as $f_{8}/f_{\pi}=1.25$ and $f_{0}/f_{\pi}=1.04\pm 0.04$.
We have used the $\eta$-$\eta'$ mixing (see Eqs. (2.6)-(2.7))
analogous to $\omega$-$\phi$ mixing
in Eq. (\ref{angle}) and observed that $\theta_{SU(3)}\approx -19^{\circ}$ 
gives a good agreement with the experimental data of decay rates.
The decay rate and the transition form factors for 
$P\to\gamma^{*}\gamma$ ($P$=$\pi,\eta$, and $\eta'$) are given 
by Eqs. (2.46) and (2.49), respectively. However, the integral formula
$I_{2}(m_M,m_q,\beta)$ in Eq. (2.50) should be changed into the following
form:
\be
I_{P\gamma^{*}\gamma}(m_q,\beta)
=\frac{\sqrt{6}}{4\pi^2}\int^{1}_{0} dx\int d^{2}k_\perp
\sqrt{\frac{\partial k_3}{\partial x}}
\frac{\phi^{C}(k^2)}{\sqrt{m^2_q + {\bf k}^{2}_{\perp}}}
\frac{(1-x)m_q}{{\bf k'}^{2}_{\perp} + m^2_q},
\ee 
where ${\bf k'}_\perp={\bf k}_\perp + (1-x){\bf q}_\perp$.  
For $\theta_{SU(3)}= -19^{\circ}$, the numerical results of the decay
widths are summarized in Table 3.3. As shown in Table 3.3,
the predictions of decay widths for $P\to\gamma\gamma$
are not much dependent on the models once the best fit parameters are 
obtained and the model predictions with $\theta_{SU(3)}= -19^{\circ}$
are in excellent agreement with the experimental data.
\begin{table}
\centering
\caption{Radiative decay widths $\Gamma(P\to\gamma\gamma)$ for
$\eta$-$\eta'$ mixing angle $\theta_{SU(3)}=-19^{\circ}$.
The results are obtained from Eq.~(\ref{pcac}).}
\begin{tabular}{|c|c|c|c|c|c|}\hline
Widths & $H$ & $J$& $PLH$& $PLJ$& Experiment\\
\hline
$\Gamma(\pi\to\gamma\gamma)$ & 7.79&7.73&7.67 &7.72&$7.8\pm 0.5$[eV]\\
\hline
$\Gamma(\eta\to\gamma\gamma)$ & 0.49&0.485&0.52 &0.52&$0.47\pm0.0 5$[keV]\\
\hline
$\Gamma(\eta'\to\gamma\gamma)$ & 4.51& 4.45&4.64&4.68&$4.3\pm 0.6$[keV] \\
\hline
\end{tabular}
\end{table}

To further justify our observation on the Jacobi factor
and the model predictions of $\eta$-$\eta'$ mixing angle, we have
calculated also the decay widths for $V(P)\to P(V)\gamma$. 
We can also use the same formula for the decay width given by 
Eq. (2.27) but the integral formula $I_{PV\gam^*}(m_M,m_q,\beta)$
in Eq. (2.30) should be changed into the following form\footnote{
Even though we considered the equal quark and antiquark masses in this
work, nevertheless, we have given the general results for the unequal 
quark and antiquark masses since we shall use it to calculate, e.g., the 
decay widths for $K^{*0}\to K^{0}\gam$ and 
$K^{*+}\to K^{+}\gam$ in the next chapter.}:
\bea
I_{PV\gam^*}(m_q,\beta)
&=&\frac{1}{2\pi}\int^{1}_{0}\int d^{2}k_{\perp}
\biggl(\frac{\partial k_3}{\partial x}\biggr)
\frac{|\phi^{C}(k^2)|^2}{x[M^{2}_{0}-(m_1-m_2)^2]} \nonumber\\
&\times&
\biggl[{\cal A} +\frac{ {\bf k}^{2}_{\perp}}{M_0 + m_1 + m_2}\biggr].
\eea 
The results are summarized in Table 3.4, where the $\omega$-$\phi$ mixing
angle $\delta_{V}=-3.3^{\circ}$ is also taken into account. 
Again, the predictions of the
model $H(PLH)$ with $\theta_{SU(3)}=-19^{\circ}$ are similar to
those of the model $J(PLJ)$.
Thus, we confirm again that the numerical results of the model $H(PLH)$
and the model $J(PLJ)$ are not much different from each other regardless
of the Jacobi factor once the best fit parameters are used.
However, the difference in the choice of radial wave function(e.g.
HO wave function versus PL wave function) is still appreciable even though
the best fit parameters are used.
The experimental data of decay widths both for $P\to\gamma\gamma$ and
$V(P)\to P(V)\gamma$ are quite consistent with the HO model
predictions for $\theta_{SU(3)}\approx -19^{\circ}$.
\begin{table}
\centering
\caption{Radiative decay widths for $V(P)\to P(V)\gamma$
transitions. We use $\eta$-$\eta'$ and $\omega$-$\phi$ mixing angles
as $\theta_{SU(3)}=-19^{\circ}$ and $\delta_{V}=-3.3^{\circ}$,
respectively.}
\label{values}
\begin{tabular}{|c|c|c|c|c|c|}\hline
Widths& $H$ & $J$& $PLH$ & $PLJ$ & Experiment[keV]\\
\hline
$\Gamma(\rho^{\pm}\to\pi^{\pm}\gamma)$ & 75 & 76&89 & 97& $68\pm 8$ \\
\hline
$\Gamma(\omega\to\pi\gamma)$ & 712 &730& 855& 938 & $717\pm 51$ \\
\hline
$\Gamma(\phi\to\pi\gamma)$ &  5.5 & 5.6&6.6& 7.2 &  $5.8\pm 0.6$ \\
\hline
$\Gamma(\rho\to\eta\gamma)$ & 59 & 59&69& 76 &  $58\pm 10$ \\
\hline
$\Gamma(\omega\to\eta\gamma)$ & 8.6 & 8.7& 10.2& 11.1 & $7.0\pm 1.8$ \\
\hline
$\Gamma(\phi\to\eta\gamma)$ & 55.9 & 55.3& 72.4& 74.2&  $56.9\pm 2.9$\\
\hline
$\Gamma(\eta'\to\rho\gamma)$ & 66.1 & 67.5&79.1&86.8 & 61 $\pm8 $ \\
\hline
$\Gamma(\eta'\to\omega\gamma)$ & 4.7 & 4.8& 5.5& 6.1 & $5.9\pm 0.9$ \\
\hline
$\Gamma(\phi\to\eta'\gamma)$ & 0.56 & 0.57& 0.71& 0.76 &  $< 1.8$\\
\hline
\end{tabular}
\end{table}
\section{Summary and Discussion}
In this work, we have first shown that the HO models $C$ and $J$ are actually
equivalent, while the model $H$ is not exactly same with the other models
$C$ and $J$ due to the Jacobi factor $\sqrt{M_{0}/4x(1-x)}$.
The effect of the presence-absence of the Jacobi factor is in principle
not negligible as we have shown in Table 3.2 and Figs. 3.1 and
3.2 [comparison between $H'(PLH')$ and $J(PLJ)$].
However, once the best fit parameters are used, the numerical results of the
physical observables from the model $H(PLH)$ are almost equivalent to those 
of the model $J(PLJ)$ regardless of the presence-absence of the Jacobi 
factor as shown in the calculations of the decay 
constants($f_{\pi},f_{\rho},f_{\omega}$, and
$f_{\phi}$), the pion form factor and the decay rates of
$V(P)\to P(V)\gamma$ and $P\to\gamma\gamma$ transitions.

In the case of best fit, the effect of the presence-absence of the Jacobi 
factor amounts to the difference in the best fit parameters as shown in 
Table 3.1.
Thus, the effect from the Jacobi factor is there no matter what we do. 
However, what we have shown in this work indicates that it is rather 
difficult to pin down a better model in the present phenomenology if the 
two models (e.g., $H$ and $J$) are differ only by the Jacobi factor. 
Nevertheless, the difference
between the HO and PL results are substantial enough to state that the 
overall agreement with the data of various radiative decay widths 
(see Table 3.4) is clearly better in the HO models than in the PL models.
We also found an excellent agreement of all three HO model($H,C$ and $J$)
predictions on the $P\to\gamma\gamma$ and $V(P)\to P(V)\gamma$ processes 
with the experimental data, if we use Eq. (\ref{pcac}) in two photon
decay processes obtained by the axial anomaly and PCAC relations and choose
the $\eta-\eta'$ mixing angle $\theta_{SU(3)}\approx -19^{\circ}$.


%% file: Mixing.tex
\newpage
\setcounter{equation}{0}
\setcounter{figure}{0}
\renewcommand{\theequation}{\mbox{4.\arabic{equation}}}
\chapter{Mixing Angles and Electromagnetic Properties of Ground
State Pseudoscalar and Vector Meson Nonets in LFQM}
It has been realized that relativistic effects are crucial to describe
the low-lying hadrons made of $u$, $d$, and $s$ quarks and
antiquarks~\cite{isgur2}. The LFQM takes advantage of the equal LF 
time ($\tau$=$t$+$z/c$) quantization and includes important relativistic
effects in the hadronic wave functions.
The distinct features of the LF equal-$\tau$ quantization
compared to the ordinary equal-$t$ quantization may be summarized as
the suppression of vacuum fluctuations with the decoupling of complicated
zero modes and the conversion of the dynamical problem from boost to
rotation. Taking advantage of this LF approach, we have shown in 
Chapters 2 and 3 that once the best fit parameters were used,
both SM and IM schemes in the LFQM provided a remarkably good agreement 
with the available experimental data for form factors,
decay constants and charge radii etc. of various light 
pseudoscalar and vector mesons as well as their radiative decay widths.

However, the radial function so far has been taken as a model wave
function rather than as a solution of the QCD-motivated dynamical
equation. Because of this, we could not calculate the mass spectra
of mesons directly from the model but assume a couple of schemes
such as SM or IM schemes.  
Even though the authors of Ref.~\cite{card} adopted the quark
potential model developed by Godfrey and Isgur~\cite{isgur2} to
reproduce the meson mass spectra, their model predictions included
neither the mixing angles of $\omega$-$\phi$ and $\eta$-$\eta'$ nor the
form factors for various radiative decay processes of pseudoscalar
and vector mesons.

In this work, we are not taking exactly the same quark potential
developed by Godfrey and Isgur~\cite{isgur2}. However, we attempt to
fill this gap between the model wave function and the QCD-motivated
potential, which includes not only the Coulomb plus confining potential
but also the hyperfine interaction, to obtain the correct $\rho$-$\pi$
splitting. For the confining potential, we take a (1) harmonic
oscillator (HO) potential and (2) linear potential and compare the numerical
results for these two cases. We use the variational principle to solve the
equation of motion. Accordingly, our analysis covers the mass spectra
of light pseudoscalar ($\pi,K,\eta,\eta'$) and
vector ($\rho,K^{*},\omega,\phi$) mesons and the mixing angles of
$\omega$-$\phi$ and $\eta$-$\eta'$ as well as other observables such as
charge radii, decay constants, radiative decay widths, etc.
We exploit the invariant meson mass scheme in this model.
We also adopt the parametrization to
incorporate the quark-annihilation diagrams~\cite{isgur1,georgi,scadron}
mediated by gluon exchanges and the SU(3) symmetry
breaking, i.e., $m_{u(d)}\neq m_{s}$, in the determination of
meson mixing angles.

This Chapter is organized as follows:
In Section 4.1, we set up a simple QCD motivated effective Hamiltonian
and use the Gaussian radial wave function as a trial function of the
variational principle. We find the optimum values of the model
parameters, quark masses ($m_{u(d)},m_{s}$) and Gaussian parameters (
$\beta_{u\bar{u}}$=$\beta_{u\bar{d}}$=$\beta_{d\bar{d}},
\beta_{u\bar{s}},\beta_{s\bar{s}}$) for the two cases of confining
potentials (1) and (2). We also analyze the meson mass spectra and
predict the mixing angles of $\omega$-$\phi$ and $\eta$-$\eta'$.
We adopt a formulation to incorporate the quark-annihilation
diagrams and the effect of SU(3) symmetry breaking on the meson
mixing angles. In Section 4.2, we calculate the decay constants, charge 
radii, form factors, and  radiative decay rates of various light pseudoscalar
and vector mesons and discuss the numerical results of the two confining
potentials (1) and (2) in comparison with the available
experimental data. A summary and discussions follow in Section 4.3.
The details of fixing the model parameters and the mixing angle
formulations are presented in Appendices D and E, respectively.

\section{Model Description}
The QCD-motivated effective Hamiltonian for a description of the meson
mass spectra is given by~\cite{card,isgur2}
\begin{eqnarray}\label{hamil}
H_{q\bar{q}}|\Psi^{SS_{z}}_{nlm}\rangle
&=& \biggl[\sqrt{m_{q}^{2}+k^{2}} + \sqrt{m_{\bar{q}}^{2}+k^{2}}
+ V_{q\bar{q}}\biggr]|\Psi^{SS_{z}}_{nlm}\rangle,\nonumber\\
&=& \biggl[H_{0} + V_{q\bar{q}}\biggr]|\Psi^{SS_{z}}_{nlm}\rangle =
M_{q\bar{q}}|\Psi^{SS_{z}}_{nlm}\rangle,
\end{eqnarray}
where $M_{q\bar{q}}$ is the mass of the meson,
$k^{2}={\bf k}^{2}_{\perp}+k^{2}_{n}$, and
$|\Psi^{SS_{z}}_{nlm}\rangle$ is the meson
wave function given in Eq. (1.5)

In this work, we use the two interaction potentials $V_{q\bar{q}}$ for 
the pseudoscalar $(0^{-+})$ and vector ($1^{--}$) mesons:(1)
Coulomb plus HO, and (2) Coulomb plus linear
confining potentials. In addition, the hyperfine interaction,
which is essential to distinguish vector from pseudoscalar mesons,
is included for both cases, viz.,
\begin{eqnarray}\label{potent}
V_{q\bar{q}}= V_{0}(r) + V_{\rm hyp}(r)
= a + {\cal V}_{\rm conf} - \frac{4\kappa}{3r}
+ \frac{2\vec{S}_{q}\cdot\vec{S}_{\bar{q}}}
{3m_{q}m_{\bar{q}}}\nabla^{2}V_{\rm Coul},
\end{eqnarray}
where ${\cal V}_{\rm conf}= br (r^{2})$ for the linear (HO)
potential and $\la\vec{S}_{q}\cdot\vec{S}_{\bar{q}}\ra$=1/4 $(-3/4)$
for the vector (pseudoscalar) meson.
Even though more realistic solution
of Eq.~(\ref{hamil}) can be obtained by expanding the radial function
$\phi_{n,l=0}(k^{2})$ onto a truncated set of HO basis
states~\cite{card,isgur2}, i.e.,
$\sum_{n=1}^{n_{\rm max}}c_{n}\phi_{n,0}(k^{2})$, our intention
in this work is to explore only the $0^{-+}$ and $1^{--}$ ground state meson
properties. Therefore, we use the $1S$ state harmonic wave function
$\phi_{10}(k^{2})$ as a trial function of the variational principle
\be\label{radial}
\phi_{10}(x,{\bf k}_{\perp})=
\biggl(\frac{1}{\pi^{3/2}\beta^{3}}\biggr)^{1/2}
\exp(-k^{2}/2\beta^{2}),
\ee
where $\phi(x,{\bf k}_{\perp})$ is normalized according to
\begin{eqnarray}
&&\sum_{\nu\bar{\nu}}\int^{1}_{0}dx\int d^{2}{\bf k}_{\perp}
|\Psi^{SS_{z}}_{100}(x,{\bf k}_{\perp},\nu\bar{\nu})|^{2}
\nonumber\\
&&\;\;\;=\int^{1}_{0}dx\int d^{2}{\bf k}_{\perp}
\biggl(\frac{\partial k_{n}}{\partial x}\biggr)
|\phi_{10}(x,{\bf k}_{\perp})|^{2}=1.
\end{eqnarray}
Because of this rather simple trial function, our results could be
regarded as crude approximations. However, we note that this choice
is consistent with the LFQM wave
function which has been quite successful in describing various meson
properties~\cite{jaus,chung2,IM,dziem1,ji,spin}.
Furthermore, Eq.~(\ref{radial}) takes the same form as the ground
state solution of the HO potential even though
it is not the exact solution for the linear potential case.
As we show in Appendix D, after fixing the parameters $a$, $b$, and
$\kappa$, the Coulomb plus HO potential
$V_{0}(r)$ in Eq.~(\ref{potent}) turns out to be very similar in the relevant
range of potential ($r\leq 2$ fm) to the Coulomb plus linear confining
potentials [see Figs. 4.1(a) and 4.1(b)] which are frequently used in the
literature~\cite{card,isgur2,lucha,Karl,gromes,isgw,isgw2}.
The details of fixing the parameters of our model, i.e., quark
masses ($m_{u(d)},m_{s}$), Gaussian parameters ($\beta_{u\bar{d}},
\beta_{u\bar{s}},\beta_{s\bar{s}}$), and potential parameters
($a,b,\kappa$) in $V_{q\bar{q}}$ given by Eq.~(\ref{potent}), 
are summarized in Appendix D.
\begin{figure}
\centerline{\psfig{figure=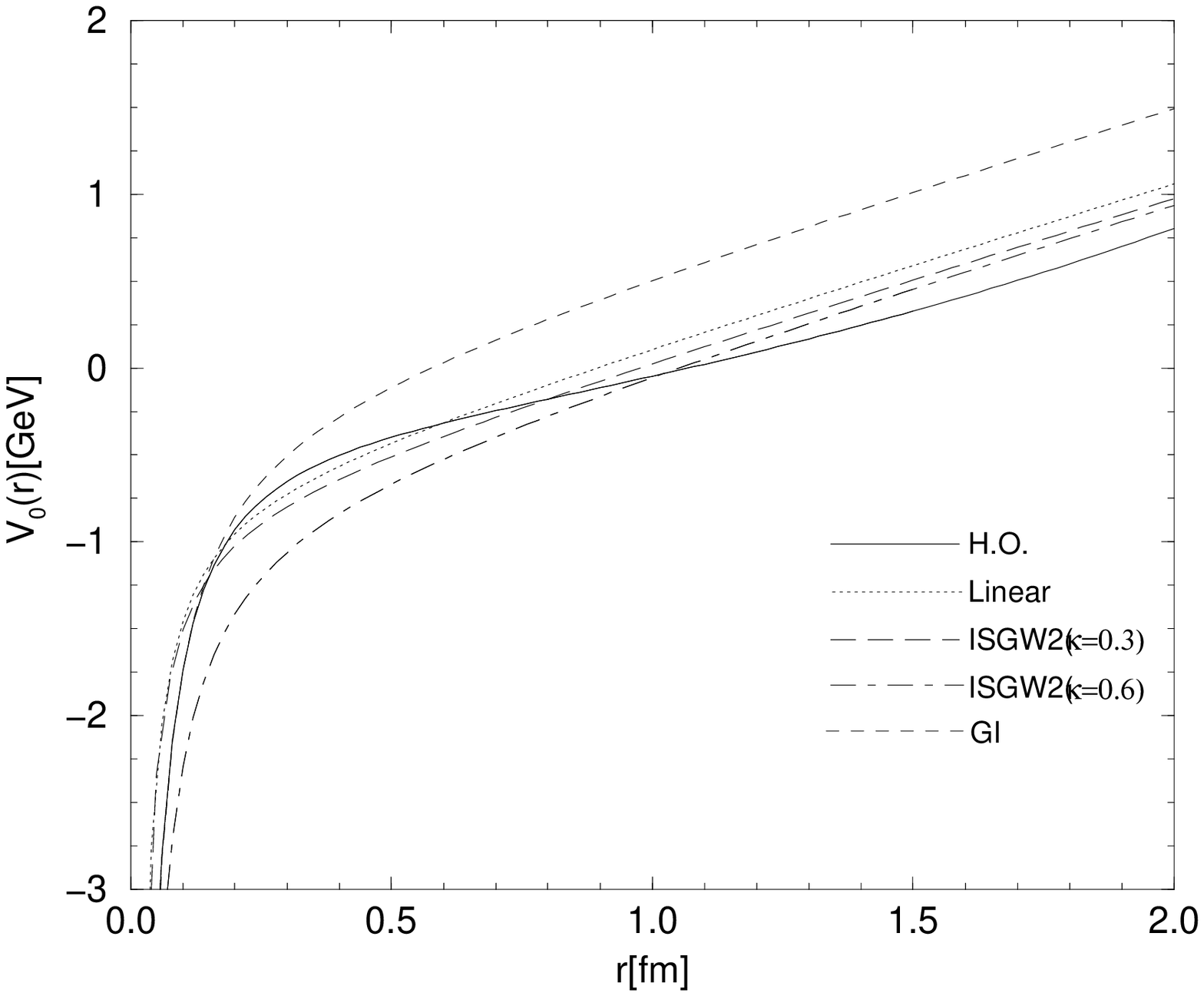,width=3.5in,height=3.2in}}
\centerline{\psfig{figure=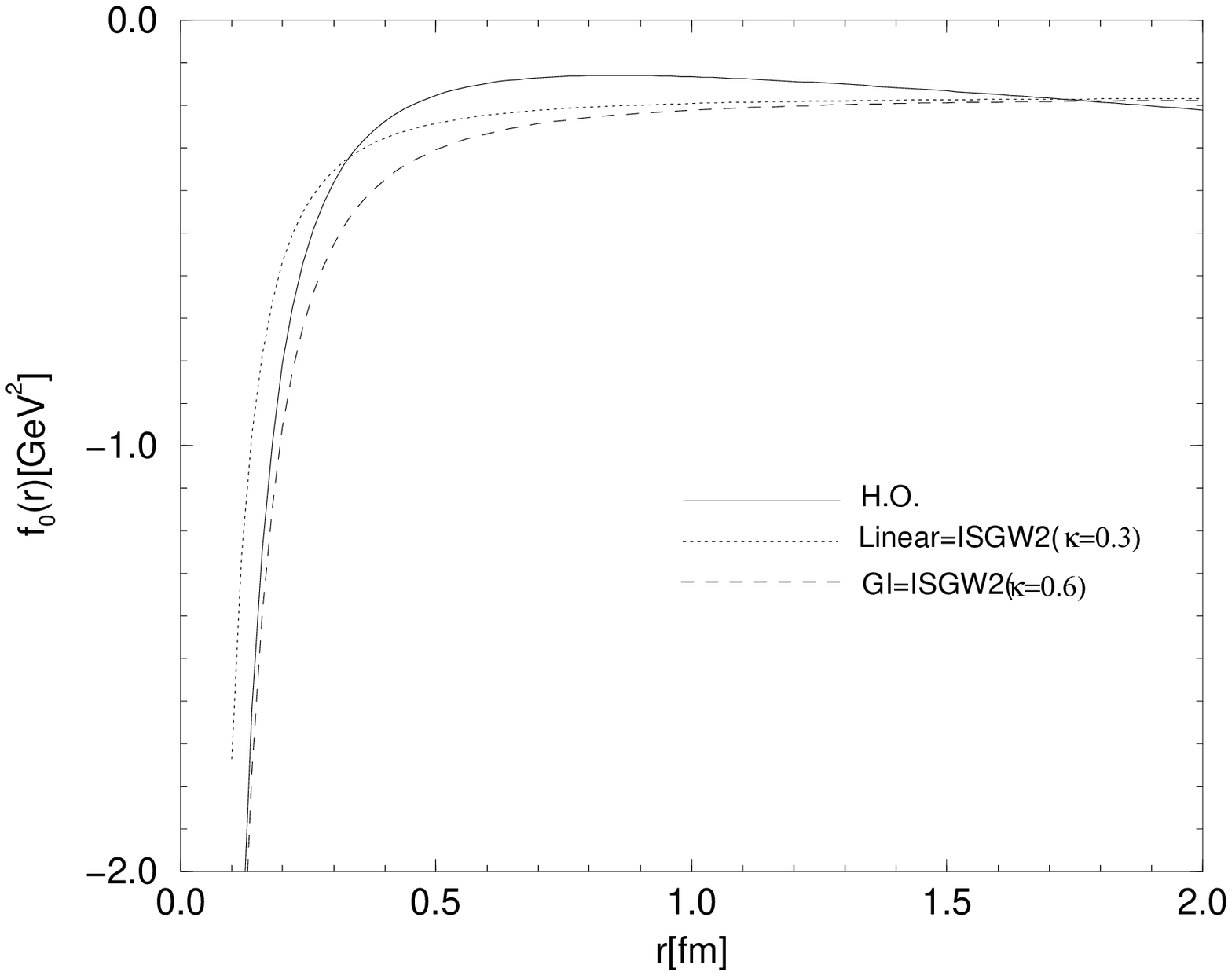,width=3.5in,height=3.2in}}
\caption{
(a) The central potential $V_{0}(r)$ versus $r$.
Our Coulomb plus HO (solid line) and linear (dotted line)
potentials are compared with the quasi-relativistic
potential of ISGW2 model with $\kappa=0.3$ (long-dashed line)
and $\kappa=0.6$ (dot-dashed line) and the relativized
potential of GI model (short-dashed line).
(b) The central force $f_{0}(r)$ versus $r$.
Our force for the linear potential is the same
as that of ISGW2~\protect\cite{isgw2} with $\kappa=0.3$ (dotted lines).
The forces of GI~\protect\cite{isgur2} and ISGW2 with $\kappa=0.6$ are the 
same as each other (dashed lines). Our force for the HO 
potential (solid line) is quite comparable with the other four forces
up to the range of $r \leq 2$ fm.}
\end{figure}

Following the procedure listed in Appendix D, our optimized model
parameters are given in Table 4.1. In fixing all of these parameters,
the variational principle [Eq.~(\ref{variation})] plays the crucial role
for $u\bar{d}$, $u\bar{s}$, and $s\bar{s}$ meson systems to share
the same potential parameters $(a,b,\kappa)$ regardless of their
quark-antiquark contents [see Figs. 4.2(a) and 4.2(b)].
\begin{table}
\centering
\caption{Optimized quark masses $(m_{q},m_{s})$ (in unit of GeV)
and the Gaussian parameters $\beta$ (in unit of GeV) for both harmonic
oscillator and linear potentials obtained from the variational principle.
The values in parentheses are results from the smearing
function in Eq.~(\ref{smear}) instead of the contact term.  
$q$=$u$ and $d$.}
\begin{tabular}{|c|c|c|c|c|c|}\hline
 Potential& $m_{q}$ & $m_{s}$ & $\beta_{q\bar{q}}$ &
$\beta_{s\bar{s}}$& $\beta_{q\bar{s}}$\\
\hline
HO& 0.25 & 0.48 & 0.3194 & 0.3681 (0.3703) & 0.3419 (0.3428)\\
\hline
Linear&0.22 & 0.45 & 0.3659 & 0.4128 (0.4132)&0.3886 (0.3887) \\
\hline
\end{tabular}
\end{table}
\begin{figure}
\centerline{\psfig{figure=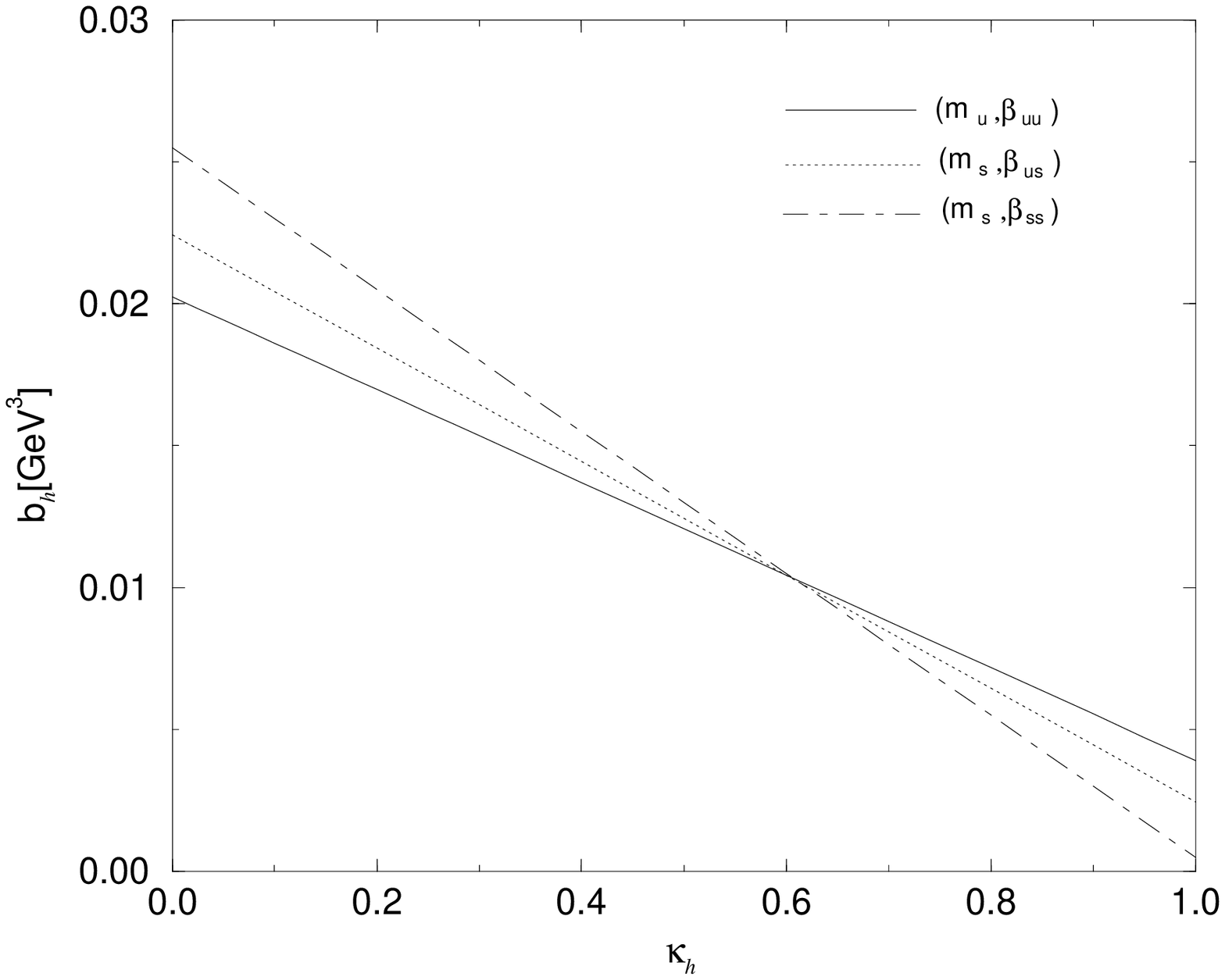,width=3.5in,height=3.2in}}
\centerline{\psfig{figure=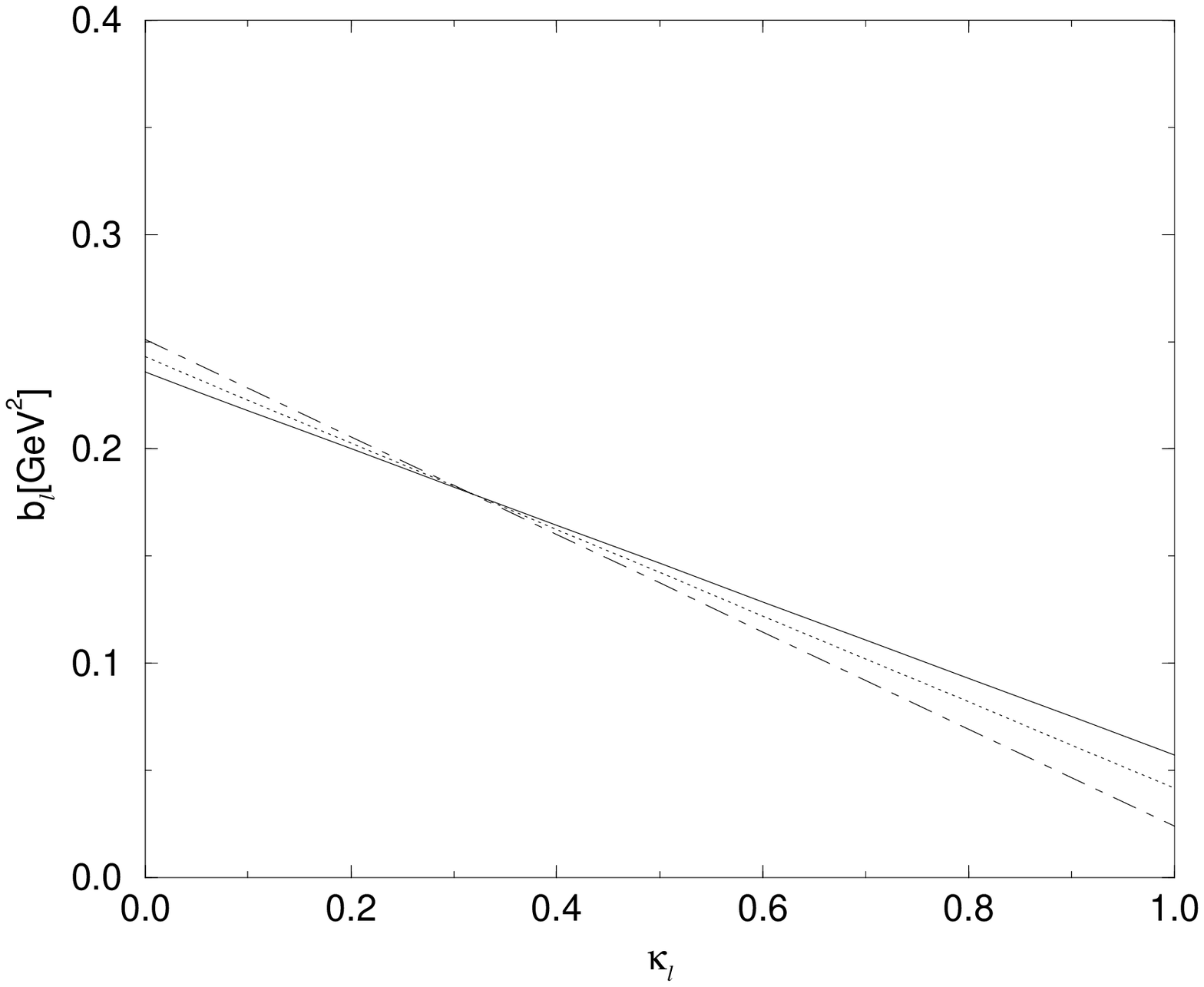,width=3.5in,height=3.2in}}
\caption{ (a) The parameters $m_q$, $m_s$, $\beta_{qq}$, 
$\beta_{qs}$, and $\beta_{ss}$ satisfying the variational principle
given by Eq. (D.2). The solid, dotted, and dot-dashed lines
are fixed by the sets of ($m_{u},\beta_{u\bar{u}}$),
($m_{s},\beta_{u\bar{s}}$), and ($m_{s},\beta_{s\bar{s}}$),
respectively. (b) The parameters $m_q$, $m_s$, $\beta_{qq}$, 
$\beta_{qs}$, and $\beta_{ss}$ satisfying the variational principle given
by Eq. (D.3). The same line codes are used as in (a).}
\end{figure}

We also determine the mixing angles from the mass spectra of
$(\omega,\phi)$ and $(\eta,\eta')$.
Identifying $(f_{1},f_{2})$=$(\phi,\omega)$ and $(\eta,\eta')$
for vector and pseudoscalar nonets, the physical meson states
$f_{1}$ and $f_{2}$ are given by
\begin{eqnarray}\label{f1f2}
&&|f_{1}\rangle = -\sin\delta|n\bar{n}\rangle -\cos\delta|s\bar{s}\rangle,
\nonumber\\
&&|f_{2}\rangle = \cos\delta|n\bar{n}\rangle -\sin\delta|s\bar{s}\rangle,
\end{eqnarray}
where $|n\bar{n}\rangle\equiv 1/\sqrt{2}|u\bar{u} + d\bar{d}\rangle$ and
$\delta=\theta_{SU(3)}-35.26^{\circ}$ is the mixing angle.
Taking into account SU(3) symmetry breaking and using the
parametrization for the (mass)$^{2}$ matrix
suggested by Scadron~\cite{scadron}, we obtain
\begin{eqnarray}\label{tan2}
\tan^{2}\delta = \frac{(M^{2}_{f_{2}} - M^{2}_{n\bar{n}})
(M^{2}_{s\bar{s}} - M^{2}_{f_{1}})}{ (M^{2}_{f_{2}}-M^{2}_{s\bar{s}})
(M^{2}_{f_{1}}-M^{2}_{n\bar{n}})},
\end{eqnarray}
which is the model-independent equation for any meson $q\bar{q}$ nonets.
The details of obtaining meson mixing angles using quark-annihilation
diagrams are summarized in Appendix E.
In order to predict the $\omega$-$\phi$ and $\eta$-$\eta'$ mixing angles,
we use the physical masses~\cite{data} of
$M_{f_{1}}=(m_{\phi},m_{\eta})$ and $M_{f_{2}}=(m_{\omega},m_{\eta'})$
as well as the masses of $M^{V}_{s\bar{s}}$= 996 (952) MeV and
$M^{P}_{s\bar{s}}$= 732 (734) MeV obtained from the expectation
value of $H_{s\bar{s}}$ in Eq.~(\ref{hamil}) for the
HO (linear) potential case (see Appendix D for more details).
Our predictions for $\omega$-$\phi$ and $\eta$-$\eta'$ mixing angles
for the HO (linear) potential are
$|\delta_{V}|\approx 4.2^{\circ} (7.8^{\circ})$ and
$\theta_{SU(3)}\approx -19.3^{\circ} (-19.6.^{\circ})$, respectively.
The mass spectra of light pseudoscalar and vector mesons
used are summarized in Table 4.2. 
Since the signs of $\delta_{V}$
for $\omega$-$\phi$ mixing are not yet
definite~\cite{isgur1,georgi,scadron,Das,Sakurai,Coleman}
in the analysis of the quark-annihilation diagram (see Appendix E),
we will keep both signs of $\delta_{V}$ when we compare various physical
observables in the next section.
\begin{table}
\centering
\caption{Fit of the ground state meson masses with
the parameters given in Table 4.1. Underlined masses are input data.
The masses of $(\eta,\eta')$ and $(\omega,\phi)$ were used to determine
the mixing angles of $\eta$-$\eta'$ and $\omega$-$\phi$, respectively.
The values in parentheses are results from the smearing function
in Eq.~(\ref{smear}) instead of the contact term.}
{\small
\begin{tabular}{|c|c|c|c|c|c|c|c|c|c|}\hline
$^{1}S_{0}$ & Expt.[MeV] & HO& Linear&
$^{3}S_{1}$ & Expt. & HO& Linear\\
\hline
$\pi$ & 135 & \underline {135}&\underline{135}
&$\rho$ & 770 & \underline {770}& \underline{770}\\
\hline
$K$ & 494 & 470 (469)& 478 (478) & $K^{*}$
& 892 & 875 (875)& 850 (850)\\
\hline
$\eta$ & 547 & \underline{547}& \underline{547}
& $\omega$ & 782 & \underline{782}& \underline{782} \\
\hline
$\eta'$ & 958 & \underline{958}& \underline{958}
& $\phi$ & 1020 & \underline{1020} &\underline{1020} \\
\hline
\end{tabular}
}
\end{table}

\section{Application}
In this section, we now use the optimum model parameters presented in
the previous section and calculate various physical observables:
(1) decay constants of light pseudoscalar and vector mesons,
(2) charge radii of pion and kaon, (3) form factors of neutral and
charged kaons, and (4) radiative decay widths for the
$V(P)\to P(V)\gamma$ and $P\to\gamma\gamma$ transitions.
These observables are calculated for the two potentials (HO and linear)
to gauge the sensitivity of our results.

Our calculation is carried out using the standard LF frame (
$q^{+}=0$) with ${\bf q}_{\perp}^{2}=Q^{2}=-q^{2}$.
We think that this is a distinct advantage in the LFQM 
because various form factor formulations are well established
in the LF quantization method using this well-known
Drell-Yan-West frame ($q^{+}=0$).
The charge form factor of the pseudoscalar meson can be expressed
for the ``+" component of the current $J^{\mu}$ as follows:
\begin{eqnarray}
F(Q^{2})&=& e_{q}I(Q^{2},m_{q},m_{\bar{q}}) +
e_{\bar{q}}I(Q^{2},m_{\bar{q}},m_{q}),
\end{eqnarray} 
where $e_{q} (e_{\bar{q}})$ is the charge of quark (antiquark) and
\bea\label{psform}
I(Q^{2},m_{q},m_{\bar{q}})
&=&\int^{1}_{0}dx\int d^{2}{\bf k}_{\perp}
\sqrt{\frac{\partial k_{n}}{\partial x}} 
\sqrt{\frac{\partial k'_{n}}{\partial x}}
\phi(x,{\bf k}_{\perp})\phi^{*}(x,{\bf k'}_{\perp})
\nonumber\\
&\times&\frac{{\cal A}^{2} + {\bf k}_{\perp}\cdot{\bf k'}_{\perp}}
{ \sqrt{{\cal A}^{2} + {\bf k}^{2}_{\perp}}
\sqrt{{\cal A}^{2} + {\bf k'}^{2}_{\perp}} },
\eea
with the definition of ${\cal A}$ and ${\bf k'}_{\perp}$ given by
\be
{\cal A}= xm_{\bar{q}} + (1-x)m_{q},\;\;
{\bf k'}_{\perp}={\bf k}_{\perp} + (1-x){\bf q}_{\perp}.
\ee
The charge radius of the meson can be calculated by
$r^{2}$=$-6dF(Q^{2})/dQ^{2}|_{Q^{2}=0}$.
Since all other formulas for the physical observables such as 
pseudoscalar and vector meson decay constants, decay rates for the
$V(P)\to P(V)\gamma$ and $P\to\gamma\gamma$ transitions, have already
been given in Chapter 3, we do not list them here again.
\begin{figure}
\centerline{\psfig{figure=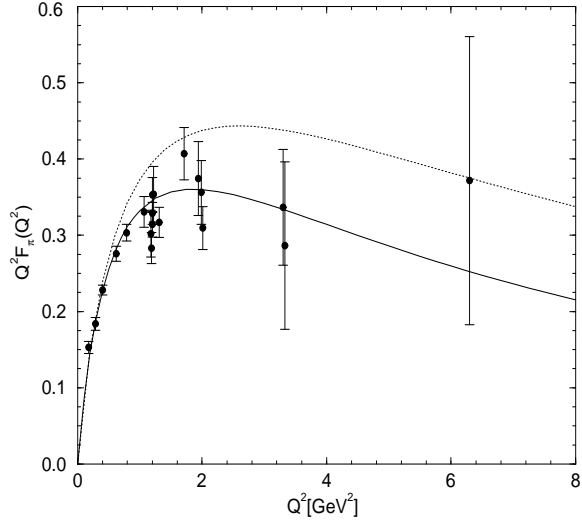,width=3.5in,height=3.2in}}
\caption{ The charge form factor for the pion
compared with data taken from Ref.~\protect\cite{Bebek}. 
The solid and dotted lines correspond to the results of HO and
linear potential cases, respectively.}
\end{figure}

In Fig. 4.3, we show our numerical results of the pion form factor
for the HO (solid line) and linear (dotted line) cases
and compare with the available experimental data~\cite{Bebek} up to
the $Q^{2}\sim 8$ GeV$^{2}$ region. Since our model parameters of
$m_{u}$= 0.25 GeV and $\beta_{u\bar{u}}$= 0.3194 GeV for the
HO case are the same as the ones used
in Refs.~\cite{jaus} and \cite{sch}, our numerical result
of the pion form factor is identical with the
Fig. 2 (solid line) in Ref.~\cite{sch}.
In Figs. 4.4(a) and 4.4(b), we show our numerical results for the form
factors of the charged and neutral kaons and compare with the results of
vector model dominance (VMD)~\cite{bell}, where a simple two-pole model of
the kaon form factors was assumed, i.e.,
$F_{K^{+}(K^{0})}(Q^{2})= e_{u(d)}m^{2}_{\omega}/(m^{2}_{\omega}+Q^{2})
+ e_{\bar{s}}m^{2}_{\phi}/(m^{2}_{\phi} + Q^{2})$.
From Figs. 4.4(a) and 4.4(b), we can see that the neutral kaon form 
factors using the model parameters obtained from
HO and linear potentials are not much different from each
other even though the charged ones are somewhat different.
\begin{figure}
\centerline{\psfig{figure=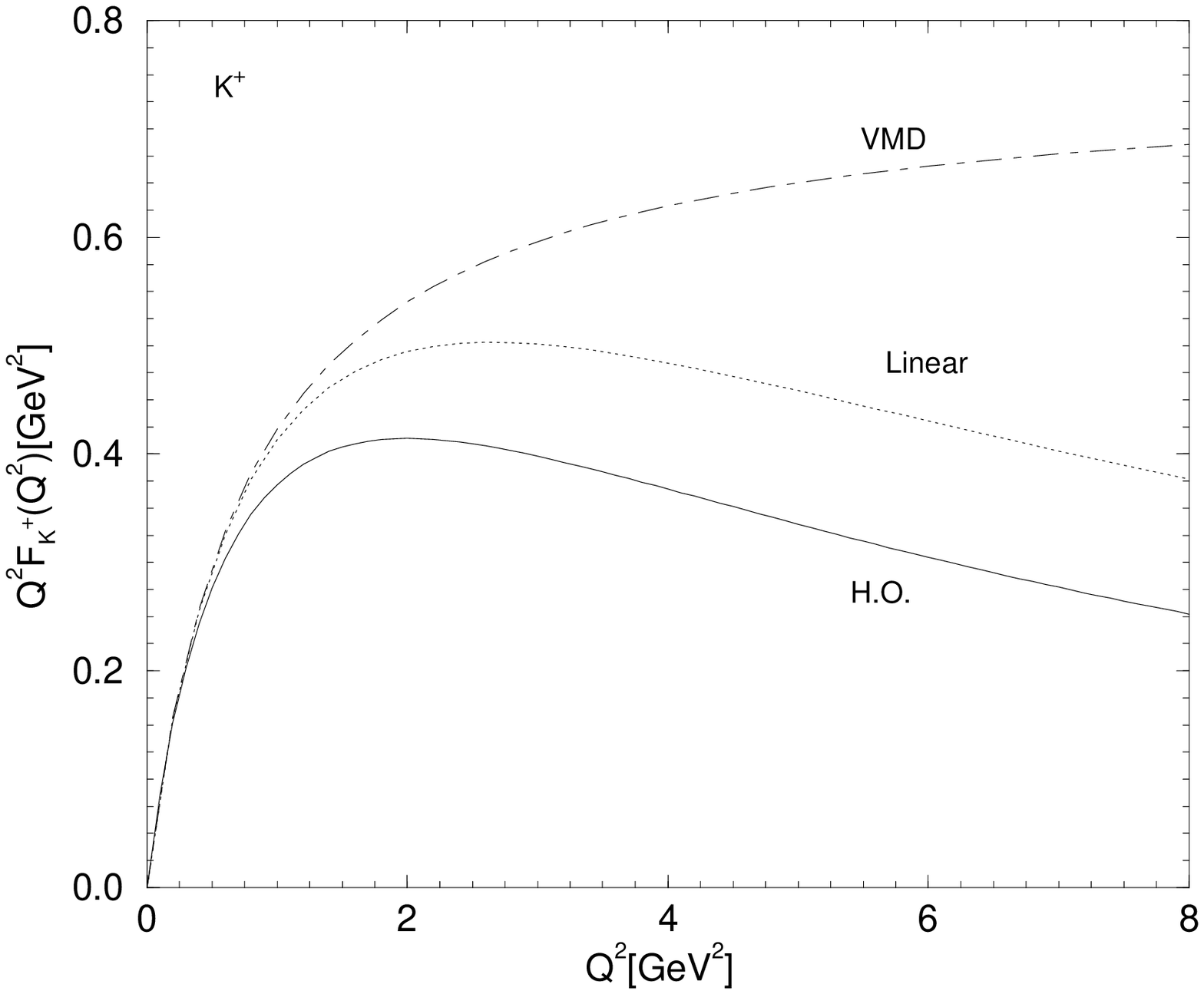,width=3.5in,height=3.2in}}
\centerline{\psfig{figure=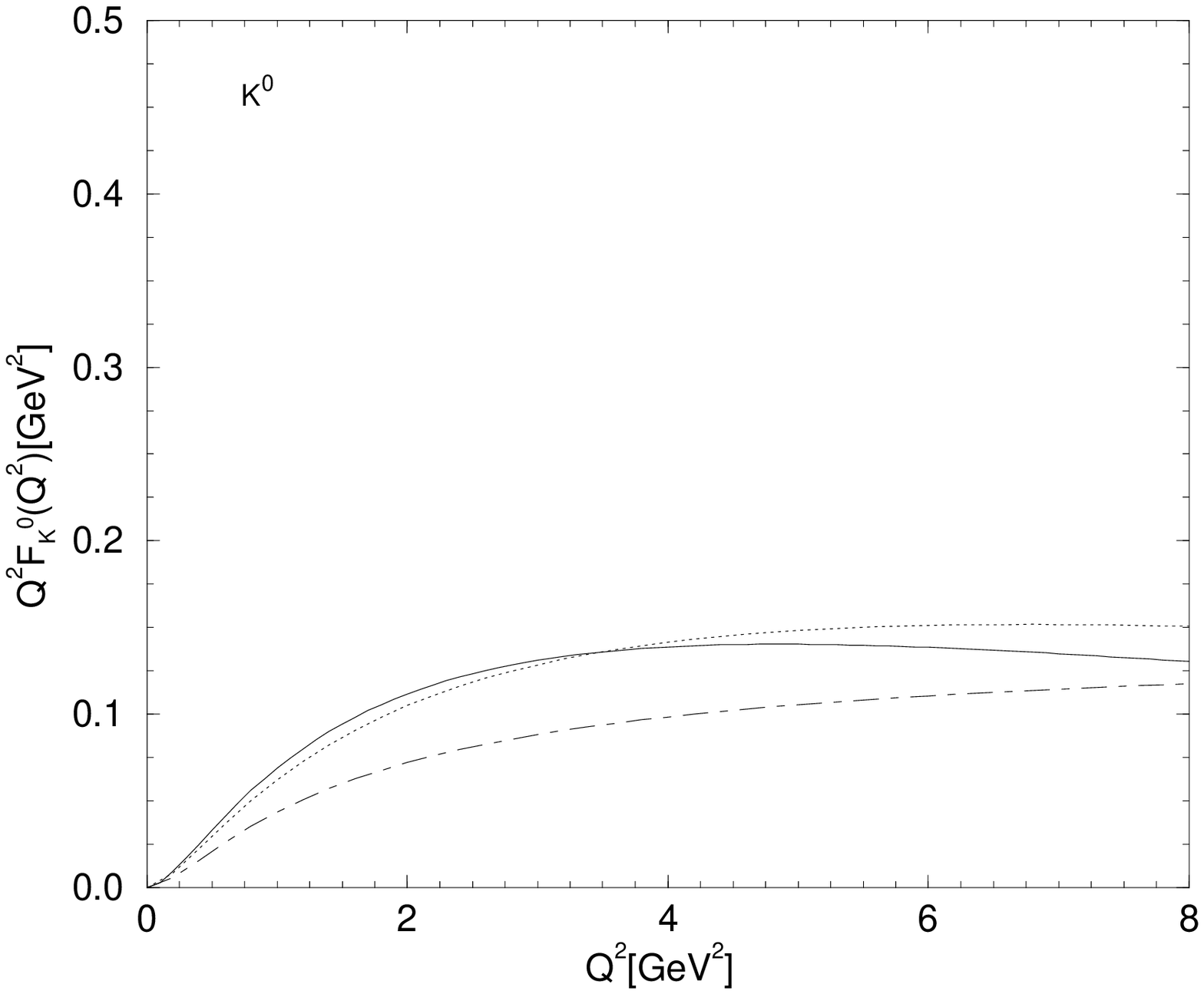,width=3.5in,height=3.2in}}
\caption{ (a) Theoretical predictions of
charged $K^{+}$ form factors using the parameters of both  
HO (solid) and linear (dotted) potentials compared with a simple
two-pole VMD model~\protect\cite{bell} (dot-dashed), 
$F^{\rm VDM}_{K^{+}(K^{0})}
= e_{u(d)}m^{2}_{\omega}/(m^{2}_{\omega} + Q^{2})
+ e_{\bar{s}}m^{2}_{\phi}/(m^{2}_{\phi} + Q^{2})$. 
(b) Theoretical predictions of neutral $K^{0}$ form factors.
The same line codes are used as in (a).} 
\end{figure}

\begin{table}
\centering
\caption{Decay constants (in unit of MeV) and charge radii 
(in unit of fm$^2$) for various pseudoscalar and
vector mesons. For comparison, we use
$|\delta_{V}|=3.3^{\circ}\pm 1^{\circ}$ for both potential cases.
The experimental data are taken from Ref.~\protect\cite{data}, 
unless otherwise noted. Note that $f_{K^{*}}$ was obtained by 
$\la 0|{\bar s}\gam^\mu u|K^{*}\ra=
2^{1/2}M_{K^{*}}f_{K^{*}}\vep^{\mu}$.}\label{dcon1}
\begin{tabular}{|c|c|c|c|c|c|}\hline
&\multicolumn{2}{|c|}{$\delta_{V}=-3.3^{\circ}\pm 1^{\circ}$}
&\multicolumn{2}{|c|}{$\delta_{V}=+3.3^{\circ}\pm 1^{\circ}$}& \\
\cline{2-5}
Observables & HO& Linear &HO & Linear  & Experiment\\
\hline
$f_{\pi}$ & 92.4 & 91.8& 92.4 & 91.8 &92.4$\pm$0.25 \\
\hline
$f_{K}$ & 109.3 & 114.1& 109.3 & 114.1& 113.4$\pm$ 1.1 \\
\hline
$f_{\rho}$ & 151.9 & 173.9 &151.9 &173.9 & 152.8$\pm$ 3.6 \\
\hline
$f_{K^{*}}$ & 157.6 & 180.8& 157.6 & 180.8 & ---\\
\hline
$f_{\omega}$ & 45.9$\pm$1.4 & 52.6$\pm$1.6 &
55.1$\pm$1.3 & 63.1$\pm$1.5 & 45.9$\pm$ 0.7 \\
\hline
$f_{\phi}$ & 82.6$\mp$ 0.8 & 94.3$\mp$0.9&
76.7$\mp$ 1.0 & 87.6$\mp$1.1 & 79.1$\pm$ 1.3  \\
\hline
$r^{2}_{\pi}$ & 0.449 & 0.425& 0.449& 0.425
& 0.432 $\pm$ 0.016~\cite{amen2} \\
\hline
$r^{2}_{K^{+}}$ & 0.384 & 0.354& 0.384 & 0.354
& 0.34$\pm$ 0.05~\cite{amen2}\\
\hline
$r^{2}_{K^{0}}$ & $-0.091$ & $-0.082$ & $-0.091$ &
$-0.082$ & $-0.054\pm$ 0.101~\cite{amen2}\\
\hline
\end{tabular}
\end{table}
The decay constants and charge radii of various pseudoscalar and
vector mesons for the two potential cases are given in Table 4.3 
and compared with experimental data\cite{data,amen2}.
While our optimal prediction of $\delta_{V}$ was
$|\delta_{V}|= 4.2^{\circ} (7.8^{\circ})$ for HO (linear) potential model,
we displayed our results for the common $\delta_{V}$ value
with a small variation (i.e.,
$|\delta_{V}|= 3.3^{\circ}\pm 1^{\circ}$) in Table 4.3 to show
the sensitivity. The results for both potentials are not
much different from each other and both results are quite comparable
with the experimental data. The decay widths of the $V(P)\to P(V)\gamma$
transitions are also given for the two different potential models in
Table 4.4. Although it is not easy to see which sign of $\delta_{V}$ for
the HO potential model is more favorable  to the
experimental data, the positive sign of $\delta_{V}$ looks a little
better than the negative one for the processes of
$\omega(\phi)\to\eta\gamma$ and $\eta'\to\omega\gamma$ transitions.
Especially, the overall predictions of the HO potential model
with positive $\delta_{V}$
seem to be in good agreement with the experimental data.
However, more observables should be compared with the data in
order to give a more definite answer for this sign issue of
$\omega$-$\phi$ mixing angle. The overall predictions of the linear potential
model are also comparable with the experimental data even though
the large variation of the mixing angle $\delta_{V}$ should be taken
into account in this case.
\begin{table}
\centering
\caption{Radiative decay widths for the $V(P)\to P(V)\gamma$ transitions.
The mixing angles, $\theta_{SU(3)}=-19^{\circ}$ for $\eta$-$\eta'$ and
$|\delta_{V}|=3.3^{\circ}\pm 1^{\circ}$ for $\omega$-$\phi$, are used
for both potential models, respectively.
The experimental data are taken from 
Ref.~\protect\cite{data}.}\label{rad1}
\begin{tabular}{|l|c|c|c|c|c|}\hline
&\multicolumn{2}{|c|}{$\delta_{V}=-3.3^{\circ}\pm 1^{\circ}$}
&\multicolumn{2}{|c|}{$\delta_{V}=+3.3^{\circ}\pm 1^{\circ}$}& \\
\cline{2-5}
Widths&HO & Linear & HO & Linear & Expt.[keV]\\
\hline
$\Gamma(\rho^{\pm}\to\pi^{\pm}\gamma)$ & 76 & 69 & 76 &69
&$68\pm 8$ \\
\hline
$\Gamma(\omega\to\pi\gamma)$ & 730$\pm$1.3 & 667$\pm$1.3
& 730$\mp$1.3 & 667$\mp$1.3 & $717\pm 51$ \\
\hline
$\Gamma(\phi\to\pi\gamma)$ & 5.6$^{-2.9}_{+3.9}$& 5.1$^{-2.6}_{+3.6}$
& 5.6$^{+3.9}_{-2.9}$ & 5.1$^{+3.6}_{-2.6}$ &  $5.8\pm 0.6$ \\
\hline
$\Gamma(\rho\to\eta\gamma)$ & 59& 54 & 59 & 54 &  $58\pm 10$ \\
\hline
$\Gamma(\omega\to\eta\gamma)$ & 8.7$\mp$ 0.3 & 7.9$\mp$0.3
& 6.9$\mp$ 0.3 & 6.3$\mp$0.3 & $7.0\pm 1.8$ \\
\hline
$\Gamma(\phi\to\eta\gamma)$ & 38.7$\pm$ 1.6  & 37.8$\pm$ 1.5
& 49.2$\pm$ 1.6 & 47.6$\pm$ 1.5 & $55.8\pm 3.3$\\
\hline
$\Gamma(\eta'\to\rho\gamma)$ & 68 & 62 & 68 & 62 & 61 $\pm8 $ \\
\hline
$\Gamma(\eta'\to\omega\gamma)$ & 4.9$\pm$ 0.4 & 4.5$\pm$ 0.4
& 7.6$\pm$ 0.4 & 7.0$\pm$ 0.4 & $6.1\pm 1.1$ \\
\hline
$\Gamma(\phi\to\eta'\gamma)$ & 0.41$\mp$0.01& 0.39$\mp$0.01
&0.36$\mp$ 0.01 & 0.34$\mp$0.01 & $< 1.8$\\
\hline
$\Gamma(K^{*0}\to K^{0}\gamma)$ & 124.5 & 116.6& 124.5& 116.6
& 117$\pm$ 10\\
\hline
$\Gamma(K^{*+}\to K^{+}\gamma)$ & 79.5 & 71.4& 79.5 & 71.4 & 50 $\pm$ 5 \\
\hline
\end{tabular}
\end{table}

In Table 4.4, we show the results
of $P(=\pi,\eta,\eta')\to\gamma\gamma$ decay widths obtained from our
two potential models with the axial anomaly plus partial conservation
of the axial vector current (PCAC) relations given
by Eq.~(\ref{pcac}). The predictions of
$\eta(\eta')\to\gamma\gamma$ decay widths using PCAC are in a
good agreement with the experimental data for both the HO
and linear potential models with $\eta$-$\eta'$ mixing angle,
$\theta_{SU(3)}=-19^{\circ}$. The predictions of the
decay constants for the octet and singlet mesons, i.e., $\eta_{8}$
and $\eta_{0}$, are $f_{8}/f_{\pi}=1.254 (1.324)$
and $f_{0}/f_{\pi}=1.127 (1.162)$ MeV for the HO (linear)
potential model, respectively.
Our predictions of $f_{8}$ and $f_{0}$ are not much different from the
predictions of chiral perturbation theory~\cite{dono}
reported as $f_{8}/f_{\pi}= 1.25$ and
$f_{0}/f_{\pi}=1.04\pm 0.04$, respectively.
Another important mixing-independent quantity related to
$f_{8}$ and $f_{0}$ is the $R$ ratio defined by
\begin{eqnarray}\label{ratio}
R&\equiv&\biggl[\frac{\Gamma(\eta\to\gamma\gamma)}{m^{3}_{\eta}}
+\frac{\Gamma(\eta'\to\gamma\gamma)}{m^{3}_{\eta'}}\biggr]
\frac{m^{3}_{\pi}}{\Gamma(\pi\to\gamma\gamma)}
= \frac{1}{3}\biggl(\frac{f^{2}_{\pi}}{f^{2}_{8}}
+ 8\frac{f^{2}_{\pi}}{f^{2}_{0}}\biggr).
\end{eqnarray}
\begin{table}
\centering
\caption{Radiative decay widths $\Gamma(P\to\gamma\gamma)$
obtained by using the axial anomaly plus PCAC
relations (see Eq.~(3.16)).
$\theta_{SU(3)}=-19^{\circ}$ for $\eta$-$\eta'$ mixing
is used for both potential cases.
The experimental data are taken from 
Ref.~\protect\cite{data}.}\label{rad2}
\begin{tabular}{|c|c|c|c|c|}\hline
Widths &  HO & Linear & Experiment\\
\hline
$\Gamma(\pi\to\gamma\gamma)$ &  7.73& 7.84 & $7.8\pm 0.5$ [eV]\\
\hline
$\Gamma(\eta\to\gamma\gamma)$ & 0.42& 0.42 & $0.47\pm0.0 5$ [keV]\\
\hline
$\Gamma(\eta'\to\gamma\gamma)$ & 4.1& 3.9 & $4.3\pm 0.6$ [keV] \\
\hline
\end{tabular}
\end{table}
\begin{figure}
\centerline{\psfig{figure=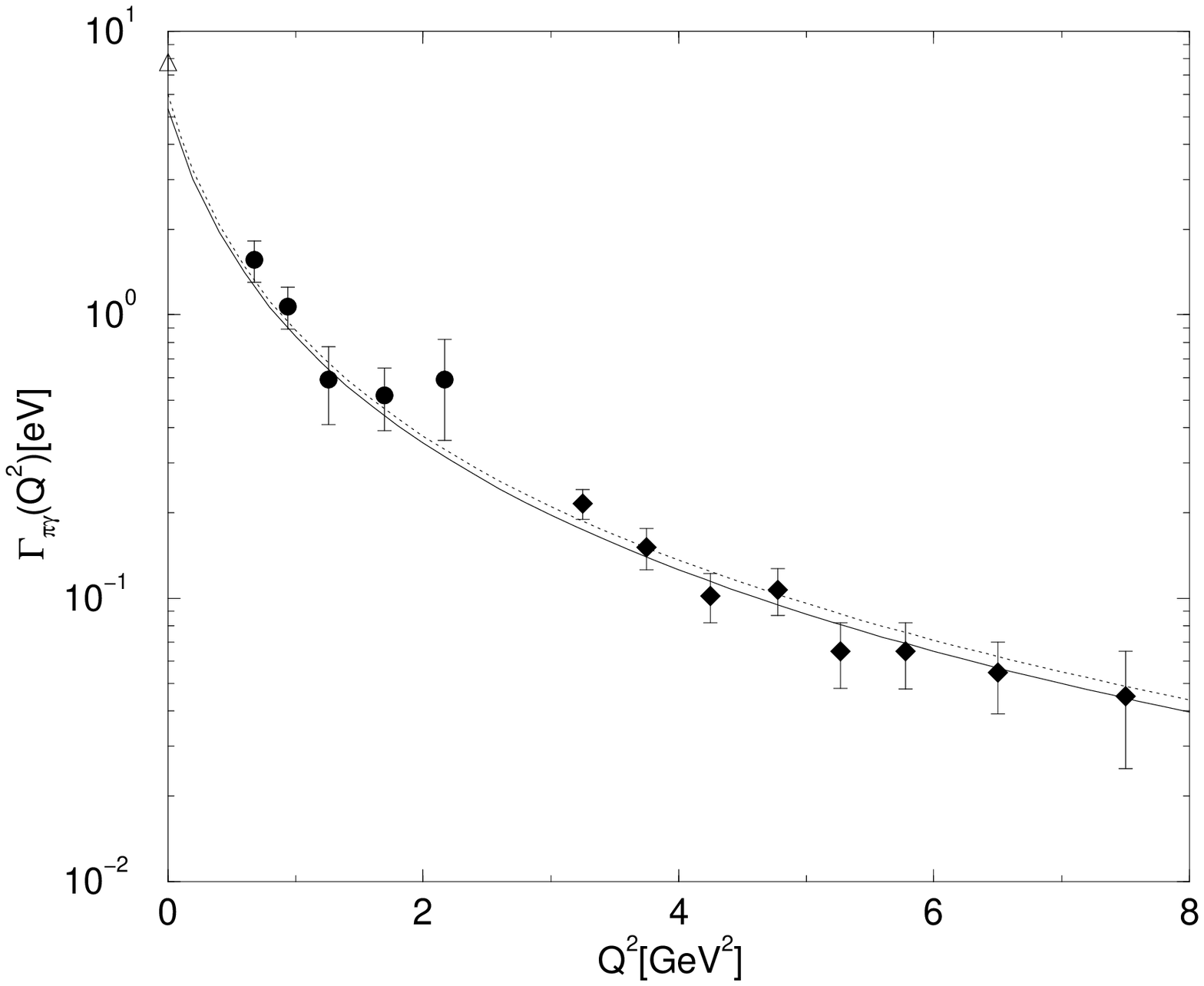,width=3.5in,height=3.2in}}
\caption{The decay rate for the
$\pi\to\gamma^{*}\gamma$  transition obtained from
the one-loop diagram. Data are taken from 
Refs.~\protect\cite{cello1,cello2}.}
\end{figure}
\begin{figure}
\centerline{\psfig{figure=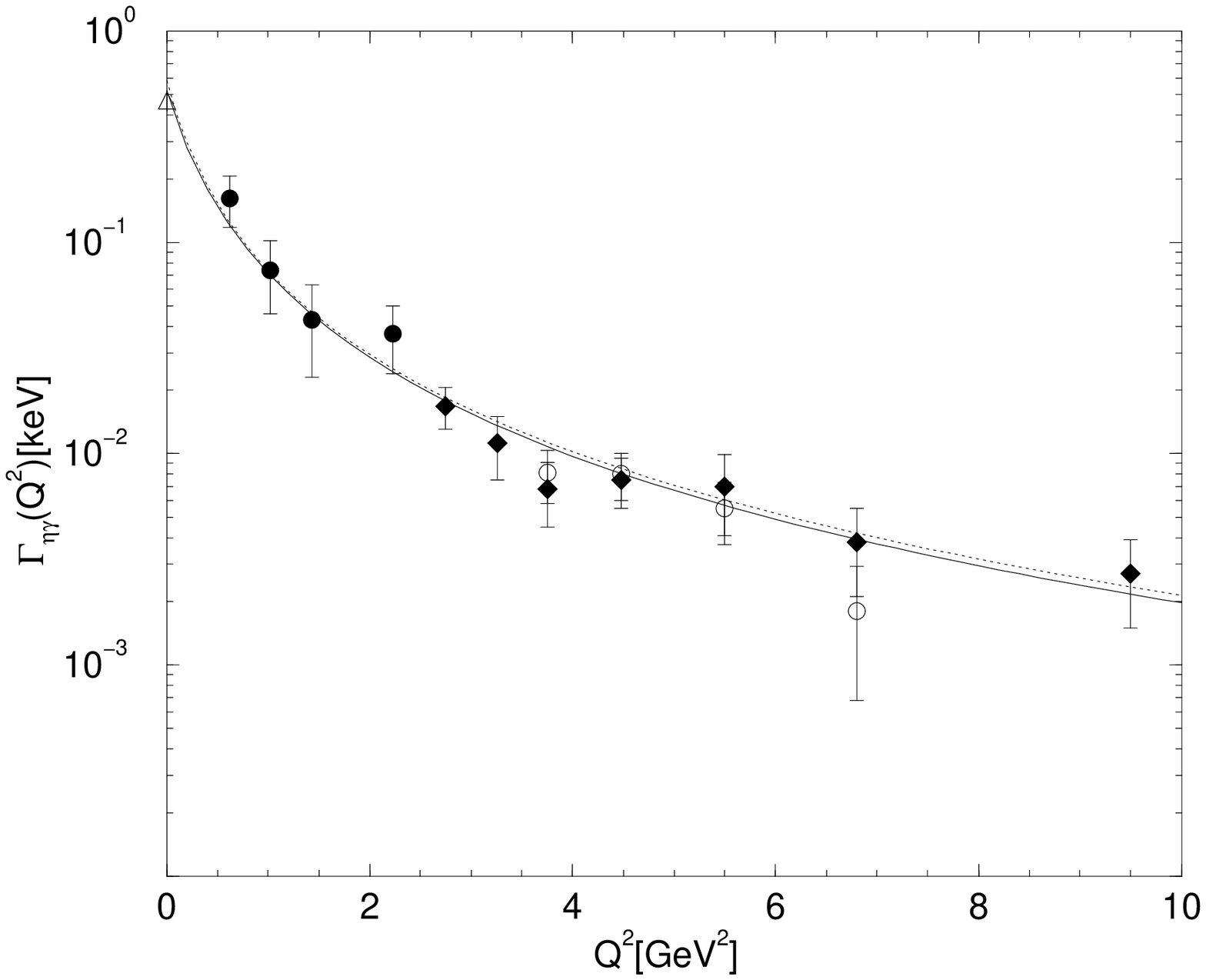,width=3.5in,height=3.2in}}
\caption{The decay rate for the
$\eta\to\gamma^{*}\gamma$ transition obtained from
the one-loop diagram. Data are taken from 
Refs.~\protect\cite{cello1,cello2,tpc}.}
\end{figure}
\begin{figure}
\centerline{\psfig{figure=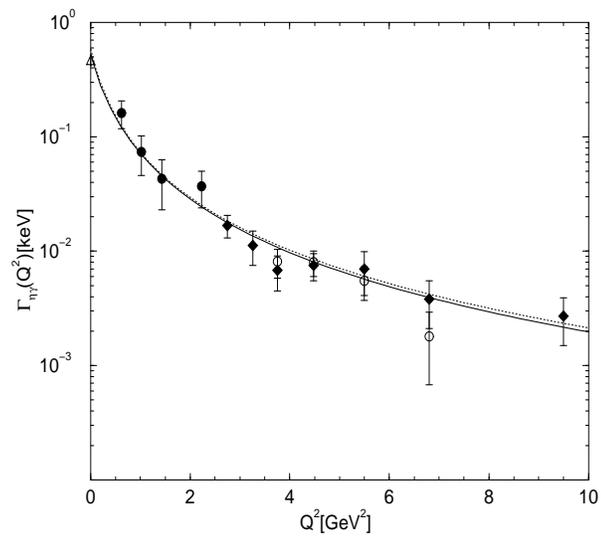,width=3.5in,height=3.2in}}
\caption{The decay rate for the
$\eta'\to\gamma^{*}\gamma$ transition obtained from
the one-loop diagram. Data are taken from 
Refs.~\protect\cite{cello1,cello2,tpc}.}
\end{figure}
Our predictions, $R=2.31$ and 2.17 for the HO and
linear potential model cases, respectively, are quite comparable to
the available experimental data~\cite{r1,r2},
$R_{\rm expt}=2.5\pm0.5({\rm stat})\pm 0.5({\rm syst})$.
Also, the $Q^{2}$-dependent
decay rates $\Gamma_{P\gamma}(Q^{2})$ are calculated from the usual
one-loop diagram~\cite{jaus,spin} and the results are shown in
Figs. 4.5-4.7. Our results for both potential models are not only very
similar to each other but also in remarkably good
agreement with the experimental data~\cite{cello1,cello2,tpc} up to
the $Q^{2}\sim 10$ GeV$^{2}$ region. We think that the reason
why our model is so successful for $P\to\gamma^{*}\gamma$ transition
form factors is because the $Q^{2}$ dependence ($\sim 1/Q^{2}$) is
due to the off-shell quark propagator in the one-loop diagram and
there is no angular condition~\cite{spin} associated with the pseudoscalar
meson.

\section{Summary and Discussion}
In the LFQM approach, we have investigated the mass spectra, mixing angles,
and other physical observables of light pseudoscalar and vector
mesons using QCD-motivated potentials given by Eq.~(\ref{potent}).
The variational principle for the effective Hamiltonian is crucial
to find the optimum values of our model parameters.
As shown in Figs. 4.1(a) and 4.1(b), we noticed that both central potentials
in Eq.~(\ref{potent}) are not only very similar to each other
but also quite close to the Isgur-Scora-Grinstein-Wise model 2
(ISGW2)~\cite{isgw2} potentials.
In Figs. 4.1(a) and 4.1(b), we have also included the Godfrey-Isgur (GI)
potential for comparison.
Using the physical masses of ($\omega,\phi$) and
($\eta,\eta'$), we were able to predict the $\omega$-$\phi$ and
$\eta$-$\eta'$ mixing angles as $|\delta_{V}|\approx
4.2^{\circ} (7.8^{\circ})$ and
$\theta_{SU(3)}\approx -19.3^{\circ} (-19.6^{\circ})$ for the
HO (linear) potential model, respectively.
We also have checked that the sensitivity of the mass spectra of
($\omega,\phi$) to $\sim 1^{\circ} (5^{\circ})$ variation of
$\delta_{V}$, i.e., from $\delta_{V}=4.2^{\circ} (7.8^{\circ})$ to
$3.3^{\circ}$ for the HO (linear) potential case,
is within the $1\% (5\%)$ level.

Then, we applied our models to compute the observables
such as charge radii, decay constants, and radiative decays of
$P(V)\to V(P)\gamma^{*}$ and $P\to\gamma\gamma^{*}$.
As summarized in Tables 4.3, 4.4, and 4.5, our numerical results for these
observables in the two cases (HO and linear)
are overall not much different from each other and are in a rather good
agreement with the available experimental data~\cite{data}. 
Furthermore, our results of the $R$ ratio presented in
Eq.~(\ref{ratio}) are in a good agreement with the experimental 
data~\cite{r1,r2}.
The $Q^{2}$ dependences of $P\to\gamma\gamma^{*}$ processes were also
compared with the experimental data up to $Q^{2}\sim 8$ GeV$^{2}$.
The $Q^{2}$ dependence for
these processes is basically given by the off-shell quark propagator
in the one-loop diagrams.
As shown in Figs. 4.5-4.7, our results are in an excellent agreement with
the experimental data~\cite{cello1,cello2,tpc}.
Both the pion and kaon form factors were also predicted
in Figs. 4.3 and 4.4, respectively.
We believe that the success of LFQM hinges upon the
suppression of complicated zero-mode contributions from the LF 
vacuum due to the rather large constituent quark masses. The
well-established formulation of form factors in the Drell-Yan-West
frame also plays an important role for our model to provide comparable
result with the experimental data. Because of these successful applications
of our variational effective Hamiltonian method, the extension
to the heavy ($b$ and $c$ quark sector) pseudoscalar and vector
mesons will be shown in Chapter 7~\cite{Semi}
and the $0^{++}$ scalar mesons is currently under
consideration.

While there have been previous LFQM results on
the observables that we calculated in this work, they were based
on the approach of modeling the wave function rather than modeling
the potential. Our approach in this work attempting to fill the gap
between the model wave function and the QCD-motivated potential has
not yet been explored to cover as many observables as we did in
this work.
Nevertheless, it is not yet clear which sign of $\omega$-$\phi$ mixing
angle should be taken, even though the overall agreement between our
HO potential model with the positive sign, i.e.,
$\delta_{V}\sim 3.3^{\circ}$, and the
available experimental data seem to be quite good.
If we were to choose the sign of $X$ as $X>0$ in Eq. (E.4), then
the fact that the mass difference $m_{\omega}-m_{\rho}$ is
positive is correlated with the sign of the $\omega$-$\phi$
mixing angle~\cite{private}. In other words, $m_{\omega}>m_{\rho}$
implies $\delta_{V}>0$ from Eqs. (E.3)-(E.5).
Perhaps, the precision measurement of $\phi\to\eta'\gamma$
envisioned in the future at TJNAF experiments might be helpful to
give a more stringent test of $\delta_{V}$.
In any case, more observables
should be compared with the experimental data to give more definite
assessment of this sign issue.


%% file: anal_final.tex
\newpage
\setcounter{equation}{0}
\setcounter{figure}{0}
\renewcommand{\theequation}{\mbox{5.\arabic{equation}}}
\chapter{Exploring the Timelike Region For the Elastic Form Factor 
in a Scalar Field Theory}
The Drell-Yan-West ($q^{+}$=$q^{0}+q^{3}$=0) frame in the LF 
quantization provided an effective formulation for the calculation 
of various form factors in the spacelike momentum transfer region
$q^{2}$=$-Q^{2}$$<0$~\cite{LB}.
In $q^{+}$=0 frame, only parton-number-conserving Fock
state (valence) contribution is needed when the ``good" components of the
current, $J^{+}$ and ${\bf J}_{\perp}$=$(J_{x},J_{y})$, are used~\cite{Kaon}.
For example, only the valence diagram shown in Fig. 1.1(a) is used in the
LFQM analysis of spacelike meson form factors.
Successful LFQM description of various hadron form
factors can be found in the literatures~\cite{jaus,chung2,Mix,card}.

However, the timelike ($q^{2}$$>0$) form factor analysis in the LFQM 
has been hindered by the fact that $q^{+}$=0 frame is defined
only in the spacelike region ($q^{2}$=$q^{+}q^{-}-q^{2}_{\perp}$$<0$).
While the $q^{+}$$\neq$0 frame can be used in principle to compute the 
timelike form factors, it is inevitable (if $q^{+}$$\neq$0) to encounter 
the nonvalence diagram arising from the quark-antiquark pair creation 
(so called ``Z-graph"). 
For example, the nonvalence diagram in the case of semileptonic
meson decays is shown in Fig. 1.1(b).
The main source of the difficulty, however, in calculating the nonvalence
diagram (see Fig. 1.1(b)) is the lack of information on the black blob which
should contrast with the white blob representing the usual LF 
valence wave function. In fact, it was reported~\cite{Kaon} that
the omission of nonvalence contribution
leads to a large deviation from the full results.
The timelike form factors associated with the hadron pair productions in
$e^{+}e^{-}$ annihilations also involve the nonvalence contributions.
Therefore, it would be very useful to avoid encountering the nonvalence
diagram and still be able to generate the results of timelike form factors.

In this work, we show an explicit example of generating the exact result
of the timelike form factor without encountering the nonvalence diagram.
This can be done by the analytic continuation from the spacelike form
factor calculated in the Drell-Yan-West ($q^{+}$=0) frame to the timelike
region. To explicitly show it, we use an exactly solvable model of ($3+1$)
dimensional scalar field theory interacting with gauge fields.
Our model is essentially the $(3+1)$ dimensional extension of Mankiewicz
and Sawicki's $(1+1)$ dimensional quantum field theory model~\cite{SM},
which was later reinvestigated by several others~\cite{zm,BH,GS,Sa,Ba}.
The starting model wave function is the solution of covariant
Bethe-Salpeter (BS) equation in the ladder approximation with a
relativistic version of the contact interaction~\cite{SM}.
Here, we do not take the Hamiltonian approach. The covariant model wave
function is a product of two free single particle propagators, the overall
momentum-conservation Dirac delta, and a constant vertex function.
Consequently, all our form factor calculations are nothing
but various ways of evaluating the Feynman perturbation-theory triangle
diagram in scalar field theory.
In this model, we calculate:
(A) the timelike process of $\gamma^{*}\to M + \bar{M}$
transition in $q^{+}$$\neq$0 ($q^{2}$$>$0) frame,
(B) the spacelike process of $M\to\gamma^{*} +M$ in $q^{+}$$\neq$0 
($q^{2}$$<$0) frame, and
(C) the spacelike process of $M\to\gamma^{*} +M$ in $q^{+}$=0 frame.
Using the analytic continuation from $q^{2}$$<$0 to $q^{2}$$>$0, we
show that the result in (C), which is obtained without encountering the
nonvalence contributions at all, exactly reproduces the result in (A).
In fact, all three results (A), (B), and (C) coincide with each other in the
entire $q^{2}$ range.
We also confirm that our results are consistent with the dispersion
relations~\cite{SDrell,Gasio,Man,Gell}. We consider not only for the equal
quark/antiquark mass case such as the pion but also for the unequal
mass cases such as $K$ and $D$.

This Chapter is organized as follows: 
In Section 5.1, we derive the timelike
electromagnetic (EM) form factor of $\gamma^{*}\to M+\bar{M}$ process in the
$q^{+}$$\neq$0 frame (A)
and discuss the singularities occuring from the on-energy shell
of quark-antiquark pair creation.
In Section 5.2, the spacelike form factor of $M\to\gamma^{*} + M$ process
is calculated both in the $q^{+}$$\neq$0 (B) and $q^{+}$=0 (C) frames.
We then analytically continue the spacelike form factors to the timelike
region. The singularities occured in the timelike region are also discussed.
In Section 5.3, for the numerical calculation of the
EM ($\pi$, $K$, and $D$) meson
form factors for three different cases (A), (B)
and (C), we use the constituent quark and antiquark masses
($m_{u}$=$m_{d}$=0.25 GeV, $m_{s}$=0.48 GeV, and 
$m_{c}$=1.8 GeV)~\cite{Mix,Kaon,Semi} and show that the form 
factors obtained from those three different cases are indeed equal to 
each other for the entire $q^{2}$ region.
The meson peaks analogous to the vector meson dominance(VMD) are also
obtained. The conclusion and discussion follows in Section 5.4.
\section{ Form Factors in the Timelike Region }
The EM local current $J^{\mu}(0)$
responsible for a virtual photon decay into two $q\bar{Q}$ bound states
in the scalar field theory can be calculated using the diagrams shown in
Fig. 5.1.
The covariant diagram shown in Fig. 5.1(a) is equivalent to the sum of
two LF time-ordered diagrams in Figs. 5.1(b) and 5.1(c).
The EM current $J^{\mu}(0)$ obtained from the covariant
diagram of Fig. 5.1(a) is given by
\bea\label{eq:51}
J^{\mu}(0)&=& e_{q}\int d^{4}k\frac{1}{ (q-k)^{2}-m_{q}^{2}+i\epsilon}
(q-2k)^{\mu}\frac{1}{ (q-k-P_{2})^{2}-m_{\bar{Q}}^{2}+i\epsilon}
\nonumber\\
&\times&\frac{1}{ k^{2}-m_{q}^{2}+i\epsilon}\;\;
+\;\; e_{\bar{Q}}(m_{q}\leftrightarrow m_{\bar{Q}}\; {\rm of\; the\;
first\; term}),
\eea
where $m_{q(\bar{Q})}$ and $e_{q(\bar{Q})}$ are the constituent
quark (antiquark) mass and charge, respectively.
The corresponding form factor $F(q^2)$ of the $q{\bar{Q}}$ bound state in
timelike($q^{2}>0$) region is defined by
\be\label{eq:52}
J^{\mu}(0)= (P_{1}-P_{2})^{\mu}F(q^{2}),
\ee
where $q=P_{1}+P_{2}$, $P^{1}_{1}=P^{2}_{2}=M^{2}$ and $M$ is the
mass of a $q\bar{Q}$ bound state scalar particle.
\begin{figure}
\centerline{\psfig{figure=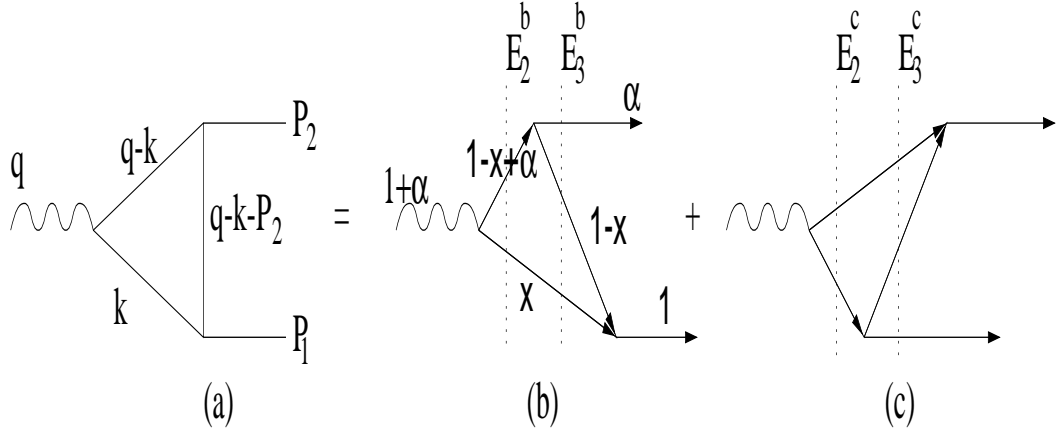,width=5.5in,height=2.2in}}
\caption{The electromagnetic decays of a photon into two-body bound
sates, i.e., $\gamma^{*}\to q\bar{q}$(or $Q\bar{Q}$)$\to
{\cal M}(q\bar{Q}){\cal M}(q\bar{Q})$,
in scalar theory: Covariant representation (a),
and the LF time ordered contributions to the decay amplitude (b)
and (c).}
\end{figure}
Using the Cauchy integration over $k^-$ in Eq.~(\ref{eq:51}), 
we can find each time-ordered contribution (Figs. 5.1(b) and 5.1(c)) 
to the timelike form factor $F(q^2>4M^2)$ in Eq.~(\ref{eq:52}).
This procedure allows us to analyze the singularity structure
of each LF time-ordered diagram as well.
For the calculation of each LF time-ordered contribution,
we take the purely longitudinal momentum frame, i.e.,
$q^{+}$$\neq$0,  ${\bf q}_{\perp}$=0 and 
${\bf P}_{1\perp}$=${\bf P}_{2\perp}$=0.
Accordingly, $q^{2}(=q^{+}q^{-}) > 4 M^2$ is given by
\be\label{eq:53}
q^{2}=M^{2}(1+\al)^{2}/\al,
\ee
where $\al$=$P^{+}_{2}/P^{+}_{1}$=$q^{+}/P^{+}_{1}-1$ is
the longitudinal momentum fraction and
the two solutions for $\al$ are given by
\be\label{alpm}
\al_{\pm}=\biggl(\frac{q^{2}}{2M^{2}}-1\biggr)\pm
\sqrt{\biggl(\frac{q^{2}}{2M^{2}}-1\biggr)^{2}-1}.
\ee
Note that both $\al_{\pm}$=1 correspond to the threshold,
$q^{2}$=$4M^{2}$. The EM form factor $F(q^{2})$ in Eq.~(\ref{eq:52})
is independent of the subscript sign of $\al$. Thus,
one can take either ${\al}_{+}$ or ${\al}_{-}$ to calculate $F(q^2)$.
Here, for convenience, we use $\al$=$\al_{-}$ which ranges from 0 to 1
for the physical momentum transfer region, i.e., $\al_{-}\to0$ as
$q^{2}\to\infty$ and 1 as $q^{2}\to 4M^2$. Of course, we can use
$\al$=$\al_{+}$ equally well and verify that the two results (${\al}_{+}$
and ${\al}_{-}$) are exactly same for the calculation of $F(q^{2})$.

Since $q^{+}>P^{+}_{1}\geq P^{+}_{2}>0$ for $\al$=$\al_{-}$,
the Cauchy integration over $k^{-}$ in Eq.~(\ref{eq:51}) has
two nonzero contributions to the residue calculations, one coming from
the interval (i) $0<k^{+}<P^{+}_{1}$ (see Fig. 5.1(b)) and the other
from (ii) $P^{+}_{1}<k^{+}<q^{+}$ (see Fig. 5.1(c)). The internal
momentum $k^{+}$ is defined by $k^{+}$=$xP^{+}_{1}$, where $x$
is the Lorentz invariant longitudinal momentum variable.
The ``good"-current $J^{+}(0)$ is used in our computation of
the two LF diagrams Figs. 5.1(b) and 5.1(c).
In the following, for simplicity, we won't explicitly write
either the obvious second term in Eq.~(\ref{eq:51}) nor the charge 
factor ($e_q$ or $e_{\bar{Q}}$).

In the region of $0<k^{+}<P^{+}_{1}$, the residue is at the
pole of $k^{-}$=$[m_{q}^{2}+k^{2}_{\perp}-i\epsilon]/k^{+}$, which is placed
in the lower half of complex-$k^{-}$ plane. Thus, the Cauchy integration of
$J^{+}$ in Eq.~(\ref{eq:51}) over $k^{-}$ in this region gives
\bea\label{eq:55}
J^{+}_{b}(0)&=&{\cal N}\int^{P^{+}_{1}}_{0}
dk^{+}d^{2}k_{\perp}\frac{q^{+}-2k^{+}}
{k^{+}(q^{+}-k^{+})(P^{+}_{1}-k^{+})}\nonumber\\
&\times& \frac{1}{\biggl[ q^{-}-(m_{q}^{2}+k^{2}_{\perp})/k^{+} -
(m_{q}^{2}+k_{\perp}^{2})/(q^{+}-k^{+})\biggr]}\\
&\times&
\frac{1}{\biggl[ q^{-}-P^{-}_{2}- (m_{q}^{2}+k^{2}_{\perp})/k^{+} -
(m_{\bar{Q}}^{2}+k_{\perp}^{2})/(P_{1}^{+}-k^{+})\biggr]},\nonumber
\eea
where the subscript $b$ in $J^{+}_{b}(0)$ implies the current in 
Fig.~5.1(b) and ${\cal N}$(=$-i\pi)$ is the normalization constant.
The last two factors in Eq.~(\ref{eq:55}) correspond to the two and 
three particle intermediate states. We represent the energy denominators 
of the two and three particle intermediate states as $E^{b}_{2}$ and
$E^{b}_{3}$ in Fig. 5.1(b), respectively.

To analyze the singularities of Eq.~(\ref{eq:55}), we further integrate 
over ${\bf k}_{\perp}$ and obtain
\be\label{eq:56}
J^{+}_{b}(0)=\pi{\cal N}\int^{1}_{0}dx
\frac{x(1+\al-2x)/(1+\al)}{{\cal E}^{b}_{3}-{\cal E}^{b}_{2}}
\ln\biggl(\frac{{\cal E}^{b}_{2}}{{\cal E}^{b}_{3}}\biggr),
\ee
where ${\cal E}^{b}_{2}$=$x(1+\al-x)M^{2}/\al- m_{q}^{2}$ and
${\cal E}^{b}_{3}$=$x(1-x)M^{2}-[xm_{q}^{2} + (1-x)m_{\bar{Q}}^{2}]$.
While ${\cal E}^{b}_{3}$ is not zero (${\cal E}^{b}_{3}\neq 0$) in general
for the entire physical region, ${\cal E}^{b}_{2}$ can be zero
when $q^{2}\geq4m^{2}_{q(\bar{Q})}$.
The singular structure of ${\cal E}^{b}_{3}-{\cal E}^{b}_{2}$ term in 
Eq.~(\ref{eq:56}) depends on whether a $q\bar{Q}$ bound state scalar 
particle is strongly bounded ($M^{2} < m^{2}_{q} + m^{2}_{\bar{Q}}$) or
weakly bounded ($M^{2} > m^{2}_{q} + m^{2}_{\bar{Q}}$).
As we will show in our numerical calculations (Section 5.4),
anomalous threshold appears for $M^{2} > m^{2}_{q} + m^{2}_{\bar{Q}}$
while only the normal threshold of bound state exists for
$M^{2} < m^{2}_{q} + m^{2}_{\bar{Q}}$.

In the region of $P^{+}_{1}<k^{+}<q^{+}$, the residue is at the pole of
$k^{-}$=$q^{-}-[m_{q}^{2}+({\bf q}_{\perp}-{\bf k}_{\perp})^{2}
-i\epsilon]/(q^{+}-k^{+})$,
which is placed in the upper half of complex-$k^{-}$ plane.
Thus, the Cauchy integration of
$J^{+}(0)$ in Eq.~(\ref{eq:51}) over $k^{-}$ in this region yields the 
result
\bea\label{eq:57}
J^{+}_{c}(0)&=& -{\cal N}\int^{q^{+}}_{P^{+}_{1}}
dk^{+}d^{2}k_{\perp}\frac{q^{+}-2k^{+}}
{k^{+}(q^{+}-k^{+})(P^{+}_{1}-k^{+})}\nonumber\\
&\times&
\frac{1}{\biggl[ q^{-}- (m_{q}^{2}+k^{2}_{\perp})/k^{+} -
(m_{q}^{2}+k_{\perp}^{2})/(q^{+}-k^{+})\biggr]}\\
&\times&
\frac{1}{\biggl[ q^{-}-P^{-}_{1}
+ (m_{\bar{Q}}^{2}+k^{2}_{\perp})/(P^{+}_{1}-k^{+})
- (m_{q}^{2}+k_{\perp}^{2})/(q^{+}-k^{+})\biggr]},\nonumber
\eea
where the subscript $c$ in $J^{+}_{c}(0)$ means the current in 
Fig. 5.1(c). After the integration over the ${\bf k}_{\perp}$ in 
Eq.~(\ref{eq:57}), we obtain
\be\label{eq:58}
J^{+}_{c}(0)= -\pi{\cal N}\int^{1}_{0} dX
\frac{\al X(1+\al-2\al X)/(1+\al)}{{\cal E}^{c}_{3}-{\cal E}^{c}_{2}}
\ln\biggl( \frac{ {\cal E}^{c}_{2}}{{\cal E}^{c}_{3}}\biggr),
\ee
where $x$=$1+\al (1-X)$, ${\cal E}^{c}_{3}$=$X(1-X)M^{2}-[X m_{\bar{Q}}^{2}
+ (1-X)m_{q}^{2}]$ and ${\cal E}^{c}_{2}$=$\al X(1+ \al -\al X)M^{2}/\al
- m_{q}^{2}$. The pole structure in Eq.~(\ref{eq:58}) is equivalent to 
that of Eq.~(\ref{eq:56}).

Consequently, the timelike form factor in Eq.~(\ref{eq:52}) is given by  
\bea\label{eq:59}
F(q^{2}) &=& \frac{\pi{\cal N}}{\al^{2}-1}\int^{1}_{0}dx
\biggl\{ \frac{x(1+\al-2x)}{{\cal E}^{b}_{3}-{\cal E}^{b}_{2}}
\ln\biggl(\frac{{\cal E}^{b}_{2}}{{\cal E}^{b}_{3}}\biggr)
\nonumber\\
&-& \frac{\al x(1+\al-2\al x)}{{\cal E}^{c}_{3}-{\cal E}^{c}_{2}}
\ln\biggl( \frac{ {\cal E}^{c}_{2}}{{\cal E}^{c}_{3}}\biggr)\biggr\},
\eea
where ${\cal E}^{b}_{2}, {\cal E}^{b}_{3}, {\cal E}^{c}_{2}$ and
${\cal E}^{c}_{3}$ are defined in Eqs.~(\ref{eq:56}) and (\ref{eq:58}).
\section{Form Factors in Spacelike Region and the Analytic Continuation
to the Timelike Region}
In this section, we calculate the EM form factor in spacelike momentum
transfer region and then analytically continue to the timelike region
to compare the result with the timelike form factor 
(i.e. Eq.~(\ref{eq:59})) that we obtained in the previous section.
The EM current of a $q\bar{Q}$ bound state in spacelike
momentum transfer region is defined by the local current $j^{\mu}(0)$;
\be\label{eq:510}
j^{\mu}(0)= (P_{1}+P_{2})^{\mu}{\cal F}(q^{2}),
\ee
where $q$=$P_{1}-P_{2}$, $q^{2}$$<$0 and ${\cal F}(q^2)$ is the spacelike
form factor.
The EM current $j^{\mu}(0)$ obtained
from the covariant triangle diagram of Fig. 5.2(a) is given by 
\be\label{eq:511} 
j^{\mu}(0)= \int d^{4}k\frac{1}{ (P_{1}-k)^{2}-m_{q}^{2}+i\epsilon}
(P_{1}+P_{2}-2k)^{\mu}\frac{1}{ (P_{2}-k)^{2}-m_{q}^{2}+i\epsilon}
\frac{1}{ k^{2}-m_{\bar{Q}}^{2}+i\epsilon}.
\ee
\begin{figure}
\centerline{\psfig{figure=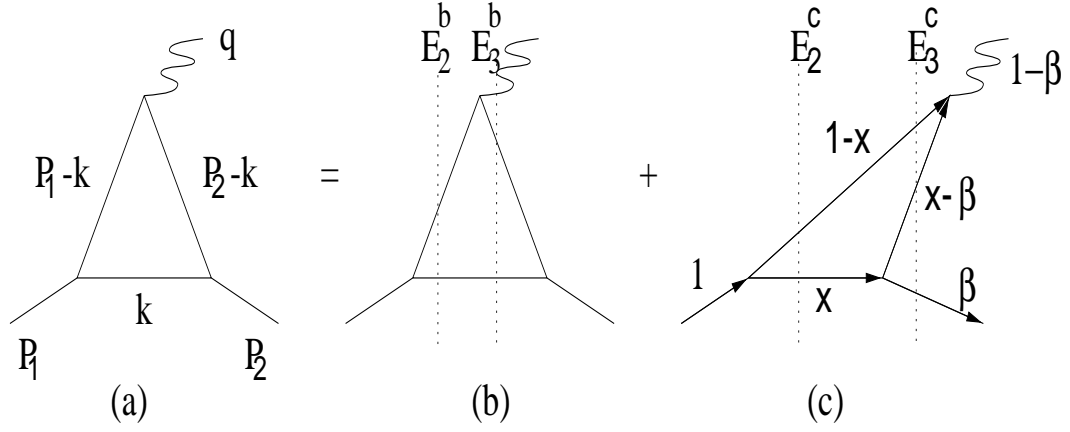,width=5.5in,height=2.2in}}
\caption{Covariant triangle diagram (a) is represented as the sum of
LF triangle diagram (b) and the LF pair-creation
diagram (c).}
\end{figure}
As in the case of the timelike form factor in Section 5.2,
the Cauchy integration of $k^{-}$ in Eq.~(\ref{eq:511}) has also two 
contributions to the residue calculations, one coming from the interval
$0<k^{+}<P^{+}_{2}$ (see Fig. 5.2(b)) and the other from
$P^{+}_{2}<k^{+}<P^{+}_{1}$ (see Fig. 5.2(c)). Again, only the
``good"-current $j^{+}(0)$ in Eq.~(\ref{eq:511}) 
is used to obtain the contributions from Figs. 5.2(b) and 5.2(c).

In the region of $0<k^{+}<P^{+}_{2}$, the residue is at the
pole of $k^{-}$=$[m_{\bar{Q}}^{2}+k^{2}_{\perp}-i\epsilon]/k^{+}$, which is
placed in the lower half of complex-$k^{-}$ plane. Thus, the Cauchy
integration of $j^{+}$ in Eq.~(\ref{eq:511}) over $k^{-}$ in this region 
yields
\bea\label{eq:512}
j^{+}_{b}(0)&=& {\cal N}\int^{P^{+}_{2}}_{0}dk^{+}d^{2}k_{\perp}
\frac{(P_{1}+P_{2}-2k)^{+}}{(P_{1}-k)^{+}(P_{1}-k-q)^{+}k^{+}}
\nonumber\\
&\times&
\frac{1}{ \biggl[P^{-}_{1}- [m_{q}^{2}+k^{2}_{\perp}]/(P^{+}_{1}-k^{+})
- [m_{\bar{Q}}^{2}+k^{2}_{\perp}]/k^{+}\biggr]}\\
&\times&
\frac{1}{\biggl[P^{-}_{1}-q^{-} -
[m_{q}^{2}+({\bf k}_{\perp}+{\bf q}_{\perp})^{2}]/(P^{+}_{1}-k^{+})
- [m_{\bar{Q}}^{2}+k^{2}_{\perp}]/k^{+}\biggr]},\nonumber
\eea
where the subscript $b$ in $j^{+}_{b}(0)$ implies the current 
in Fig. 5.2(b).

In the region of $P^{+}_{2}<k^{+}<P^{+}_{1}$, the residue
is at the pole of $k^{-}$=$P^{-}_{1} -
[m_{q}^{2}+k^{2}_{\perp}-i\epsilon]/(P_{1}^{+}-k^{+})$, which is
placed in the upper half of complex-$k^{-}$ plane. Thus, the Cauchy
integration of $j^{+}$ in Eq.~(\ref{eq:511}) over $k^{-}$ in this region 
becomes
\bea\label{eq:513}
j^{+}_{c}(0)&=& {\cal N}\int^{P^{+}_{1}}_{P^{+}_{2}}dk^{+}d^{2}k_{\perp}
\frac{(P_{1}+P_{2}-2k)^{+}}{(P_{1}-k)^{+}(P_{1}-k-q)^{+}k^{+}}
\nonumber\\
&\times&
\frac{1}{\biggl[P^{-}_{1}-[m_{q}^{2}+k_{\perp}^{2}]/(P^{+}_{1}-k^{+})
- [m_{\bar{Q}}^{2}+k^{2}_{\perp}]/k^{+}\biggr]}
\\
&\times& \frac{1}{ \biggl[P^{-}_{2}- P^{-}_{1} +
[m_{q}^{2}+k_{\perp}^{2}]/(P^{+}_{1}-k^{+})
 - [m_{q}^{2}+({\bf q}_{\perp}+{\bf k}_{\perp})^{2}]/(P^{+}_{2}
-k^{+})\biggr]}.
\nonumber
\eea
As one can see from Eq.~(\ref{eq:513}), the nonvalence contribution  
(Fig. 5.2(c)) vanishes only in $q^{+}$=0 frame.
In the following, we investigate the spacelike form factor
${\cal F}(q^{2})$ given in Eq.~(\ref{eq:510}) using both $q^{+}$$\neq$0 
and $q^{+}$=0 frames. We then analytically continue to the timelike region
in order to compare the result with the direct calculation of the timelike
form factor $F(q^2)$ presented in the previous section.
\subsection{The $q^{+}\neq0$ and ${\bf q}_{\perp}=0$ frame}
In the purely longitudinal momentum frame $q^{+}$$\neq$0, 
${\bf q}_{\perp}$=0, and ${\bf P}_{1\perp}$=${\bf P}_{2\perp}$=0, 
the momentum transfer 
$q^{2}$=$q^{+}q^{-}$ can be written in terms of the longitudinal momentum 
fraction $\beta$=$P^{+}_{2}/P^{+}_{1}= 1-q^{+}/P^{+}_{1}$; 
\be\label{eq:514}
q^{2}=-M^{2}(1-\beta)^{2}/\beta\leq 0,
\ee
where the two solutions of $\beta$ are given by 
\be\label{beta}
\beta_{\pm}= \biggl(1 -\frac{q^{2}}{2M^{2}}\biggr)\pm
\sqrt{ \biggl(1-\frac{q^{2}}{2M^{2}}\biggr)^{2} -1}.
\ee  
The form factor ${\cal F}(q^{2})$ in Eq.~(\ref{eq:510}) is also independent 
of the subscript sign of $\beta$. However, $P_{2}^{+}\leq P^{+}_{1}$ was 
used in obtaining Eqs.~(\ref{eq:512}) and (\ref{eq:513}) and thus here we 
use $\beta$=$\beta_{-}$ ($0\leq\beta\leq 1$) in spacelike region. 
As shown in Eqs.~(\ref{eq:511})-(\ref{eq:513}), the sum of valence 
(Fig. 5.2(b)) and nonvalence (Fig. 5.2(c)) diagrams is equivalent to the 
covariant triangle diagram in Fig. 5.2(a).

For the analysis of singularity structures, we integrate over
${\bf k}_{\perp}$ and obtain from the valence contribution(Fig. 5.2(b));
\be\label{eq:516}
j^{+}_{b}(0)= \pi{\cal N}\int^{1}_{0}dx
\frac{\beta x(1+\beta-2\beta x)}{\tilde{{\cal E}}_{3}^{b}
-\tilde{{\cal E}}_{2}^{b}}
\ln\biggl(\frac{\tilde{{\cal E}}_{2}^{b}}{\tilde{{\cal E}}_{3}^{b}}\biggr),
\ee
where $\tilde{{\cal E}}_{3}^{b}$=$x(1-x)M^{2} -[xm_{q}^{2} +
(1-x)m_{\bar{Q}}^{2}]$
and $\tilde{{\cal E}}_{2}^{b}$=$\beta x(1-\beta x)M^{2}
- [\beta xm_{q}^{2} + (1-\beta x)m_{\bar{Q}}^{2}]$.
It turns out that Eq.~(\ref{eq:516}) has no singularities because
$\tilde{{\cal E}}_{3}^{b}$$\neq$0, $\tilde{{\cal E}}_{2}^{b}$$\neq$0,
and $\tilde{{\cal E}}_{3}^{b}-\tilde{{\cal E}}_{2}^{b}$$\neq$0
for the entire $q^{2}$ region.
On the other hand, the $k_{\perp}$ integration for the
current $j^{+}_{c}(0)$ in Eq.~(\ref{eq:513}) yields
\be\label{eq:517}
j^{+}_{c}(0)= \pi{\cal N}\int^{1}_{0}d{\cal X} 
\frac{(1-\beta)^{2}{\cal X}(2{\cal X}-1)}{\tilde{{\cal E}}_{2}^{c}
-\tilde{{\cal E}}_{3}^{c}}
\ln\biggl(\frac{\tilde{{\cal E}}_{3}^{c}}{\tilde{{\cal E}}_{2}^{c}}\biggr),
\ee
where $x$=$1-(1-\beta){\cal X}$,
$\tilde{{\cal E}}_{3}^{c}$=$(1-\beta){\cal X}[1-(1-\beta){\cal X}]M^{2}-
[(1-\beta){\cal X}m_{\bar{Q}}^{2} + (1 - (1-\beta){\cal X})m_{q}^{2}]$ and
$\tilde{{\cal E}}_{2}^{c}$=$-(1-\beta)^{2}{\cal X}(1-{\cal X})M^{2}/\beta
-m_{q}^{2}$.
While $\tilde{{\cal E}}_{2}^{c}$ corresponding to the energy denominator
of the two particle intermediate state does not vanish both in spacelike 
and timelike region, $\tilde{{\cal E}}_{3}^{c}$ of the
three-particle intermediate state is not zero only for the
spacelike momentum transfer region. For the timelike region,
$\tilde{{\cal E}}_{3}^{c}$ can be zero so that
the singularities start at $q^{2}_{\rm min}$=$4m^{2}_{q(\bar{Q})}$ for
$\gamma^{*} q\bar{q}(\gamma^{*} Q\bar{Q})$ vertex. The singularity 
structure of $\tilde{{\cal E}}_{2}^{c}-\tilde{{\cal E}}_{3}^{c}$ 
in Eq.~(\ref{eq:517}) is the same as in the case of timelike form factor 
(Section 5.4), following the condition of a $q\bar{Q}$ bound state.

The EM form factor ${\cal F}(q^{2})$ in Eq.~(\ref{eq:510}) of a 
$q\bar{Q}$ bound state in $q^{+}$$\neq$0 frame is then obtained by
\bea\label{eq:518}
{\cal F}(q^{2},q^{+}\neq0)&=&
\frac{\pi{\cal N}}{1+\beta}\int^{1}_{0}dx
\biggl\{\frac{\beta x(1+\beta-2\beta x)}{\tilde{{\cal E}}_{3}^{b}
-\tilde{{\cal E}}_{2}^{b}}
\ln\biggl(\frac{\tilde{{\cal E}}_{2}^{b}}{\tilde{{\cal E}}_{3}^{b}}\biggr)
\nonumber\\
&+& \frac{(1-\beta)^{2}x(2x-1)}{\tilde{{\cal E}}_{2}^{c}
-\tilde{{\cal E}}_{3}^{c}}
\ln\biggl(\frac{\tilde{{\cal E}}_{3}^{c}}{\tilde{{\cal E}}_{2}^{c}}\biggr)
\biggr\},
\eea
where ${\tilde{{\cal E}}_{2}^{b}},{\tilde{{\cal E}}_{3}^{b}},
{\tilde{{\cal E}}_{2}^{c}}$ and ${\tilde{{\cal E}}_{2}^{c}}$ are defined
in Eqs.~(\ref{eq:516}) and (\ref{eq:517}). Here, $\beta$ is a function of 
$q^2$.  According to the analytic continuation, the sign of $q^{2}$
in Eq.~(\ref{eq:518}) must be changed from $-$ to $+$ for the timelike region.
\subsection{ The Drell-Yan-West ($q^{+}=0$) frame}
In Drell-Yan-West frame, $q^{+}=0$, $q^{2}=-q^{2}_{\perp}$, 
and ${\bf P}_{1\perp}$=${\bf P}_{2\perp}$=0, the `$+$' component of the 
current has only the valence contribution, i.e., $j^{+}_{b}(0)$ in
Eq.~(\ref{eq:512}). The current $j^{+}_{b}(0)$ in $q^{+}$=0 frame is given by
\be\label{eq:519}
j^{+}_{b}(0)= {\cal N}\int^{1}_{0}dx d^{2}\ell_{\perp}
\frac{2x(1-x)}{({\cal A}-\ell^{2}_{\perp}-\xi^{2})^{2}
-4\xi^{2}\ell^{2}_{\perp}\cos^{2}\phi},
\ee
where ${\bf \ell}_{\perp}$=${\bf k}_{\perp}+x{\bf q}_{\perp}/2$,
${\cal A}$=$x(1-x)M^{2}-[xm^{2}_{q} + (1-x)m^{2}_{\bar{Q}}]$, and
$\xi^{2}$=$x^{2}q^{2}_{\perp}/4$. The angle $\phi$ ($0\leq\phi\leq 2\pi$) is
defined by ${\bf\ell}_{\perp}\cdot{\bf q}_{\perp}$=
$|{\bf\ell}_{\perp}||{\bf q}_{\perp}|\cos\phi$.

Integrating Eq.~(\ref{eq:519}) over $\phi$ and ${\bf\ell}_{\perp}$,
we obtain
\bea\label{eq:520}
j^{+}_{b}(0)&=&
-\frac{8\pi{\cal N}}{\sqrt{q^{2}_{\perp}(4M^{2}+q^{2}_{\perp})}}
\int^{1}_{0}dx\frac{(1-x)}{\sqrt{(x-x_{+})(x-x_{-})}}\nonumber\\
&\times&
\tanh^{-1}\biggl[\sqrt{\frac{q^{2}_{\perp}}{4M^{2}+q^{2}_{\perp}}}
\frac{x}{\sqrt{(x-x_{+})(x-x_{-})}}\biggr],
\eea 
where
\be\label{eq:521}
x_{\pm}=\frac{2(M^{2}-m_{q}^{2}+m_{\bar{Q}}^{2})}{4M^{2}+q^{2}_{\perp}}\pm
\sqrt{\frac{4(M^{2} - m_{q}^{2} + m_{\bar{Q}}^{2})^{2}}
{(4M^{2}+q^{2}_{\perp})^{2}} -
\frac{4m_{\bar{Q}}^{2}}{(4M^{2}+q^{2}_{\perp})} }.
\ee
The analytic continuation from spacelike to timelike region
in the $q^{+}$=0 frame requires the change of
$q_{\perp}$ to $iq_{\perp}$ in Eqs.~(\ref{eq:520}) and (\ref{eq:521}).
We note from Eqs.~(\ref{eq:520}) and (\ref{eq:521}) that the result of
the timelike region exhibits the same singularity structure
as the direct analyses in $q^{+}$$\neq$0 frame, 
i.e., Eqs.~(\ref{eq:56}) and (\ref{eq:58}), even though the nonvalence 
contribution in Fig. 5.2(c) is absent here. 

After some manipulation, we obtain the EM form factor of a
$q\bar{Q}$ bound state in the $q^{+}$=0 frame as follows 
\bea\label{eq:522}
{\cal F}(q^{2},q^{+}=0)&=&-\frac{4\pi{\cal N}}{\sqrt{q^{2}(q^{2}-4M^{2})}}
\int^{\arcsin(\frac{1-a}{b})}_{\arcsin(-\frac{a}{b})}
d\theta (1-a-b\sin\theta)\nonumber\\
&\times&
\tanh^{-1}\biggl[\sqrt{\frac{q^{2}}{q^{2}-4M^{2}}}
\frac{a+b\sin\theta}{ib\cos\theta}\biggr],
\eea 
where $a=(x_{+}+x_{-})/2$, $b=(x_{+}-x_{-})/2$, and 
$q^{2}$=$-q^{2}_{\perp}$.  While the representations 
in Eqs.~(\ref{eq:518}) and (\ref{eq:522}) look apparently different,
it is amazing to realize that the two formulas 
(Eqs.~(\ref{eq:518}) and (\ref{eq:522})) turn
out to be actually identical. As we will show explicitly in the next
section of numerical calculations, all three results of 
Eqs.~(\ref{eq:59}), (\ref{eq:518}) and (\ref{eq:522})
indeed coincide exactly in the entire $q^2$ range.
\section{Numerical Results}
For our numerical analysis of $\pi$, $K$, and $D$ meson form factors,
we use the physical meson masses together with the following constituent
quark and antiquark masses: $m_{u}$=$m_{d}$=0.25 GeV, $m_{s}$=0.48 GeV, and
$m_{c}$=1.8 GeV~\cite{Mix,Kaon,Semi}.
Since our numerical results of the EM form factors obtained from 
Eqs.~(\ref{eq:59}), (\ref{eq:518}) and (\ref{eq:522}) 
turn out to be exactly same with each other for the entire
$q^{2}$ region, only a single line is depicted in 
Figs. 5.3, 5.4 and 5.5 for the form factor calculations of
$\pi,K$, and $D$ mesons, respectively.

In should be noted from our constituent masses
that $M^{2}< m_{q}^{2} + m_{\bar{Q}}^{2}$ for $\pi$ and $K$ and
$M^{2}> m_{q}^{2} + m_{\bar{Q}}^{2}$ for $D$ meson cases.
As discussed in Ref.~\cite{Gasio} for the analysis of the
one-particle matrix element of a scalar current,
the sigularity for
$M^{2}> m_{q}^{2} + m_{\bar{Q}}^{2}$ case starts at
\be\label{eq:523}
q^{2}_{\rm min}=\frac{1}{m^{2}_{\bar{Q}(q)}}[ m^{2}_{q(\bar{Q})} -
(M - m_{\bar{Q}(q)})^{2}][ (M + m_{\bar{Q}(q)})^{2}
- m^{2}_{q(\bar{Q})}],
\ee
for $\gamma^{*} q\bar{q}(\gamma^{*} Q\bar{Q})$ vertex,
while the singularity for
$M^{2}< m_{q}^{2} + m_{\bar{Q}}^{2}$ case starts on the positive $q^{2}$-axis
at the threshold point $q^{2}_{\rm min}$=$4m_{q(\bar{Q})}^{2}$ for
$\gamma^{*} q\bar{q}(\gamma^{*} Q\bar{Q})$ vertex.
Our numerical results exhibit all of these threshold
behaviors coming from the normal ($\pi,K$) and anomalous ($D$) cases. 
As a consistency check, we also compare 
our numerical results of the form factor $F(q^{2})= {\rm Re}\;
F(q^{2}) + i\;{\rm Im}\; F(q^{2})$ with the dispersion relations given by
\begin{eqnarray}\label{dispersion}
{\rm Re}\;F(q^{2})&=&\frac{1}{\pi}P\int^{\infty}_{-\infty}
\frac{{\rm Im}\;F(q'^{2})}{q'^{2}-q^{2}}dq'^{2},\\
{\rm Im}\;F(q^{2})&=&-\frac{1}{\pi}P\int^{\infty}_{-\infty}
\frac{{\rm Re}\;F(q'^{2})}{q'^{2}-q^{2}}dq'^{2},
\end{eqnarray}
where $P$ indicates the Cauchy principal value.

\begin{figure}
\centerline{\psfig{figure=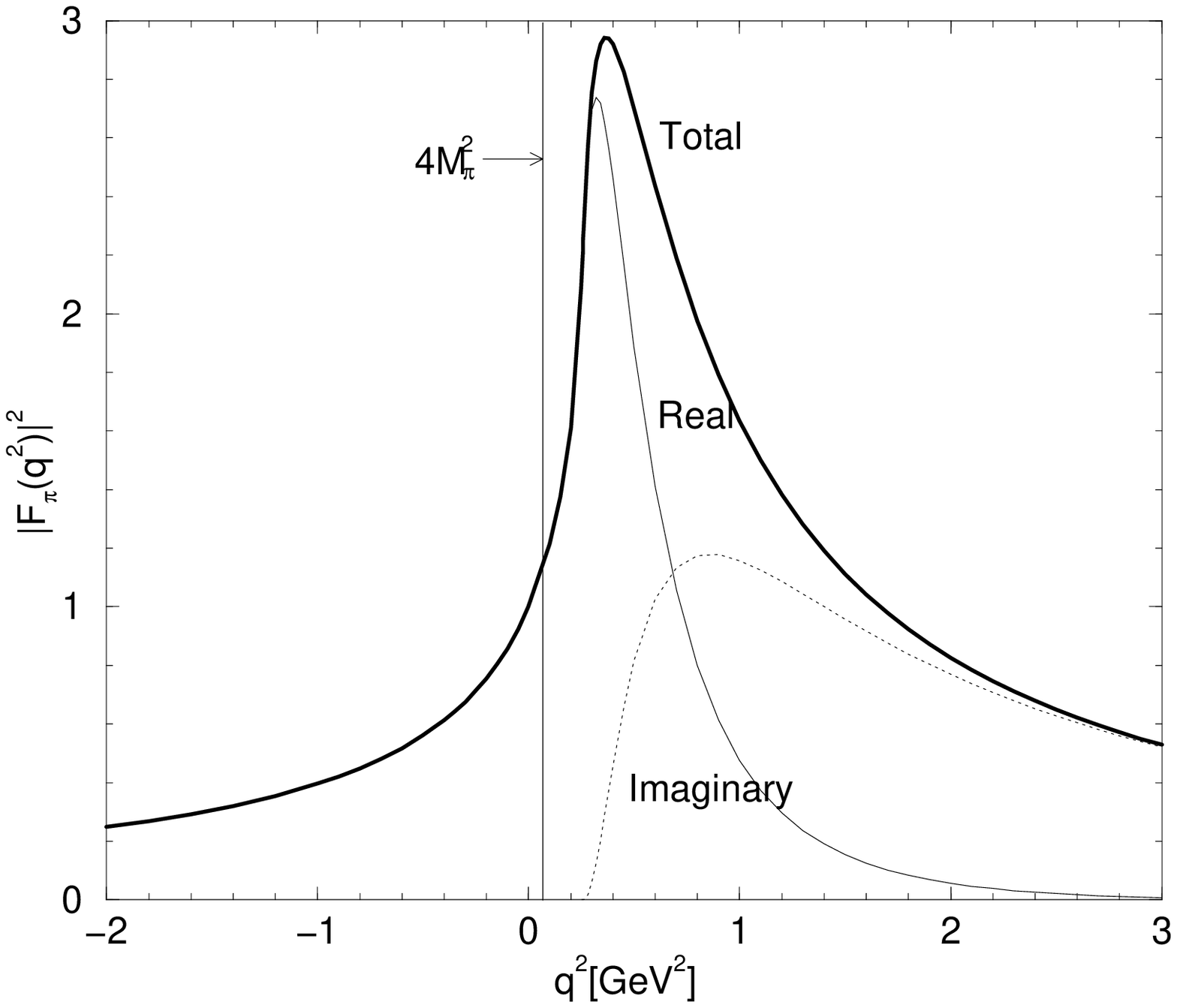,width=3.5in,height=3.2in}}
\centerline{\psfig{figure=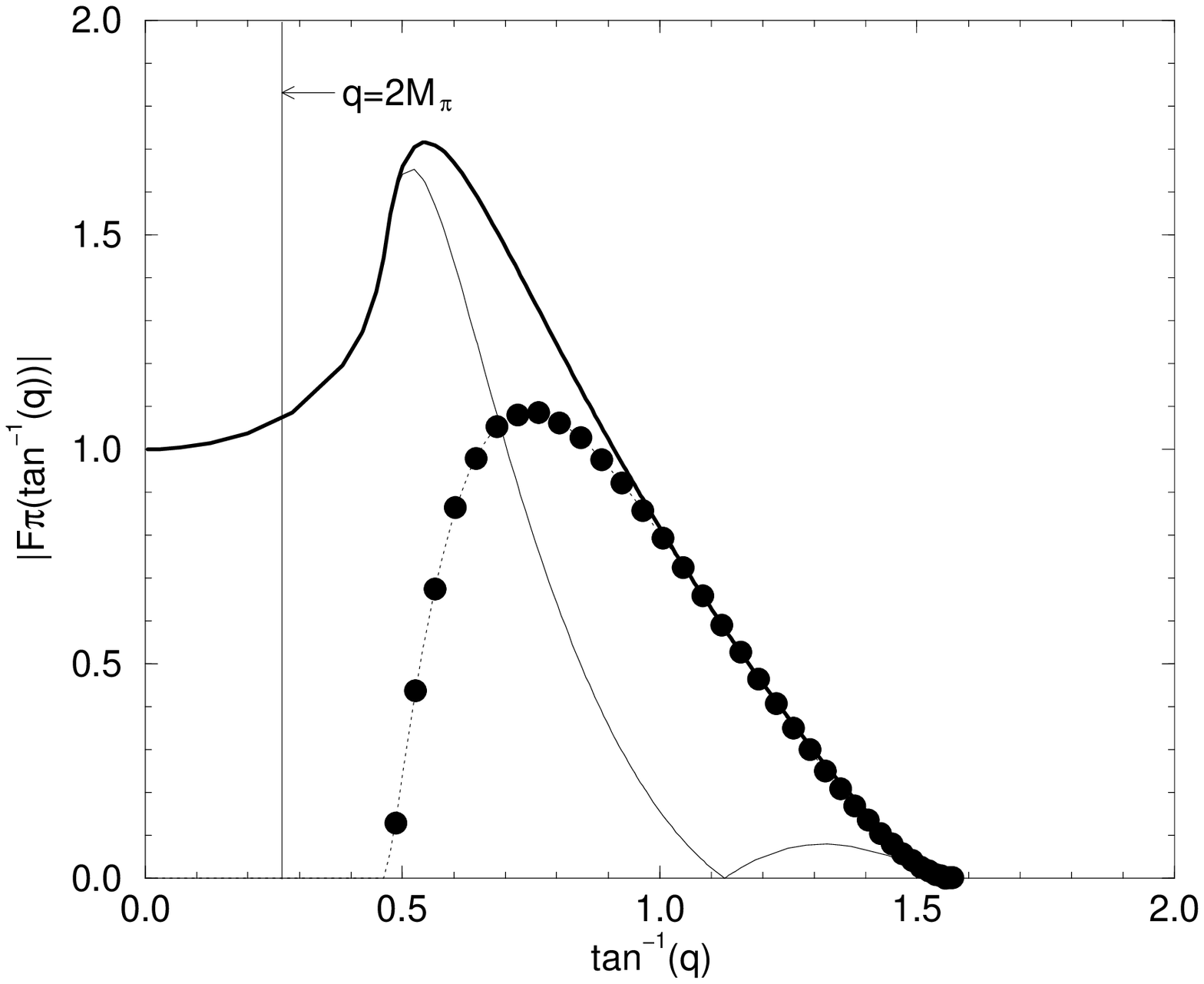,width=3.5in,height=3.2in}}
\caption{ (a) The electromagnetic form fator of the pion in $(3+1)$ 
dimensional scalar field theory for $-2\leq q^{2}\leq 3$ GeV$^{2}$. The 
total, real, and imaginary parts of $|F_{\pi}(q^{2})|^{2}$ are represented
by thick solid, solid, and dotted lines, respectively.
(b) The electromagnetic form fator of the pion in $(3+1)$ dimensional
scalar field theory for the entire timelike region compared to the
dispersion relations (data of black dots) given by Eq.~(5.25).
The same line code as in (a) is used.}
\end{figure}
In Fig. 5.3(a), we show the EM form factor of the pion for
$-2\;{\rm GeV}^{2}\leq q^{2}\leq 3\;{\rm GeV}^{2}$.
The imaginary part (the dotted line)
of the form factor starts at $q^{2}_{\rm min}$=$4m^{2}_{u(d)}$=0.25
GeV, which is consistent with the condition for
$M^{2}< m_{q}^{2} + m_{\bar{Q}}^{2}$ case. It is interesting to note that
the square of the total form factor $|F_{\pi}(q^{2})|^{2}$ (thick solid line)
produces a $\rho$ meson-type peak near $q^{2}\sim M^{2}_{\rho}$.
However, it is not yet clear if this model indeed reproduces all the
features of the vector meson dominance (VMD) phenomena. Even though
the generated position of peak is consistent with VMD, the final state
interaction is not included in this simple model calculation.
We believe that much more complex mechanisms may be necessary to
reproduce the realistic VMD phenomena. More detailed analysis along this
line is under consideration. Nevertheless, it is remarkable that this
simple model is capable of generating the peak and the position
of peak is quite consistent with the VMD.

In Fig. 5.3(b), we show the timelike form factor of the pion for the entire
$q^{2}>0$ region and compare the imaginary part of our direct calculations
(dotted line) obtained from 
Eqs.~(\ref{eq:59}), (\ref{eq:518}), and (\ref{eq:522}) with the result
(data of black dots) obtained from the dispersion relations given by
Eq.~(5.25). Our direct calculation is in an excellent agreement
with the solution of the dispersion relations. Our result for the real part
are also confirmed to be in complete agreement with the dispersion
relations. For high $q^{2}$ region, the
imaginary part of the form factor is dominant over the real part
(thin solid line).

\begin{figure}
\centerline{\psfig{figure=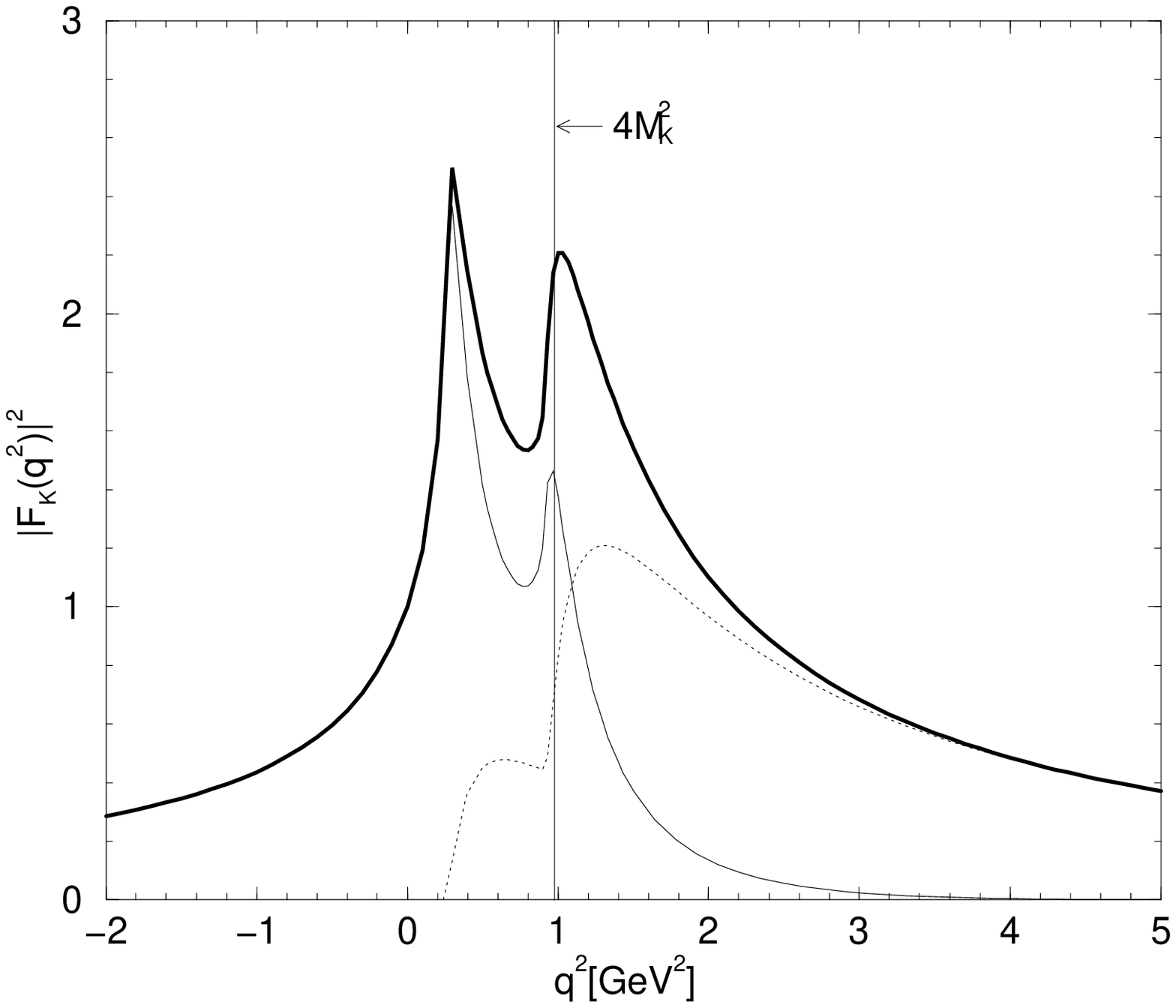,width=3.5in,height=3.2in}}
\centerline{\psfig{figure=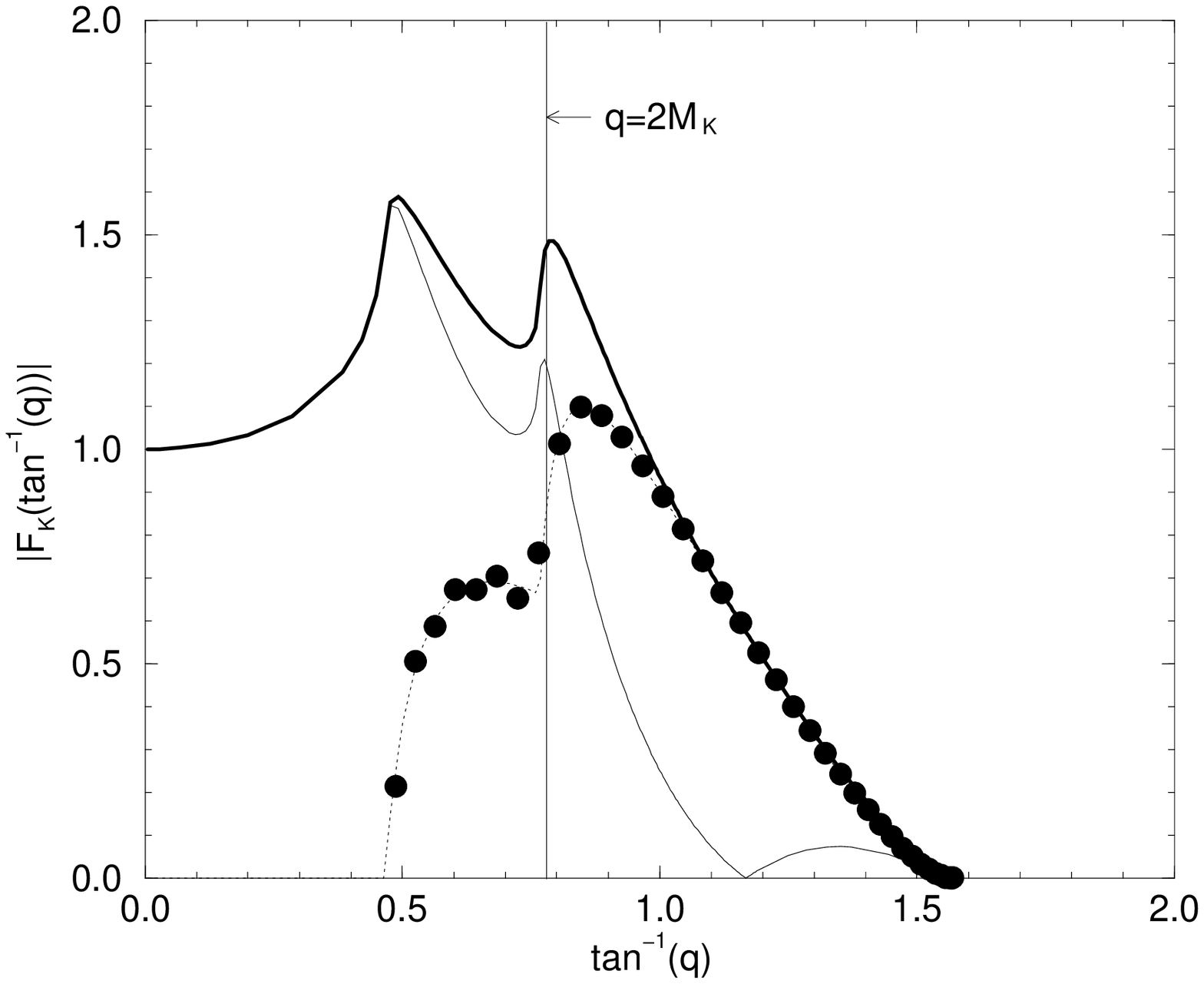,width=3.5in,height=3.2in}}
\caption{The electromagnetic form fator of the kaon in $(3+1)$ dimensional
scalar field theory for $-2\leq q^{2}\leq 5$ GeV$^{2}$. 
The same line code as in Fig. 5.3(a) is used.
(b) The electromagnetic form fator of the kaon in $(3+1)$ dimensional
scalar field theory for the entire timelike region compared to the
dispersion relations (data of black dots) given by Eq.~(5.25).
The same line code as in Fig. 5.3(a) is used.}
\end{figure}
In Fig. 5.4(a), we show the kaon form factor for
$-2\;{\rm GeV}^{2}\leq q^{2}\leq 5\;{\rm GeV}^{2}$. The kaon also has
the normal singularity.
However, it has two thresholds for the imaginary parts; one is
$q^{2}_{\rm min}$=$4m^{2}_{u}$ and the other is
$q^{2}_{\rm min}$=$4m^{2}_{s}$.
These lead to the humped shape (dotted line) of the
imaginary part shown in Fig. 5.4(a).
While we have in principle two vector-meson-type peaks (i.e.
$\rho$ and $\phi$), one can see in Fig. 5.4(a) only
$\phi$ meson-type peak for the timelike kaon EM form factor above
the physical threshold at $q^{2}_{\rm min}$=$4M^{2}_{K}$.
We also show in Fig. 5.4(b) the imaginary part from our direct calculation 
is in an excellent agreement with the result (data of black dots)
from the dispersion relations
for the entire timelike $q^{2}$ region. Again, the imaginary part is
predominant for high $q^{2}$ region.

\begin{figure}
\centerline{\psfig{figure=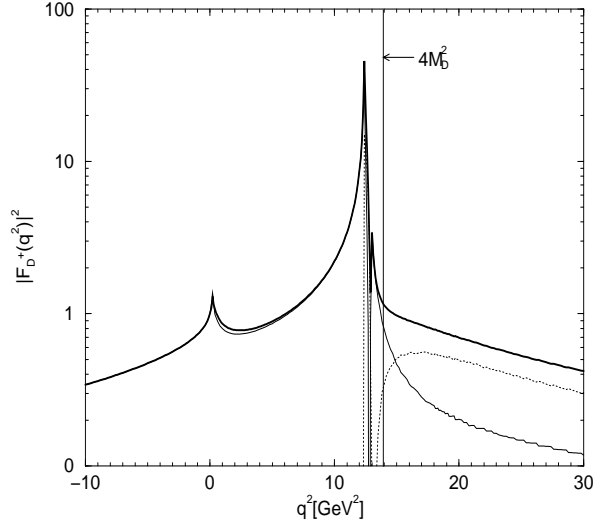,width=3.5in,height=3.2in}}
\caption{The electromagnetic form fator of the $D$ meson in $(3+1)$ 
dimensional scalar field theory for $-10\leq q^{2}\leq 30$ GeV$^{2}$. 
The same line code as in Fig. 5.3(a) is used.}
\end{figure}
In Fig. 5.5, we show the $D$ meson form factor for
$-10\;{\rm GeV}^{2}\leq q^{2}\leq 30\;{\rm GeV}^{2}$.
Unlike the normal threshold of $\pi$ and $K$ form factor calculations,
the $D$ meson form factor shows anomalous thresholds according to 
Eq.~(\ref{eq:523}), i.e., $q^{2}_{\rm min}\sim 0.24$ GeV$^{2}$ 
(compared to $4m^{2}_{d}$=0.25 GeV$^{2}$ for normal case) and 
$q^{2}_{\rm min}\sim 12.4$ GeV$^{2}$
(compared to $4m^{2}_{c}$=12.96 GeV$^{2}$ for normal case) for the
$\gamma^{*}-\bar{d}$ and $\gamma^{*}-c$ vertices, respectively.
Similar to the kaon case in Fig. 5.4, we also have
two unphysical peaks, i.e., $\rho$ and $J/\psi(1S)$ meson type peaks due
to $\bar{d}$ and $c$ quarks, respectively.
However, the timelike form factor of $D$ meson has no pole
structure for the the physical $q^{2}\geq 4M^{2}_{D}$ region.
In all of these figures (Figs. 5.3-5.5), it is astonishing that the
numerical result of Eq.~(\ref{eq:522})
obtained from $q^{+}$=0 frame without encountering the nonvalence diagram
coincides exactly with the numerical results of 
Eqs.~(\ref{eq:59}) and (\ref{eq:518}) obtained from $q^{+}$$\neq$0 frame.

\section{Summary and Discussion}
In this chapter, we presented the investigation on the EM form factors 
of various mesons
both in spacelike and timelike regions using an exactly solvable model
of $(3+1)$ dimensional scalar field theory interacting with gauge fields.
Our calculations demonstrated that one can compute the timelike
form factor without encountering the nonvalence contributions.
We calculated the form factor in spacelike region using the Drell-Yan-West
($q^+$=0) frame and showed that its analytic continuation to the
timelike region reproduces exactly the direct result of
timelike form factor obtained in the longitudinal momentum 
($q^{+}$$\neq$0 and ${\bf q}_{\perp}$=0) frame. 
It is remarkable that the analytic
continuation of the result in Drell-Yan-West frame to the timelike
region automatically generate the effect of the nonvalence contributions
to the timelike form factor.
Another interesting result in our model calculations is that
the peaks analogous to the VMD were generated and the position
of peaks were indeed quite consistent with the VMD.
Even though much more detailed analyses including the final state
interaction may be necessary to reproduce the entire feature of VMD, our
results seem pretty encouraging for further investigations.
Using the dispersion relations, we have also confirmed that
our numerical results of imaginary parts start at
$q^{2}_{\rm min}$=$4m_{q(\bar{Q})}$ and
the normal thresholds appear for
$\pi$ and $K$ ($M^{2}<m^{2}_{q} +m^{2}_{\bar{Q}}$) systems
while the anomalous threshold exists for $D$ ($M^{2}>m^{2}_{q}
+m^{2}_{\bar{Q}}$) system.
Thus, it is hopeful that one can use the same technique of analytic
continuation and calculate the timelike form factors in more
realistic models. Detailed anlysis along this line is underway.
Applications to the semileptonic decay processes in $(3+1)$ dimensional
scalar field model are also in progress.


%% file: zm.tex
\newpage
\setcounter{equation}{0}
\setcounter{figure}{0}
\renewcommand{\theequation}{\mbox{6.\arabic{equation}}}
\chapter{Nonvanishing Zero Modes in the Light-Front Current}
One of the distinguishing features in LF quantization is the
rational energy-momentum dispersion relation which gives a sign
correlation between the LF energy($P^{-}$) and the LF 
longitudinal momentum($P^{+}$). In the old-fashioned time-ordered perturbation
theory~\cite{Brodsky}, this sign correlation allows one to remove the 
so-called ``Z-graphs" such as the diagram of particle-antiparticle pair
creation(annihilation) from(to) the vacuum.
As an example, in the theory of scalar fields interacting with gauge
fields~\cite{GS,Sa}, the covariant triangle diagram shown in Fig. 5.2(a)
corresponds to only two LF time-ordered diagrams shown in 
Figs. 5.2(b) and 5.2(c), while in the ordinary time-ordered perturbation 
theory, Fig. 5.2(a) would correspond to the six time-ordered diagrams 
including the ``Z-graphs". Furthermore, the Drell-Yan-West ($q^{+}$=0) frame
may even allow one to remove the diagram shown in Fig. 5.2(c) 
because of the same reasoning from the energy-momentum dispersion relation 
and the conservation of the LF longitudinal momenta at the vertex 
of the gauge field and the two scalar fields.

Based on this idea, the Drell-Yan-West ($q^{+}$=0) frame is frequently used
for the bound-state form factor calculations.
Taking advantage of $q^{+}$=0 frame, one may need to
consider only the valence diagram shown in Fig. 5.2(b), where the 
three-point scalar vertices should be replaced by the LF bound-state
wavefunction.

In this work, however, we point out that even at $q^{+}$=0 frame
one should not overlook the possibility of non-zero contribution from the
nonvalence( pair creation or annihilation) diagram shown in Fig. 5.2(c).
As we will show explicitly in the simple $(1+1)$-dimensional scalar field
theory interacting with gauge fields, the current $J^{-}$ is not immune to
the zero mode contribution shown in Fig. 5.2(c) at $q^{+}$=0. While
the current $J^{+}$ does not have any zero mode contribution from
Fig. 5.2(c), the processes that involve more than one form factor, e.g.,
semileptonic decay processes, require the calculations of more components
of the current other than $J^{+}$ in order to find all the necessary
form factors in $q^{+}$=0 frame.
For instance, in the analysis of the semileptonic decays between
two pseudoscalar mesons, two form factors, $f_{\pm}(q^{2})$, are involved
and one has to use not only $J^{+}$ but also $J^{-}$(or $J^{\perp}$
in $3+1$ dimensions) to obtain both form factors in $q^{+}$=0 frame.
Thus, the zero mode contribution is crucial to obtaining the correct 
results of electroweak form factors.
Only a brief exactly solvable model calculation is provided here. A full,
detailed treatment of $(3+1)$ dimensional semileptonic decay processes
such as $K\to\pi$, $B\to\pi$, $B\to D$ etc. will be presented in a
separate communication.

This Chapter is organized as follows: In Section 6.1,
we describe the general formalism of the semileptonic decay form
factors for non-zero momentum transfer in ($1+1$)-dimensions and then
discuss the zero mode problem in the limiting cases of the form factors
as $q^{+}\to 0$. The numerical results of the
zero mode contributions for $K\to\pi$, $B\to\pi$, and $B\to D$ transitions
are given in Section 6.2. We also briefly discuss in Section 6.2 the zero
modes in the electromagnectic form factor calculations. 
The summary and discussion follow in Section 6.3.

\section{Zero modes in $0^-\to0^-$ Transitions}
The semileptonic decay of a $Q_{1}\bar{q}$ bound state into
another $Q_{2}\bar{q}$ bound state is governed by the weak current, viz.,
\be\label{eq:61}
J^{\mu}(0)=\la P_{2}|\bar{Q_{2}}\gamma^{\mu}Q_{1}|P_{1}\ra=
f_{+}(q^{2})(P_{1}+P_{2})^{\mu} + f_{-}(q^{2})(P_{1}-P_{2})^{\mu},
\ee
where $P_{2}$=$P_{1}-q$ and the non-zero momentum transfer square
$q^{2}$=$q^{+}q^{-}$ is timelike, i.e., $q^{2}$=$[0,(M_{1}-M_{2})^{2}]$.
One can easily obtain $q^{2}$ in terms of the fraction $\alpha$ as follows
\be\label{eq:62}
q^{2}= (1-\alpha)(M^{2}_{1}-\frac{M^{2}_{2}}{\alpha}),
\ee
where $\alpha$=$P^{+}_{2}/P^{+}_{1}$=$1-q^{+}/P^{+}_{1}$.
Accordingly, the two solutions for $\alpha$ are given by
\be\label{eq:63}
\alpha_{\pm}=\frac{M_{2}}{M_{1}}\biggl[
\frac{ M^{2}_{1}+M^{2}_{2}-q^{2}}{2M_{1}M_{2}}
\pm \sqrt{\biggl(\frac{ M^{2}_{1}+M^{2}_{2}-q^{2}}
{2M_{1}M_{2}}\biggr)^{2}-1} \biggr].
\ee
The $+(-)$ sign in Eq.~(6.3) corresponds to the daughter meson
recoiling in the positive(negative) $z$-direction relative to
the parent meson. At zero recoil($q^{2}$=$q^{2}_{\rm max}$) and
maximum recoil($q^{2}$=0), $\alpha_{\pm}$ are given by
\bea\label{eq:64}
\alpha_{+}(q^{2}_{\rm max})&=&
\alpha_{-}(q^{2}_{\rm max})=\frac{M_{2}}{M_{1}},
\nonumber\\
\alpha_{+}(0) &=& 1,\;\;
\alpha_{-}(0)=\biggl(\frac{M_{2}}{M_{1}}\biggr)^{2}.
\eea 
In order to obtain the form factors $f_{\pm}(q^{2})$ which are
independent of $\alpha_{\pm}$, we can define
\begin{eqnarray}
\la P_{2}|\bar{Q_{2}}\gamma^{\mu}Q_{1}|P_{1}\ra|_{\alpha
=\alpha_{\pm}}&\equiv&
2P_{1}^{+}H^{+}(\alpha_{\pm})\hspace{.2cm}{\rm for}\hspace{.2cm}\mu=+,\\
&\equiv& 2\biggl(\frac{M^{2}_{1}}{P^{+}_{1}}\biggr)H^{-}(\alpha_{\pm})
\hspace{.2cm}{\rm for}\hspace{.2cm}\mu=-,
\end{eqnarray}
and obtain from Eq.~(6.1)
\begin{eqnarray}
f_{\pm}(q^{2})&=&\pm \frac{(1\mp \alpha_{-})H^{+}(\alpha_{+}) -
(1\mp \alpha_{+})H^{+}(\alpha_{-})}{\alpha_{+}-\alpha_{-}}
\hspace{.2cm}{\rm for}\hspace{.2cm}\mu=+,\\
&=&\pm \frac{(1\mp \beta_{-})H^{-}(\alpha_{+}) -
(1\mp \beta_{+})H^{-}(\alpha_{-})}{\beta_{+}-\beta_{-}}
\hspace{.2cm}{\rm for}\hspace{.2cm}\mu=-,
\end{eqnarray}
where $\beta_{\pm}$=$\alpha_{-}(0)/\alpha_{\pm}$.

Now, the current $J^{\mu}(0)$ obtained from the covariant triangle
diagram of Fig.~5.2(a) is given by
\be
J^{\mu}(0)=\int d^{2}k\frac{1}{ (P_{1}-k)^{2}-m^{2}_{1}+i\epsilon}
(P_{1}+P_{2}-2k)^{\mu}\frac{1}{ (P_{2}-k)^{2}-m^{2}_{2}+i\epsilon}
\frac{1}{ k^{2}-m^{2}_{\bar{q}}+i\epsilon}.
\ee
From this, we obtain for the ``$\pm$"-components of the current $J^{\mu}(0)$ 
as
\be
J^{\pm}(0)=-2\pi i(I^{\pm}_{1}+I^{\pm}_{2}),
\ee
where $I^{\pm}_{1}$ and $I^{\pm}_{2}$ corresponding to diagrams 
Figs.~5.2(b) and 5.2(c), respectively, are given by
\be
I^{+}_{1}(\alpha)=\int^{\alpha}_{0}dx\frac{1-2x+\alpha}{x(1-x)(\alpha-x)
\biggl(M^{2}_{1}-\frac{m^{2}_{1}}{1-x}-\frac{m^{2}_{\bar{q}}}{x}
\biggr)\biggl(\frac{M^{2}_{2}}{\alpha}-\frac{m^{2}_{2}}{\alpha-x}-
\frac{m^{2}_{\bar{q}}}{x}\biggr )}, 
\ee
\be
I^{+}_{2}(\alpha)=\int^{1}_{\alpha}dx \frac{1-2x+\alpha}{x(1-x)(\alpha-x)
\biggl(M^{2}_{1}-\frac{m^{2}_{1}}{1-x}-\frac{m^{2}_{\bar{q}}}{x}
\biggr) \biggl(\frac{M^{2}_{2}}{\alpha}+\frac{m^{2}_{1}}{1-x}-
\frac{m^{2}_{2}}{\alpha-x}-M^{2}_{1}\biggr )},
\ee
and
\be
I^{-}_{1}(\alpha)=\int^{\alpha}_{0}dx \frac{M^{2}_{1}+M^{2}_{2}/\alpha
-2m^{2}_{\bar{q}}/x}{x(1-x)(\alpha-x)
\biggl(M^{2}_{1}-\frac{m^{2}_{1}}{1-x}-\frac{m^{2}_{\bar{q}}}{x}
\biggr)\biggl(\frac{M^{2}_{2}}{\alpha}-\frac{m^{2}_{2}}{\alpha-x}-
\frac{m^{2}_{\bar{q}}}{x}\biggr)},
\ee
\be
I^{-}_{2}(\alpha)=\int^{1}_{\alpha}dx \frac{M^{2}_{2}/\alpha -M^{2}_{1}
+2m^{2}_{1}/(1-x)}{x(1-x)(\alpha-x)\biggl(M^{2}_{1}-\frac{m^{2}_{1}}{1-x}-
\frac{m^{2}_{\bar{q}}}{x}\biggr)\biggl(\frac{M^{2}_{2}}{\alpha}
+ \frac{m^{2}_{1}}{1-x}- \frac{m^{2}_{2}}{\alpha-x} - M^{2}_{1}
\biggr)}.
\ee
Note that at zero momentum transfer limit, $q^{2}$=$q^{+}q^{-}\to 0$,
the contributions of $I^{\pm}_{2}(\alpha)$
come from either $lim_{q^{+}\to0}I^{\pm}_{2}(\alpha)$=$
I^{\pm}_{2}(\alpha_{+}(0))$ or $lim_{q^{-}\to0}I^{\pm}_{2}(\alpha)$=$
I^{\pm}_{2}(\alpha_{-}(0))$.
It is crucial to note in $q^{+}$=0 frame that while
$I^{+}_{2}(\alpha_{+}(0))$ vanishes, $I^{-}_{2}(\alpha_{+}(0))$
does not vanish because the integrand has a singularity even though
the region of integration shrinks to zero.
Its nonvanishing term is thus given by
\be\label{zero}
I^{-}_{2}(\alpha_{+}(0))= -\frac{2}{m^{2}_{1}-m^{2}_{2}}{\rm ln}\biggl(
\frac{m^{2}_{2}}{m^{2}_{1}}\biggr).
\ee
This nonvanishing term is ascribed to the term proportional to
$k^{-}$=$P^{-}_{1}-m^{2}_{1}/(P^{+}_{1}-k^{+})$ in Eq.~(6.14), which prevents
Eq.~(6.14) from vanishing in the limit, $\alpha\to 1$.
This is precisely the contribution from ``zero mode" at $q^{+}$=0 frame.
$I^{-}_{2}(\alpha_{+}(0))$ should be distingushed from the other
nonvanishing pair-creation diagrams at
$q^{-}$=0 frame, i.e., $I^{\pm}_{2}(\alpha_{-}(0))$.
Some relevant but different applications of zero modes were discussed
in the literature~\cite{zero,BH,Fred}.

\begin{table}
\centering
\caption{Form factors of $f_{\pm}(0)$ obtained for different zero-momentum
transfer limit, $q^{+}$=0 and $q^{-}$=0. The notations of 
$\alpha_{\pm}$, $\alpha_{p}$, and $\alpha_{m}$
used in table are defined as $\alpha_{\pm}=\alpha_{\pm}(0)$,
$\alpha_{p}$=$1+\alpha_{-}(0)$, and
$\alpha_{m}$=$1-\alpha_{-}(0)$, respectively.
$I^{\pm}_{i}(\alpha_{\pm})$ implies 
$\sum^{2}_{i=1}I^{\pm}_{i}(\alpha_{\pm})$.}
\begin{tabular}{|c|c|c|}\hline
Form factor & $q^{+}=0$ & $q^{-}=0$\\ \hline
$f_{+}(0)$ & $I^{+}_{1}(\alpha_{+})/2$
& $I^{-}_{i}(\alpha_{-})/2M^{2}_{1}$\\ \hline
$f_{-}(0)$ & $[I^{-}_{i}(\alpha_{+})/M^{2}_{1}
-\alpha_{p}I^{+}_{1}(\alpha_{+})/2]/\alpha_{m}$
& $[I^{+}_{i}(\alpha_{-})
-\alpha_{p}I^{-}_{i}(\alpha_{-})/2M^{2}_{1}]/\alpha_{m}$  \\
\hline
\end{tabular}
\end{table}
In Table 6.1, we summarized the form factors $f_{\pm}(0)$
obtained from both currents, $J^{+}$ and $J^{-}$, for different
zero momentum transfer limit, i.e., $q^{+}$=0 or $q^{-}$=0.
As shown in Table 6.1, the nonvalence contributions,
$I^{\pm}_{2}(\alpha_{\pm}(0))$, are separated from the
valence contributions, $I^{\pm}_{1}(\alpha_{\pm}(0))$.
Of special interest, we observed that the form factor $f_{-}(0)$ at
$q^{+}$=0 is no longer free from the zero mode, $I^{-}_{2}(\alpha_{+}(0))$.

\begin{table}
\centering
\caption{ Zero-mode(Z.M.) and nonvalence(N.V.) contributions to the exact 
form factors of $f_{\pm}(0)$ for the semileptonic decays of $K(B)\to\pi$ 
and $B\to D$ in $(1+1)$ dimensions.
We distinguished the zero mode contribution at $q^{+}$=0 from the usual
nonvalence one at $q^{-}$=0.}
\begin{tabular}{|c|c|c|c|c|c|}\hline
Frame& $f^{\rm N.V.(Z.M.)}_{\pm}(0)$/$f^{\rm full}_{\pm}(0)$
& N.V.(Z.M.) factor & $K\to\pi$
& $B\to\pi$ & $B\to D$\\ \hline
$q^{+}=0$& $f^{\rm Z.M.}_{+}(0)/f^{\rm full}_{+}(0)$ & None & 0 & 0 & 0 \\
\cline{2-6}
& $f^{\rm Z.M.}_{-}(0)/f^{\rm full}_{-}(0)$ 
&$\propto I^{-}_{2}(\alpha_{+}(0))$ & 6.9 & 0.03 & 0.1 \\ \hline
$q^{-}=0$& $f^{\rm N.V}_{+}(0)/f^{\rm full}_{+}(0)$ &
$\propto I^{-}_{2}(\alpha_{-}(0))$ & 2.8& 1.3& 0.05 \\ \cline{2-6}
& $f^{\rm N.V1.}_{-}(0)/f^{\rm full}_{-}(0)^{[a]}$ &
$\propto I^{+}_{2}(\alpha_{-}(0))$  & 3.8 & 3.8 & 0.6 \\ \cline{2-6}
& $f^{\rm N.V2.}_{-}(0)/f^{\rm full}_{-}(0)^{[a]}$ &
$\propto I^{-}_{2}(\alpha_{-}(0))$ & $-11.1$ & $-4.0$ & $-1.1$ \\
\hline
\end{tabular}

$^{[a]}$ We show the separate contributions of the nonvalence
terms proportional to
$I^{+}_{2}(\alpha_{-}(0))$ and $I^{-}_{2}(\alpha_{-}(0))$ to the
exact form factor of $f_{-}(0)$ at $q^{-}$=0.
\end{table}
\section{Numerical Results}
To give some quantitative idea how much these nonvalence contributions
$I^{\pm}_{2}(\alpha_{\pm}(0))$ are for a few different decay processes,
we performed model calculations for $K\to\pi$, $B\to\pi$, and $B\to D$
transitions in $(1+1)$ dimensions using rather widely used constituent
quark masses, $m_{u(d)}$=0.25 GeV, $m_{c}$=1.8 GeV,
and $m_{b}$=5.2 GeV. Numerically, we first verified that the form factors,
$f_{+}(0)$ and $f_{-}(0)$, obtained from the $q^{+}$=0 frame are
in fact exactly the same with $f_{+}(0)$ and $f_{-}(0)$ obtained from the
$q^{-}$=0 frame, respectively, once the nonvalence contributions
(including zero mode) are added.
The nonvalence contributions to the form factors
of $f_{\pm}(0)$ at $q^{-}$=0 are also shown in Table 6.2.
In Figs.~6.1a(b)-6.3a(b), the effects of pair-creation (nonvalence)
diagram to the exact form factors are shown for the non-zero momentum
transfer region for the above three decay processes.
Especially, the zero mode contributions $I^{-}_{2}(\alpha_{+}(0))$
to the exact solutions for the $f_{-}(0)$ at
$q^{+}$=0, i.e., $f^{\rm Z.M.}_{-}(0)/f^{\rm full}_{-}(0)$, are
estimated as $6.9$ for $K\to\pi$, $0.03$ for $B\to\pi$, and $0.12$ for
$B\to D$ decays.
The zero mode contributions on $f_{-}(0)$ at $q^{+}$=0
frame are drastically reduced from the light-to-light meson transition to
the heavy-to-light and heavy-to-heavy ones.
This qualitative feature of zero mode effects
on different initial and final states are expected to remain same
even in $(3+1)$ dimensional case, even though the actual quantitative values
must be different from $(1+1)$ dimensional case.
\begin{figure}
\centerline{\psfig{figure=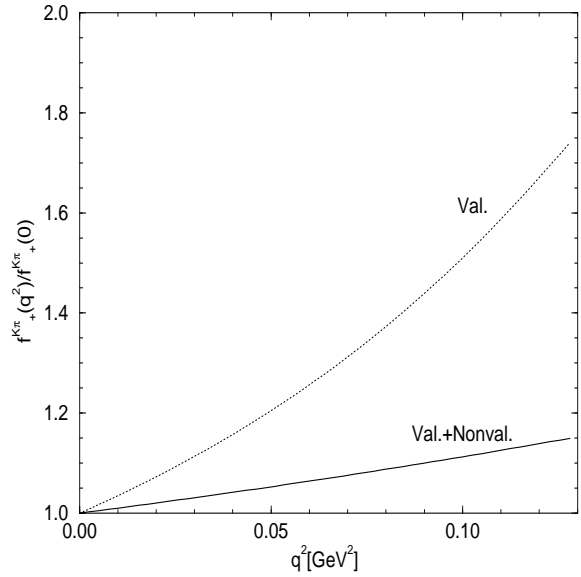,width=3.5in,height=3.5in}}
\centerline{\psfig{figure=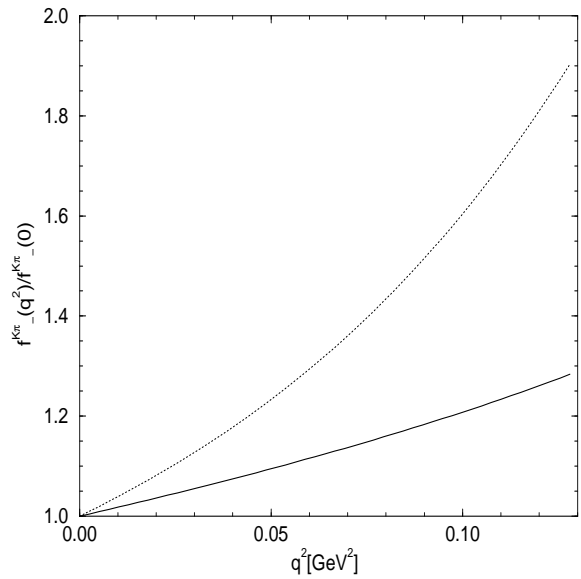,width=3.5in,height=3.5in}}
\caption{(a) Normalized form factor of $f_{+}(q^{2})$ for $K\to\pi$
in $(1+1)$ dimension. The solid line is the result from the
valence plus nonvalence contributions.
The dotted line is the result from the valence contribution.
(b) Normalized form factor of $f_{-}(q^{2})$ for $K\to\pi$
in $(1+1)$ dimension. The same line code as in (a) is used.}
\end{figure}
\begin{figure}
\centerline{\psfig{figure=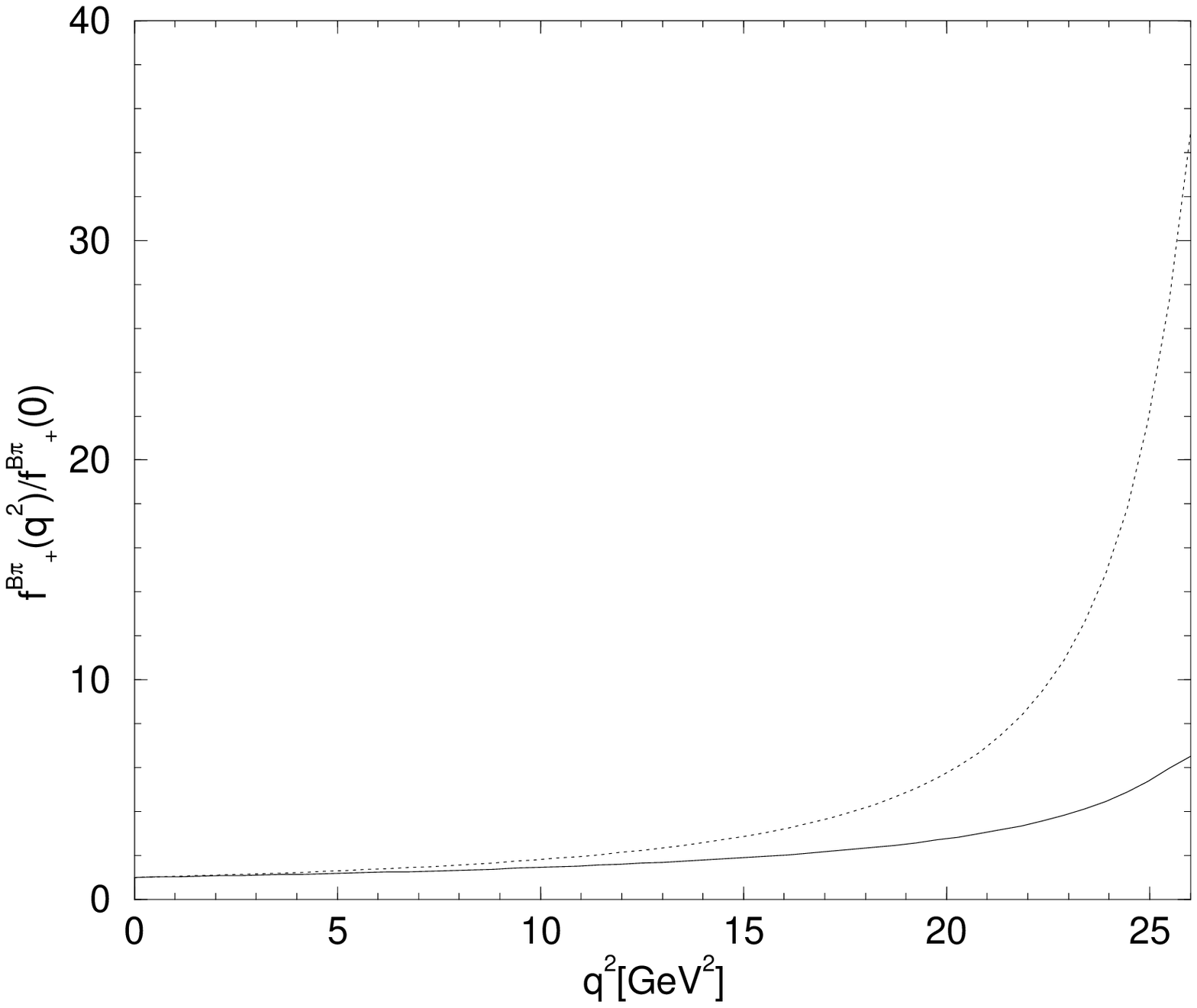,width=3.5in,height=3.5in}}
\centerline{\psfig{figure=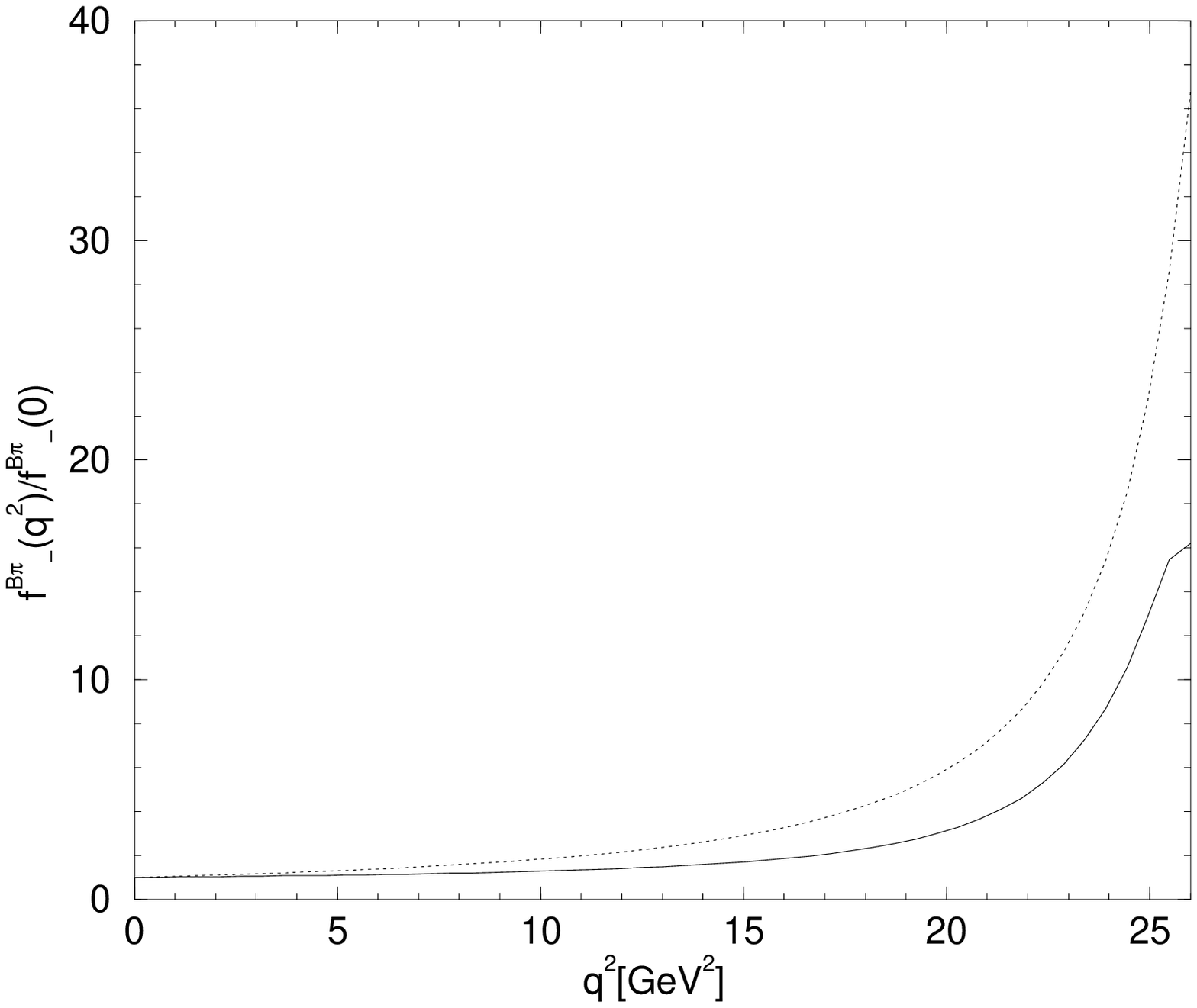,width=3.5in,height=3.5in}}
\caption{(a) Normalized form factor of $f_{+}(q^{2})$ for $B\to\pi$
in $(1+1)$ dimension. The same line code as in Fig. 6.1(a) is used.
(b) Normalized form factor of $f_{-}(q^{2})$ for $B\to\pi$
in $(1+1)$ dimension. The same line code as in Fig. 6.1(a) is used.}
\end{figure}
\begin{figure}
\centerline{\psfig{figure=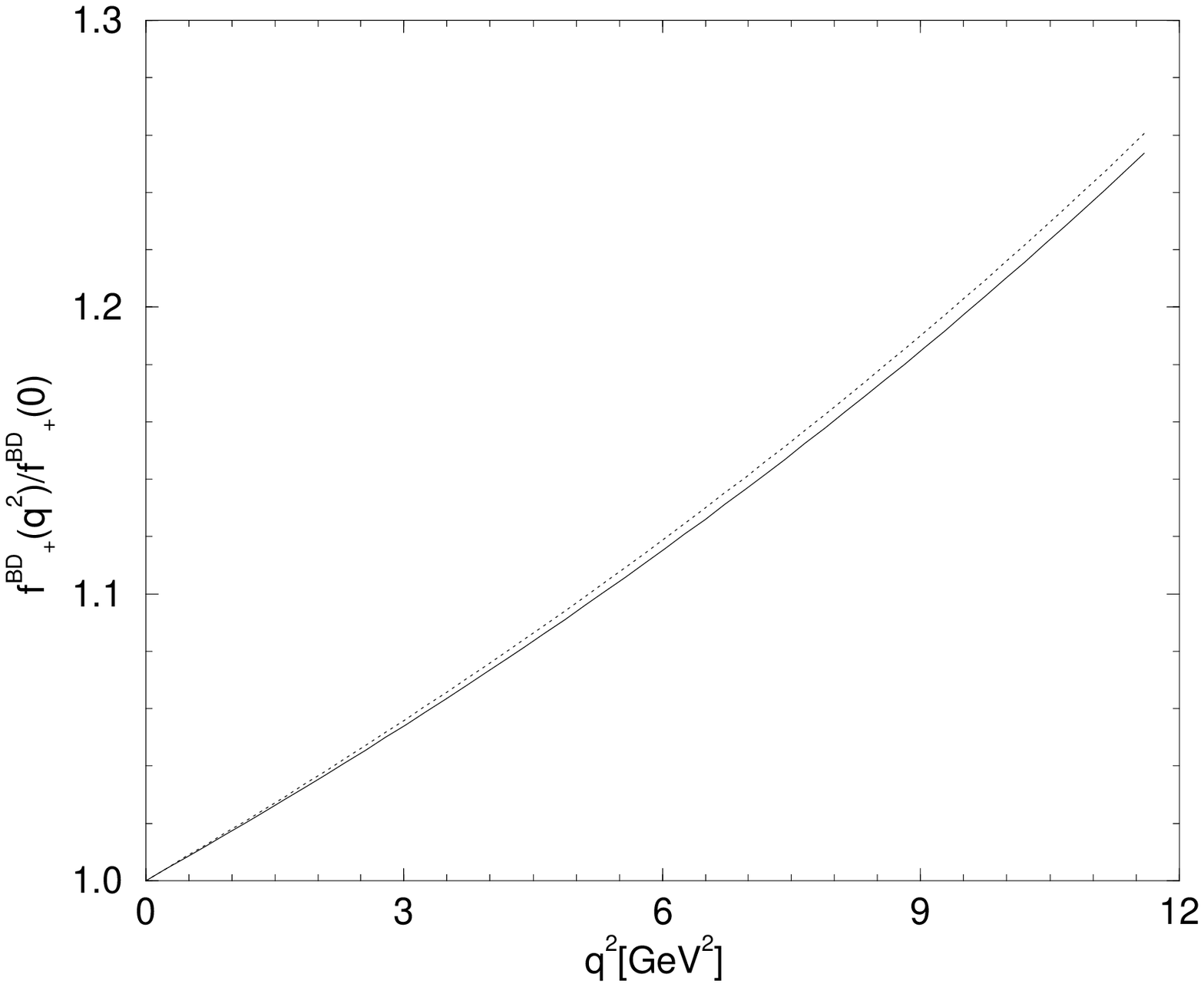,width=3.5in,height=3.5in}}
\centerline{\psfig{figure=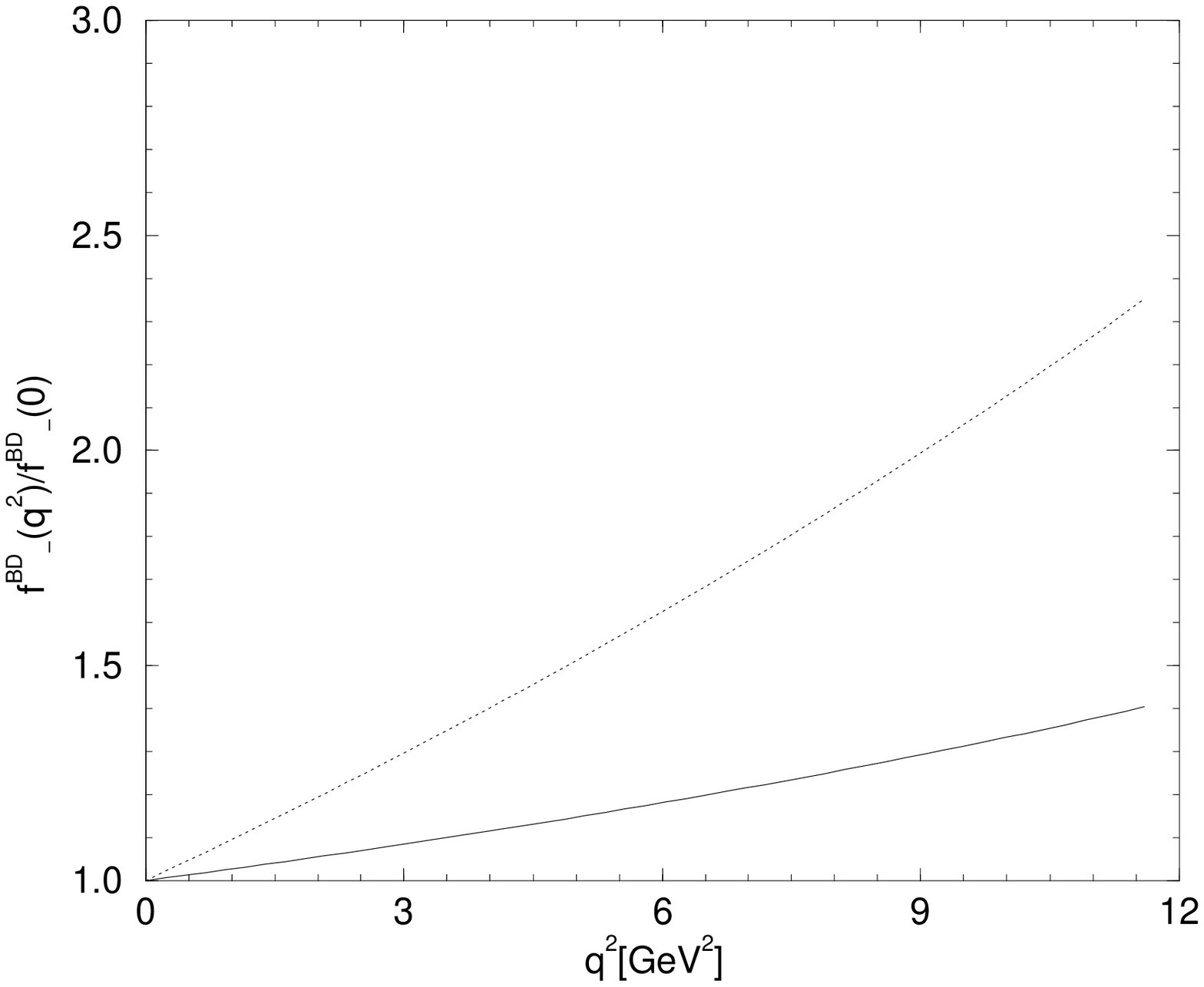,width=3.5in,height=3.5in}}
\caption{(a) Normalized form factor of $f_{+}(q^{2})$ for $B\to D$
in $(1+1)$ dimension. The same line code as in Fig. 6.1(a) is used.
(b) Normalized form factor of $f_{-}(q^{2})$ for $B\to D$
in $(1+1)$ dimension. The same line code as in Fig. 6.1(a) is used.}
\end{figure} 

Furthermore, we have found the effect of zero mode to the EM form factor:
\be
J^{\mu}(0)= (2P_{1}-q)^{\mu}F_{M}(Q^{2}).
\ee
The EM form factor at $q^{+}$=0 using $J^{-}(0)$ current is obtained by
\be\label{emzero}
F_{M}(0)= N\biggl\{\int^{1}_{0}dx\frac{M^{2}-m^{2}_{\bar{q}}/x}
{x(1-x)^{2}\biggl(M^{2}-\frac{m^{2}_{q}}{1-x}-
\frac{m^{2}_{q}}{x}\biggr)^{2}}+ 1/m^{2}_{q}\biggr\},
\ee
where $N$ is the normalization constant and
the $1/m^{2}_{q}$ in Eq.~(\ref{emzero}) is the ``zero mode" term. 
Numerically, using the previous quark masses, the effects of zero modes on 
the form factors of $F_{\pi}(0)$ and $F_{B}(0)$, i.e.,
$F^{\rm Z.M.}_{\pi}(0)/F^{\rm full}_{\pi}(0)$ and
$F^{\rm Z.M.}_{B}(0)/F^{\rm full}_{B}(0)$, are  estimated as
$16.9$ and $0.75$, respectively.
Again, the zero mode contribution is drastically reduced for the heavy
meson form factor. However, it gives a very large effect on the light
meson form factors.
The similar observation on the EM form factor was made in the Breit
frame recently~\cite{Fred}. In ($3+1$) dimensions, however, we note that
the relation between the Breit frame and the Drell-Yan-West frame involves
the transverse rotation in addition to the boost and therefore the results
obtained from the Breit frame cannot be taken as the same with those obtained
from the Drell-Yan-West frame or vice versa.
\section{Summary and Discussion}
In conclusion, we investigated the zero mode effects on the form factors
of semileptonic decays as well as the electromagnetic transition in
the exactly solvable model. Our main observation was the nonvanishing
zero mode contribution to the $J^{-}$ current and our results are
directly applicable to the real $(3+1)$ dimensional calculations.
The effect of zero mode to the $f_{-}(0)$ form factor
is especially important in the application for the physical semileptonic
decays in the Drell-Yan-West ($q^{+}$=0) frame.
To the extent that the zero modes have a significant contribution to
some physical observables as shown in this work, one may even conjecture
that the condensation of zero modes could lead to the nontrivial
realization of chiral symmetry breaking in the LF quantization
approach. The work along this line is in progress.


%% file: Semi.tex
\newpage
\setcounter{equation}{0}
\setcounter{figure}{0}
\renewcommand{\theequation}{\mbox{7.\arabic{equation}}}
\chapter{LFQM Analysis of Exclusive
$0^{-}$$\to$$0^{-}(1^-)$ Semileptonic Meson Decays}
In recent years, the exclusive semileptonic decay processes generated a
great excitement not only in measuring the most accurate values of the
Cabbibo-Kobayashi-Maskawa (CKM) matrix elements but also in testing diverse
theoretical approaches to describe the internal structure of hadrons.
Especially, due to the anticipated abundance of accurate experimental data
from the $B$-factories (e.g. HERA-B at HERA, BaBar at SLAC and Belle at KEK),
the heavy-to-heavy and heavy-to-light meson decays such as 
$B$$\to$$D(D^*)$, $B$$\to$$\pi(\rho)$, $D$$\to$$\pi(\rho)$ etc.
become invaluable processes deserving thorough analysis.
While the available experimental data of heavy meson branching ratios
have still rather large uncertainties~\cite{data}, various theoretical methods
have been applied to calculate the weak decay processes, e.g., lattice
QCD~\cite{Flynn,Flynn2,UK2,ALL,LMMS,Bern2,UKQCD,Bernard,Bowler,Abada}, 
QCD sum rules~\cite{BB1,BB2,Ball,Nar2,BBD}, 
Heavy quark effective theory~\cite{IW}, and quark models
~\cite{Kaon,Semi,isgw,isgw2,Wirbel,Dem,Cheng,OD1,OD2,Melikhov,Mel2,Jaus3}.
In particular, the weak transition form factors determined by the lattice
QCD~\cite{Bernard} provided a useful guidance for the model building of 
hadrons, making definitive tests on existing models, even though the current 
error bars in the lattice data are yet too large to pin down the best 
phenomenological model of hadrons. These weak form factors, however, are 
the essential informations of the strongly interacting quark/gluon structure 
inside hadrons and thus it is very important to analyze these processes 
with the viable model that has been very successful in analyzing other 
processes.

In this work, we report the analysis of exclusive semileptonic decays
of $0^-\to 0^-$ and $0^-\to 1^-$ heavy meson decays  
using our LFQM~\cite{Mix} which has been quite successful 
in the analysis of EM form factors and radiative decays. 
In addition to these heavy meson semileptonic decays, the light-to-light
weak form factor analysis such as $K_{\ell3}$ decays will be disscussed by
comparing with the experiment~\cite{data} as well as many other theoretical
models, e.g., chiral perturbation theory(CPT)~\cite{Roos,Gasser},
the effective chiral Lagrangian approach~\cite{CL}, vector meson
dominance~\cite{VMD}, the extended Nambu-Jona-Lasino model~\cite{Andrei},
Dyson-Schwinger approach~\cite{Yuri} and other quark
models~\cite{isgw2}.

The LFQM takes advantage of the equal LF time ($\tau$=$t+z/c$)
quantization~\cite{BPP} and includes important relativistic effects
in the hadronic wave functions. The distinguished feature of the LF
equal-$\tau$ quantization compared to the ordinary equal-$t$ quantization is
the rational energy-momentum dispersion relation~\cite{spin} which leads to 
the suppression of vacuum fluctuations with the decoupling of complicated
zero modes~\cite{zm,zero,BH,Fred} and the conversion of the dynamical 
problem from boost to rotation~\cite{JS}. 
Moreover, one of the most distinctive advantages in the
LFQM has been the utility of the well-established Drell-Yan-West
($q^{+}$=0) frame for the calculation of various form 
factors~\cite{LB}. By taking the ``good" components of the current 
($J^{+}$ and ${\bf J}_{\perp}$), one can get rid of the zero 
mode~\cite{Kaon} problem and compute the full theoretical prediction for
the spacelike form factors in $q^{+}$=0 frame. The weak
transition form factors that we are considering, however, are the
timelike $q^{2}>0$ observables.
Our method is to rely on the analytic continuation
from the spacelike region to the timelike region calculating the ``good"
components of the current in the $q^{+}$=0 frame~\cite{Kaon}.
If we were to take the $q^{+}$$\neq$0 frame, then we must take
into account the higher Fock-state (nonvalence) contributions
arising from quark-antiquark pair creation (so called ``Z-graph") as well as
the valence configuations. In fact, we notice that a few previous
analyses~\cite{Dem} were performed in the $q^{+}$$\neq$0 frame without 
taking into account the nonvalence contributions.
We find that such omission leads to a large deviation from the full 
results~\cite{Kaon}. Our method is to rely on the analytic continuation
from the spacelike region to the timelike region calculating the ``good"
components of the current in the $q^{+}$=0 frame.

The key idea in our LFQM~\cite{Mix} for mesons is to treat
the radial wave function as a trial function for the variational
principle to the QCD-motivated Hamiltonian saturating
the Fock state expansion by the \underline {constituent} quark and antiquark.
The spin-orbit wave function is uniquely determined by the Melosh
transformation (see Appendix A).
We take the same QCD-motivated effective Hamiltonian given by
Eqs.~(4.1) and (4.2) and the Gaussian radial wave function
$\phi(k^{2})$ given by Eq.~(4.3) as our trial wave function 
to minimize the central Hamiltonian~\cite{Mix}.
The model parameters of heavy-quark sector ($c$ and $b$) such as 
$m_{c}$, $m_{b}$, $\beta_{uc}$, $\beta_{ub}$, etc. are then uniquely 
determined by the same procedure as the light-quark analysis~\cite{Mix}
discussed in Chapter 4. The procedure of determining model parameters
constrained by the variational principle (see Eqs~(D.1)-(D.3) in Appendix
D) is shown in Fig. 7.1, where the lines of $qq$ and $qc$ ($q$=$u$ and $d$) 
etc. represent the sets of $\{m_{q},m_{q},\beta_{qq}\}$ and
$\{m_{q},m_{c},\beta_{qc}\}$, respectively, etc.
\begin{figure}
\centerline{\psfig{figure=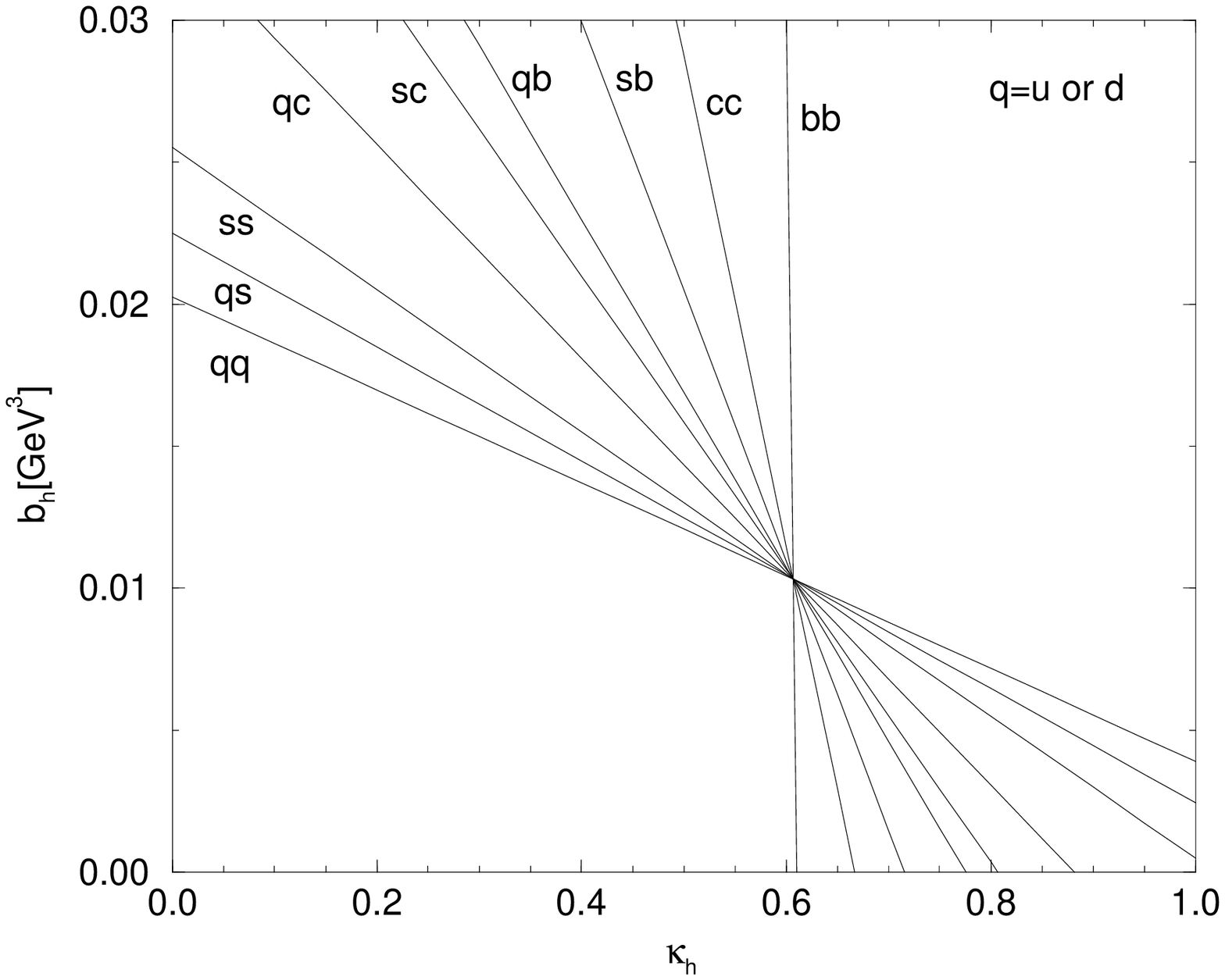,width=3.5in,height=3.5in}}
\centerline{\psfig{figure=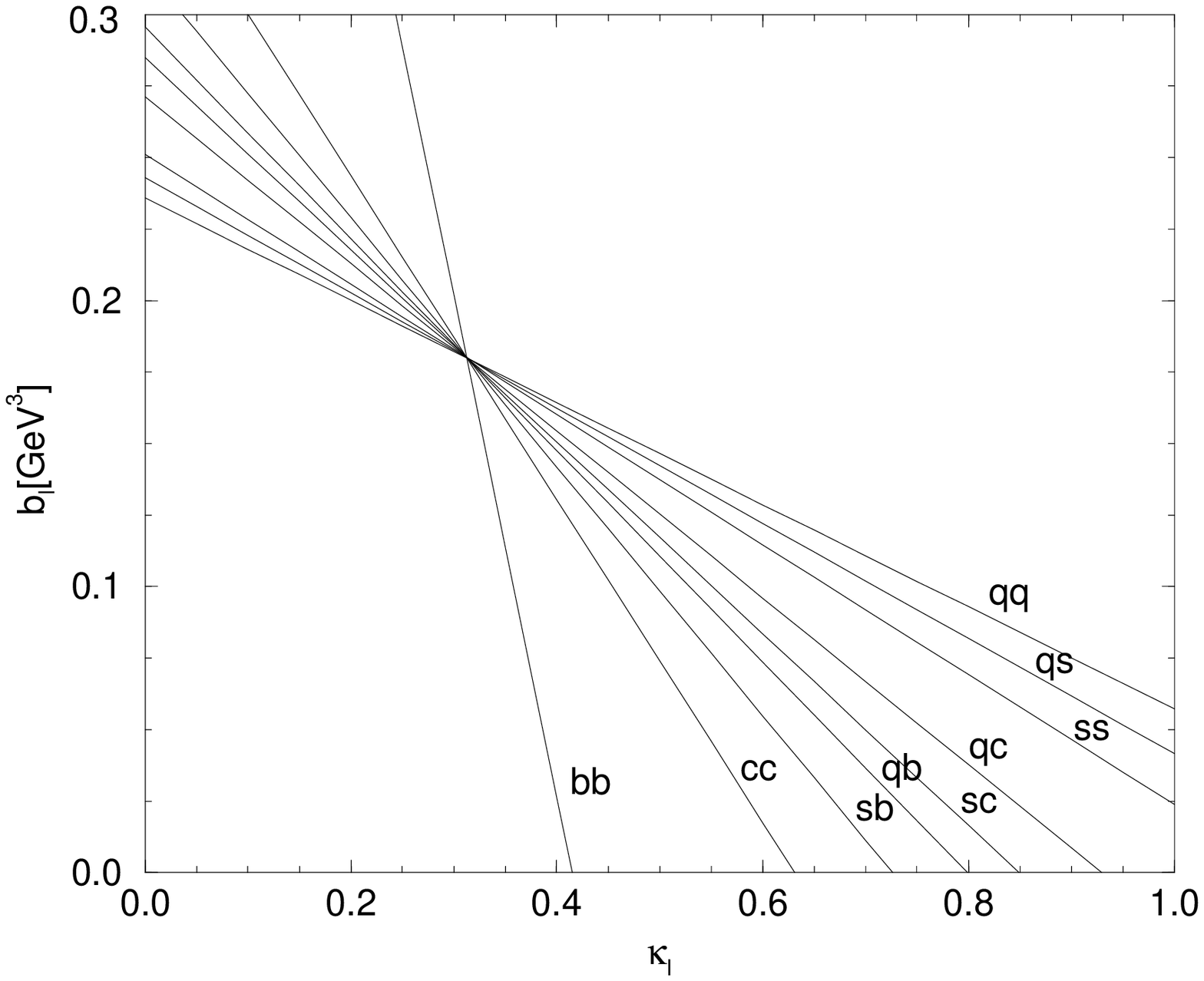,width=3.5in,height=3.5in}}
\caption{ The parameters $m_{s}$, $m_{c}$, $m_{b}$, $\beta_{qs}$, 
$\beta_{qc}$, etc. satisfying variational principle with (a) HO and 
(b) Linear potential models. The $qq$ and $qc$ etc. represents the sets 
of $(m_{q},m_{q},\beta_{qq})$ and $(m_{q},m_{c},\beta_{qc})$ etc.,
respectively.}
\end{figure}
Because all the lines in Fig. 7.1 should go through the same point of
($b$=0.18 GeV$^{2}$,$\kappa=0.313$), the parameters of $m_{c}$, $m_{b}$,
$\beta_{uc}$, $\beta_{ud}$, etc. are all automatically determined  without
any adjustment. Our model parameters obtained by the variational principle
are summarized in Tables~\ref{t71} and~\ref{t72}.
\begin{table}[p]
\centering
\caption{Optimized quark masses $m_{q}$[GeV] for both HO
and linear potentials obtained
from the variational principle. We also include the results from
the smearing function(SF) instead of the
Breit-Fermi contact term. $q$=$u$ and $d$.}\label{t71}
\begin{tabular}{|c|c|c|c|c|}\hline
Potential & $m_{q}$ & $m_{s}$ & $m_{c}$ & $m_{b}$ \\
\hline
HO& 0.25 & 0.48 & 1.8 & 5.2 \\
\hline
SF& 0.25 & 0.48 & 1.8 & 5.2  \\
\hline
Linear& 0.22 & 0.45 & 1.8 & 5.2 \\
\hline
SF & 0.22 & 0.45 & 1.8 & 5.2 \\
\hline
\end{tabular}
\end{table}

\begin{table}
\centering
\caption{Optimized Gaussian parameters $\beta$[GeV] for both HO
and linear potentials obtained
from the variational principle. We also include the results from
the smearing function(SF) instead of the Breit-Fermi contact term. 
}\label{t72}
{\footnotesize 
\begin{tabular}{|c|c|c|c|c|c|c|c|c|c|}\hline
Potential & $\beta_{q\bar{q}}$ &
$\beta_{s\bar{s}}$& $\beta_{q\bar{s}}$ & $\beta_{q\bar{c}}$ &
$\beta_{s\bar{c}}$ & $\beta_{c\bar{c}}$& $\beta_{q\bar{b}}$&
$\beta_{s\bar{b}}$ & $\beta_{b\bar{b}}$ \\
\hline
HO& 0.3194 & 0.3681 & 0.3419
& 0.4216 & 0.4686 & 0.6998 & 0.4960 & 0.5740 & 1.8025\\
\hline
SF& 0.3194 & 0.3703 & 0.3428
& 0.4280 & 0.4782 & 0.7278 & 0.5122 & 0.5980 & 1.9101 \\
\hline
Linear & 0.3659 & 0.4128 & 0.3886
& 0.4679 & 0.5016 & 0.6509 & 0.5266 & 0.5712 & 1.1452 \\
\hline
SF & 0.3659 & 0.4132 & 0.3887
& 0.4697 & 0.5042 & 0.6548 & 0.5307 & 0.5767 & 1.1806\\
\hline
\end{tabular}
}
\end{table}

\begin{table}
\centering
\caption{Fit of the ground state meson masses with
the parameters given in Tables~\protect\ref{t71} and~\protect\ref{t72}. 
Underline masses are input data.
The masses of ($\omega$-$\phi$) and ($\eta$-$\eta'$) were used to
determine the mixing angles of $\omega$-$\phi$ and $\eta$-$\eta'$,
respectively.}\label{t73}
{\footnotesize
\begin{tabular}{|c|c|c|c|c|c|c|c|}\hline
$^{1}S_{0}$ & Expt. [MeV] & HO (SF) & Linear (SF) &
$^{3}S_{1}$ & Expt. [MeV] & HO (SF) & Linear (SF) \\
\hline
$\pi$ & 135$\pm$0.00035 & \underline{135 (135)} &\underline{135 (135)} &
$\rho$ & 770$\pm$ 0.8 & \underline{770 (770)} & \underline{770 (770)}\\
$K$ & 494$\pm$ 0.016 & 470 (469) & 478 (478) &
$K^{*}$ & 892$\pm$ 0.24 & 875 (875) & 850 (850) \\
$\eta$ & 547$\pm$ 0.19 & \underline{547 (547)}&\underline{547 (547)} &
$\omega$ & 782$\pm$ 0.12 & \underline{782 (782)}&\underline{782 (782)}\\
$\eta'$ & 958$\pm$0.14 & \underline{958 (958)}&\underline{958 (958)} &
$\phi$ & 1020$\pm$0.008 &\underline{1020 (1020)}&\underline{1020 (1020)}\\
$D$ & 1869$\pm$0.5 & 1821 (1821) & 1836 (1840) &
$D^{*}$ & 2010$\pm$ 0.5 & 2024 (2026) &1998 (1997)\\
$D_{s}$ & 1969$\pm$0.6 & 2005 (2004) & 2011 (2014) &
$D^{*}_{s}$ &2112$\pm$0.7 & 2150 (2150)& 2109 (2108) \\
$\eta_{c}$ & 2980$\pm$ 2.1 & 3128 (3123) & 3171 (3173) &
$J/\psi$ & 3097$\pm$0.04 & 3257 (3244) &3225 (3221)\\
$B$ & 5279$\pm$ 1.8 & 5235 (5231) & 5235 (5237) &
$B^{*}$ & 5325$\pm$ 1.8& 5349 (5349) & 5315 (5314)\\
$B_{s}$ & 5369$\pm$2.0 & 5378 (5372) & 5375 (5376) &
$(b\bar{s})$ & -- & 5471 (5466) & 5424 (5423)\\
$(b\bar{b})$ & -- & 9295 (9353) & 9657 (9651) &
$\Upsilon$ & 9460$\pm$ 0.21 & 9558 (9459) & 9691 (9675)\\
\hline
\end{tabular}
}
\end{table}
Our predictions of the ground state meson mass spectra and the
decay constants
are summarized in Tables~\ref{t73} and~\ref{dcon2}, 
respectively, and compared with the experimental data~\cite{data}
and the lattice QCD results~\cite{Flynn,Bern2}. 
Our predictions of ground state 
meson mass spectra agree with the experimental data~\cite{data}
within 6$\%$ error. Furthermore, our model predicts the two unmeasured mass
spectra of $^{1}S_{0}(b\bar{b})$ and $^{3}S_{1}(b\bar{s})$ systems as
$M_{b\bar{b}}$=9295 (9657) MeV and $M_{b\bar{s}}$=5471 (5424) MeV 
for the HO (linear) potential, respectively.
Our predictions for the decay constants of heavy mesons are quite
comparable with the lattice QCD~\cite{Flynn,Bern2} anticipating  
future accurate experimental data.
\begin{table}
\caption{Decay constants~[MeV] and charge radii~[fm$^{2}$]
for various heavy pseudoscalar and vector mesons.
Note that the $f_P$ and $f_V$ are obtained by
$\la 0|{\bar q}_2\gam^\mu\gam_5q_1|P\ra=if_{P}P^{\mu}$
and
$\la 0|{\bar q}_2\gam^\mu q_1|V\ra=M_{V}f_{V}\vep^{\mu}$,
respectively, to compare with the lattice results. In the above 
definitions of the decay constants, our $f_\pi$, for example,
obtained in Chapter 4 becomes 130.7 (129.8) for the HO (linear) 
parameters.}\label{dcon2}
\centering
{\scriptsize
\begin{tabular}{|r|c|c|c|c|c|c|}\hline
References& $f_{D}$ & $f_{D^{*}}$& $f_{D_{s}}$& $f_{B}$& $f_{B^{*}}$
&$f_{B_{s}}$\\
\hline
HO & 179.7 & 211.6 & 218.6 &160.9 &173.0 &207.0\\
Linear & 196.9 & 238.9 & 233.1 &171.4 & 185.8 & 203.9 \\
\hline
LAT~\cite{Flynn}& 200$\pm$30 & $\cdots$& 220$\pm$30 &
170$\pm$35 & $\cdots$ & 195$\pm$35\\
~\cite{Bern2} & $195\pm11^{+15+15}_{-8-0}$& $\cdots$ 
&$213\pm9^{+23+17}_{-9-0}$ & $159\pm11^{+22+21}_{-9-0}$ & $\cdots$ &
$175\pm10^{+28+25}_{-10-1}$\\
\hline
Expt.~\cite{data} & $<219$ & &$137-304$ & & & \\
\hline
References& $r^{2}_{D^{+}}$ & $r^{2}_{D^{0}}$ & $r^{2}_{D^{+}_{s}}$ &
$r^{2}_{B^{+}}$ & $r^{2}_{B^{0}}$ & $r^{2}_{B^{0}_{s}}$\\
\hline
HO & 0.182 & $-0.309$ & 0.106 & 0.420 & $-0.208$& $-0.081$\\
Linear & 0.176 & $-0.301$ &0.101 & 0.438 & $-0.217$ & $-0.083$\\
\hline
\end{tabular}
}
\end{table}

This Chapter is organized as follows:
In Section 7.1, we calculate the EM form factors of $D$ and $B$ mesons 
as well as $\pi$ and $K$ mesons in the Drell-Yan-West ($q^+$=0) frame and 
compare the results with those obtained from the valence
contributions in $q^+$$\neq$0 frame.
Comparing the form factors between the two reference frames, 
we quantify the nonvalence contributions from $q^+$$\neq$0 frame.
In Section 7.2, we discuss the formalism of the form factors for
various $0^{-}\to 0^{-}$ and $0^-\to 1^-$ semileptonic decays.
We calculate the weak form factors in $q^+$=0 frame (i.e. spacelike
$q^2<0$ region) and then analytically continue the form factors
to the timelike $q^{2}$ region by changing $q_{\perp}$ to $iq_{\perp}$ 
in the form factors. 
In Section 7.3, our numerical results of the observables for
various $0^{-}\to 0^{-}$ and $0^-\to 1^-$ semileptonic decays 
are presented and compared with the available experimental data 
as well as other theoretical results.
Of special interest, we compare our analytic solutions of the
form factors with the simple pole approximation motivated by
the vector dominance model. The monople-type form factors 
turn out to be good approximations to our analytic solutions for
most decay processes except heavy-to-light decays, e.g., 
$B\to\pi$ and $B\to\rho$. We also quantify for $0^-\to 0^-$ decays the
nonvalence contributions from $q^{+}$$\neq$0 frame as in the 
case of the EM form factor analyses.
Summary and discussion of our main results follow in Section 7.4.
In the Appendix F, we show the derivation of the matrix element of
the form factors for $0^{-}\to 0^{-}$ semileptonic decays
both in the $q^{+}$=0 and $q^+$$\neq$0 frames.

\section{EM Form Factors of $B$ and $D$ Mesons}
The EM form factor of a pseudoscalar meson is defined by the the relation
\be\label{eq71}
\la P_{2}|J^{\mu}_{\rm em}|P_{1}\ra=
(P^{\mu}_{1}+P^{\mu}_{2})F(Q^{2}),
\ee
where $J^{\mu}_{\rm em}$ is the EM current for the quarks
and $P_{1}(P_{2})$ is the four-momenta of the initial(final) meson.
In the calculations of the hadronic matrix elements in the LF 
frame, one usually use $q^{+}$=$P^{+}_{1}-P^{+}_{2}$=0 frame. This leads
to $q^{2}=q^{+}q^{-} - q^{2}_{\perp}= -q^{2}_{\perp}\equiv -Q^{2}$
implying a spacelike momentum transfer.

In this work, however, we calculate the valence contributions
from both $q^{+}$=0 and $q^{+}$$\neq$0 frames and compare the
difference in spacelike region so that we can indirectly estimate
the nonvalence contributions from $q^{+}$$\neq$0 frame.
The form factor from the
valence contribution in $q^{+}$=0 frame is an exact solution
since the ``Z"-graph does not contribute at all in this $q^{+}$=0
frame as far as the ``good" components of the currents are concerned,
i.e., $J^{+}$ or ${\bf J}_{\perp}$~\cite{zm,Kaon}.
In the standard $q^{+}$=0 frame, the quark momentum variables
are given by
In the standard $q^{+}$=0 frame, the quark momentum variables
are given by
\begin{eqnarray}\label{qq_var}
p^{+}_{1}&=&(1-x)P^{+}_{1},\;\;
p^{+}_{\bar{q}}=xP^{+}_{1},\nonumber\\
{\bf p}_{1\perp}&=&(1-x){\bf P}_{1\perp} + {\bf k}_{\perp},\;\;
{\bf p}_{\bar{q}\perp}= x{\bf P}_{1\perp} - {\bf k}_{\perp},
\nonumber\\
p^{+}_{2}&=&(1-x)P^{+}_{2},\;\;
p'^{+}_{\bar{q}}=xP^{+}_{2},\nonumber\\
{\bf p}_{2\perp}&=&(1-x){\bf P}_{2\perp} +
{\bf k'}_{\perp},\;\;
{\bf p'}_{\bar{q}\perp}= x{\bf P}_{2\perp} - {\bf k'}_{\perp},
\end{eqnarray}
which requires that $p^{+}_{\bar{q}}$=$p'^{+}_{\bar{q}}$
and ${\bf p}_{\bar{q}\perp}$=${\bf p'}_{\bar{q}\perp}$.

Taking ${\bf P}_{1\perp}$=0, ${\bf P}_{2\perp}$=$-{\bf q}_{\perp}$,
we obtain from the matrix element of the $``+"$ component of the current 
$J^{+}$:
\begin{eqnarray}\label{EMD}
F(Q^{2}) &=& e_{q}\int^{1}_{0}dx\int d^{2}k_{\perp}
\sqrt{\frac{\partial k_{1n}}{\partial x}}
\sqrt{\frac{\partial k'_{2n}}{\partial x}}
\phi^{*}_{2}(x,{\bf k'}_{\perp})\phi_{1}(x,{\bf k}_{\perp})\nonumber\\
&\times&\frac{{\cal A}^{2}+{\bf k}_{\perp}\cdot{\bf k'}_{\perp}}
{ \sqrt{ {\cal A}^{2}+ k^{2}_{\perp}}\sqrt{ {\cal A}^{2}+
k^{'2}_{\perp}} }
+ e_{\bar{q}}(q\leftrightarrow\bar{q}\hspace{.2cm}
{\rm of\hspace{.1cm}the\hspace{.1cm}first\hspace{.1cm}term}),
\end{eqnarray}
where ${\bf k'}_{\perp}$=${\bf k}_{\perp} - x{\bf q}_{\perp}$,
${\cal A}$=$xm_{q} + (1-x)m_{\bar{q}}$ and $e_{q} (e_{\bar q})$
is the charge factor of the quark (antiquark).

The EM form factor of a pseudoscalar meson obtained from the valence
contributions in $q^{+}$$\neq$0 frame is given in appendix F.
Because of the deficit of the nonvalence (or ``Z-graph")
contributions in $q^{+}$$\neq$0 frame, the EM form
factors in $q^{+}$=0 [Eq.~(\ref{EMD})] and $q^{+}$$\neq$0
frames [Eq.~(F.10)] are different with each other
for the non-zero momentum transfer.
Thus, we can estimate quantitatively the effect of nonvalence
contributions by comparing Eq.~(\ref{EMD}) and Eq.~(F.10).
In Figs.~7.2  and 7.3, we show the EM form factors of the $D$ and $B$
mesons and compare the results of $q^{+}$=0 frame with those of
$q^{+}$$\neq$0 frame, respectively. 
For comparison, we also calculated~\cite{Kaon} the EM form 
factors $F_{\pi}(q^{2})$ and $F_{K}(q^{2})$ in the spacelike region using
both $q^{+}$=0 and $q^{+}$$\neq$0 frames to estimate the nonvalence
contributions in $q^{+}$$\neq$0 frame.
As shown in Figs.~7.4 and 7.5 for $F_{\pi}(q^{2})$ and $F_{K}(q^{2})$,
respectively, our predictions in $q^{+}$=0 frame are in a very good
agreement with the available data~\cite{amendolia,amen2} while the results
for $q^{+}$$\neq$0 frame deviate from the data significantly. The deviations
represent the nonvalence contributions in $q^{+}$$\neq$0
frame.  However, the deviations are clearly
reduced for $F_{K}(q^{2})$ because of the large suppression
from the energy denominator shown in Fig.~1.1 in Chapter 1
for the nonvalence contributions. 
The suppressions are much bigger for the heavier mesons
such as $D$ (see Fig.~7.2) and $B$ (see Fig.~7.3). 
Especially, for the $B$ meson case, the
nonvalence contributions are almost negligible
up to $Q^{2}=-q^{2}\sim 10$ GeV$^{2}$.
\begin{figure}
\centerline{\psfig{figure=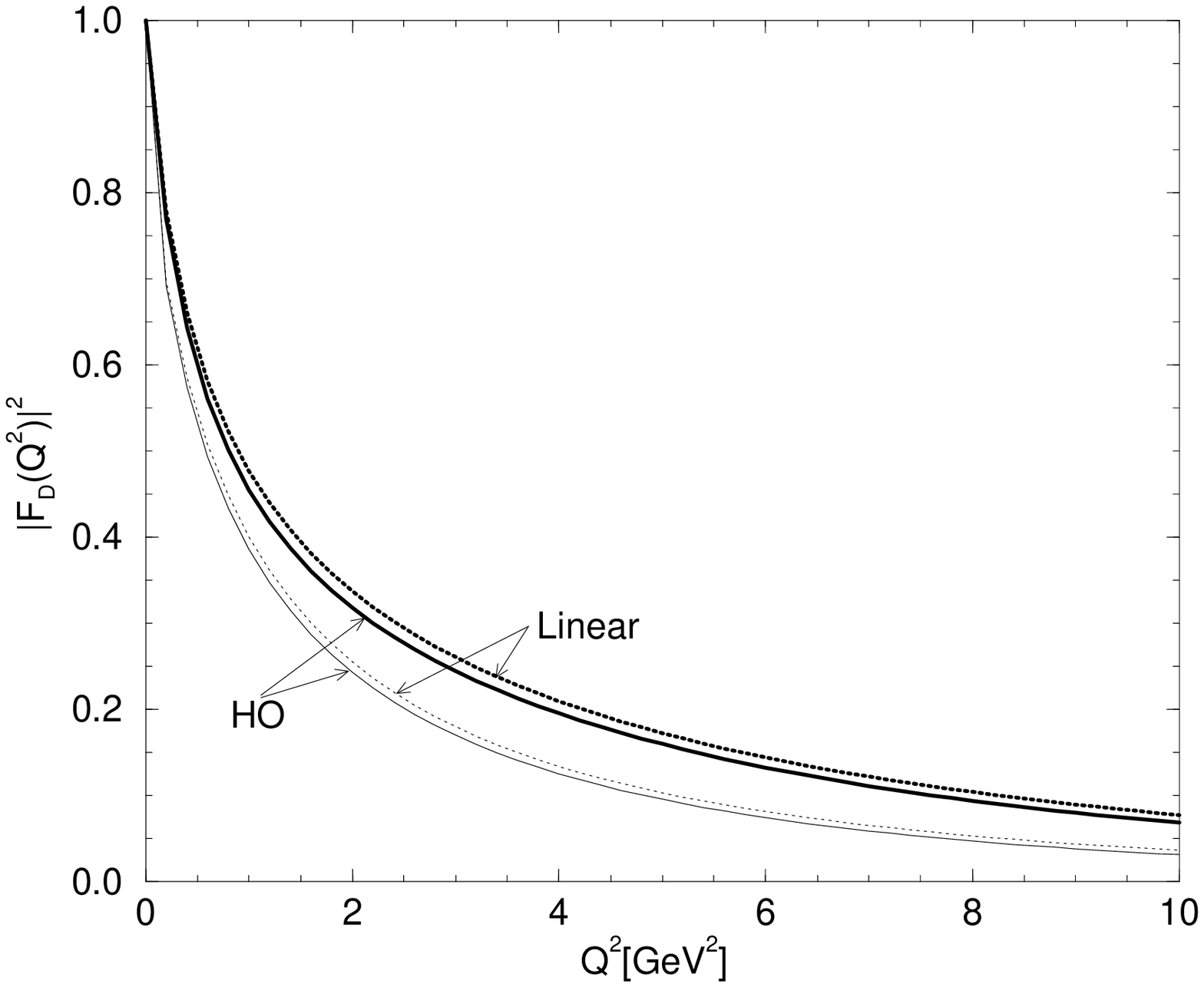,width=3.5in,height=3.2in}}
\centerline{\psfig{figure=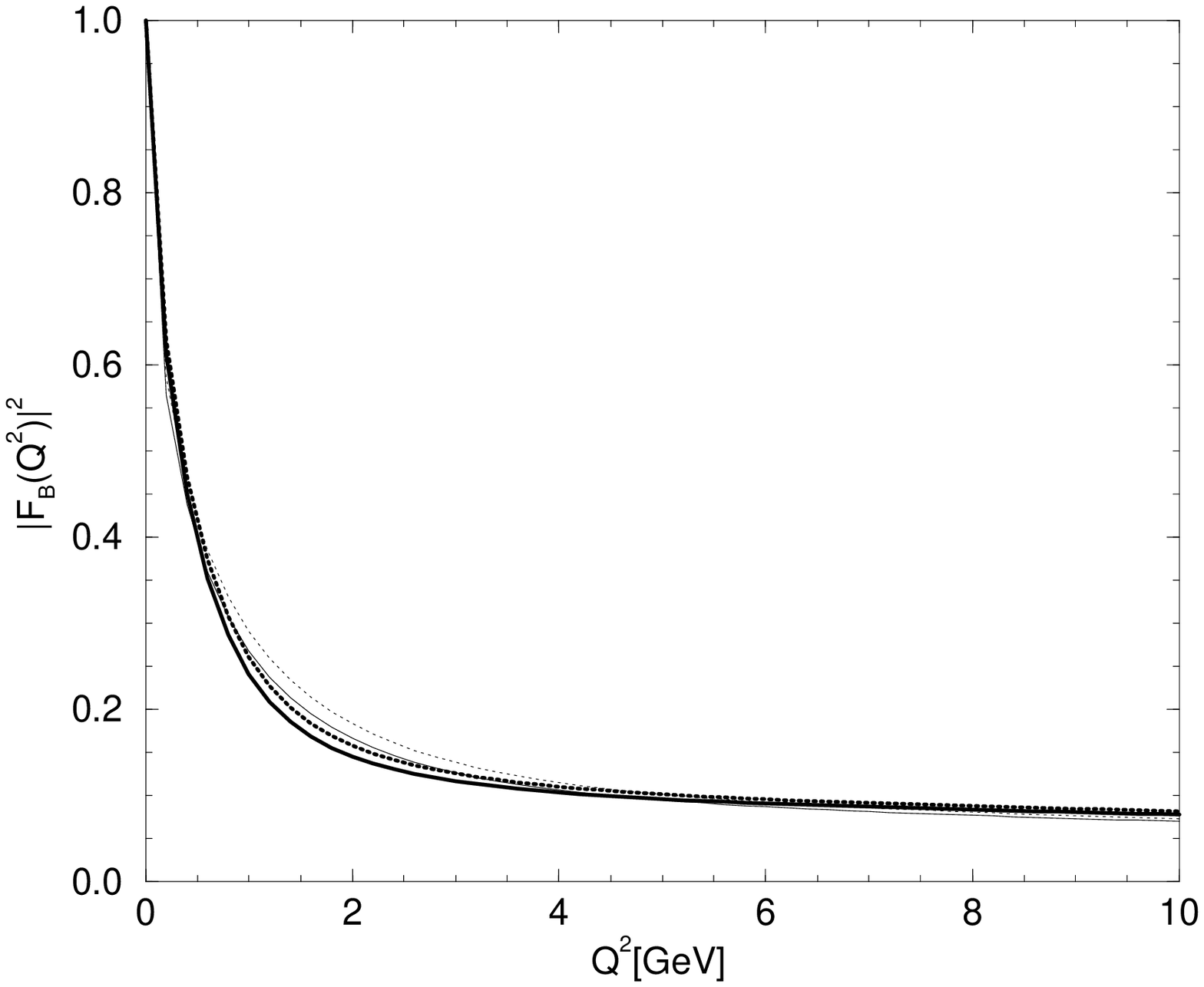,width=3.5in,height=3.2in}}
\caption{
The EM form factor of $D$ meson for the
spacelike $Q^{2}=-q^{2}$ region. The
thick solid (light solid) and thick dotted (light dotted) lines are
the results from the $q^{+}$=0 ($q^{+}$$\neq$0) frame for the HO and linear
potential parameters, respectively.
7.3:
The EM form factor of $B$ meson for the spacelike $Q^{2}=-q^{2}$ region. 
The same line code as in Fig.~7.2 is used.
}
\end{figure}
\setcounter{figure}{3}
\begin{figure}
\centerline{\psfig{figure=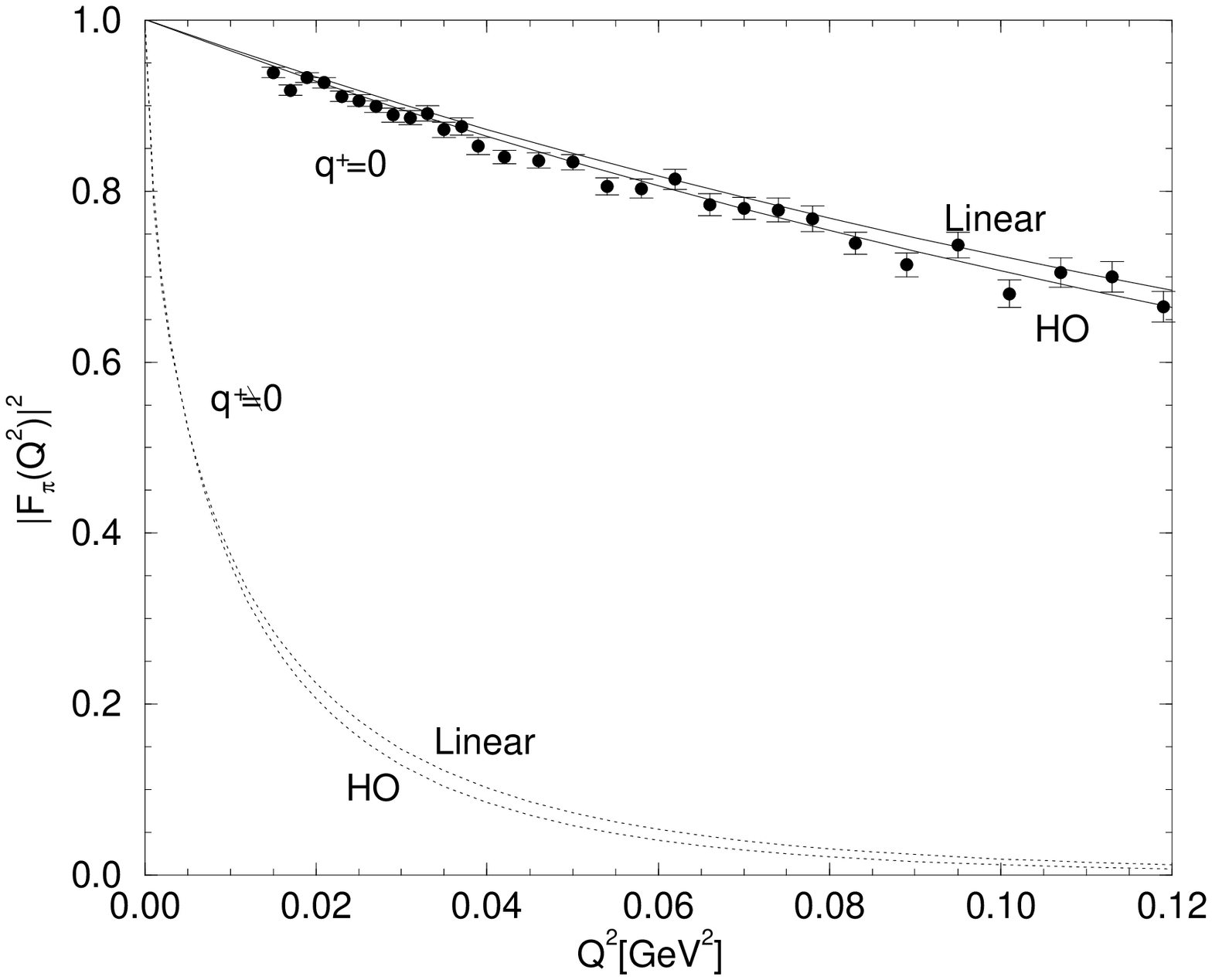,width=3.5in,height=3.2in}}
\centerline{\psfig{figure=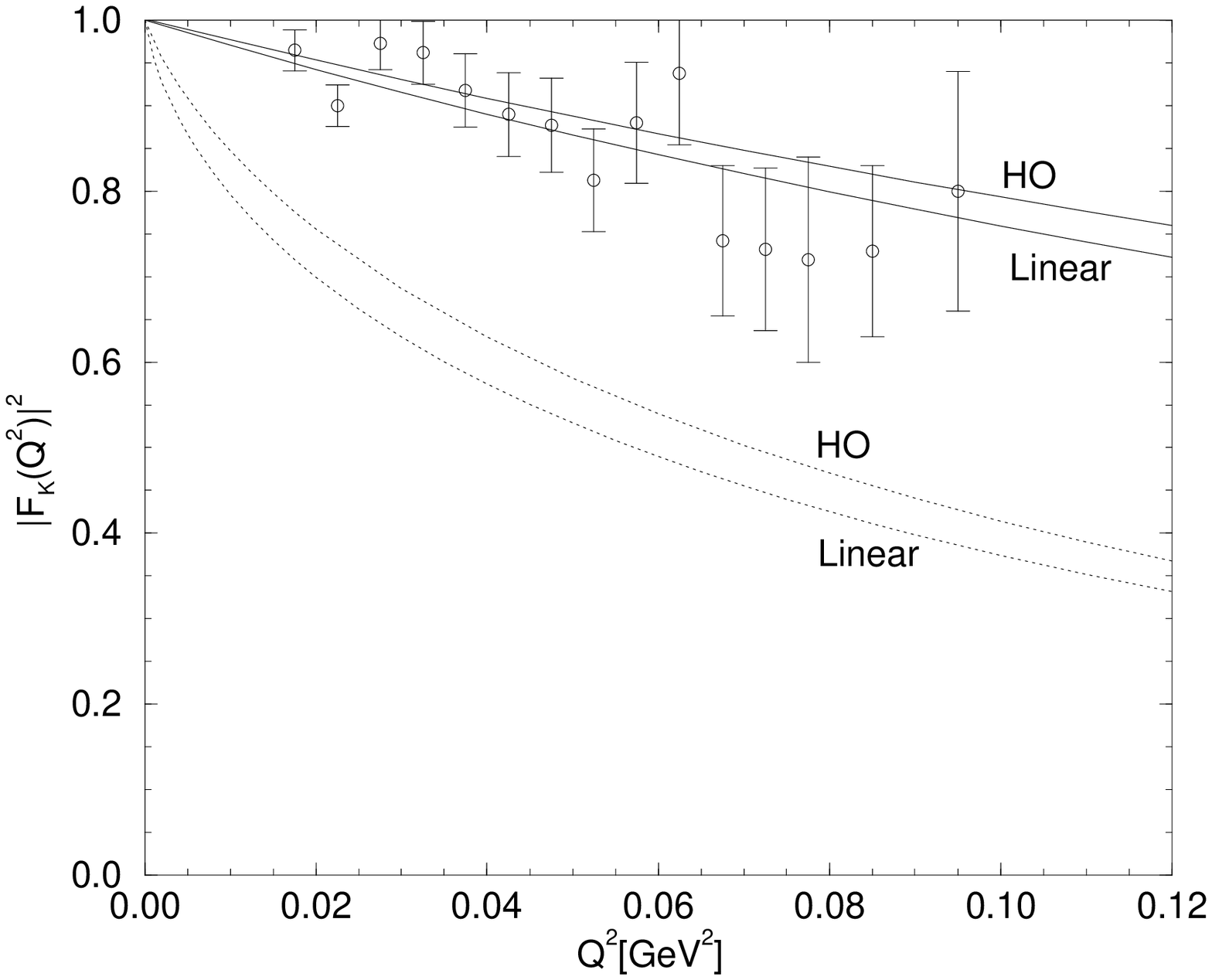,width=3.5in,height=3.2in}}
\caption{
The EM form factor of pion for low
$Q^{2}=-q^{2}$ compared with data~\protect\cite{amendolia}.
7.5:
The EM form factor of kaon compared with data~\protect\cite{amen2}. 
The same line code as in Fig.~7.4 is used.}
\end{figure}
This seems quite natural because it is much harder to create the
heavy quark pair than the light one in ``Z-graph".

\section{Weak Form Factors for $0^{-}\to 0^{-}$ Transitions}
The matrix element of the current $J^\mu$=$\bar{q}_2\gam^\mu Q_1$
for $0^{-}(Q_{1}\bar{q})\to 0^{-}(q_{2}\bar{q})$ decay is given
by two weak form factors $f_+$ and $f_{-}$, viz., 
\be\label{PP_form}
\la P_{2}|\bar{q}_{2}\gamma^{\mu}Q_{1}|P_{1}\ra=
f_{+}(q^{2})(P_{1}+P_{2})^{\mu} + f_{-}(q^{2})q^{\mu},
\ee
where $q^\mu$=$(P_1-P_2)^\mu$ is the four-momentum transfer
to the lepton and $m^{2}_{\ell}\leq q^2\leq (M_1-M_2)^2$.
Equation~(\ref{PP_form}) 
can also be parametrized into two other different ways
\begin{eqnarray}
\la P_{2}|\bar{q}_{2}\gamma^{\mu}Q_{1}|P_{1}\ra&=&
\sqrt{M_{1}M_{2}}\biggl[h_{+}(q^{2})(v_{1}+v_{2})^{\mu}
+ h_{-}(q^{2})(v_{1}-v_{2})^{\mu}\biggr],\nonumber\\
&=& f_{+}(q^{2})\biggl[(P_{1}+P_{2})^{\mu}
-\frac{M^{2}_{1}-M^{2}_{2}}{q^{2}}q^{\mu}\biggr]\nonumber\\
&+& f_{0}(q^{2})\frac{M^{2}_{1}-M^{2}_{2}}{q^{2}}q^{\mu},
\end{eqnarray}
where $v_{i}=P_{i}/M_{i}$ and satisfy the following relations:
\be
f_{+}(0)= f_{0}(0),\;\;
f_{0}(q^{2})= f_{+}(q^{2})
+ \frac{q^{2}}{M^{2}_{1}-M^{2}_{2}}f_{-}(q^{2}).
\ee
The second parametrization uses the helicity basis, where $f_{+}$
is related to the exchange of a vector particle and $f_{0}$ to
the exchange of a scalar particle.
In the heavy-quark limit $M_{1,2}\to\infty$, heavy-quark symmetry
requires that~\cite{Neubert}
\be\label{heavy_h}
h_{+}(q^{2})=\xi(v_{1}\cdot v_{2}),\;\;
h_{-}(q^{2})=0,
\ee
where $\xi(v_{1}\cdot v_{2})$ is the universal Isgur-Wise (IW)
function\footnote{ From Eqs.~(\ref{PP_form})-(\ref{heavy_h}), 
one can easily find that the
form factors $f_{+}(q^{2})$ and $f_{0}(q^{2})$ are related to the
IW function via $\xi(v_{1}\cdot v_{2})$=$\frac{2\sqrt{M_{1}M_{2}}}
{M_{1}+M_{2}}f_{+}(q^{2})$= $\frac{2\sqrt{M_{1}M_{2}}}{M_{1}+M_{2}}
\frac{f_{0}(q^{2})}{[1-q^{2}/(M_{1}+M_{2})^{2}]}$. That means the $q^{2}$
dependence of $f_{+}$ is different from that of $f_{0}$ by an additional
factor.} normalized to unity at the point of equal velocities, $\xi(1)$=1. 

Since the lepton mass is small except in the case of the $\tau$
lepton, one may safely neglect the lepton mass in the decay rate
calculation of the heavy-to-heavy and heavy-to-light transitions.
However, for $K_{\ell 3}$ decays, the muon($\mu$) mass is not negligible,
even though electron mass can be neglected. Thus, including nonzero
lepton mass, the formula for the decay rate of $K_{\ell 3}$ is given
by~\cite{Kaon}
\begin{eqnarray}\label{PP_rate}
\frac{d\Gamma}{dq^{2}}&=& \frac{G^{2}_{F}}{24\pi^{3}}
|V_{Q_{1}\bar{Q}_{2}}|^{2} K_{f}(q^{2})
\biggl(1-\frac{m^{2}_{\ell}}{q^{2}}\biggr)^{2}
\nonumber\\
&\times&\biggl\{
[K_{f}(q^{2})]^{2}\biggl(1+\frac{m^{2}_{\ell}}{2q^{2}}\biggr)
|f_{+}(q^{2})|^{2}
+ M^{2}_{1}\biggl(1-\frac{M^{2}_{2}}{M^{2}_{1}}\biggr)^{2}
\frac{3}{8}\frac{m^{2}_{\ell}}{q^{2}}|f_{0}(q^{2})|^{2}\biggr\},
\end{eqnarray}
where $G_{F}$ is the Fermi constant, $V_{Q_{1}\bar{Q}_{2}}$ is
the element of the CKM mixing matrix and
the kinematic factor $K_{f}(q^{2})$ (=$|{\bf P}_{2}|$) in the decaying
meson rest frame (${\bf P}_{1}$=0) is given by
\be\label{KF}
K_{f}(q^{2})= \frac{1}{2M_{1}}
\sqrt{ (M_{1}^{2}+M_{2}^{2}-q^{2})^{2}-4M_{1}^{2}M_{2}^{2} }.
\ee
Note that the form factor $f_{0}(q^{2})$ does not contribute to
the decay rate in the limit of massless leptons.
Since our analysis will be performed in the isospin symmetry ($m_{u}$=$m_{d}$)
but $SU_{f}(3)$ breaking ($m_{s}$$\neq$$m_{u(d)}$) limit, we do not
discriminate between the charged and neutral kaon weak decays,
i.e., $f^{K^{0}}_{\pm}$=$f^{K^{+}}_{\pm}$.
For $K_{l3}$ decays, the three form factor parameters,
i.e., $\lambda_{+}, \lambda_{0}$ and $\xi_{A}$,
have been measured using the following linear parametrization~\cite{data}:
\be\label{K_lam}
f_{\pm}(q^{2})= f_{\pm}(q^{2}=m^{2}_{l})\biggl(
1 + \lambda_{\pm}\frac{q^{2}}{M^{2}_{\pi^{+}}}\biggr),
\ee
where $\lambda_{\pm,0}$ is the slope of $f_{\pm,0}$ evaluated at
$q^{2}$=$m^{2}_{l}$ and $\xi_{A}$=$f_{-}/f_{+}|_{q^{2}=m^{2}_{l}}$.

In LFQM, the matrix element of the weak vector current can be obtained
by the convolution of initial and final LF meson wave functions in 
$q^{+}$=0 frame where the decaying hadron is at rest: 
\begin{eqnarray}\label{PP_m}
& &\la P_{2}|\bar{q}_{2}\gamma^{\mu}Q_{1}|P_{1}\ra\nonumber\\
& &= -\int^{1}_{0}dx\int d^{2}{\bf k}_{\perp}
\sqrt{\frac{\partial k_{1n}}{\partial x}}
\sqrt{\frac{\partial k'_{2n}}{\partial x}}
\frac{\phi^{*}_{2}(x,{\bf k'}_{\perp})\phi_{1}(x,{\bf k}_{\perp})}
{2(1-x)\prod^{2}_{i=1}\sqrt{ M^{2}_{i0} -(m_{i}-m_{\bar{q}})^{2} }}
\nonumber\\
& &\;\;\;\;\times{\rm Tr} \biggl[\gamma_{5}({\not\! p}_{2}+m_{2})
\gamma^{\mu}({\not\!p}_{1}+m_{1})\gamma_{5}
({\not\! p}_{\bar{q}}-m_{\bar{q}})\biggr],
\end{eqnarray}
where $M^{2}_{i0}$=$(k_{\perp}^{2}+m^{2}_{i})/(1-x)
+ (k_{\perp}^{2}+m^{2}_{\bar{q}})/x$.
Using the matrix element of the ``$+$" component of the current, $J^{+}$,
we obtain from Eqs.~(\ref{PP_form}) and (\ref{PP_m})
the form factor $f_{+}(q_{\perp}^{2})$ as follows
\be\label{fp}
f_{+}(q_{\perp}^{2})=\int^{1}_{0}dx\int d^{2}{\bf k}_{\perp}
\sqrt{\frac{\partial k_{1n}}{\partial x}}
\sqrt{\frac{\partial k'_{2n}}{\partial x}}
\phi^{*}_{2}(x,{\bf k'}_{\perp})\phi_{1}(x,{\bf k}_{\perp})
\frac{{\cal A}_{1}{\cal A}_{2}+{\bf k}_{\perp}\cdot{\bf k'}_{\perp}}
{ \sqrt{ {\cal A}_{1}^{2}+ k^{2}_{\perp}}\sqrt{ {\cal A}_{2}^{2}+
k^{'2}_{\perp}} },
\ee
where $q_{\perp}^{2}$=$-q^{2}$, ${\cal A}_{i}$=$m_{i}x + m_{\bar{q}}(1-x)$
and ${\bf k'}_{\perp}$=${\bf k}_{\perp}-x{\bf q}_{\perp}$. As we discussed
in the introduction, we need the ``$\perp$" component of the current,
${\bf J}_{\perp}$, to obtain the form factor $f_{-}(q_{\perp}^{2})$ 
in Eq.~(\ref{PP_form}), viz.,
\be\label{Jperp1}
\la P_{2}|\bar{q}_{2}({\bf q}_{\perp}\cdot\vec{\gamma}_{\perp})Q_{1}|P_{1}\ra
= q^{2}_{\perp}[f_{-}(q_{\perp}^{2}) -f_{+}(q_{\perp}^{2})],
\ee
after multiplying ${\bf q}_{\perp}$ on both sides of Eq.~(\ref{PP_form}).
The left-hand side (LHS) of Eq.~(\ref{Jperp1}) is given by
\begin{eqnarray}\label{Jperp2}
\la P_{2}|\bar{q}_{2}({\bf q}_{\perp}\cdot\vec{\gamma}_{\perp})Q_{1}|P_{1}\ra
&=& -\int^{1}_{0}dx\int d^{2}{\bf k}_{\perp}
\sqrt{\frac{\partial k_{1n}}{\partial x}}
\sqrt{\frac{\partial k'_{2n}}{\partial x}}
\frac{x\phi^{*}_{2}(x,{\bf k'}_{\perp})\phi_{1}(x,{\bf k}_{\perp})}
{ 2\sqrt{ {\cal A}_{1}^{2}+ k^{2}_{\perp}}\sqrt{ {\cal A}_{2}^{2}+
k_{\perp}^{'2} } }\nonumber\\
&\times&{\rm Tr} \biggl[\gamma_{5}({\not\! p}_{2}+m_{2})
({\bf q}_{\perp}\cdot\vec{\gamma}_{\perp})({\not\! p}_{1}+m_{1})\gamma_{5}
({\not\! p}_{\bar{q}}-m_{\bar{q}})\biggr].
\end{eqnarray}
Using the quark momentum variables given in Eq.~(\ref{qq_var}), 
we obtain the trace term in Eq.~(\ref{Jperp2}) as follows:
\begin{eqnarray}
&& {\rm Tr} \biggl[\gamma_{5}({\not\! p}_{2}+m_{2})
({\bf q}_{\perp}\cdot\vec{\gamma}_{\perp})({\not\! p}_{1}+m_{1})\gamma_{5}
({\not\! p}_{\bar{q}}-m_{\bar{q}})\biggr]\nonumber\\
&&= -2\biggl\{ \frac{({\cal A}^{2}_{1}+ k_{\perp}^{2})}{x(1-x)}
({\bf k}_{\perp}-{\bf q}_{\perp})\cdot{\bf q}_{\perp} +
\frac{({\cal A}^{2}_{2}+ k_{\perp}^{'2})}{x(1-x)}
{\bf k}_{\perp}\cdot{\bf q}_{\perp}
\nonumber\\
&&\;\;\;\; + [ (m_{1}-m_{2})^{2}
+ q_{\perp}^{2}]{\bf k}_{\perp}\cdot{\bf q}_{\perp}\biggr\}.
\end{eqnarray}
The more detailed derivation of Eqs.~(\ref{fp}) and (\ref{Jperp2}) are 
presented in Appendix F. 
Since both sides of Eq.~(\ref{Jperp1}) vanish as $q^{2}\to 0$, one has to
be cautious for the numerical computation of $f_{-}$ at $q^{2}$=0. Thus,
for the numerical computation at $q^{2}$=0, we need to find an analytic
formula for $f_{-}(0)$. In order to obtain the analytic formula for the form
factor $f_{-}(0)$, we make a low $q^{2}_{\perp}$ expansion to extract the
overall $q^{2}_{\perp}$ from Eq.~(\ref{Jperp2}).
Then, the form factor $f_{-}(0)$ is obtained as follows:
\begin{eqnarray}\label{f0}
f_{-}(0)&=& f_{+}(0) + \int^{1}_{0}dx\int d^{2}{\bf k}_{\perp}
\sqrt{\frac{\partial k_{1n}}{\partial x}}
\sqrt{\frac{\partial k_{2n}}{\partial x}}
\frac{x\phi^{*}_{2}(x,{\bf k}_{\perp})\phi_{1}(x,{\bf k}_{\perp})}
{ \sqrt{ {\cal A}_{1}^{2}+ k^{2}_{\perp}}\sqrt{ {\cal A}_{2}^{2}+
k_{\perp}^{2} } }\nonumber\\
&\times&\biggl\{ \biggl[C_{T1}(C_{J1}-C_{J2}+C_{M}+C_{R}) + C_{T2}\biggr ]
k^{2}_{\perp}\cos^{2}\phi + C_{T3}\biggr\},
\end{eqnarray}
where the angle $\phi$ is defined by ${\bf k}_{\perp}\cdot{\bf q}_{\perp}$
=$|{\bf k}_{\perp}||{\bf q}_{\perp}|\cos\phi$ and the terms of $C's$ are
given by
\begin{eqnarray}
C_{Ji}&=&
\frac{2\beta^{2}_{i}}{(1-x)(\beta^{2}_{1}+\beta^{2}_{2})M^{2}_{i0}}
\biggl[\frac{1}{1 -[(m^{2}_{i}-m^{2}_{\bar{q}})/M^{2}_{i0}]^{2}}
-\frac{3}{4}\biggr], \nonumber\\
C_{M}&=&\frac{1}{(1-x)(\beta^{2}_{1}+\beta^{2}_{2})}
\biggl[\frac{\beta^{2}_{2}}{M^{2}_{20}-(m_{2}-m_{\bar{q}})^{2}}
- \frac{\beta^{2}_{1}}{M^{2}_{10}-(m_{1}-m_{\bar{q}})^{2}}\biggr],\nonumber\\
C_{R}&=&\frac{-1}{4(1-x)(\beta^{2}_{1}+\beta^{2}_{2})}
\biggl[ \biggl(\frac{m^{2}_{2}-m^{2}_{\bar{q}}}{M^{2}_{20}}\biggr)^{2}
- \biggl(\frac{m^{2}_{1}-m^{2}_{\bar{q}}}{M^{2}_{10}}\biggr)^{2}\biggr],
\nonumber\\
C_{T1}&=& \frac{1}{x(1-x)}({\cal A}_{1}^{2}+ {\cal A}_{2}^{2}+
2k^{2}_{\perp}) + (m_{1}-m_{2})^{2},\nonumber\\
C_{T2}&=& \frac{2(\beta^{2}_{1}-\beta^{2}_{2})}
{(1-x)(\beta^{2}_{1}+\beta^{2}_{2})},\hspace{.1cm}
C_{T3}= \frac{x\beta^{2}_{1}}{\beta^{2}_{1}+\beta^{2}_{2}}C_{T1}
-\frac{{\cal A}^{2}_{1}+k^{2}_{\perp}}{x(1-x)}.
\end{eqnarray}
The form factors $f_{+}$ and $f_{-}$ can be analytically continued to
the timelike $q^{2}$$>$0 region by replacing $q_{\perp}$ by $iq_{\perp}$ 
in Eqs.~(\ref{fp}) and (\ref{Jperp1}). Since $f_{-}(0)$ in Eq.~(\ref{f0}) 
is exactly zero in the $SU_{f}(3)$ symmetry~\cite{Yuri}, i.e., 
$m_{u(d)}$=$m_{s}$ 
and $\beta_{u\bar{d}}$=$\beta_{u\bar{s}}$=$\beta_{s\bar{s}}$, 
one can get $f_{+}(q_{\perp}^{2})$=$F_{\pi}(q_{\perp}^{2})$ for the 
$\pi^{+}\to\pi^{0}$ weak decay($\pi_{e3}$), where $F_{\pi}(q_{\perp}^{2})$
is the EM form factor of pion, and $f_{-}(q^{2})=0$ because of
the isospin symmetry. 

\section{Weak Form Factors for $0^{-}\to 1^{-}$ Transitions}
The form factors for the semileptonic $P(Q_{1}\bar{q})\to V(q_{2}\bar{q})$
decay of a pseudoscalar to a vector meson is given by
\begin{eqnarray}\label{PV_m1}
&&\la V(P_{2},\vep)|\bar{q}_{2}\gamma^{\mu}(1-\gamma_{5})Q_{1}|P(P_{1})\ra
\nonumber\\
&&= 2i\vep_{\mu\nu\rho\sigma}\vep^{*\nu}P^{\rho}_{1}P^{\sigma}_{2}
\frac{V(q^{2})}{M_{1} + M_{2}}
\nonumber\\
&&\;\;-\vep^{*\mu}(M_{1} + M_{2})A_{1}(q^{2})
+ (P_{1} + P_{2})^{\mu}(\vep^{*}\cdot q)
\frac{A_{2}(q^{2})}{M_{1} + M_{2}}\nonumber\\
&&\;\; +q^{\mu}(\vep^{*}\cdot q)\frac{2M_{2}}{q^{2}}[A_{3}(q^{2})
- A_{0}(q^{2})],
\end{eqnarray}
where 
\be
A_{3}(q^{2})=\frac{M_{1} + M_{2}}{2M_{2}}A_{1}(q^{2}) -
\frac{M_{1}-M_{2}}{2M_{2}}A_{2}(q^{2}).
\ee
Alternatively, one can also use another set of the form factors in
the following way:
\begin{eqnarray}\label{PV_m2}
&&\la V(P_{2},\vep)|\bar{q}_{2}\gamma^{\mu}(1-\gamma_{5})Q_{1}|P(P_{1})\ra
\nonumber\\
&&= i\vep_{\mu\nu\rho\sigma}\vep^{*\nu}P^{\rho}_{1}P^{\sigma}_{2}g(q^{2})
\nonumber\\
&&\;\;+ \vep^{*\mu}f(q^{2}) + P^{\mu}(\vep^{*}\cdot P)a_{+}(q^{2})
+ q^{\mu}(\vep^{*}\cdot P)a_{-}(q^{2}),
\end{eqnarray}
where $P$=$P_{1} + P_{2}$.
The relationship between the two sets of the form factors
is given by
\begin{eqnarray}\label{fag}
&&V(q^{2})=-(M_{1}+M_{2})g(q^{2}),\nonumber\\
&&A_{0}(q^{2})=-\frac{1}{2M_{2}}[ f(q^{2})
+ (M^{2}_{1}-M^{2}_{2})a_{+}(q^{2}) + q^{2}a_{-}(q^2) ],\nonumber\\
&&A_{1}(q^{2})=-\frac{f(q^{2})}{M_{1}+M_{2}},\nonumber\\
&&A_{2}(q^{2})= (M_{1}+M_{2})a_{+}(q^{2}).
\end{eqnarray}
The form factors proportional to $q^{\mu}$, i.e., $A_{0},A_{3}$ and
$a_{-}$, do not contribute to the decay rate in the limit of massless
leptons and the corresponding differential decay rate for
$P\to V\ell\nu_{\ell}$ is given by
\be
\frac{d\Gamma}{dq^{2}}=\frac{G^{2}_{F}}{96\pi^{3}M^{2}_{1}}
|V_{Q_{1}Q_{2}}|^{2}K_{f}(q^{2})q^{2}
\biggl[ |H_{+}(q^{2})|^{2} + |H_{-}(q^{2})|^{2}
+ |H_{0}(q^{2})|^{2} \biggr],
\ee
where $K_{f}(q^2)$ is the kinematic factor given in Eq.~(\ref{KF})
and the helicity amplitudes are given by
\begin{eqnarray}
&&H_{0}=\frac{1}{2M_{2}\sqrt{2}}\biggl[
(M^{2}_{1}-M^{2}_{2}-q^{2})(M_{1}+M_{2})A_{1}(q^{2})
-4\frac{M^{2}_{1}K^{2}_{f}(q^{2})}{M_{1}+M_{2}}A_{2}(q^{2})\biggr],
\nonumber\\
&&H_{\pm}= (M_{1} + M_{2})A_{1}(q^{2})\mp
2\frac{M_{1}K_{f}(q^{2})}{M_{1}+M_{2}}V(q^{2}).
\end{eqnarray}
Note that the form factor $A_{1}(q^{2})$ dominates at large $q^{2}$
all three of the helicity amplitudes because of $K_{f}(q^{2})$=0 at
$q^{2}_{\rm max}$=$(M_{1}-M_{2})^{2}$.
Thus, to obtain the decay rates for $0^-\to1^-$ processes
in the massless lepton limit,
we need to calculate the form factors of $V,A_1$, and $A_2$ 
(or $g,f$, and $a_+$).  
The form factors $g,a_+$, and $f$ can be calculated using
the `+'-component of the current in $q^+=0$ frame, i.e.,
the transverse decay modes of the vector and axial-vector
currents determine $g$\footnote{ The use of the longitudinal polarization
vector gives zero on both sides of Eq.~(\ref{PV_m1}).}
and $a_+$ and the longitudinal mode
of the axial-vector current determines $f$.

The explicit expressions of the form factors $g,a_+$ and 
$f$ are given  by 
\begin{eqnarray}\label{gq}
&&g(q^{2})=\int^{1}_{0}dx\int d^{2}k_{\perp}
\sqrt{\frac{\partial k_{1n}}{\partial x}}
\sqrt{\frac{\partial k'_{2n}}{\partial x}}
\frac{x\phi^{*}_2(x,{\bf k'}_{\perp})\phi_1(x,{\bf k}_{\perp})}
{\sqrt{ {\cal A}_{2}^{2} +  k^{'2}_{\perp} }
\sqrt{ {\cal A}_{1}^{2} + k^{2}_{\perp} } }
\\
&&\hspace{1cm}
\;\;\times\biggl\{ {\cal A}_{1}
- \frac{(m_{1}-m_{2})}{q^{2}_{\perp}}{\bf k}_{\perp}\cdot{\bf q}_{\perp}
+ \frac{2}{M_{20} + m_{2} + m_{\bar{q}}}\biggl[ k^{2}_{\perp}
-\frac{({\bf k}_{\perp}\cdot{\bf q}_{\perp})^{2}}{q^{2}_{\perp}}\biggr]
\biggr\},\nonumber\\
&&a_+(q^{2})=\int^{1}_{0}dx\int d^{2}k_{\perp}
\sqrt{\frac{\partial k_{1n}}{\partial x}}
\sqrt{\frac{\partial k'_{2n}}{\partial x}}
\frac{x\phi^{*}_2(x,{\bf k'}_{\perp})\phi_1(x,{\bf k}_{\perp})}
{\sqrt{ {\cal A}_{2}^{2} +  k^{'2}_{\perp} }
\sqrt{ {\cal A}_{1}^{2} + k^{2}_{\perp} } } \nonumber\\
&&\hspace{1cm}\;\;\;\;
\times\biggl\{ (1-2x){\cal A}_{1}
-\frac{{\bf k}_{\perp}\cdot{\bf q}_{\perp}}{xq^{2}_{\perp}}
[(1-2x){\cal A}_{1} - {\cal A}_{2} ]\nonumber\\
&&\hspace{1.5cm}\;\;\;\;\;
-2\frac{(1- {\bf k}_{\perp}\cdot{\bf q}_{\perp}/xq^{2}_{\perp})}
{M_{20} + m_{2} + m_{\bar{q}}} 
({\bf k'}_{\perp}\cdot{\bf k}_{\perp}+{\cal A}_{1}{\cal B}_{2})
\biggr\},
\end{eqnarray}
and 
\begin{eqnarray}\label{fq}
&&f(q^2)= (M^2_2-M^2_1+q^2)a_{+}(q^2) 
\nonumber\\
&&\hspace{1cm}\;\;\;-2M_2\int^{1}_{0}dx\int d^{2}k_{\perp}
\sqrt{\frac{\partial k_{1n}}{\partial x}}
\sqrt{\frac{\partial k'_{2n}}{\partial x}}
\frac{\phi^{*}_2(x,{\bf k'}_{\perp})\phi_1(x,{\bf k}_{\perp})} 
{\sqrt{ {\cal A}_{2}^{2} +  k^{'2}_{\perp} }
\sqrt{ {\cal A}_{1}^{2} + k^{2}_{\perp} } }
\\
&&\hspace{1cm}\;\;\;\;\times\biggl\{ 2x(1-x){\cal A}_{1}M_{20}
+\frac{ (1-2x)M_{20} + m_{2}-m_{\bar{q}}}{M_{20} + m_{2} + m_{\bar{q}}}
[{\bf k'}_{\perp}\cdot{\bf k}_{\perp} + {\cal A}_{1}{\cal B}_{2} ]
\biggr\},\nonumber
\end{eqnarray}
where ${\cal B}_{i}$=$-xm_{i} + (1-x)m_{\bar{q}}$.
Accordingly, the form factors $V(q^2)$, $A_{1}(q^2)$ and $A_{2}(q^2)$
can be obtained from Eq.~(\ref{fag}). 
As in the case of $0^-\to 0^-$ transitions, we now analytically
continue the form factors $V(q^2)$, $A_{1}(q^2)$ and $A_{2}(q^2)$ 
to the timelike $q^2$ region by changing $q_\perp$ to $iq_\perp$
in Eqs.~(\ref{gq})-(\ref{fq}).
Of special interest, we also compare our analytic solutions with
the following form factor parametrization~\cite{jaus1,Melikhov}:
\be\label{ansatz}
{\cal F}(q^2)=\frac{F(0)}{1-q^2/\Lambda^{2}_1 
+ s_2q^4/\Lambda^{4}_2},
\ee 
where $s_2$=$\pm1$ and the parameters $\Lambda_i$ are determined by 
the calculation of the appropriate derivatives of ${\cal F}(q^2)$ at 
$q^2$=0 using Eqs.~(\ref{fp}) and (\ref{gq})-(\ref{fq}).

\section{Numerical Results}
For the numerical calculations for various $0^{-}\to 0^{-}(1^-)$
semileptonic decays, we used the quark model parameters given in
Tables~\ref{t71} and~\ref{t72}. Since the smearing effect 
is very small, we do not include the
results obtained from the parameter sets of the SF in our numerical
analyses. 
\subsection{$K_{\ell 3}$ decays}
Our predictions of the parameters for $K_{\ell 3}$ decays in $q^{+}$=0
frame, i.e., $f_{+}(0)$, $\lambda_{+}$, $\lambda_{0}$,
$\la r^{2}\ra_{K\pi}$=$6f'_{+}(0)/f_{+}(0)$=$6\lambda_{+}/M^{2}_{\pi^{+}}$, 
and $\xi_{A}$=$f_{-}/f_{+}|_{q^{2}=m^{2}_{l}}$ given by Eq.~(\ref{K_lam}),
are summarized in Table~\ref{t75}. We do not distinguish $K_{e3}$
from $K_{\mu3}$ in the calculation of the above parameters
since the slopes of $f_{\pm}$ are
almost constant in the range of $m^{2}_{e}\leq q^{2}\leq m^{2}_{\mu}$.
However, the decay rates should be different due to the phase space
factors given by Eq.~(\ref{PP_rate}) and our numerical results for 
$\Gamma(K_{e3})$ and $\Gamma(K_{\mu3})$ in $q^{+}$=0 frame are also 
presented in Table~\ref{t75}.
\begin{table}
\centering
\caption{
Model predictions for the parameters of $K_{\ell 3}$ decay form factors
obtained from $q^{+}=0$ frame. As a sensitivity check,
we include the results in square brackets by changing $m_{s}$=0.48 to
0.43 GeV for the HO parameters. The CKM matrix used in the calculation of
the decay width (in units of $10^{6}$ s$^{-1}$) is
$|V_{us}|=0.2205\pm0.0018$~\protect\cite{data}.}\label{t75}
{\footnotesize
\begin{tabular}{|l|c|c|c|r|}\hline
Observables & HO & Linear & Other models & Expt.~\cite{data}\\
\hline
$f_{+}(0)$ & 0.961 & 0.962 & 0.961$\pm$0.008$^{a}$,
$0.952^{e}$,$0.98^{f}$,$0.93^{g}$ & \\
 & [0.974] & & & \\
\hline
$\lambda_{+}$& 0.025 & 0.026
& 0.031$^{b}$,0.033$^{c}$,0.025$^{d}$
& 0.0286$\pm$0.0022$[K^{+}_{e3}]$\\
 & [0.029] & &0.028$^{e}$,0.018$^{f}$,0.019$^{g}$ & \\
& & & & 0.0300$\pm$0.0016$[K^{0}_{e3}]$\\
\hline
$\lambda_{0}$& $-0.007$ & $-0.009$ & 0.017$\pm$0.004$^{b}$
0.013$^{c}$,0.0$^{d}$ & 0.004$\pm$0.007$[K^{+}_{\mu3}]$\\
 & $[+0.0027]$ & &0.0026$^{e}$,$-0.0024^{f}$,$-0.005^{g}$ &
0.025$\pm$0.006$[K^{0}_{\mu3}]$\\
\hline
$\xi_{A}$& $-0.38$& $-0.41$ & $-0.164$$\pm$0.047$^{b}$
$-0.24^{c}$,$-0.28^{d}$ & $-0.35$$\pm$0.15$[K^{+}_{\mu3}]$\\
 & $[-0.31]$ & 
&$-0.28^{e}$,$-0.25^{f}$,$-0.28^{g}$ & $-0.11$$\pm$0.09$[K^{0}_{\mu3}]$\\
\hline
$\la r\ra_{\pi K}$(fm) & 0.55 & 0.56 &
0.61$^{b}$,0.57$^{e}$, 0.47$^{f}$,0.48$^{g}$ & \\
 & [0.59] & & & \\
\hline
$\Gamma(K^{0}_{e3})$ & 7.30$\pm$0.12
& 7.36$\pm$0.12 & &7.7$\pm$0.5$[K^{0}_{e3}]$\\
 & [7.60$\pm$0.12] & & & \\ 
\hline
$\Gamma(K^{0}_{\mu3})$& 4.57$\pm$0.07& 4.56$\pm$0.07 &
&5.25$\pm$0.07$[K^{0}_{\mu3}]$\\
 & [4.84$\pm$0.08] & & & \\
\hline
\end{tabular}
$^{a}$ Ref.~\cite{Roos},$^{b}$ Ref.~\cite{Gasser},$^{c}$ Ref.~\cite{CL},
$^{d}$ Ref.~\cite{VMD}, $^{e}$ Ref.~\cite{Andrei}.
$^{f}$ Ref.~\cite{Yuri},$^{g}$ Ref.~\cite{isgw2}.
}
\end{table}
Our results for the form factor $f_{+}$ at zero momentum
transfer, $f_{+}(0)$=0.961 [0.962] for the HO [linear] parameters, 
are consistent with the Ademollo-Gatto theorem~\cite{Gatto} and also in 
an excellent agreement with the result of chiral perturbation 
theory~\cite{Roos},
$f_{+}(0)$=0.961$\pm$0.008. Our results for other observables such as
$\lambda_{+}$, $\xi_{A}$, and $\Gamma(K_{\ell 3})$
are overall in a good agreement with the experimental data~\cite{data}.
We have also investigated the sensitivity of our results by
varying quark masses. For instance, the results\footnote{ Even though we
show the results only for the HO parameters, we find the similar variations 
for the linear parameters; i.e., the posivive sign of $\lambda_{0}$ can be 
obtained when $m_{s}/m_{u}$$\leq$1.8 for both HO and linear paramters. 
In addition to the observables
in this work, our predictions for $f_{K}$, $r^{2}_{K^{+}}$, and
$r^{2}_{K^{0}}$ in~\cite{Mix} are changed to 108 MeV ($1\%$ change),
0.385 fm$^{2}$ ($0.3\%$ change), and $-0.077$ fm$^{2}$ (15$\%$),
respectively.} obtained by changing the strange quark mass from
$m_{s}$=0.48 GeV to 0.43 GeV ($10\%$ change) for the HO parameters are 
included in Table~\ref{t75}. As one can see in Table~\ref{t75}, 
our model predictions are quite stable for the
variation of $m_{s}$ except $\lambda_{0}$, which changes its sign from
$-0.007$ to $+0.0027$. The large variation of $\lambda_{0}$
is mainly due to the rather large sensitivity of $f_{-}(0)$ (18$\%$
change) to the variation of $m_{s}$.
Similar observation regarding on the large sensitivity for $\lambda_{0}$
compared to other observables has also been reported in Ref.~\cite{Andrei}
for the variation of quark masses.
As discussed in Refs.~\cite{Yuri} and~\cite{IS}, $f_{-}(0)$ is sensitive
to the nonperturbative enhancement of the SU(3) symmetry breaking mass
difference $m_{s}-m_{u(d)}$ since $f_{-}(0)$ depends on the ratio of $m_{s}$
and $m_{u(d)}$.

Of special interest, we also observed that the nonvalence contributions
from $q^{+}$$\neq$0 frame are clearly visible for $\lambda_{+}$,
$\lambda_{0}$ and $\xi_{A}$ even though it may not be quite significant
for the decay rate $\Gamma(K_{\ell3})$.
Our predictions with only the valence contributions in $q^{+}$$\neq$0
frame are $f_{+}(0)$=0.961 [0.962], $\lambda_{+}$=0.081 [0.083],
$\lambda_{0}$=$-0.014$ $[-0.017]$, $\xi_{A}$=$-1.12$ $[-1.10]$,
$\Gamma(K_{e3})$=(8.02[7.83]$\pm$0.13)$\times10^{6}$ s$^{-1}$ and
$\Gamma(K_{\mu3})$=(4.49[4.36]$\pm$ 0.13)$\times10^{6}$ s$^{-1}$ for the
HO [linear] parameters. 
Even though the form factor $f_{+}(0)$ in $q^{+}$$\neq$0 frame is free 
from the nonvalence contributions, its derivative at $q^{2}$=0,
i.e., $\lambda_{+}$, receives the nonvalence contributions. Moreover,
the form factor $f_{-}(q^{2})$ in $q^{+}$$\neq$0 frame is not immune to
the nonvalence contributions even at $q^{2}$=0~\cite{zm}.
Unless one includes the nonvalence contributions in the $q^{+}$$\neq$0 frame,
one cannot really obtain reliable predictions for the observables such as
$\lambda_{+}, \lambda_{0}$ and $\xi_{A}$ for $K_{\ell3}$ decays.

\setcounter{figure}{5}
\begin{figure}[t]
\centerline{\psfig{figure=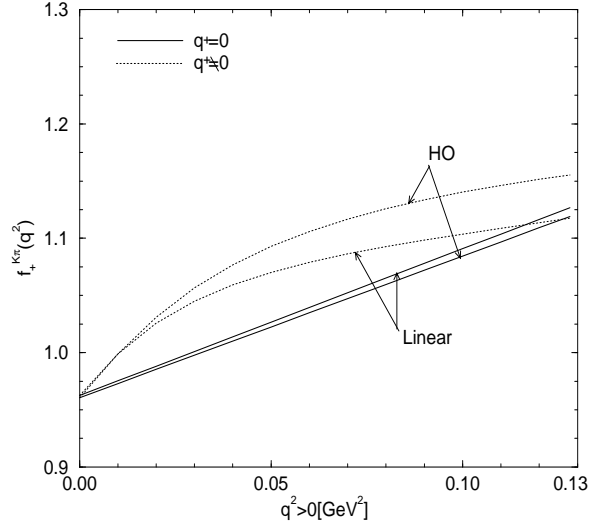,width=3.5in,height=3.2in}}
\caption{The form factors $f_{+}(q^{2})$ for the
$K\to\pi$ transition in timelike momentum transfer $q^{2}>0$. The solid
and dotted lines are the results from the $q^{+}$=0 and $q^{+}$$\neq$0
frames for the HO and linear parameters, respectively.
The differences of the results between the two frames are the measure of
the nonvalence contributions from $q^{+}$$\neq$0 frame.} 
\end{figure}
\begin{figure}
\centerline{\psfig{figure=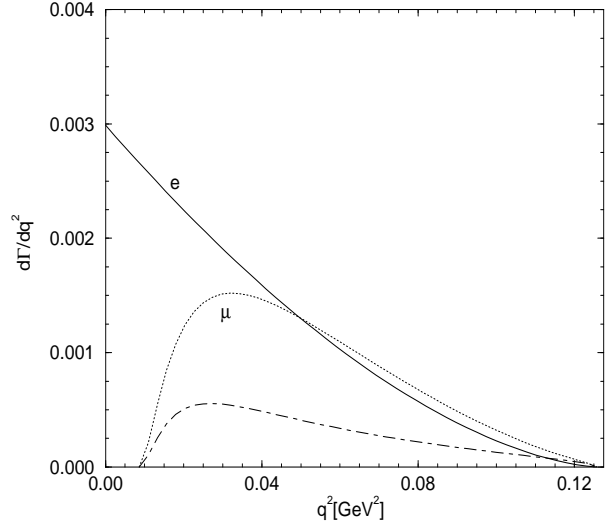,width=3.5in,height=3.2in}}
\caption{ The decay rates $d\Gamma/dq^{2}$ of
$K_{e3}$(solid line) and $K_{\mu3}$(dotted line) for the HO parameters
in $q^{+}$=0 frame. The dot-dashed line is the contribution 
from the term proportional to $f_{0}$ in Eq.~(\ref{PP_rate})
for $K_{\mu 3}$ decay. The results for the linear parameters are not much 
different from those for the HO parameters.}
\end{figure}
In Fig.~7.6, we show the form factors $f_{+}$ obtained from both $q^{+}$=0
and $q^{+}$$\neq$0 frames for 0$\leq$$q^{2}$$\leq$$(M_{K}-M_{\pi})^{2}$ region.
As one can see in Fig.~7.6, the form factors $f_{+}$
obtained from $q^{+}$=0 frame (solid lines) for both HO and linear parameters 
appear to be linear functions of $q^{2}$ justifying Eq.~(\ref{K_lam}) 
usually employed in the analysis of experimental data~\cite{data}. 
Note, however, that the curves without the nonvalence contributions in
$q^{+}$$\neq$0 frame (dotted lines) do not exhibit the same behavior.
In Fig.~7.7, we show $d\Gamma/dq^{2}$ spectra for $K_{e3}$ (solid line) and
$K_{\mu3}$ (dotted line) obtained from $q^{+}$=0 frame.
While the term proportional to $f_{0}$ in Eq.~(\ref{PP_rate}) is negligible 
for $K_{e3}$ decay rate, its contribution for $K_{\mu3}$ decay rate is quite
substantial (dot-dashed line). Also, we show in Fig.~7.8 the form factors
$f_{+}(q^{2})$ (solid and dotted lines for the HO and linear
parameters, respectively)
at spacelike momentum transfer region and compare with the theoretical
prediction from Ref.~\cite{Andrei} (dot-dashed line).
The measurement of this observable in $q^{2}<0$ region is anticipated
from TJNAF~\cite{Andrei}.
\begin{figure}
\centerline{\psfig{figure=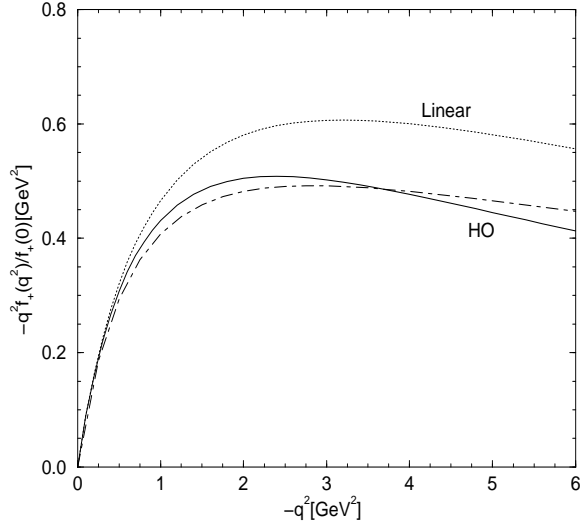,width=3.5in,height=3.2in}}
\caption{ The form factors $f_{+}(q^{2})$ for the
$K\to\pi$ transition in spacelike momentum transfer $-q^{2}<0$. The
solid and dotted lines are the results from the HO and linear parameters,
respectively. The dot-dashed line is the result from 
Ref.~\protect\cite{Andrei}.}
\end{figure} 

We have also estimated the zero-mode contribution by calculating
the $``-"$ component of the current. Our observation in an exactly
solvable scalar field theory was presented in Chapter 6~\cite{zm}.
Using the light-front bad current $J^{-}$ in $q^{+}$=0 frame,
we obtained $f_{-}(0)$=12.6 (18.6) for the HO (linear) parameters.
The huge ratio of $f_{-}(0)|_{J^{-}}/f_{-}(0)|_{J_{\perp}}$$\approx$
$-36$ ($-48$) for the HO (linear) parameters 
is consistent with our observation in Chapter 6~\cite{zm}.
We also found that the zero-mode contribution is highly
suppressed as the quark mass increases.
\subsection{$D$ and $D_s$ decays}
(1) {\bf $D\to\pi^{-}e^{+}\nu_{e}$}:
Our predicted decay rate for $D\to\pi$ in $q^{+}$=0 frame is
$\Gamma(D^{0}\to\pi^{-}e^{+}\nu_{e})$=0.110 (0.113)$|V_{cd}|^{2}$
ps$^{-1}$. Using the lifetime $\tau_{D^{0}}=0.415\pm 0.004$ ps and 
$|V_{cd}|$=0.224$\pm$0.016~\cite{data},
we obtain the following branching ratio
\begin{eqnarray}
{\rm Br}(D^{0}\to\pi^{-}e^{+}\nu_{e})&=& (2.30\pm 0.33)\times 10^{-3}
\;\;{\rm (HO)},\nonumber\\
&=&(2.36\pm 0.34)\times 10^{-3} \;\;{\rm (Linear)},
\end{eqnarray}
while the experimental data~\cite{data} reported 
Br$_{\rm exp.}$$(D^{0}\to\pi^{-}e^{+}\nu_{e})$=
$(3.9\pm^{+2.3}_{-1.1}\pm0.4)\times 10^{-3}$.
Even though our predicted decay rates are rather smaller than the 
central value of the data, one shoud note that the number of events
for the $D\to\pi$ data is currently very small compared to other
processes~\cite{data}. We also found that
the decay rate obtained from the valence contributions in
$q^{+}$$\neq$0 frame is about 18 (17)$\%$ larger for the HO (linear)
parameters than that obtained from $q^{+}$=0 frame.
 
\begin{figure}
\centerline{\psfig{figure=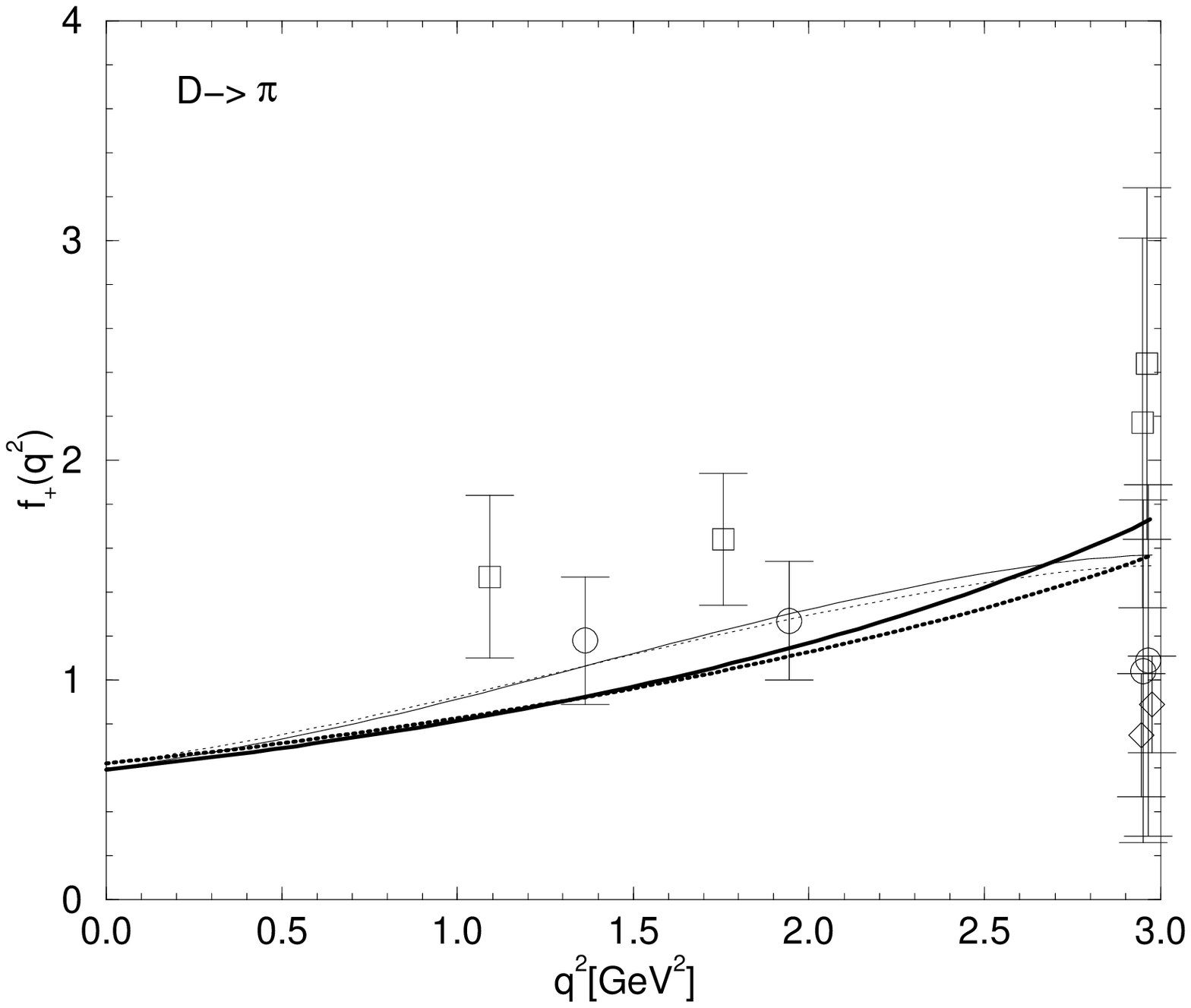,width=3.5in,height=3.5in}}
\centerline{\psfig{figure=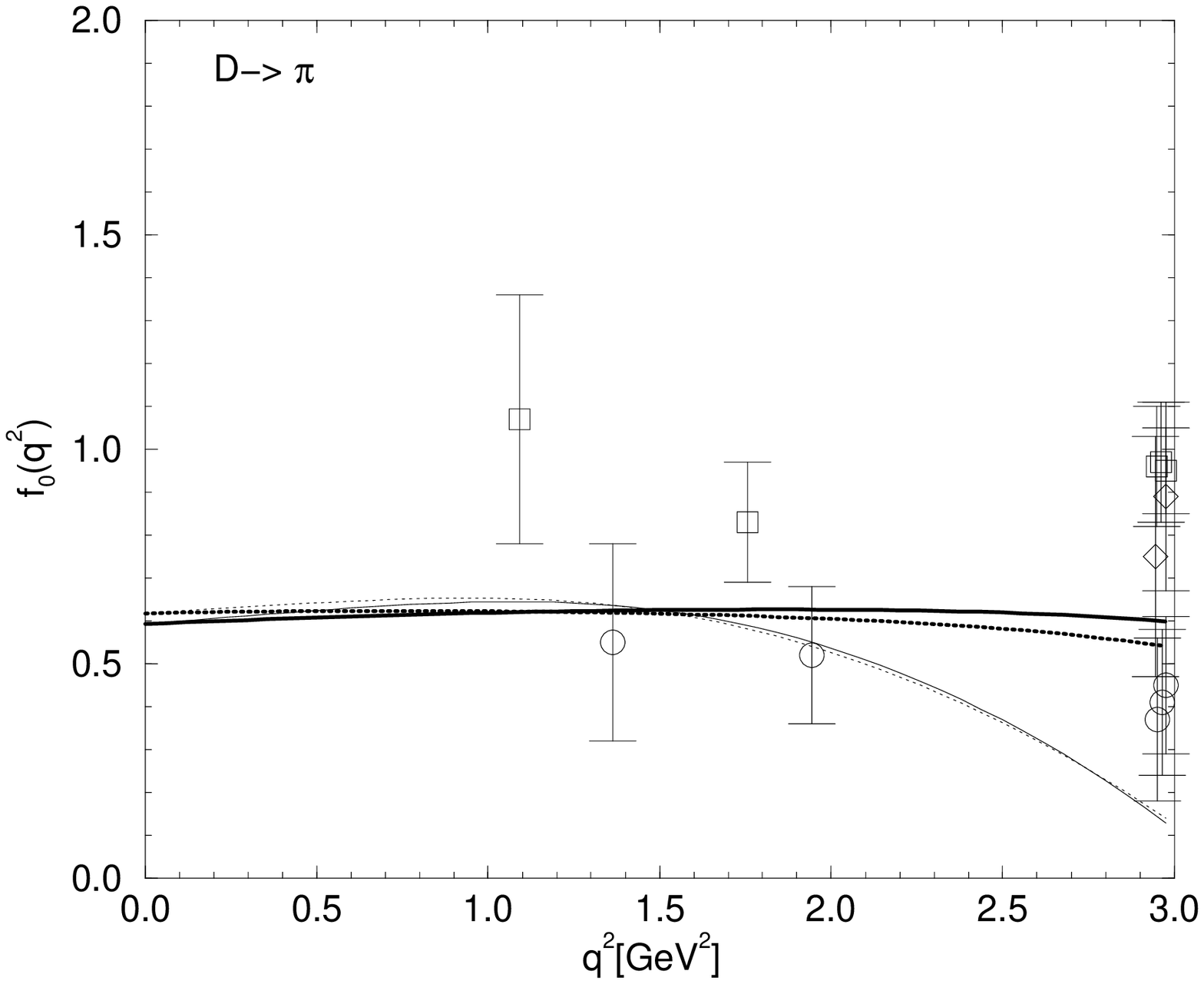,width=3.5in,height=3.5in}}
\caption{ (a) The form factor $f_{+}(q^{2})$ for
the $D\to\pi$  transition compared with a lattice
calculation~\protect\cite{Bernard}. 
The same line code as in Fig.~7.2 is used for our results. 
(b) The form factor $f_{0}(q^{2})$ for the $D\to\pi$ transition.
The same line code is used as in (a).}
\end{figure}
In Figs.~7.9(a) and 7.9(b), we present the $q^{2}$
dependence of the form factors (thick lines) $f^{D\pi}_{+}(q^{2})$ 
and $f^{D\pi}_{0}(q^{2})$ obtained from $q^+$=0 frame 
(thick solid lines for HO paramters and dotted lines for linear
parameters), respectively. For comparison, we also show the
results (thin lines) obtained from the valence contributions in
$q^+$$\neq$0 frame as well as the lattice calculations~\cite{Bernard}.
As on can see in Figs.~7.9(a) and 7.9(b), while the differences from the 
choice of the model parameters (i.e., HO or linear) are not significant,
the ones between the two frames (i.e., $q^+$=0 and $q^+$$\neq$0 frames),
which are the measure of the nonvalence contributions from $q^+$$\neq$0
frame, are quite substantial, especially for the form factor 
$f^{D\pi}_{0}(q^{2})$ case. 
Our predictions in $q^+$=0 frame are also comparable to the
the lattice results~\cite{Bernard}.

\begin{table}
\centering
\caption{ Summary of the parameters for the various form factors
for $D$$\to$$\pi$ $[K]$ and $D$$\to$$\rho$ $[K^*]$ transitions.
The experimental
data are taken from the survey in Ref.~\protect\cite{data2}.}\label{t76}
{\scriptsize
\begin{tabular}{|r|c|c|c|c|c|}\hline
   &\multicolumn{2}{|c|}{$D$$\to$$\pi$ $[K]$}
   &\multicolumn{2}{|c}{$D$$\to$$\rho$ $[K^*]$}& \\
\cline{2-6}
   & $F$     &  $f_+$    & $V$  & $A_1$    & $A_2$  \\
\hline
HO &$F(0)$ &0.593 [0.724]& 0.709 [0.809]& 0.510 [0.640]& 0.342 [0.450]\\
\cline{2-6}
   & $F(q^{2}_{\rm max})$ & 1.73 [1.25]& 1.006 [1.029]& 0.572 [0.698]
   & 0.419 [0.534] \\
\cline{2-6}
   & $\Lambda_1$ [GeV]
   & 1.95 [2.06] & 1.94 [2.06]& 3.37 [3.41] & 2.45 [2.41] \\
\cline{2-6}
   & $s_2\Lambda_2$ [GeV]
   & 2.88 [3.04] & 2.86 [3.04]& $-5.05$ [$-5.05$] & 3.18 [3.43] \\
\cline{2-6}
   & ${\cal F}_{\Lambda_1}(q^{2}_{\rm max})$
   & 2.71 [1.30]& 1.039 [1.029]
   & 0.570 [0.698] & 0.428 [0.534] \\
\cline{2-6}
   & ${\cal F}_{\Lambda_1\Lambda_2}(q^{2}_{\rm max})$ 
   & 1.71 [1.21]& 1.007 [1.042]
   & 0.571 [0.697] & 0.420 [0.538] \\
\hline
Linear& $F(0)$ &0.618 [0.736]& 0.739 [0.822]& 0.517 [0.638]& 0.311 [0.414]\\
\cline{2-6}
   & $F(q^{2}_{\rm max})$ & 1.57 [1.23]& 1.002 [1.024]& 0.563 [0.683]
   & 0.357 [0.475] \\
\cline{2-6}
   & $\Lambda_1$ [GeV]
   & 2.06 [2.16] & 2.05 [2.15]& 3.99 [3.85] & 2.92 [2.67] \\
\cline{2-6}
   & $s_2\Lambda_2$ [GeV] 
   & 3.17 [3.29] & 3.26 [3.32]& $-4.62$ [$-4.61$] & 3.27 [3.76] \\
\cline{2-6}
   & ${\cal F}_{\Lambda_1}(q^{2}_{\rm max})$
   & 2.10 [1.23]& 1.036 [1.033]
   & 0.559 [0.683]  & 0.362 [0.475] \\
\cline{2-6}
   & ${\cal F}_{\Lambda_1\Lambda_2}(q^{2}_{\rm max})$ 
   & 1.61 [1.18]& 1.018 [1.024]
   & 0.561 [0.681]  & 0.357 [0.477] \\
\hline
LAT~\cite{Flynn}
       & $F(0)$ & 0.65(10)[0.73(7)]& 1.1(2)[1.2(2)] & 0.65(7)[0.70(7)]
       & 0.55(10)[0.6(1)]\\
~\cite{ALL}& $F(0)$ & $\cdots$ [0.78$\pm$0.08] & $\cdots$ [1.08$\pm$0.22]
           & $\cdots$ [0.67$\pm$0.11] & $\cdots$ [0.49$\pm$0.34] \\
~\cite{LMMS}& $F(0)$ & 0.58(9) [0.63(8)]& 0.78(12) [0.86(10)]
            & 0.45(4) [0.53(3)] & 0.02(26) [0.19(21)] \\
~\cite{Bowler}& $F(0)$ & $0.61(^{12}_{11})[0.67(^{7}_{8})]$ 
              & $0.95(^{29}_{14})[1.01(^{30}_{13})]$ 
              & $0.63(^{6}_{9})[0.70(^{7}_{10})]$ 
              & $0.51(^{10}_{15})[0.66(^{10}_{15})]$ \\ 
\hline
SR~\cite{Ball}&$F(0)$ & 0.5$\pm$0.1 [$\cdots$]& 1.0$\pm$0.2 [$\cdots$]
              & 0.5$\pm$0.2 [$\cdots$] & 0.4$\pm$0.1 [$\cdots$]\\
~\cite{BBD}& $F(0)$ & $\cdots$ [0.60$\pm$0.15]& $\cdots$ [1.10$\pm$0.25]
             & $\cdots$ [0.50$\pm$0.15]& $\cdots$ [0.60$\pm$0.15]\\
\hline
QM~\cite{Wirbel}& $F(0)$ & 0.69 [0.76]& 1.23 [1.23]& 0.78 [0.88]
                & 0.92 [1.15]\\
~\cite{Jaus3}& $F(0)$ & 0.67 [0.78] & 0.93 [1.04] & 0.58 [0.66]
               & 0.42 [0.43]\\
\hline
Expt.~\cite{data2}
       & $F(0)$ & $\cdots$ [0.76$\pm$0.03] & $\cdots$ [1.07$\pm$0.09]
       & $\cdots$ [0.58$\pm$0.03]& $\cdots$ [0.41$\pm$0.05]\\
\hline
\end{tabular}
}
\end{table}
In Table~\ref{t76}, we compare our form factor $f^{D\pi}_{+}(q^{2})$ 
at both $q^2$=0 and
$q^2$=$q^{2}_{\rm max}$ with the simple pole parametrization given
by Eq.~(\ref{ansatz}) as well as other theoretical predictions.
It is quite interesting to note that our analytic solution 
$f_{+}(q^{2}_{\rm max})$= 1.73 (1.57) for the HO (linear) parameters
is well approximated by Eq.~(\ref{ansatz}), 
${\cal F}_{\Lambda_1\Lambda_2}(q^{2}_{\rm max})$= 1.71 (1.61), 
but not fitted by the simple pole approximation, 
${\cal F}_{\Lambda_1}(q^2)$=$F(0)/(1-q^2/\Lambda^{2}_1)$= 2.71 (2.10).
We also show in Table~\ref{t77} the form factors $f^{D\pi}_{-}(0)$ 
obtained from both $q^+$=0 and $q^+$$\neq$0 frame. 
As one can see in Table~\ref{t77}, the form factor $f^{D\pi}_{-}(0)$ 
obtained from $q^{+}$$\neq$0 frame for the HO (linear) parameters
is about 40 (38)$\%$ larger than that obtained from $q^{+}$=0 frame.
\begin{table}
\centering
\caption{Comparison of the form factors $f_{-}(q^{2})$
at $q^{2}$=0 between $q^{+}$=0 and $q^{+}$$\neq$0 frames.
The differences between the two frames are the measure of the
nonvalence contributions in $q^+$$\neq$0 frame.
Note that $c_\eta$=$|\cos\delta_P|$ and $c_{\eta'}$=$|\sin\delta_P|$.
}\label{t77}
\begin{tabular}{|c|c|c|c|c|c|}\hline
Models& &$D\to\pi$& $D\to K$&$D_{s}\to\eta$&
$D_{s}\to\eta'$ \\
\hline
HO & $q^{+}$=0 & $-0.428$ & $-0.400$ & $-0.416c_{\eta}$&
$-0.416c_{\eta'}$ \\
& $q^{+}$$\neq$0 & $-0.600$ & $-0.723$ & $-0.764c_{\eta}$&
$-0.431c_{\eta'}$ \\
\hline
Linear& $q^{+}$=0 & $-0.454$ & $-0.422$ & $-0.460c_{\eta}$&
$-0.460c_{\eta'}$ \\
& $q^{+}$$\neq$0 & $-0.625$ & $-0.719$ & $-0.779c_{\eta}$&
$-0.449c_{\eta'}$ \\
\hline
\end{tabular}
\end{table}

(2) {\bf $D^{0}\to K^{-}\ell^{+}\nu$}:
Our predicted decay rate for $D\to K$ in $q^{+}$=0 frame is
$\Gamma(D^{0}\to K^{-}e^{+}\nu_{e})$= 
8.26 (8.36)$|V_{cs}|^{2}\times 10^{-2}$ ps$^{-1}$ and 
$\Gamma(D^{0}\to K^{-}\mu^{+}\nu_{\mu})$=
6.40 (6.43)$|V_{cs}|^{2}\times 10^{-2}$ ps$^{-1}$
for the HO (linear) parameters.
Using $|V_{cs}|$=1.04$\pm$0.16~\cite{data}, our predictions for the 
branching ratio are given by
\begin{eqnarray}
{\rm Br}(D^{0}\to K^{-}e^{+}\nu_{e})&=& (3.71\pm 1.14)\%\hspace{0.2cm}
{\rm (HO)},\nonumber\\
&=&(3.75\pm 1.16)\%\hspace{0.2cm}{\rm (Linear)},\nonumber\\
{\rm Br}(D^{0}\to K^{-}\mu^{+}\nu_{\mu})&=& (2.87\pm 0.88)\%\hspace{0.2cm}
{\rm (HO)},\nonumber\\
&=&(2.89\pm 0.90)\%\hspace{0.2cm}{\rm (Linear)},
\end{eqnarray}
while the experimental data are
${\rm Br}_{\rm exp.}(D^{0}\to K^{-}e^{+}\nu_{e})$=(3.66$\pm$0.18)$\%$ and
${\rm Br}_{\rm exp.}(D^{0}\to K^{-}\mu^{+}\nu_{\mu})$=(3.23$\pm$0.17)$\%$, 
respectively.
Our results for the electron ($e$) decay mode are in a good agreement
with the experimental data for both HO and linear parameters. For the
muon ($\mu$) decay mode, our values are rather smaller than the central
value of the experimental data, neverthless, are quite comparable with
the data within the given uncertainties of CKM matrix, $V_{cs}$.
The result of the decay rate obtained from the valence contributions in
$q^{+}$$\neq$0 frame is about 16 (15)$\%$ larger for the HO (linear)
parameters than that obtained from $q^{+}$=0 frame.
Our value of $f^{DK}_{+}(0)$=0.724 (0.736) for the HO (linear)
parameters is very close to the available experimental data~\cite{data},
$f^{\rm Expt.}_{+}(0)$=0.7$\pm$0.1. Unlike the $D\to\pi$ transition,
our analytic solution $f^{DK}_{+}(q^{2}_{\rm max})$= 1.25 (1.23) for
the HO (linear) parameters is well approximated by both 
Eq.~(\ref{ansatz}), 
${\cal F}_{\Lambda_1\Lambda_2}(q^{2}_{\rm max})$= 1.21 (1.18), 
and the simple pole approximation, 
${\cal F}_{\Lambda_1}(q^{2}_{\rm max})$= 1.30 (1.23).
This implies that the simple pole approximation is enough to fit the form 
factor $f^{DK}_{+}(q^2)$ and one may neglect the $\Lambda_2$ and higher 
order contributions in Eq.~(\ref{ansatz}). 
Moreover, our simple pole masses $\Lambda_1$=2.06 and 2.16 GeV for both 
HO and linear parameters, respectively, are in
good agreement with the value of 2.11(=$D^{*\pm}_s$) GeV expected from 
the closest resonance with the proper quantum number $J^P$=$1^{-}$.
As one can see in Table~\ref{t77}, the form factor $f^{DK}_{-}(0)$
obtained from $q^{+}$$\neq$0 frame for the HO (linear) parameters
is about 80 (70)$\%$ larger than that obtained from $q^{+}$=0 frame.

\begin{figure}
\centerline{\psfig{figure=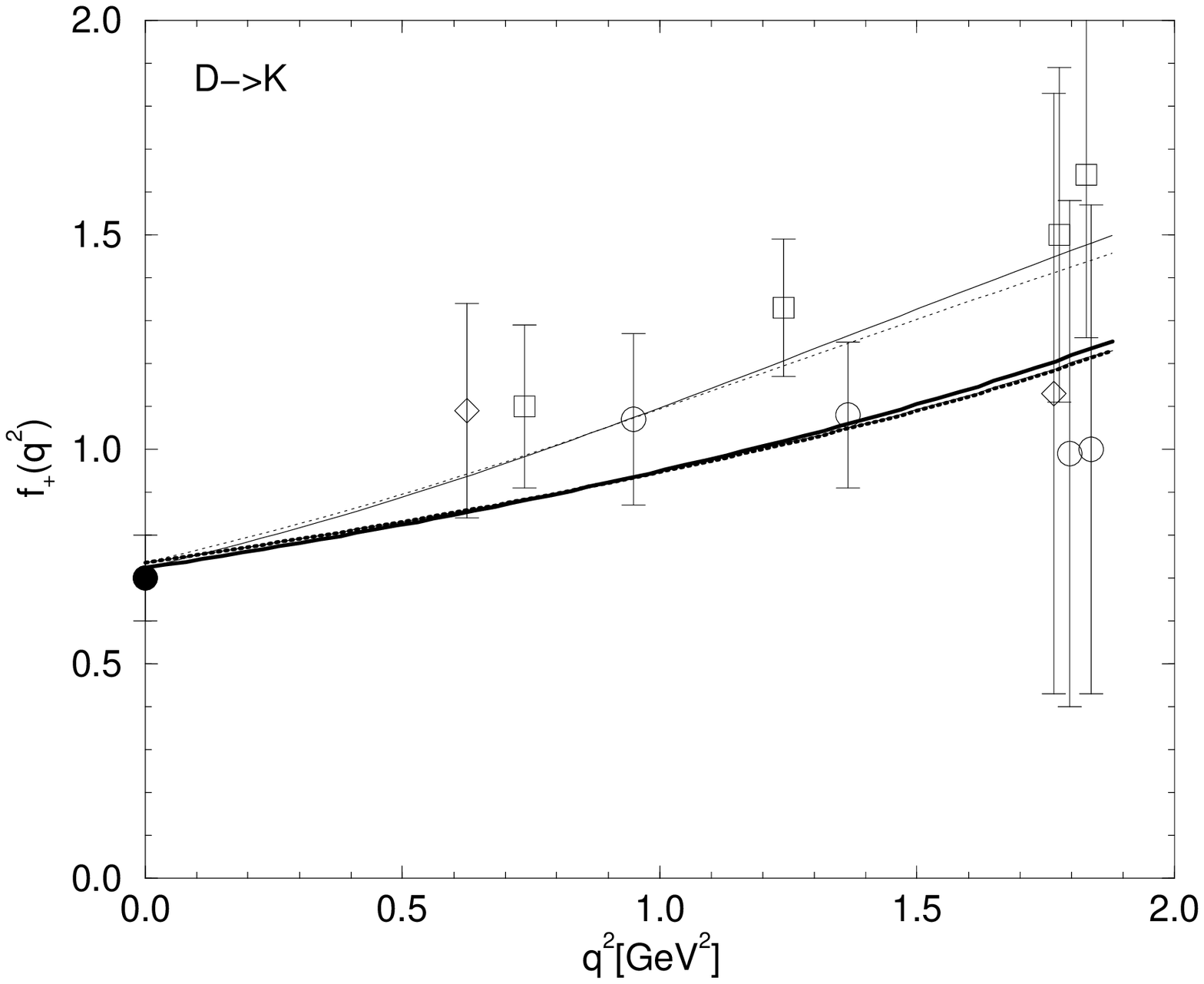,width=3.5in,height=3.5in}}
\centerline{\psfig{figure=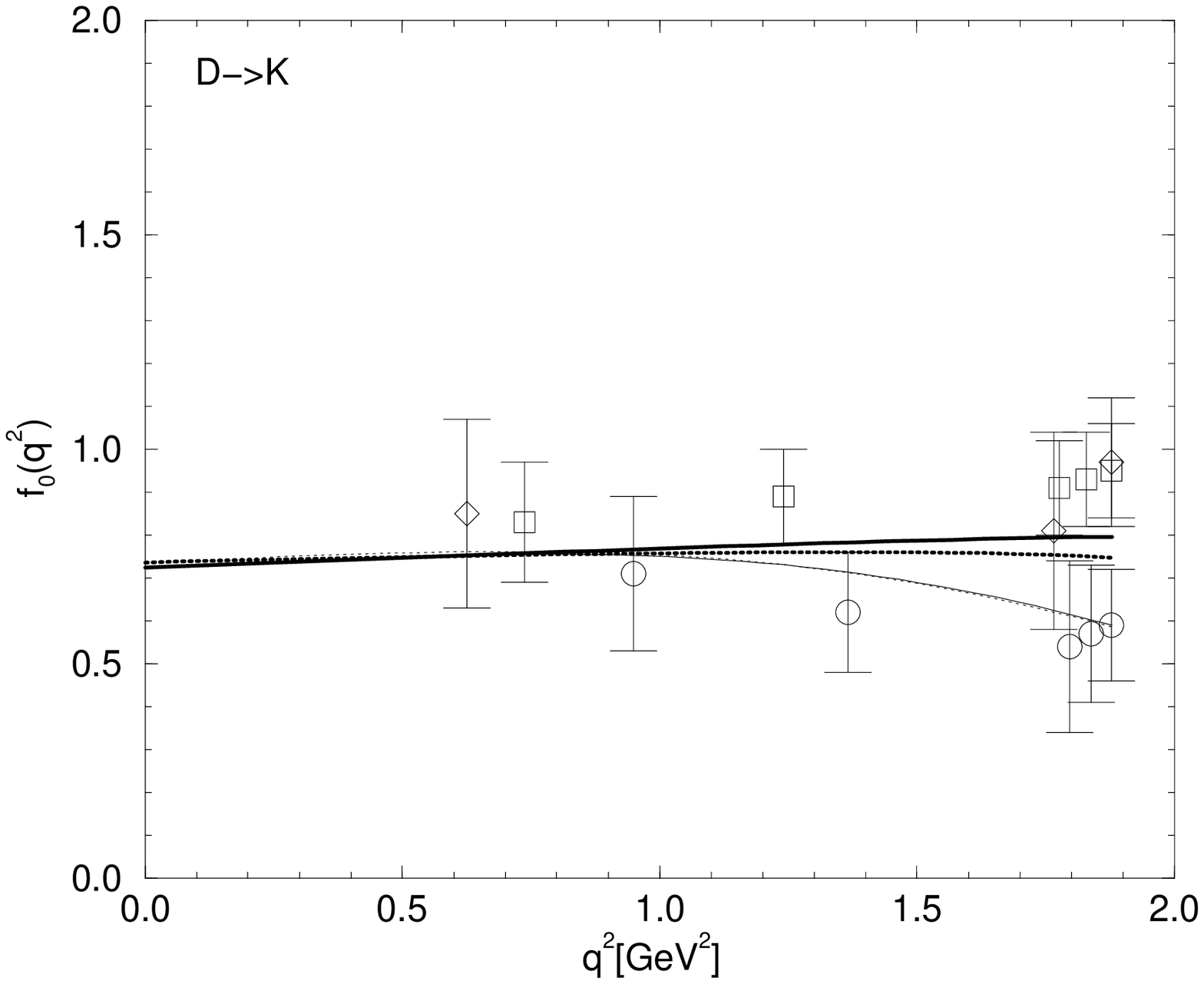,width=3.5in,height=3.5in}}
\caption{ (a) The form factor $f_{+}(q^{2})$ for the $D\to K$
transition compared with the experimental
data ($\bullet$)~\protect\cite{data}
and the lattice calculation~\protect\cite{Bernard}.
The same line code as in Fig.~7.2 is used for
our results. (b) The form factor $f_{0}(q^{2})$ for the $D\to K$ transition.
The same line code is used as in (a).}
\end{figure}
In Figs.~7.10(a) and 7.10(b), we present the $q^{2}$ dependence of the 
form factors $f^{DK}_{+}(q^{2})$ and $f^{DK}_{0}(q^{2})$
obtained from both $q^+$=0 (thick lines) and $q^{+}$$\neq$0 (thin lines) 
frames, respectively, and compare with the available 
experimental data ($\bullet$) as well as
the lattice QCD calculations~\cite{Bernard}.
Again, the nonvalence contributions from $q^+$$\neq$0 frame are
quite sizeable especially for $f^{DK}_{0}(q^{2})$ case.
Our results for both
$f^{DK}_{+}(q^{2})$ and $f^{DK}_{0}(q^{2})$ in $q^{+}$=0 frame 
are overall in a good agreement with the lattice calculations 
in Ref.~\cite{Bernard}.

\begin{figure}
\centerline{\psfig{figure=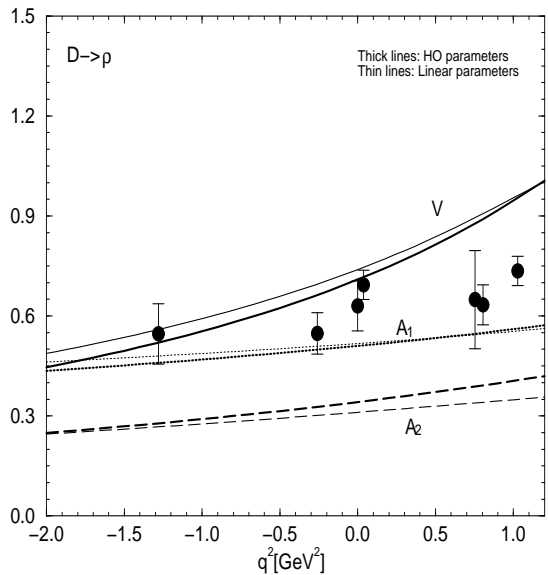,width=3.5in,height=3.5in}}
\centerline{\psfig{figure=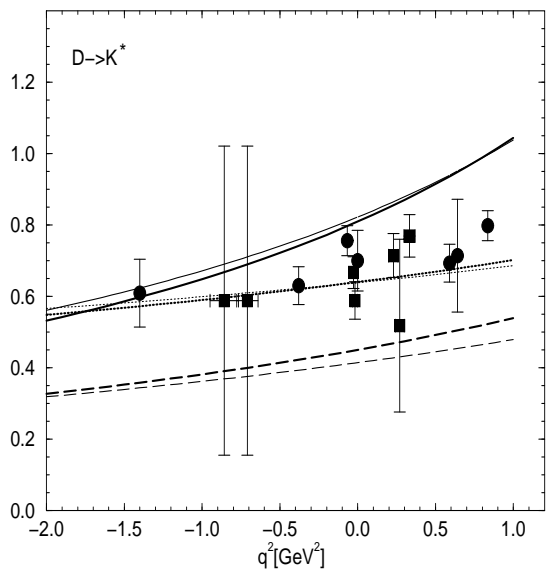,width=3.5in,height=3.5in}}
\caption{The form factors $V(q^2)$ (solid line), 
$A_1(q^2)$ (dotted line) and $A_2(q^2)$ (long-dashed line),
where the thick (thin) lines represent the HO (linear) parameters, 
for $D\to\rho$ transition compared with the lattice 
calculation~\protect\cite{Bowler} of $A_1(q^2)$ (data).
7.12: The form factors $V(q^2)$, $A_1(q^2)$ and $A_2(q^2)$
for $D\to K^{*}$ transition compared with the lattice 
calculations~\protect\cite{ALL} (square) and~\protect\cite{Bowler} 
(circle) of $A_1(q^2)$.
The same line code as in Fig.~7.11 is used.}
\end{figure} 
(3) $D^0\to\rho^-\ell^{+}\nu_{\ell}$: 
Our predicted decay rate for $D\to\rho$ in $q^+$=0 frame is 
$\Gamma(D^0\to\rho^-\ell^{+}\nu_{\ell})$=
6.38 (6.61)$|V_{cd}|^2\times 10^{-2}$ ps$^{-1}$ 
for the HO (linear) parameters. We also obtain the ratio 
$\Gamma_L/\Gamma_T$, where the subscript $L$ represents the longitudinal 
polarization and $T$ transverse polarization, for the 
HO (linear) parameters as 1.39 (1.48), which is quite comparable
to the QCD sum rule result: 1.31$\pm$0.11 in Ref.~\cite{Ball}. 
Our predictions for the branching ratio given by
\bea
{\rm Br}(D^0\to\rho^-\ell^{+}\nu_{\ell})
&=& (1.33\pm 0.19)\%\;\; {\rm (HO)},\nonumber\\
&=& (1.38\pm 0.19)\%\;\; {\rm (Linear)},
\eea
are overall in a good agreement with the available experimental 
data~\cite{data}: ${\rm Br}_{\rm exp.}=(1.87\pm 0.9)\%$.  
We show in Table~\ref{t76} our values of $V$, $A_1$ and $A_2$ at both
$q^2$=0 and $q^2$=$q^{2}_{\rm max}$ and compare with Eq.~(\ref{ansatz})
as well as other theoretical predictions.
As one can see in Table~\ref{t76}, our analytic solutions for the form
factors $V(q^{2}_{\rm max})$= 1.006 (1.002), 
$A_1(q^{2}_{\rm max})$=0.572 (0.563) and 
$A_2(q^{2}_{\rm max})$=0.419 (0.357) for the HO (linear) parameters
are well approximated by Eq.~(\ref{ansatz}),
$V_{\Lambda_1\Lambda_2}(q^{2}_{\rm max})$=1.007 (1.018),
$A_{1\Lambda_1\Lambda_2}(q^{2}_{\rm max})$=0.571 (0.561) and
$A_{2\Lambda_1\Lambda_2}(q^{2}_{\rm max})$=0.420 (0.357). Our
analytic solutions are also well fitted by the simple pole
formular,
$V_{\Lambda_1}(q^{2}_{\rm max})$=1.039 (1.036),
$A_{1\Lambda_1}(q^{2}_{\rm max})$=0.570 (0.559) and
$A_{2\Lambda_1}(q^{2}_{\rm max})$=0.428 (0.362). 
This again means that the simple pole formular may be
a good approximation to fit the form factors $V$, $A_1$ and $A_2$. 
In Fig.~7.11, we present our $q^{2}$ dependence of the form factors
$V(q^2)$ (solid line), $A_1(q^2)$ (dotted line) and 
$A_2(q^2)$ (long-dashed line) obtained from $q^+$=0 frame, 
where the thick and thin lines represent
the HO and linear parameters, respectively. 
We also compare our $A_1(q^2)$ with 
the available lattice result~\cite{Bowler} (data) and show they are 
overall in a good agreement with each other.

(4) $D^0\to K^{*-}\ell^{+}\nu_{\ell}$: 
Our predicted decay rate for $D\to K^{*}$ in $q^+$=0 frame is 
$\Gamma(D^0\to K^{*-}\ell^{+}\nu_{\ell})$=5.86 (5.85)$|V_{cs}|^2\times 
10^{-2}$ ps$^{-1}$ for the HO (linear) parameters.
We obtain the ratio $\Gamma_L/\Gamma_T$ as 1.37 (1.42) for the HO (linear) 
parameters, while the available experimental data reported  
as 1.18$\pm$0.18$\pm$0.08 (E653)~\cite{E653} and 
1.20$\pm$0.13$\pm$0.13 (E687)~\cite{E687}. 
Our predictions for the branching ratio given by
\bea
{\rm Br}(D^0\to K^{*-}\ell^{+}\nu_{\ell})
&=& (2.63\pm 0.81)\%\;\; {\rm (HO)},\nonumber\\
&=& (2.62\pm 0.81)\%\;\; {\rm (Linear)},
\eea
are quite comparable to the available experimental data~\cite{data}:
${\rm Br}_{\rm exp.}$=$(2.02\pm 0.33)\%$.
We summarize in Table~\ref{t76} our results of the form factors $V$, 
$A_1$ and $A_2$ at both $q^2$=0 and $q^2$=$q^{2}_{\rm max}$ and compare 
with the available experimental data~\cite{data2} as well as 
Eq.~(\ref{ansatz}) and other theoretical results.
While our form factors $A_{1}(0)$=0.640 (0.638) and $A_{2}(0)$=0.450 
(0.414) for the HO (linear) parameters are quite comparable with the
data, $A_{1}(0)=0.58\pm0.03$ and $A_{2}(0)$=$0.41\pm0.05$~\cite{data2}.
our form factor $V(0)$=0.809 (0.822) underestimates the current data, 
$V(0)$=$1.07\pm0.09$~\cite{data2}. As one can see in Table~\ref{t76},  
our analytic solutions for the form factors 
$V(q^{2}_{\rm max})$=1.029 (1.024), 
$A_1(q^{2}_{\rm max})$=0.698 (0.683) and 
$A_2(q^{2}_{\rm max})$=0.534 (0.475) for the HO (linear) parameters
are well approximated not only by Eq.~(\ref{ansatz}),
$V_{\Lambda_1\Lambda_2}(q^{2}_{\rm max})$=1.029 (1.024),
$A_{1\Lambda_1\Lambda_2}(q^{2}_{\rm max})$=0.698 (0.683) and
$A_{2\Lambda_1\Lambda_2}(q^{2}_{\rm max})$=0.534 (0.475),
but also by the simple pole formular,
$V_{\Lambda_1}(q^{2}_{\rm max})$=1.042 (1.033),
$A_{1\Lambda_1}(q^{2}_{\rm max})$=0.697 (0.681) and
$A_{2\Lambda_1}(q^{2}_{\rm max})=0.538 (0.477)$.
Our simple pole masses $\Lambda_1$=2.06 and 2.15 GeV (see Table~\ref{t76})
for both HO and linear parameters, respectively, are also in
good agreement with the value of $D^{*\pm}_s$=2.11 GeV.
We also present in Fig.~7.12 our $q^{2}$ dependence of the form factors
$V(q^2)$, $A_1(q^2)$ and $A_2(q^2)$ and compare our $A_1(q^2)$ with
the available lattice result~\cite{Bowler} (data).
Our results of $A_1(q^2)$ for both HO and linear parameter cases are
in a good agreement with that of Ref.~\cite{Bowler}.
 
(5) $D_{s}\to\eta(\eta')\ell^{+}\nu_{\ell}$: These two semileptonic
decays are very interesting processes to check our LFQM predictions of
$\eta$-$\eta'$ mixing angle. In our previous analysis of 
quark potential model~\cite{Mix} in Chapter 4, we
predicted the $\eta$-$\eta'$ mixing angle for the
HO (linear) potential model as
$\theta^{\eta-\eta'}_{SU(3)}$=$-19.3^{\circ} (-19.6^{\circ})$.

Our predicted decay rates for $D_{s}\to\eta$ and $D_{s}\to\eta'$ in
$q^{+}$=0 frame are
$\Gamma(D_{s}\to\eta)$=
0.100 (0.104)$\cos^{2}\delta|V_{cs}|^{2}$ ps$^{-1}$ and
$\Gamma(D_{s}\to\eta')$=
0.026 (0.028)$\sin^{2}\delta|V_{cs}|^{2}$ ps$^{-1}$,
respectively, for the HO (linear) parameters.
Using the lifetime $\tau_{D_{s}}$=0.467$\pm$0.017 ps, we obtain
the branching ratio for $\theta_{SU(3)}$=$-19^{\circ}$ as
follows:\footnote{In numerical calculations, we use the common
$\eta$-$\eta'$ mixing angle for both the HO and linear potentials.}
\begin{eqnarray}
&&{\rm Br}(D_{s}\to\eta\ell^{+}\nu_{\ell})= (1.7\pm 0.5)\%,\nonumber\\
&&{\rm Br}(D_{s}\to\eta'\ell^{+}\nu_{\ell})= (8.7\pm 2.7)\times 10^{-3}
\hspace{0.2cm}{\rm (HO)},\nonumber \\
&&{\rm Br}(D_{s}\to\eta\ell^{+}\nu_{\ell})= (1.8\pm 0.6)\%,\nonumber\\
&&{\rm Br}(D_{s}\to\eta'\ell^{+}\nu_{\ell})= (9.3\pm 2.9)\times 10^{-3}
\hspace{0.2cm}{\rm (Linear)}. 
\end{eqnarray}
While the experimental data~\cite{data} are
${\rm Br}_{\rm exp.}(D_{s}\to\eta\ell^{+}\nu_{\ell})$=(2.5$\pm$0.7)$\%$ 
and
${\rm Br}_{\rm exp.}(D_{s}\to\eta'\ell^{+}\nu_{\ell})$=(8.8$\pm$3.4)$\times 
10^{-3}$, respectively.
Our results for the $\eta'$ decay mode are in a good agreement
with the experimental data for both HO and linear parameters. For the
$\eta$ decay mode, our values are rather smaller than the central
value of the experimental data, neverthless, are quite comparable with
the data within the given uncertainties of CKM matrix, $V_{cs}$.

\begin{table}
\centering
\caption{Summary of the parameters for the $D_s\to\eta(\eta')$ and
$D_s\to\phi$ transitions.
Note that $c_\eta$=$|\cos\delta_P|$ and $c_{\eta'}$=$|\sin\delta_P|$.
}\label{t78}
{\footnotesize
\begin{tabular}{|r|c|c|c|c|c|}\hline
   &\multicolumn{2}{|c|}{$D_s$$\to$$\eta$ $[\eta']$}
   &\multicolumn{2}{|c}{$D_s$$\to$$\phi(s\bar{s})$}& \\
\cline{2-6}
   & $F$     &  $f_+$    & $V$  & $A_1$    & $A_2$  \\
\hline
HO & $F(0)$ & 0.692$c_\eta$ [0.692$c_{\eta'}$] & 0.830 & 0.597 & 0.409 \\
\cline{2-6}
   & $F(q^{2}_{\rm max})$ & 1.31$c_\eta$ [0.93$c_{\eta'}$]
   & 1.070 & 0.672 & 0.492\\
\cline{2-6}
   & $\Lambda_1$ [GeV] & 1.94 [1.94] & 1.94 & 2.83 & 2.25 \\
\cline{2-6}
   & $s_2\Lambda_2$ [GeV] & 2.68 [2.68] & 2.69 & 4.80 & 3.00 \\
\cline{2-6}
   & ${\cal F}_{\Lambda_1}(q^{2}_{\rm max})$
   & 1.49$c_\eta$ [0.95$c_{\eta'}$] & 1.091 & 0.673 & 0.498 \\
\cline{2-6}
   & ${\cal F}_{\Lambda_1\Lambda_2}(q^{2}_{\rm max})$ 
   & 1.27$c_\eta$ [0.92$c_{\eta'}$] & 1.069 & 0.672 & 0.491 \\
\hline
Linear & $F(0)$ & 0.721$c_\eta$ [0.721$c_{\eta'}$] & 0.861 
       & 0.608 & 0.385 \\
\cline{2-6}
   & $F(q^{2}_{\rm max})$ & 1.28$c_\eta$ [0.94$c_{\eta'}$]
   & 1.080 & 0.668 & 0.445\\
\cline{2-6}
   & $\Lambda_1$ [GeV] & 2.06 [2.06] & 2.05 & 3.22 & 2.53 \\
\cline{2-6}
   & $s_2\Lambda_2$ [GeV] & 2.94 [2.94] & 2.94 & -12.91 & 3.31 \\
\cline{2-6}
   & ${\cal F}_{\Lambda_1}(q^{2}_{\rm max})$
   & 1.37$c_\eta$ [0.95$c_{\eta'}$] & 1.095 & 0.666 & 0.448 \\
\cline{2-6}
   & ${\cal F}_{\Lambda_1\Lambda_2}(q^{2}_{\rm max})$
   & 1.24$c_\eta$ [0.93$c_{\eta'}$] & 1.079 & 0.666 & 0.445 \\
\hline
LAT~\cite{LMMS} & $F(0)$ 
   & 0.67(7)$c_\eta$ [0.67(7)$c_{\eta'}$]
   & 0.86$\pm$0.10 & 0.52$\pm$0.03 & 0.17$\pm$0.17\\
~\cite{Bernard} & $F(0)$ & $\cdots$ [$\cdots$] 
   & 1.30$\pm$0.32(43) & 0.73$\pm$0.12(24)
   & 0.55$\pm$0.10(24)\\
\hline
\end{tabular}
}
\end{table}
\setcounter{figure}{12}
\begin{figure}
\centerline{\psfig{figure=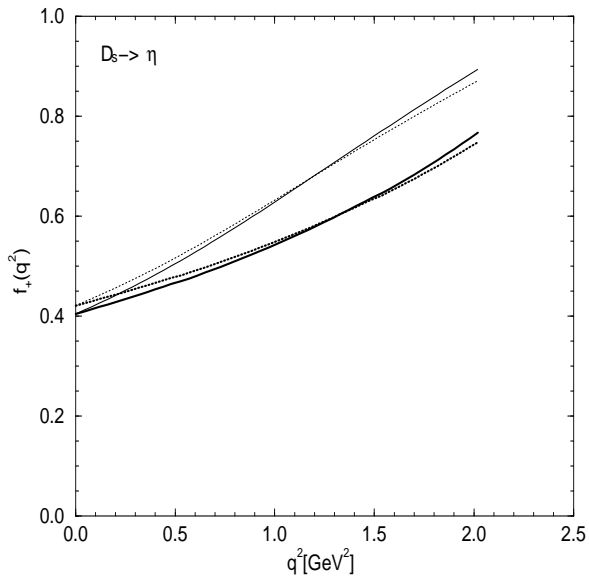,width=3.5in,height=3.5in}}
\centerline{\psfig{figure=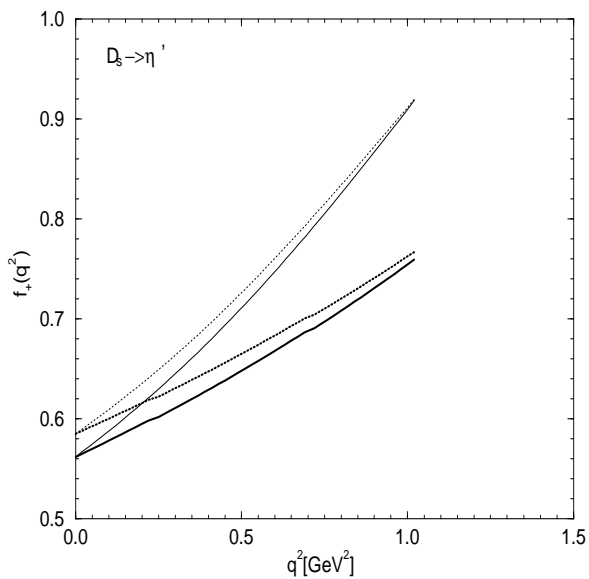,width=3.5in,height=3.5in}}  
\caption{ The form factor $f_{+}(q^{2})$ for
the $D_{s}\to\eta$ transition with the $\eta$-$\eta'$ mixing angle
$\theta_{SU(3)}$=$-19^{\circ}$. The same line code as in Fig.~7.2 is used
for our results. 
7.14: The form factor $f_{+}(q^{2})$ for
the $D_{s}\to\eta'$ transition with the $\eta$-$\eta'$ mixing angle
$\theta_{SU(3)}$=$-19^{\circ}$. The same line code as in Fig.~7.2 is used
for our results.}
\end{figure}
In Table~\ref{t78}, we compare our form factors $f^{D_s\eta}_{+}(q^{2})$
and $f^{D_s\eta'}_{+}(q^{2})$ at both $q^2$=0 and
$q^2$=$q^{2}_{\rm max}$ with Eq.~(\ref{ansatz}) and the lattice 
calculation in Ref.~\cite{LMMS}. Our predictions of 
$f^{D_s\eta}_{+}(0)$=0.692 (0.721)$c_\eta$ and 
$f^{D_s\eta'}_{+}(0)$=0.692 (0.721)$c_{\eta'}$ are in good  
agreement with the lattice results, 
$f^{D_s\eta}_{+}(0)$=(0.67$\pm$ 0.07)$c_\eta$ and
$f^{D_s\eta'}_{+}(0)$=(0.67$\pm$ 0.07)$c_\eta$~\cite{LMMS}, 
respectively. 
As one can see in Table~\ref{t78}, our analytic solutions
$f^{D_s\eta}_{+}(q^{2}_{\rm max})$= 1.31 (1.28)$c_\eta$  and
$f^{D_s\eta'}_{+}(q^{2})$= 0.93 (0.94)$c_{\eta'}$ for the HO (linear) 
parameters are well approximated by Eq.~(\ref{ansatz}),
${\cal F}^{D_s\eta}_{\Lambda_1\Lambda_2}(q^{2}_{\rm max})$=1.27 
(1.24)$c_\eta$ and
${\cal F}^{D_s\eta}_{\Lambda_1\Lambda_2}(q^{2}_{\rm max})$=0.92
(0.93)$c_{\eta'}$, respectively.
From the results of the simple pole approximation, 
${\cal F}^{D_s\eta}_{\Lambda_1}(q^2)$=1.49 (1.37)$c_\eta$
and 
${\cal F}^{D_s\eta'}_{\Lambda_1}(q^2)$=0.95 (0.95)$c_{\eta'}$,
one can find that $\Lambda_i(i\geq2)$ can be neglected 
for the $D_s\to\eta'$ transition but not for the
$D_s\to\eta$ case.
For comparison, we also present in Table~\ref{t79} the results obtained 
from $\theta_{SU(3)}$=$-10^{\circ}$ and $-23^{\circ}$ in $q^{+}$=0 frame.
Although it is not easy to conclude which mixing angle is the best
from Table~\ref{t79}, we can see at least that our predictions with
the mixing angle of $\theta^{\eta-\eta'}_{SU(3)}$$\approx$$-19^{\circ}$
are overall in a good agreement with the available experimental data.
The difference of the decay rates between the $q^{+}$=0 and 
$q^{+}$$\neq$0 frames is larger for the $\eta$ decay mode (19$\%$ for HO 
and 18$\%$ for linear) than for
the $\eta'$ decay mode (12$\%$ for HO and 11$\%$ for linear).
Likewise, comparing the form factors $f_{-}(0)$ for the
two decay modes given by Table~\ref{t77}, one can see that the difference
between the $q^{+}$=0 and  $q^{+}$$\neq$0 frames is much larger for
the $\eta$ decay mode (80$\%$ for HO and 70$\%$ for linear) than for
the $\eta'$ decay mode (4$\%$ for HO and 2$\%$ for linear).
In Figs.~7.13 and 7.14, we present the $q^{2}$ dependence of the
form factors $f_{+}(q^{2})$ for $D_{s}\to\eta$ and
$D_{s}\to\eta'$ with the mixing angle
$\theta_{SU(3)}$=$-19^{\circ}$, respectively.
\begin{table}
\centering
\caption{ Branching ratio of the $D_{s}\to\eta,\eta'$ transitions
for various $\eta-\eta'$ mixing angles, $\theta_{SU(3)}=\delta_{P}
+ 35.26^{\circ}$.}\label{t79}
{\footnotesize
\begin{tabular}{|c|c|c|c|c|c|}\hline
Processes& &$\theta_{SU(3)}=-10^{\circ}$& $\theta_{SU(3)}=-19^{\circ}$&
$\theta_{SU(3)}=-23^{\circ}$& Expt.\\ \hline
$D_{s}\to\eta$ &HO& $(2.5\pm 0.8)\%$ &$(1.7\pm 0.5)\%$
& $(1.4\pm 0.4)\%$ & $(2.5\pm0.7)\%$\\
&Linear& $(2.6\pm 0.8)\%$ & $(1.8\pm 0.6)\%$
& $(1.5\pm 0.5)\%$ &\\
\hline
$D_{s}\to\eta'$ &HO& $(6.7\pm 2.1)\cdot 10^{-3}$
& $(8.7\pm 2.7)\cdot 10^{-3}$ & $(9.6\pm 3.0)\cdot 10^{-3}$ &
$(8.8\pm3.4)\cdot 10^{-3}$\\
&Linear& $(7.1\pm 2.2)\cdot 10^{-3}$ & $(9.3\pm 2.9)\cdot 10^{-3}$
& $(10.0\pm 3.1)\cdot 10^{-3}$ & \\
\hline
\end{tabular}
}
\end{table}

(6) $D_s\to\phi\ell^{+}\nu_{\ell}$:
Our predicted decay rate for $D_s\to\phi(s\bar{s})$ in $q^+$=0 frame is 
$\Gamma(D_s\to\phi\ell^{+}\nu_{\ell})$=
4.99 (5.14)$|V_{cs}|^2\times 10^{-2}$ ps$^{-1}$ 
for the HO (linear) parameters.
We obtain the ratio $\Gamma_L/\Gamma_T$ as 1.33 (1.37) for the HO (linear)
parameters, while the lattice QCD calculation in Ref.~\cite{LMMS} 
reported 1.49$\pm$0.19.
Our predictions for the branching ratio are given by
\bea\label{Dsphi}
{\rm Br}(D_s\to\phi\ell^{+}\nu_{\ell})
&=& (2.51\pm 0.77)\%\;\; {\rm (HO)},\nonumber\\
&=& (2.59\pm 0.80)\%\;\; {\rm (Linear)},
\eea
where we used $\delta_V=-3.3^{\circ}$ for $\omega$-$\phi$ mixing 
angle. Our results in Eq.~(\ref{Dsphi}) are quite comparable to the 
available experimental data~\cite{data},
${\rm Br}_{\rm exp.}$=$(2.0\pm 0.5)\%$, within the given uncertainties
of the CKM matrix, $V_{cs}$.
\setcounter{figure}{14}
\begin{figure}
\centerline{\psfig{figure=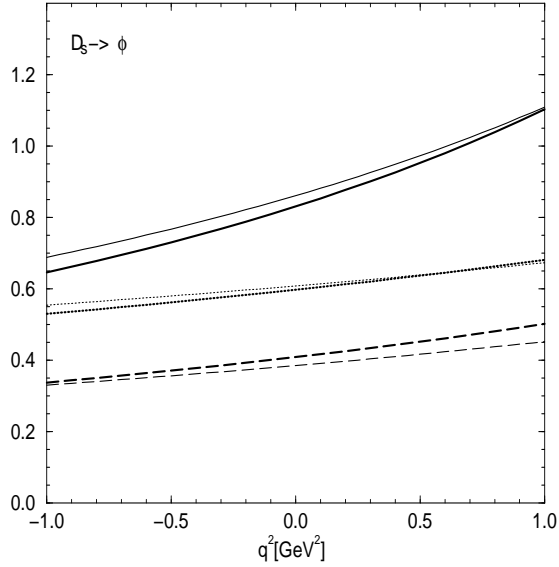,width=3.5in,height=3.5in}}
\caption{ The form factors $V(q^2)$, $A_1(q^2)$ and $A_2(q^2)$
for $D_s\to\phi(s\bar{s})$ transition.
The same line code as in Fig.~7.11 is used.}
\end{figure} 
We summarize in Table~\ref{t78} our results of the form factors $V$,
$A_1$ and $A_2$ at both $q^2$=0 and $q^2$=$q^{2}_{\rm max}$ and compare
with Eq.~(\ref{ansatz}) and the lattice results~\cite{LMMS,Bernard}.
Our results of the form factors $V$, $A_{1}$ and $A_{2}$ at $q^2$=0
are quite comparable with the lattice calculations in 
Refs.~\cite{LMMS,Bernard}. As one can see in Table~\ref{t78},
our analytic solutions for the form factors
$V(q^{2}_{\rm max})$=1.070 (1.080),
$A_1(q^{2}_{\rm max})$=0.672 (0.666) and
$A_2(q^{2}_{\rm max})$=0.492 (0.445) for the HO (linear) parameters
are well approximated not only by Eq.~(\ref{ansatz}),
$V_{\Lambda_1\Lambda_2}(q^{2}_{\rm max})$=1.069 (1.079),
$A_{1\Lambda_1\Lambda_2}(q^{2}_{\rm max})$=0.672 (0.666) and
$A_{2\Lambda_1\Lambda_2}(q^{2}_{\rm max})$=0.491 (0.445),
but also by the simple pole formular,
$V_{\Lambda_1}(q^{2}_{\rm max})$=1.091 (1.095),
$A_{1\Lambda_1}(q^{2}_{\rm max})$=0.673 (0.666) and
$A_{2\Lambda_1}(q^{2}_{\rm max})=0.498 (0.448)$. 
The simple pole approximation is already enough good approximation
for $D_s\to\phi$ transition. 
We also present in Fig.~7.15 our $q^{2}$ dependence of the form factors
$V(q^2)$, $A_1(q^2)$ and $A_2(q^2)$.

\subsection{$B$ and $B_s$ decays} 
(1) {\bf $B^{0}\to\pi^{-}\ell^{+}\nu_{\ell}$}:
Our predicted decay rate for $B\to\pi$ in $q^{+}=0$ frame is
$\Gamma(B^{0}\to\pi^{-}\ell^{+}\nu_{\ell})=
7.06 (8.16)|V_{ub}|^{2}$ ps$^{-1}$ for the HO (linear) parameters.
Using the lifetime $\tau_{B_{0}}=1.56\pm 0.04$ ps and
$|V_{ub}|=(3.3\pm 0.4\pm0.7)\times 10^{-3}$\cite{Cleo},
our predictions for the branching ratio given by
\begin{eqnarray}
{\rm Br}(B^{0}\to\pi^{-}\ell^{+}\nu_{\ell})&=&
(1.20\pm 0.29)\times 10^{-4}\hspace{.2cm} {\rm (HO)},\nonumber\\
&=& (1.40\pm 0.34)\times 10^{-4}\hspace{.2cm} {\rm (Linear)},
\end{eqnarray}
are quite comparable wiht the recent experimental data~\cite{Cleo},
${\rm Br}(B\to\pi^{-}\ell^{+}\nu_{\ell})$=$(1.8\pm0.6)\times 10^{-4}$,
within the given error range. We also performed the calculations
of the decay rate using the valence contributions from $q^+$$\neq$0
frame and found that the decay rate obtained from $q^{+}$$\neq$0 frame
is about 9 (7)$\%$ smaller for the HO (linear)
parameters than the result obtained from $q^{+}$=0 frame.

\begin{table}
\centering
\caption{Summary of the parameters for the $B\to\pi$ and
$B\to\rho$ transitions. The APE~\protect\cite{ALL} and 
ELC~\protect\cite{Abada} results are from their method `b',
which uses the heavy quark scaling laws to extrapolate from
$D$- to $B$-mesons at fixed $\omega$.}\label{t710}
{\normalsize
\begin{tabular}{|r|c|c|c|c|c|}\hline
   &\multicolumn{2}{|c|}{$B\to\pi$}
   &\multicolumn{2}{|c}{$B\to\rho$}& \\
\cline{2-6}
   & $F$     &  $f_+$    & $V$  & $A_1$    & $A_2$  \\
\hline
HO & $F(0)$ & 0.234 & 0.273 & 0.216 & 0.196 \\
\cline{2-6}
   & $F(q^{2}_{\rm max})$ & 3.34 & 0.703 & 0.449 & 0.857\\
\cline{2-6}
   & $\Lambda_1$ [GeV] & 4.37 & 4.35 & 6.91 & 4.70 \\
\cline{2-6}
   & $s_2\Lambda_2$ [GeV] & 4.74 & 5.84 & 11.01 & 6.40 \\
\cline{2-6}
   & ${\cal F}_{\Lambda_1}(q^{2}_{\rm max})$
   & $-0.61$ & $-3.66$ & 0.376 & 2.43 \\
\cline{2-6}
   & ${\cal F}_{\Lambda_1\Lambda_2}(q^{2}_{\rm max})$
   & 1.60 & 0.972 & 0.359 & 0.599 \\
\hline
Linear & $F(0)$ & 0.273 & 0.324 & 0.249 & 0.220 \\
\cline{2-6}
   & $F(q^{2}_{\rm max})$ & 2.80 & 0.688 & 0.461 & 0.851\\
\cline{2-6}
   & $\Lambda_1$ [GeV] & 4.59 & 4.57 & 8.06 & 5.09 \\
\cline{2-6}
   & $s_2\Lambda_2$ [GeV] & 6.53 & 7.11 & 14.62 & 7.02 \\
\cline{2-6}
   & ${\cal F}_{\Lambda_1}(q^{2}_{\rm max})$
   & $-1.07$ & 13.00 & 0.362 & 1.026 \\
\cline{2-6}
   & ${\cal F}_{\Lambda_1\Lambda_2}(q^{2}_{\rm max})$
   & 2.16 & 1.736 & 0.358 & 0.573 \\
\hline
LAT~\cite{UK2} & $F(0)$ & 0.27$\pm$0.11 
               & 0.35$^{+0.06}_{-0.05}$ & 0.27$^{+0.05}_{-0.04}$
               & 0.26$^{+0.05}_{-0.03}$\\ 
               & $F(q^{2}_{\rm max})$ & $\cdots$ 
               & 2.07$^{+0.11}_{-0.06}$ & 0.46$^{+0.02}_{-0.01}$
               & 0.88$^{+0.05}_{-0.03}$\\
\cline{2-6}
~\cite{ALL}    & $F(0)$ & 0.35$\pm$0.08 
               & 0.53$\pm$0.31 & 0.24$\pm$0.12 & 0.27$\pm$0.80\\
~\cite{Abada}  & $F(0)$ & 0.30$\pm$0.14(5) 
               &0.37$\pm$0.11 & 0.22$\pm$0.05 & 0.49$\pm$0.21(5)\\
\hline
SR~\cite{BB1}  & $F(0)$ & $\cdots$ 
               & 0.34$\pm$0.05 & 0.26$\pm$0.04 & 0.22$\pm$0.03\\
~\cite{BB2}    & $F(0)$ & $\cdots$ 
               & 0.35$\pm$0.07 & 0.27$\pm$0.05 & 0.28$\pm$0.05 \\
\hline 
\end{tabular}
}
\end{table}
\begin{table}
\centering
\caption{Comparison of the form factors $f_{-}(q^{2})$
at $q^{2}$=0 between $q^{+}$=0 and $q^{+}$$\neq$0 frames.
The differences between the two frames are the measure of the
nonvalence contributions in $q^+$$\neq$0 frame.}\label{t711}
\begin{tabular}{|c|c|c|c|c|}\hline
Models& & $B\to\pi$ & $B\to D$& $B_{s}\to D_{s}$ \\
\hline
HO & $q^{+}$=0 & $-0.203$ & $-0.318$ & $-0.305$ \\
& $q^{+}$$\neq$0 & $-0.235$ & $-0.478$ & $-0.518$ \\
\hline
Linear& $q^{+}$=0 & $-0.238$ & $-0.328$ & $-0.325$ \\
& $q^{+}$$\neq$0 &  $-0.273$ & $-0.499$ & $-0.533$ \\
\hline
\end{tabular}
\end{table}
We summarize in Table~\ref{t710} our analytic solutions of $f^{B\pi}_+$ 
obtained at both $q^2$=0 and $q^2$=$q^{2}_{\rm max}$ and compare with
Eq.~(\ref{ansatz}) as well as other theoretical results.
As one can see in Table~\ref{t710}, our results of $f^{B\pi}_{+}(0)$
for both HO and linear parameters are quite comparable with the
lattice QCD calculations~\cite{UK2,ALL,Abada}. Comparing with
the pole dominace formular given by Eq.~(\ref{ansatz}), we found that
our solution of $f^{B\pi}_{+}(q^{2}_{\rm max})$=3.34 (2.80) for the
HO (linear) parameters are neither fitted by Eq.~(\ref{ansatz}),
${\cal F}_{\Lambda_1\Lambda_2}(q^{2}_{\rm max})$=1.60 (2.16), nor
by the simple pole approximation,
${\cal F}_{\Lambda_1}(q^{2}_{\rm max})$=$-0.61$ $(-1.07)$.
Even though we do not know yet how many terms of $\Lambda_i(i\geq3)$
are needed to obtain the well approximated solutions motivated by
the pole dominace model, it seems quite clear that the pole
approximation up to $\Lambda_2$ given by Eq.~(\ref{ansatz}) does
not work well for the heavy-to-light transitions.
In Table~\ref{t711}, we also summarize the results of the form factor 
$f_{-}(0)$ at $q^{2}$=0 obtained from both $q^{+}$=0 and $q^{+}$$\neq$0 
frames. The difference, i.e., nonvalence contribution from $q^{+}$$\neq$0 
frame, of $f^{B\pi}_{-}(0)$ between the two frames is
about 16 (15)$\%$ for the HO (linear) parameters.

In Figs.~7.16(a) and 7.16(b), we present the $q^{2}$ dependence of the
form factors $f^{B\pi}_{+}(q^{2})$ and $f^{B\pi}_{0}(q^{2})$
obtained from both $q^+$=0 (thick lines) and $q^{+}$$\neq$0 (thin lines)
frames, respectively, and compare with the available
lattice QCD calculations~\cite{UKQCD}.
Again, the nonvalence contributions from $q^+$$\neq$0 frame are
quite sizeable especially for $f^{DK}_{0}(q^{2})$ case.
Our analytic solutions of $f^{B\pi}_{+}(q^{2})$ for both HO and linear 
parameters are overall in a good agreement with the lattice calculations
in Ref.~\cite{UKQCD}.
\begin{figure}
\centerline{\psfig{figure=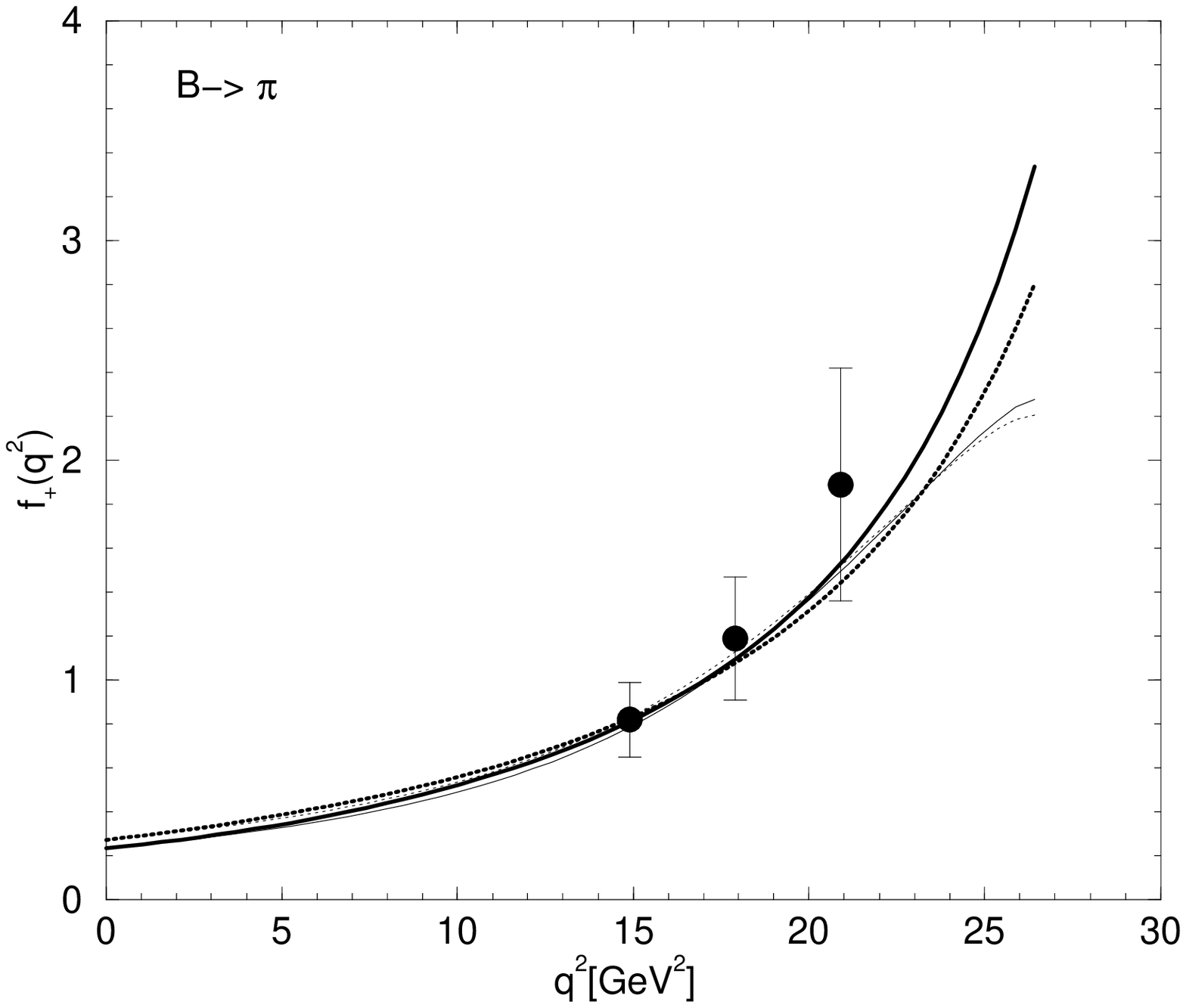,width=3.5in,height=3.5in}}
\centerline{\psfig{figure=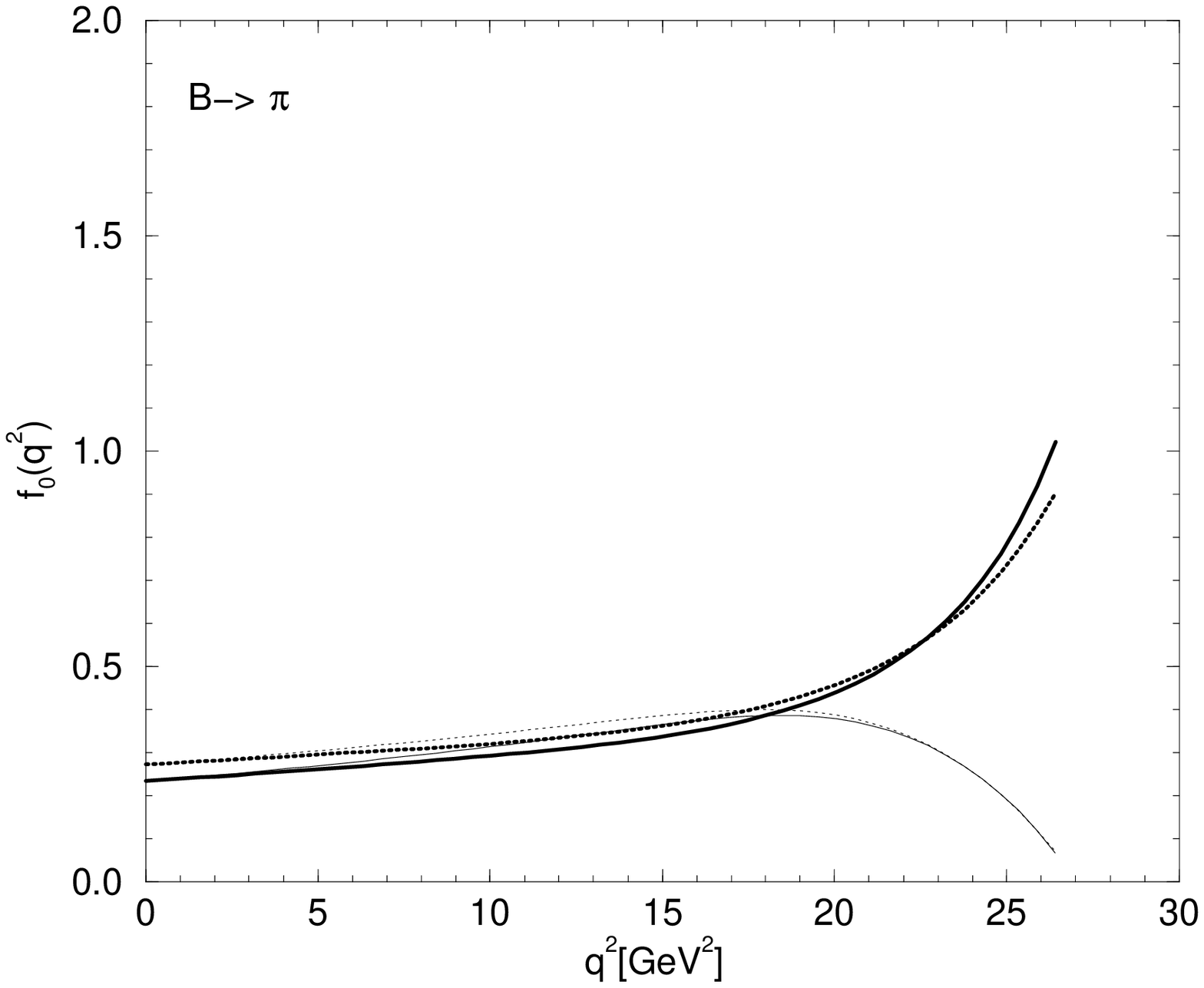,width=3.5in,height=3.5in}}
\caption{ (a) The form factor $f_{+}(q^{2})$ for the $B\to\pi$
transition compared with the lattice
calculation~\protect\cite{UKQCD} (data).
The same line code as in Fig.~7.2 is used for
our results. (b) The form factor $f_{0}(q^{2})$ for the $B\to\pi$ transition.
The same line code is used as in (a).}
\end{figure}
\begin{figure}
\centerline{\psfig{figure=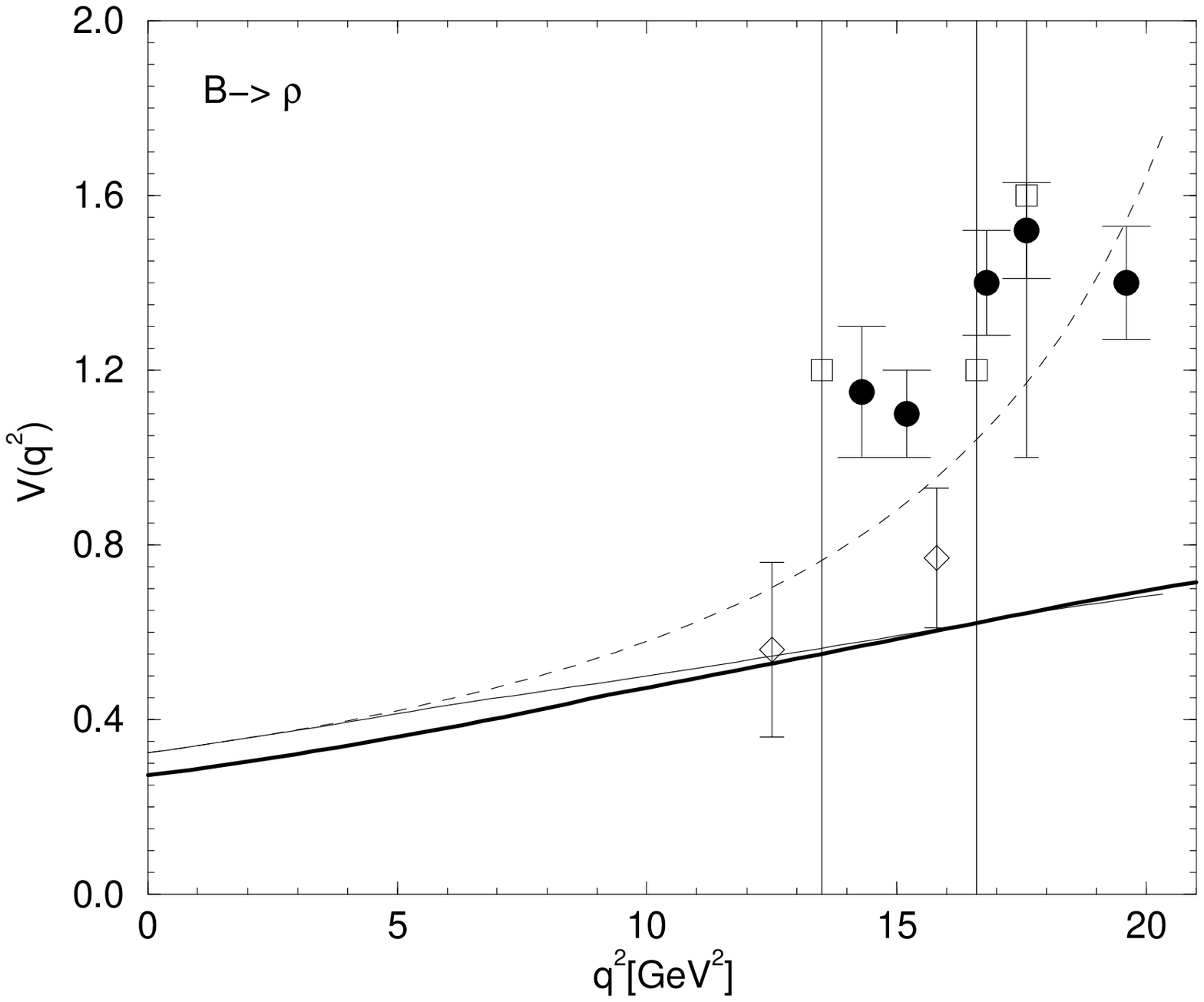,width=3.5in,height=3.5in}}
\centerline{\psfig{figure=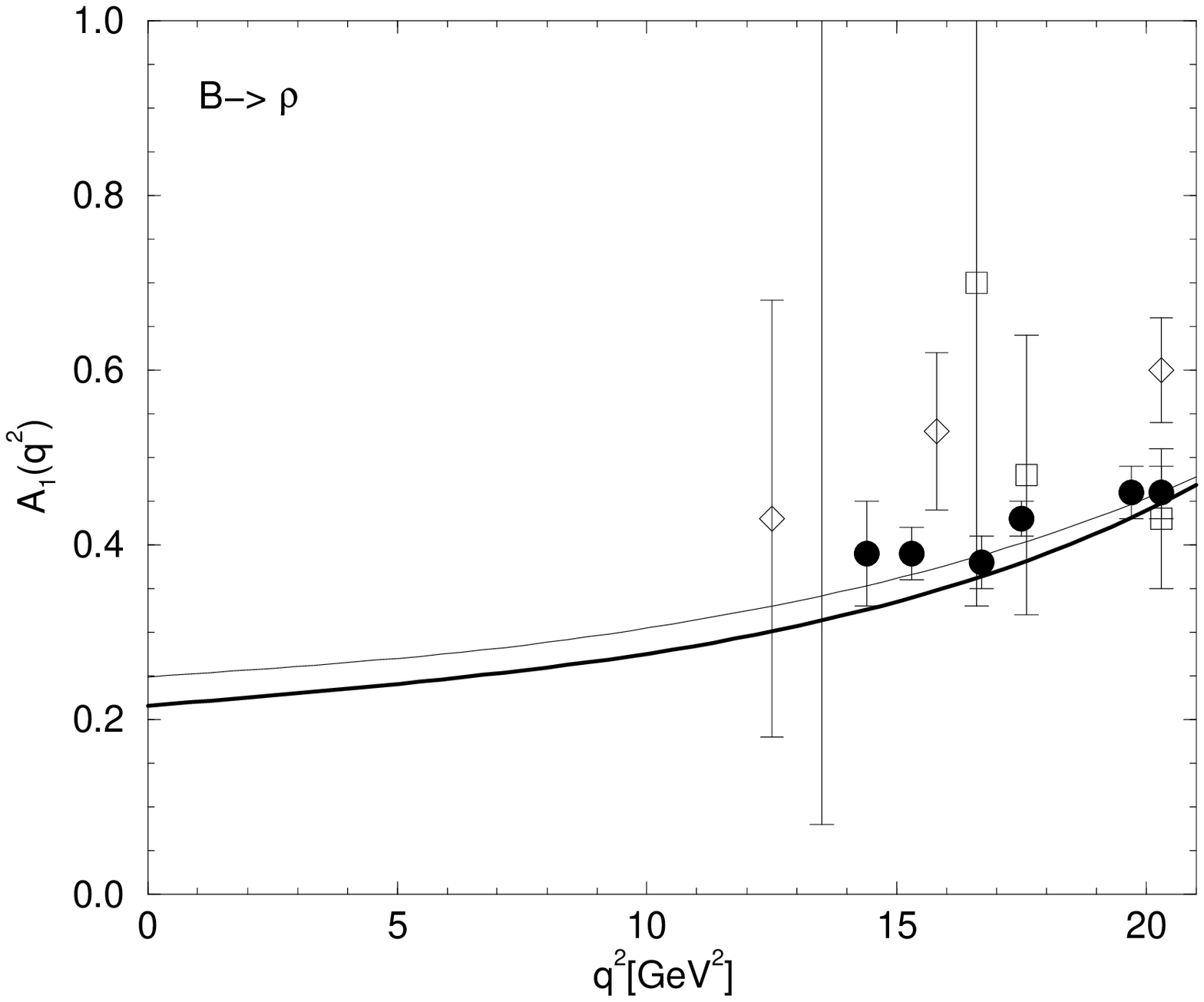,width=3.5in,height=3.5in}}
\caption{ (a) The form factor $V(q^2)$ for the $B^{0}\to\rho$ transition.
The thick (thin) solid line corresponds to our HO (linear) parameters.
For comparison, we include the results of
${\cal F}_{\Lambda_1\Lambda_2}(q^2)$ (dashed line) given by
Eq.~(\ref{ansatz}) and lattice QCD from
UKQCD~\protect\cite{Flynn2} ($\bullet$), APE~\protect\cite{ALL} (square)
and ELC~\protect\cite{Abada} ($\diamond$).
(b) The form factor $A_1(q^2)$ for the $B^{0}\to\rho$ transition.
The same line code as in (a) is used.}
\end{figure}
\setcounter{figure}{16}
\begin{figure}
\centerline{\psfig{figure=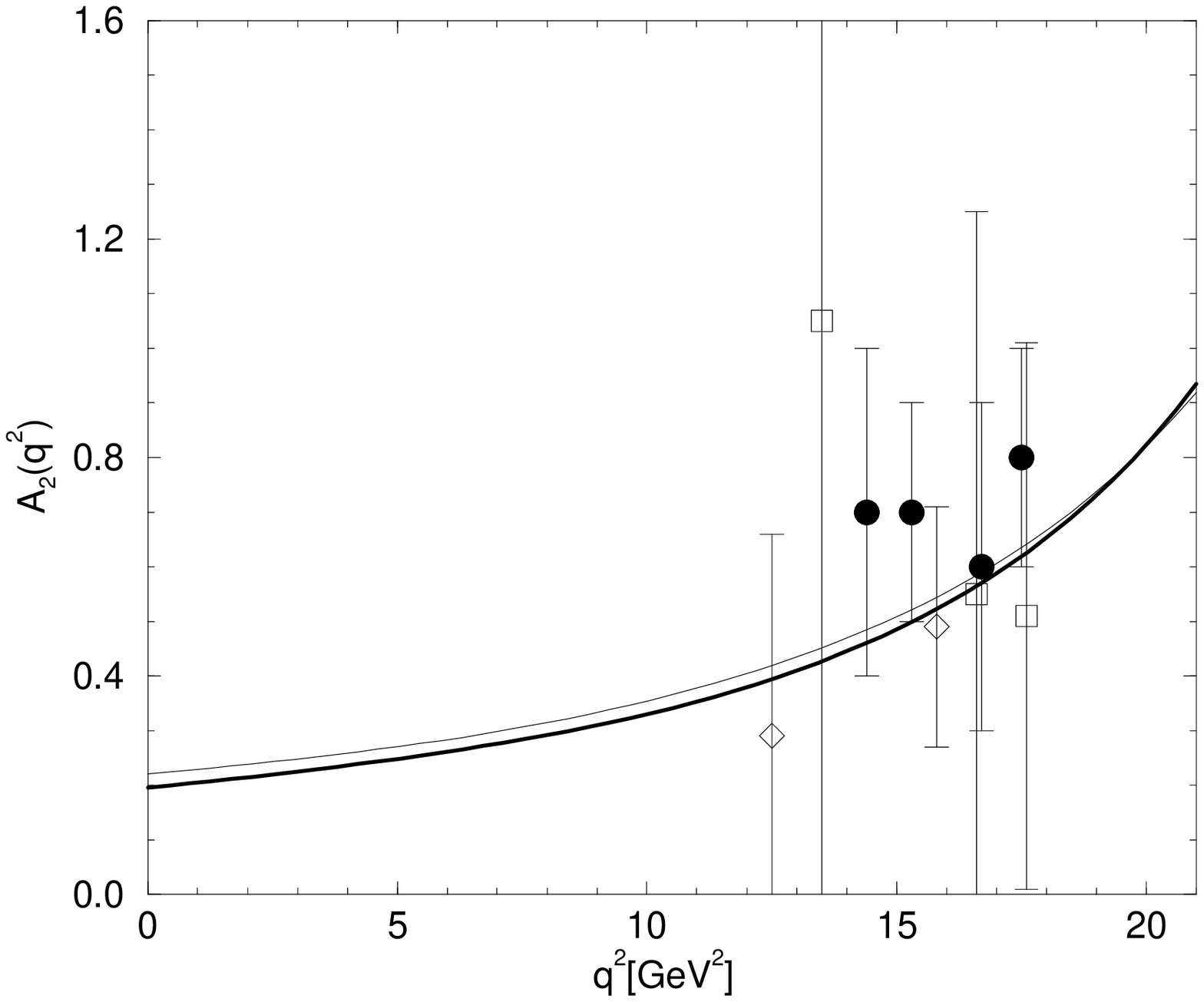,width=3.5in,height=3.5in}}
\caption{ (c) The form factor $A_2(q^2)$ for the $B^{0}\to\rho$
transition.  The same line code as in (a) is used.}
\end{figure}

(2) {\bf $B^{0}\to\rho^{-}\ell^{+}\nu_{\ell}$}:
Our predicted decay rate for $B\to\rho$ in $q^+$=0 frame is 
$\Gamma(B^0\to\rho^-\ell^{+}\nu_{\ell})$=
11.44 (14.25)$|V_{ub}|^2$ ps$^{-1}$ for the HO (linear) parameters.
We obtain the ratio $\Gamma_L/\Gamma_T$ as 1.07 (1.19) for the HO (linear)
parameters, while other QM calculations predicted as
0.3~\cite{isgw2}, 1.34~\cite{Wirbel} and 1.13~\cite{Melikhov}. The
lattice QCD of Ref.~\cite{UK2} and the QCD sum rule of 
Ref.~\cite{BB2} also calculated the ratio as 0.80$^{+0.04}_{-0.03}$
and 0.52$\pm$0.08, respectively.  
Our predictions for the branching ratio are given by
\bea\label{Brho}
{\rm Br}(B^0\to\rho^-\ell^{+}\nu_{\ell})
&=& (1.94^{+1.51}_{-1.08})\times 10^{-4}\;\; {\rm (HO)},\nonumber\\
&=& (2.36^{+1.84}_{-0.52})\times 10^{-4}\;\; {\rm (Linear)}.
\eea
While the central value of linear parameters is quite comparable to the
available experimental data~\cite{data},
${\rm Br}_{\rm exp.}$=$(2.5^{+0.8}_{-1.0})\times 10^{-4}$, the 
central value of the HO parameters underestimates the data. 

We summarize in Table~\ref{t710} our results of the form factors $V$,
$A_1$ and $A_2$ at both $q^2$=0 and $q^2$=$q^{2}_{\rm max}$ and compare
with Eq.~(\ref{ansatz}) and  other theoretical results.
Our results of the form factors $V$, $A_{1}$ and $A_{2}$ at $q^2$=0
are in good agreement with the lattice calculations in
Refs.~\cite{UK2,ALL,Abada} as well as the QCD sum rule in
Refs.~\cite{BB1,BB2}. 
Moreover, our results of the axial vector form factors 
$A_1(q^{2}_{\rm max})$=0.449 (0.461) and 
$A_2(q^{2}_{\rm max})$=0.857 (0.851) for the HO (linear) parameters
are in excellent agreement with the lattice results in Ref.~\cite{UK2}:
$A_1(q^{2}_{\rm max})$=0.46$^{+0.02}_{-0.01}$ and
$A_2(q^{2}_{\rm max})$=0.88$^{+0.05}_{-0.03}$, respectively.
Our prediction of the vector form factor 
$V(q^{2}_{\rm max})$=0.703 (0.688) for the HO(linear) parameters, however, 
disagree with the lattice result in Ref.~\cite{UK2},
$V(q^{2}_{\rm max})$=2.07$^{+0.11}_{-0.06}$.
As one can see in Table~\ref{t710}, our analytic solutions 
for the form factors $V(q^{2}_{\rm max})$ and $A_2(q^{2}_{\rm max})$ 
are neither approximated by Eq.~(\ref{ansatz}) nor the simple
pole formular, i.e., ${\cal F}_{\Lambda_1}(q^{2}_{\rm max})$.
Only the form factor of $A_1$ is reasonably approximated (but still
$\sim 20\%$ difference from our analytic solution) by the
simple pole dominace, i.e.,  
$A_{1\Lambda_1}(q^{2}_{\rm max})$=0.376 (0.362) for the HO (linear)
parameters with the pole mass $\Lambda_1$= 6.91 (8.06) GeV.
It is interesting to note that UKQCD~\cite{Flynn2} in their analysis
showed the simple pole behavior for $A_1(q^2)$ is preferred with
$m_{\rm pole}=7^{+2}_{-1}$ GeV, which is quite compatible with our
values of $\Lambda_1$.
   
In Figs.~7.17(a), 17(b) and 17(c), we show the $q^2$ dependence 
of the form factors $V$, $A_1$ and $A_2$, respectively, and 
compare with the available lattice results~\cite{Flynn2,ALL,Abada}.
For comparison, we also include in Fig.~7.17(a) the result (dashed line)
of Eq.~(\ref{ansatz}), i.e., ${\cal F}_{\Lambda_1\Lambda_2}(q^2)$ obtained 
from the linear parameters.   
As one can see in Figs.~7.17(a) and 17(b), our analytic solutions
for the $A_1(q^2)$ and $A_2(q^2)$ are in excellent agreement with
the lattice calculations~\cite{Flynn2,ALL,Abada}.
For the vector form factor $V(q^2)$ case, our analytic solution
gives reasonable agreement with the ELC result~\cite{Abada} ($\diamond$)
but underestimates the result of UKQCD~\cite{Flynn2} ($\bullet$).     

(3) $B(B_s)\to D(D_s)\ell\nu_{\ell}$: 
Our predicted decay rates for $B\to D$ and $B_{s}\to D_{s}$ in
$q^{+}$=0 frame are
$\Gamma(B^{0}\to D^{-}\ell^{+}\nu_{\ell})$= 
8.93 (9.39)$|V_{cb}|^{2}$ ps$^{-1}$ and 
$\Gamma(B_{s}\to D_{s}^{-}\ell^{+}\nu_{\ell})$= 
8.80 (9.30)$|V_{cb}|^{2}$  ps$^{-1}$, respectively, 
for the HO (linear) parameters.
Using $\tau_{B_{s}}$=(1.54$\pm$0.07) ps and
$|V_{bc}|$=0.0395$\pm$0.003~\cite{data},
we obtain the branching ratio as follows:
\begin{eqnarray}
&&{\rm Br}(B^{0}\to D^{-}\ell^{+}\nu_{\ell})= (2.17\pm 0.19)\%,\nonumber\\
&&{\rm Br}(B_{s}\to D_{s}^{-}\ell^{+}\nu_{\ell})= (2.11\pm 0.18)\%
\hspace{0.2cm} {\rm (HO)},\nonumber \\
&&{\rm Br}(B^{0}\to D^{-}\ell^{+}\nu_{\ell})= (2.28\pm 0.20)\%,\nonumber\\
&&{\rm Br}(B_{s}\to D_{s}^{-}\ell^{+}\nu_{\ell})= (2.23\pm 0.20)\%
\hspace{0.2cm} {\rm (Linear)}.
\end{eqnarray}
Our predictions of the $B\to D$ transition for both HO and linear cases
are in a good agreement with the experimental data~\cite{data},
${\rm Br}(B^{0}\to D^{-}\ell^{+}\nu_{\ell})$=(2.00$\pm$0.25)$\%$.
The decay rates of $B\to D$ and $B_{s}\to D_{s}$ obtained from the
valence contributions in $q^{+}\neq 0$ frame for the HO (linear)
parameters are about 6 (6)$\%$ and 8 (8)$\%$ larger
than those obtained from $q^{+}$=0 frame, respectively.

\begin{table}
\centering
\caption{Summary of the parameters for the $B\to D(B_s\to D_s)$ and
$B\to D^*(B_s\to D^*_s)$ transitions. The experimental data for
$B\to D^*$ transition was obtained
by assuming an exponential dependence of the form factors on $q^2$
according to the formalism outline in Ref.~\protect\cite{isgw}.
}\label{t712}
{\footnotesize
\begin{tabular}{|r|c|c|c|c|c|}\hline
   &\multicolumn{2}{|c|}{$B\to D(B_s\to D_s)$}
   &\multicolumn{2}{|c}{$B\to D^*(B_s\to D^*_s)$}& \\
\cline{2-6}
   & $F$     &  $f_+$    & $V$  & $A_1$    & $A_2$  \\
\hline
HO & $F(0)$ & 0.686 (0.664)& 0.711 (0.692)& 0.652 (0.619)
   & 0.581 (0.541)\\
\cline{2-6}
   & $F(q^{2}_{\rm max})$ & 1.11 (1.13)& 1.10 (1.11)& 0.80 (0.80)
   & 0.86 (0.84)\\
\cline{2-6}
   & $\Lambda_1$ [GeV] & 5.22 (4.95)& 5.25 (4.94)& 7.74 (6.77) 
   & 5.49 (5.15)\\
\cline{2-6}
   & $s_2\Lambda_2$ [GeV] & 7.35 (6.64)& 7.45 (6.62)&$-19.17$ (10.07) 
   & 7.81 (6.88)\\
\cline{2-6}
   & ${\cal F}_{\Lambda_1}(q^{2}_{\rm max})$
   & 1.20 (1.26)& 1.16 (1.22)& 0.79 (0.81)& 0.90 (0.90)\\
\cline{2-6}
   & ${\cal F}_{\Lambda_1\Lambda_2}(q^{2}_{\rm max})$
   & 1.11 (1.11)& 1.10 (1.11)& 0.79 (0.79)& 0.86 (0.83)\\
\hline
Linear & $F(0)$ & 0.709 (0.689) & 0.727 (0.717)& 0.667 (0.638)
       & 0.580 (0.549)\\
\cline{2-6}
   & $F(q^{2}_{\rm max})$ & 1.12 (1.14) & 1.10 (1.13)& 0.80 (0.80)
   & 0.83 (0.82)\\
\cline{2-6}
   & $\Lambda_1$ [GeV] & 5.38 (5.09)& 5.39 (5.08)& 8.24 (7.16)
   & 5.70 (5.35)\\
\cline{2-6}
   & $s_2\Lambda_2$ [GeV] & 7.77 (6.93)& 7.82 (6.91)& $-13.68$ (11.46)
   & 8.29 (7.24)\\
\cline{2-6}
   & ${\cal F}_{\Lambda_1}(q^{2}_{\rm max})$
   & 1.19 (1.24)& 1.15 (1.22)& 0.79 (0.80)& 0.86 (0.87)\\
\cline{2-6}
   & ${\cal F}_{\Lambda_1\Lambda_2}(q^{2}_{\rm max})$
   & 1.12 (1.13)& 1.10 (1.12)& 0.79 (0.80)& 0.83 (0.82)\\
\hline
ISGW~\protect\cite{isgw} & $F(q^{2}_{\rm max})$ & $\cdots$ 
  & 1.20 ($\cdots$) & 0.94 ($\cdots$) & 1.08 ($\cdots$)\\
BSW~\protect\cite{Wirbel} & $F(q^{2}_{\rm max})$ & $\cdots$ 
  & 0.97 ($\cdots$) & 0.85 ($\cdots$) & 0.90 ($\cdots$) \\
\hline
EXP.~\protect\cite{San}& $F(q^{2}_{\rm max})$ & $\cdots$ 
  & 0.91$\pm$0.49$\pm$0.12 & 0.85$\pm$0.07$\pm$0.11
  & 0.87$\pm$0.22$\pm$0.10 \\
\hline
\end{tabular}
}
\end{table}
We summarize in Table~\ref{t712} our analytic solutions of $f^{BD}_+$ and
$f^{B_sD_s}_+$ obtained at both $q^2$=0 and $q^2$=$q^{2}_{\rm max}$ 
and compare with Eq.~(\ref{ansatz}).
As one can see in Table~\ref{t712}, our results of 
$f^{BD}_{+}(q^{2}_{\rm max})$=1.11 (1.12) and 
$f^{B_sD_s}_+(q^{2}_{\rm max})$=1.13 (1.14) 
for the HO (linear) parameters are well approximated by Eq.~(\ref{ansatz}),  
${\cal F}^{BD}_{\Lambda_1\Lambda_2}(q^{2}_{\rm max})$=1.11 (1.12)
and 
${\cal F}^{B_sD_s}_{\Lambda_1\Lambda_2}(q^{2}_{\rm max})$=1.11 (1.13).
The results of the monopole approximations,  
${\cal F}^{BD}_{\Lambda_1}(q^{2}_{\rm max})$=1.20 (1.19)
and
${\cal F}^{B_sD_s}_{\Lambda_1}(q^{2}_{\rm max})$=1.26 (1.24)
are also quite comparable to our analytic solutions.
The form factors $f^{BD}_{-}(0)$ and
$f^{B_{s}D_{s}}_{-}(0)$ obtained from $q^{+}$=0 frame 
are also summarized in Table~\ref{t711} and compared with those obtained
from $q^{+}$$\neq$0 frame. As one can see in Table~\ref{t711},
the results of $f^{BD}_{-}(0)$ and $f^{B_{s}D_{s}}_{-}(0)$
in $q^{+}$$\neq$0 frame are about 50 (52)$\%$ and
70 (64)$\%$ larger for the HO (linear) parameters than 
those in $q^{+}$=0 frame, respectively.
\begin{figure}
\centerline{\psfig{figure=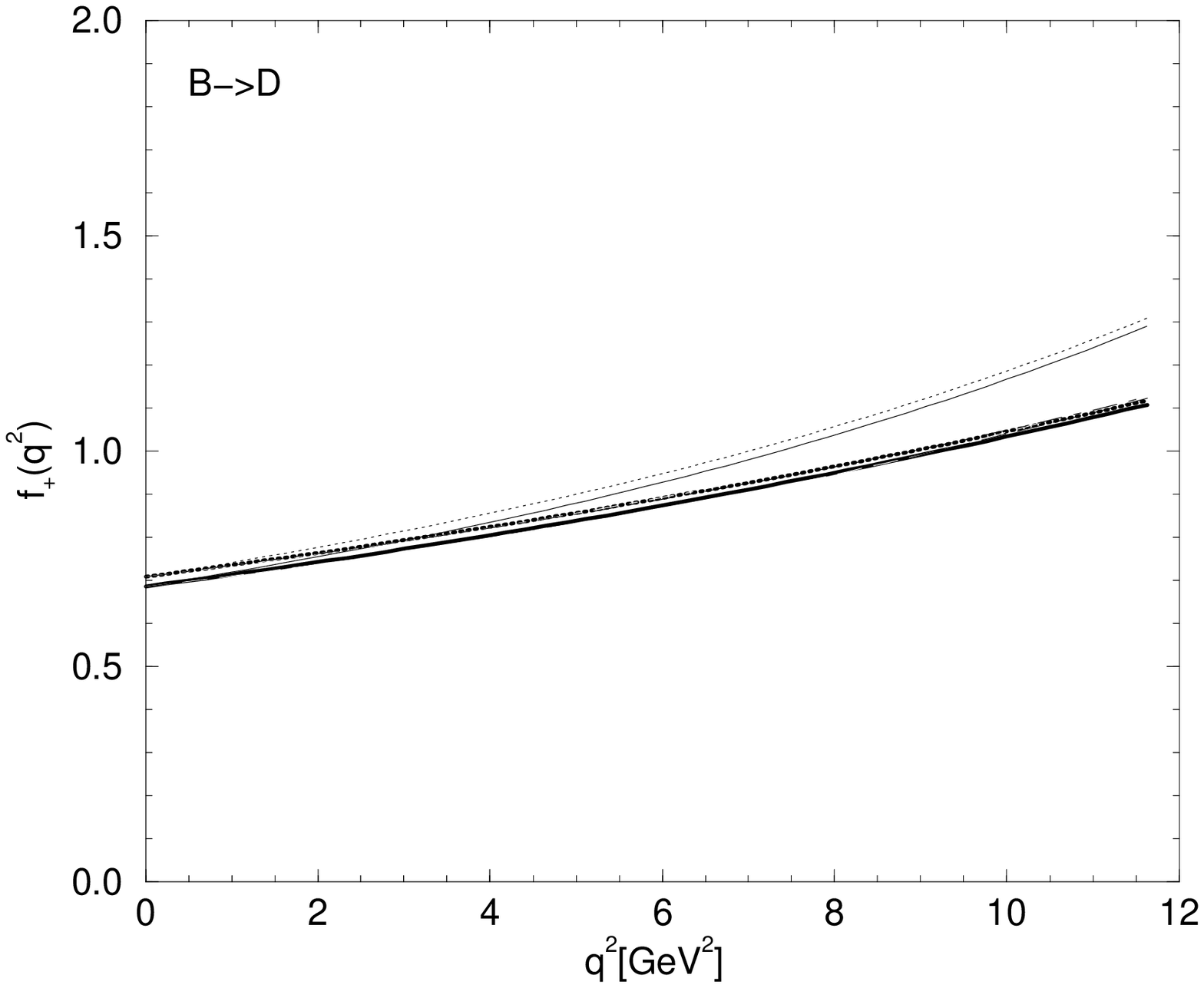,width=3.5in,height=3.5in}}
\centerline{\psfig{figure=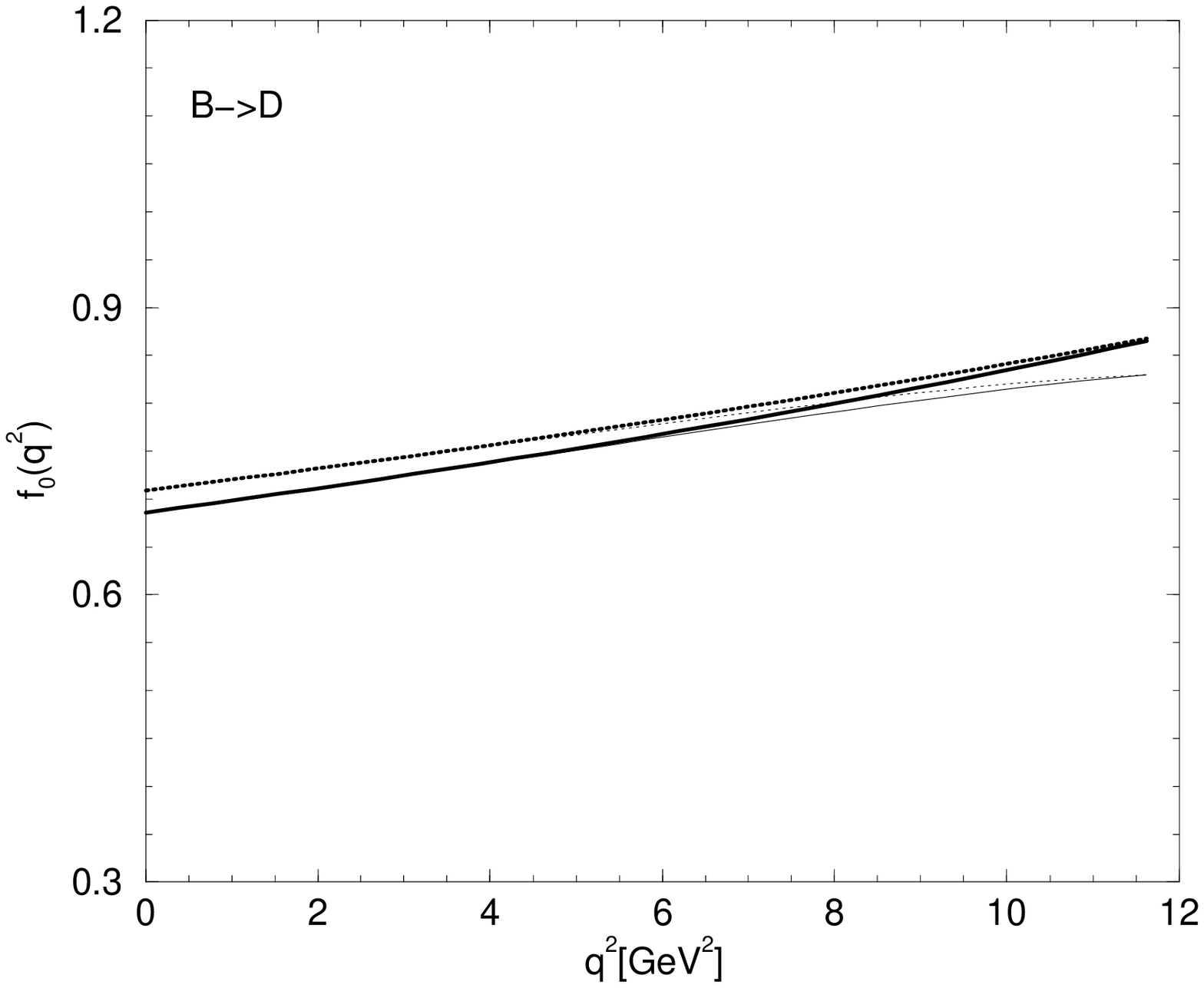,width=3.5in,height=3.5in}}
\caption{ (a) The form factor $f_+(q^2)$ for $B$$\to$$D$ transition.
The same line code as in Fig.~7.2 is used.
(b) The form factor $f_0(q^2)$ for $B$$\to$$D$ transition.
The same line code as in Fig.~7.2 is used. }
\end{figure}
\begin{figure}
\centerline{\psfig{figure=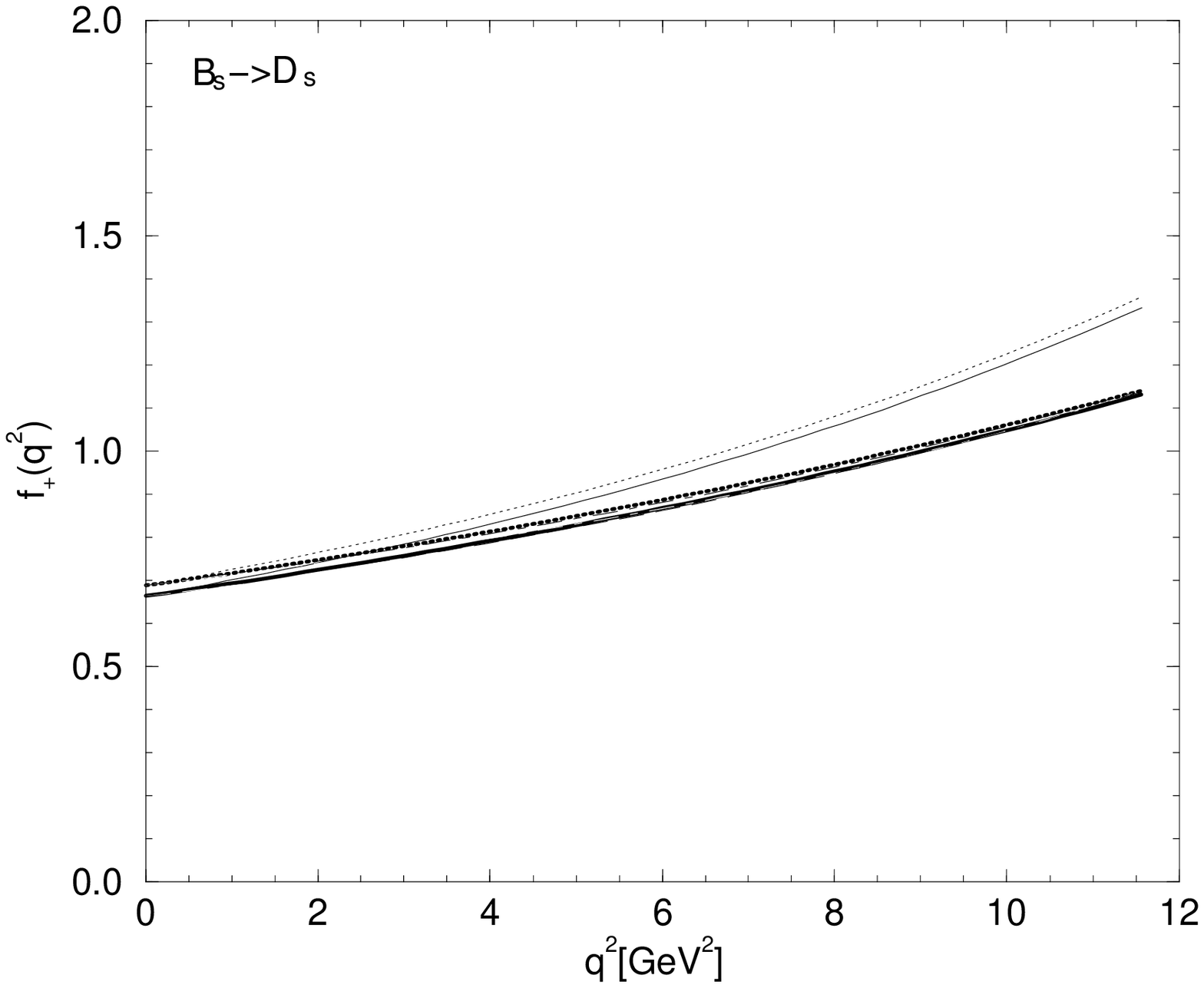,width=3.5in,height=3.5in}}
\centerline{\psfig{figure=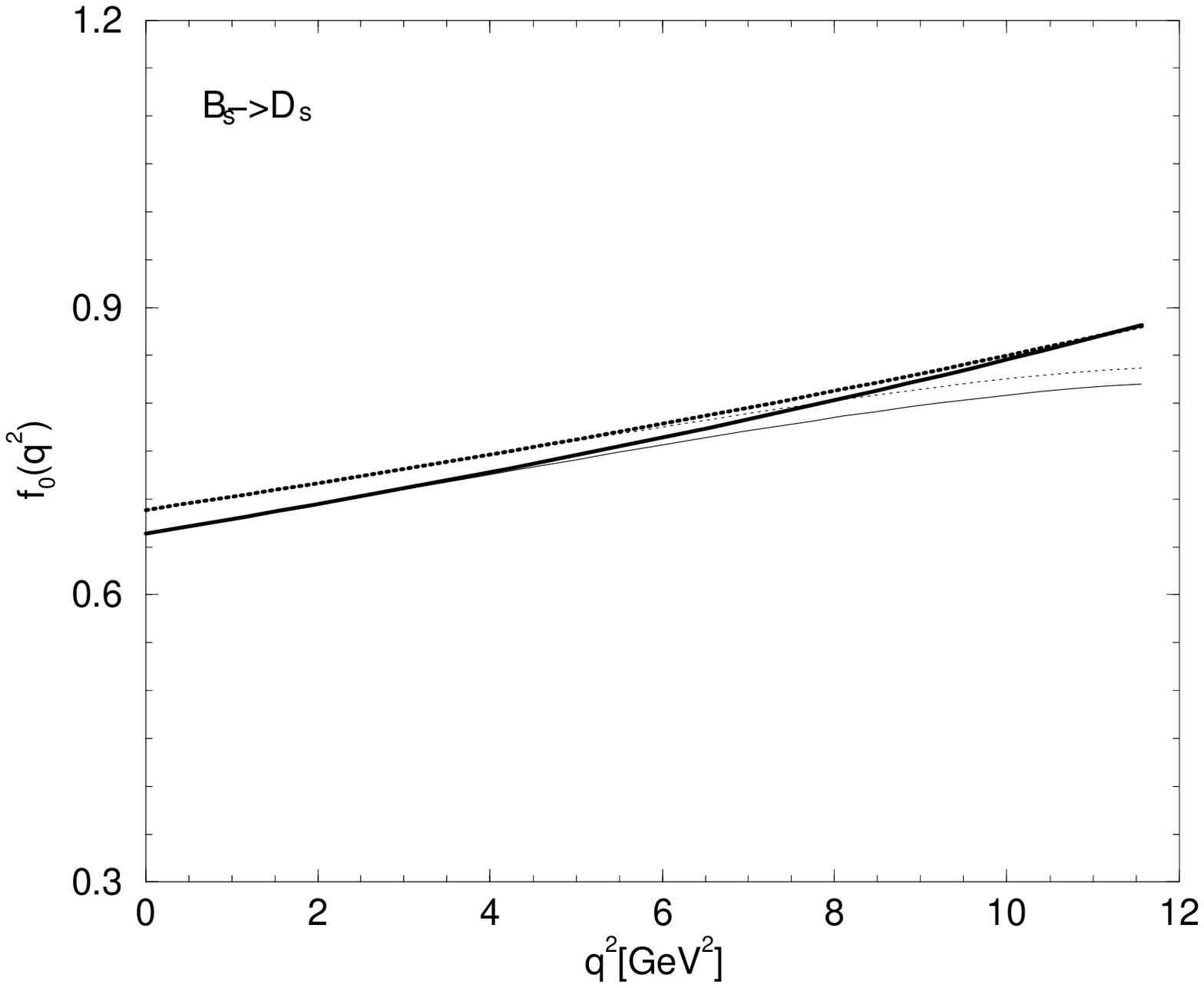,width=3.5in,height=3.5in}}
\caption{ (a) The form factor $f_+(q^2)$ for $B_s$$\to$$D_s$ transition.
The same line code as in Fig.~7.2 is used.
(b) The form factor $f_0(q^2)$ for $B_s$$\to$$D_s$ transition.
The same line code as in Fig.~7.2 is used. }
\end{figure}
In Figs.~7.18 and 7.19, we show the $q^2$ dependence behaviors
of the form factors $f_+$ and $f_0$ for $B\to D$ and
$B_s\to D_s$ transitions, respectively. For comparison, we also
include the results of the valence contributions in $q^+$$\neq$0 frame.
For these heavy-to-heavy transition cases, we can easily see the
nonvalence contributions (i.e., the difference of the results between
the two frames) are much suppressed compared to the previous light-to-light
and heavy-to-light transition cases.

These heavy-to-heavy transitions are also used to investigate model
reliability by checking the universal IW function given by
Eq.~(\ref{heavy_h}) in heavy-quark symmetry.
The slope $\rho^{2}$ of the IW function at the zero-recoil point is
defined as
\be\label{slope}
\xi(w)= 1-\rho^{2}(w-1),
\ee
where $w=v_{1}\cdot v_{2}=(M^2_1+M^2_2-q^2)/(2M_1M_2)$.
Our predictions of the slope $\rho^2=0.80\leq \rho^{2}\leq
0.92$ for possible combinations\footnote{
The slopes of the IW function for the HO (linear) 
parameters are $\rho^2$=0.85 (0.80) for $\beta_B=\beta_D$=0.496 (0.5266) GeV, 
0.92 (0.86) for $\beta_B=\beta_D$=0.4216 (0.4679) GeV 
and 0.90 (0.85) for $\beta_B\neq\beta_D$ case as usual,respectively.} 
of $\beta_{D}$ and $\beta_{B}$
are quite comparable with the current world average
$\rho^2_{\rm avg.}$=0.66$\pm$0.19~\cite{data} extracted from exclusive 
semileptonic ${\bar B}\to D\ell{\bar\nu}$ decay as well as other  
theoretical estimates, $0.7\leq \rho^{2}\leq0.88$ in~\cite{Melikhov} 
and $0.87$ in~\cite{Ivan}.
In Fig.~7.20, our predictions of the IW function 
$\xi^{BD}(v_{1}\cdot v_{2})$ for the
HO (solid) with the common gaussian $\beta=\beta_B$=0.496
and linear (dotted line) parameters with the common
$\beta=\beta_B$=0.5266, respectively, are compared with the 
available experimental data~\cite{Argus,Athanas} and show good agreement 
with the data. Other combinations of the gaussian $\beta$ paramters 
give very similar $q^2$ behavior to those as shown in Fig.~7.20. 
\begin{figure}[t]
\centerline{\psfig{figure=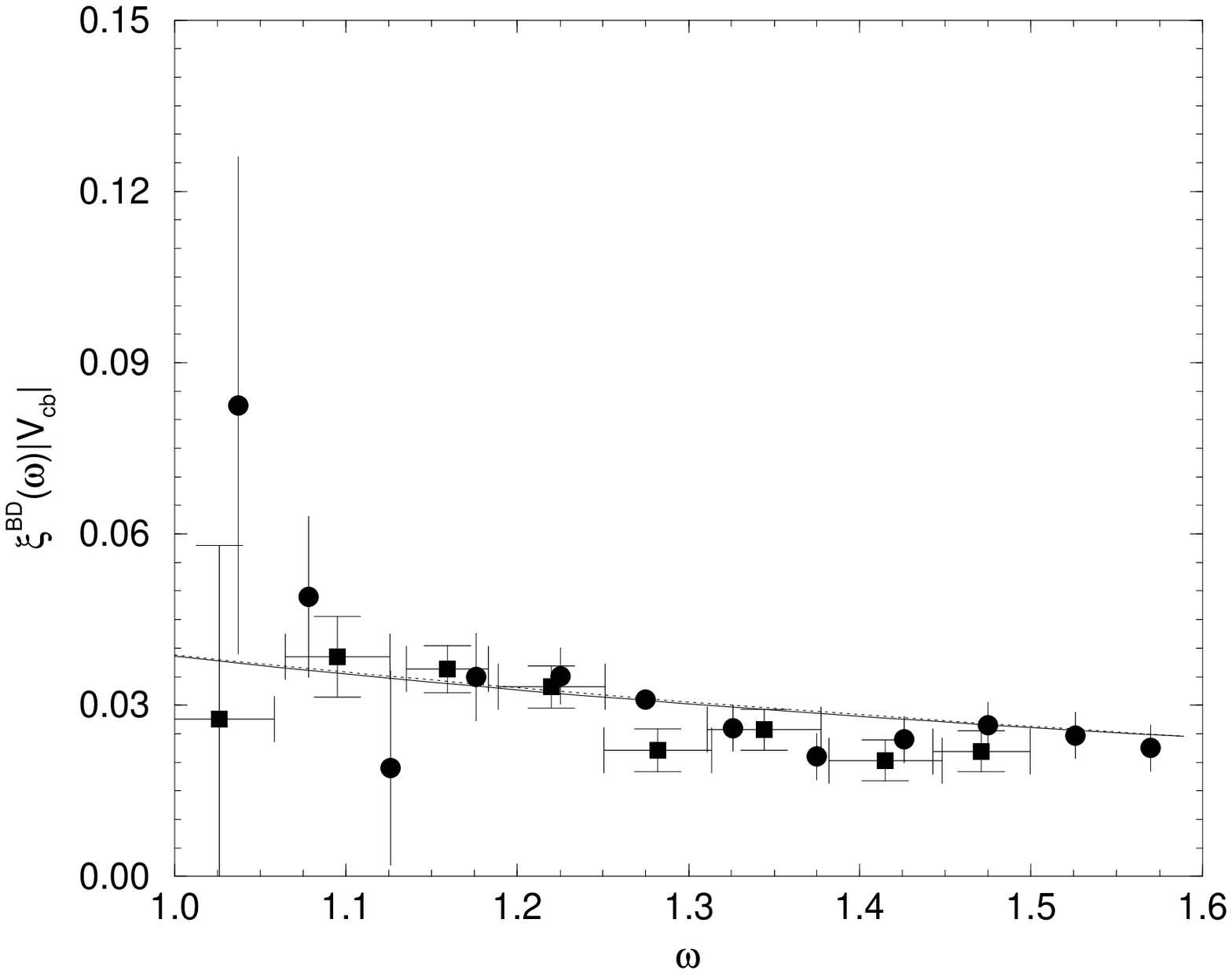,width=3.5in,height=3.5in}}
\caption{ The IW function $\xi^{BD}(v_{1}\cdot v_{2})$ for the
HO (solid) with the common gaussian $\beta=\beta_B$=0.496
and linear (dotted line) parameters with the common
$\beta=\beta_B$=0.5266, respectively. We compare our results with the
experimental data of ARGUS~\protect\cite{Argus}
(square) and CLEO~\protect\cite{Athanas} (circle).}
\end{figure}

(2) $\bar{B}^{0}(\bar{B}^{0}_s)\to 
D^{*+}(D^{*+}_s)\ell^{-}\bar{\nu}_{\ell}$:
Our predicted decay rates for $B\to D^*$ and $B_s\to D_s$ in $q^+$=0 frame
are 
$\Gamma(\bar{B}^0\to D^{*+}\ell^{-}\bar{\nu}_{\ell})$=
22.13 (22.90)$|V_{bc}|^2$ ps$^{-1}$ and 
$\Gamma(\bar{B}^{0}_s\to D^{*+}_s\ell^{-}\bar{\nu}_{\ell})$=
21.34 (22.25)$|V_{bc}|^2$ ps$^{-1}$ 
for the HO (linear) parameters, respectively.
The ratio $\Gamma_L/\Gamma_T$ for $B\to D$ and $B_s\to D_s$ decays
are obtained as 1.20 (1.23) and 1.19 (1.21) for the HO (linear)
parameters, respectively. While the experimental data for 
$B\to D^*$ reported as 0.85$\pm$0.45~\cite{ARLT} and 
0.82$\pm$0.36~\cite{CLLT}.

\begin{figure}[t]
\centerline{\psfig{figure=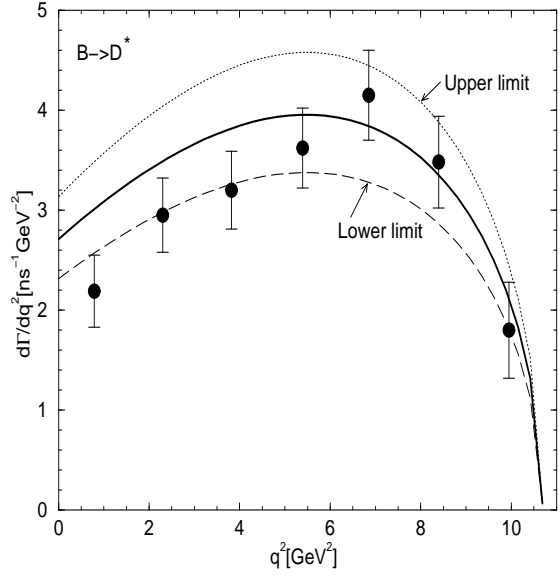,width=3.5in,height=3.5in}}
\caption{ The $d\Gamma/dq^2$ distribution for
$\bar{B}\to D^*\ell\bar{\nu}$ decays using the linear parameters.
The thick solid, dotted, and long-dashed lines are the results from the
central value, upper limit, and the lower limit of the CKM matrix
element, $|V_{bc}|$=0.0395$\pm$0.003, respectively. The experimental
data are taken from Ref.~\protect\cite{Barish}. }
\end{figure}
Our predictions for the branching ratio are given by
\bea\label{BDs}
&&{\rm Br}(\bar{B}^0\to D^{*+}\ell^{-}\bar{\nu}_{\ell})= 
(5.39\pm 0.82)\%,
\nonumber\\
&&{\rm Br}(\bar{B}^{0}_s\to D^{*+}_s\ell^{-}\bar{\nu}_{\ell})= 
(5.13\pm0.75)\%
\hspace{0.2cm} {\rm (HO)},\nonumber \\
&&{\rm Br}(\bar{B}^0\to D^{*+}\ell^{-}\bar{\nu}_{\ell})= 
(5.58\pm 0.85)\%,
\nonumber\\
&&{\rm Br}({\bar B}^{0}_s\to D^{*+}_s\ell^{-}\bar{\nu}_{\ell})= 
(5.35\pm0.78)\%
\hspace{0.2cm} {\rm (Linear)}.
\eea
Even though the current average value for $B^0\to D^*$ decays,
${\rm Br}_{\rm exp.}(\bar{B}^0\to D^{*+}\ell^{-}\bar{\nu}_{\ell})$
=$(4.60\pm0.27)\%$, is smaller than our predictions, it is interesting 
to note that the recent experiments~\cite{OPAL,ALEPH,DELPHI} reported
results close to our predictions, i.e.,
$(5.08\pm0.21\pm0.66)\%$~\cite{OPAL},
$(5.53\pm0.26\pm0.52)\%$~\cite{ALEPH}, and
$(5.52\pm0.17\pm0.68)\%$~\cite{DELPHI}, respectively.
In Fig.~7.21, we present the differential rate for
$B\to D^*$ decays using the linear parameters and compare with the 
available experimental data~\cite{Barish}. Within the uncertainties
of the CKM matrix element $|V_{bc}|$=0.0395$\pm$0.003, our results are 
overall in good agreement with the data. The results for the HO parameters
are not much different from those for the linear parameters.

We summarize in Table~\ref{t712} our analytic solutions of $V$, $A_1$ and
$A_2$ obtained at both $q^2$=0 and $q^2$=$q^{2}_{\rm max}$
and compare with Eq.~(\ref{ansatz}) as well as the available experimental
data~\cite{San} and other theoretical results~\cite{isgw,Wirbel}.
As one can see in Table~\ref{t712}, our results of the three form factors
for $B\to D^*$ transition, i.e., $V(q^{2}_{\rm max})$=1.10 (1.10), 
$A_1(q^{2}_{\rm max})$=0.80 (0.80) and $A_2(q^{2}_{\rm max})$=0.86 (0.83)
for the HO (linear) parameters are in good agreement with the experimental 
data~\cite{San}, $V_{\rm exp.}(q^{2}_{\rm max})$=0.91$\pm$0.49$\pm$0.12,
$A_{1{\rm exp.}}(q^{2}_{\rm max})$=0.85$\pm$0.07$\pm$0.11 and
$A_{2{\rm exp.}}(q^{2}_{\rm max})$=0.87$\pm$0.22$\pm$0.10, respectively.  
Our analytic solutions are also well approximated by not only  
Eq.~(\ref{ansatz}) but also the monopole form factor 
${\cal F}_{\Lambda_1}$ (see Table~\ref{t712}).  
In Figs.~7.22 and 7.23, we show the $q^2$ dependence behaviors
of the three form factors for $B\to D^*$ and $B_s\to D^{*}_s$ decays,
respectively.
\begin{figure}
\centerline{\psfig{figure=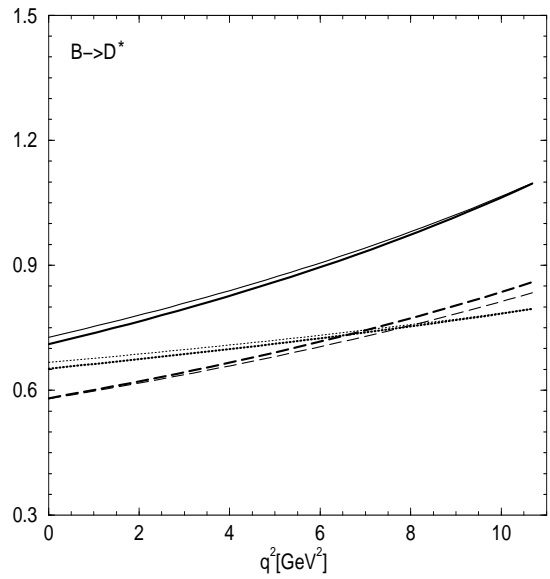,width=3.5in,height=3.5in}}
\centerline{\psfig{figure=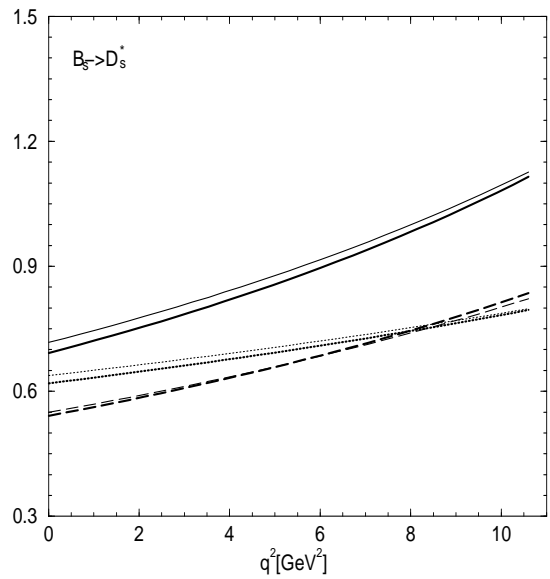,width=3.5in,height=3.5in}}
\caption{ The form factors $V(q^2)$, $A_1(q^2)$ and $A_2(q^2)$ for
$B\to D^*$ transition. The same line code as in Fig.~7.11 is used.
7.23: The form factors $V(q^2)$, $A_1(q^2)$ and $A_2(q^2)$ for
$B\to D^{*}_s$ transition. The same line code as in Fig.~7.11 is used.}
\end{figure}
 
Experimentally, two form-factor ratios for $B\to D^*$ decays
defined by~\cite{Neubert,Dubo}
\bea\label{R12}
&&R_1(q^2)=\biggl[1-\frac{q^2}{(M_B+M_{D^*})^2}\biggl]
\frac{V(q^2)}{A_1(q^2)},\nonumber\\
&&R_2(q^2)=\biggl[1-\frac{q^2}{(M_B+M_{D^*})^2}\biggl]
\frac{A_2(q^2)}{A_1(q^2)},
\eea
have been measured by CLEO~\cite{Dubo} as follows:
$R_1(q^{2}_{\rm max})=1.24\pm0.26\pm0.12$ and 
$R_2(q^{2}_{\rm max})=0.72\pm0.18\pm0.07$.
From Table~\ref{t712}, we obtain the ratios as 
$R_1(q^{2}_{\rm max})$=1.10 (1.11) and
$R_1(q^{2}_{\rm max})$=0.86 (0.84) for the HO (linear) parameters, 
which are in good agreement with the data.     
Our results are also agree with the predictions of Neubert~\cite{Neubert},
$R_1(q^{2}_{\rm max})$=1.35 and $R_2(q^{2}_{\rm max})$=0.79, and
the ISGW2 model~\cite{isgw2},
$R_1(q^{2}_{\rm max})$=1.27 and $R_2(q^{2}_{\rm max})$=1.01.
In Fig.~7.24, we present the two form-factor ratios $R_1$ and
$R_2$ as a function of $w$ and compare with 
the data~\cite{Dubo} at $q^2=q^{2}_{\rm max}$(or $w$=1) 
anticipating more data in the entire physical $q^2$ range. 
Our results show almost $w$ independent 
behavior as indicated by Ref.~\cite{Neubert}.  
\setcounter{figure}{23}
\begin{figure}
\centerline{\psfig{figure=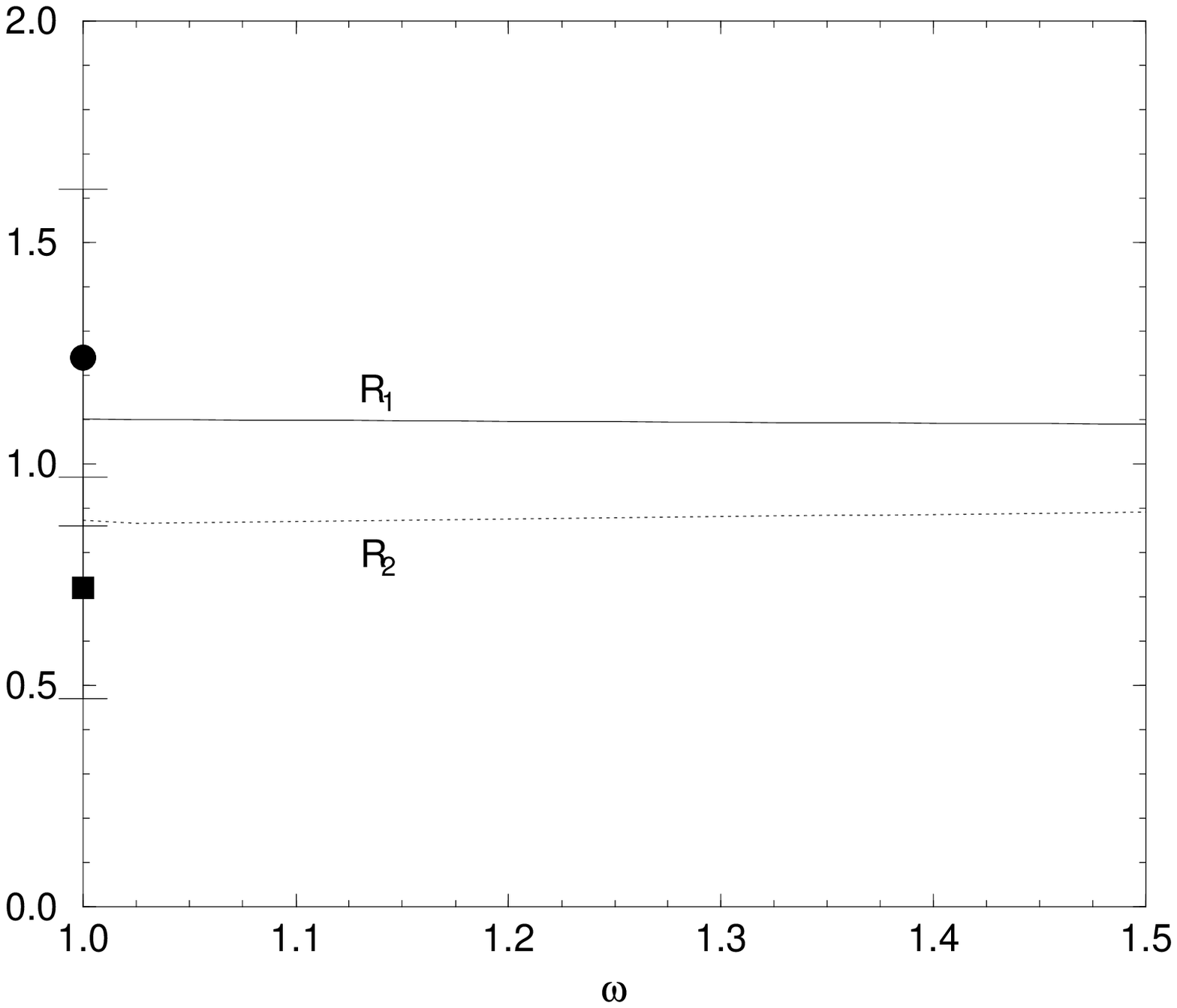,width=3.5in,height=3.5in}}
\centerline{\psfig{figure=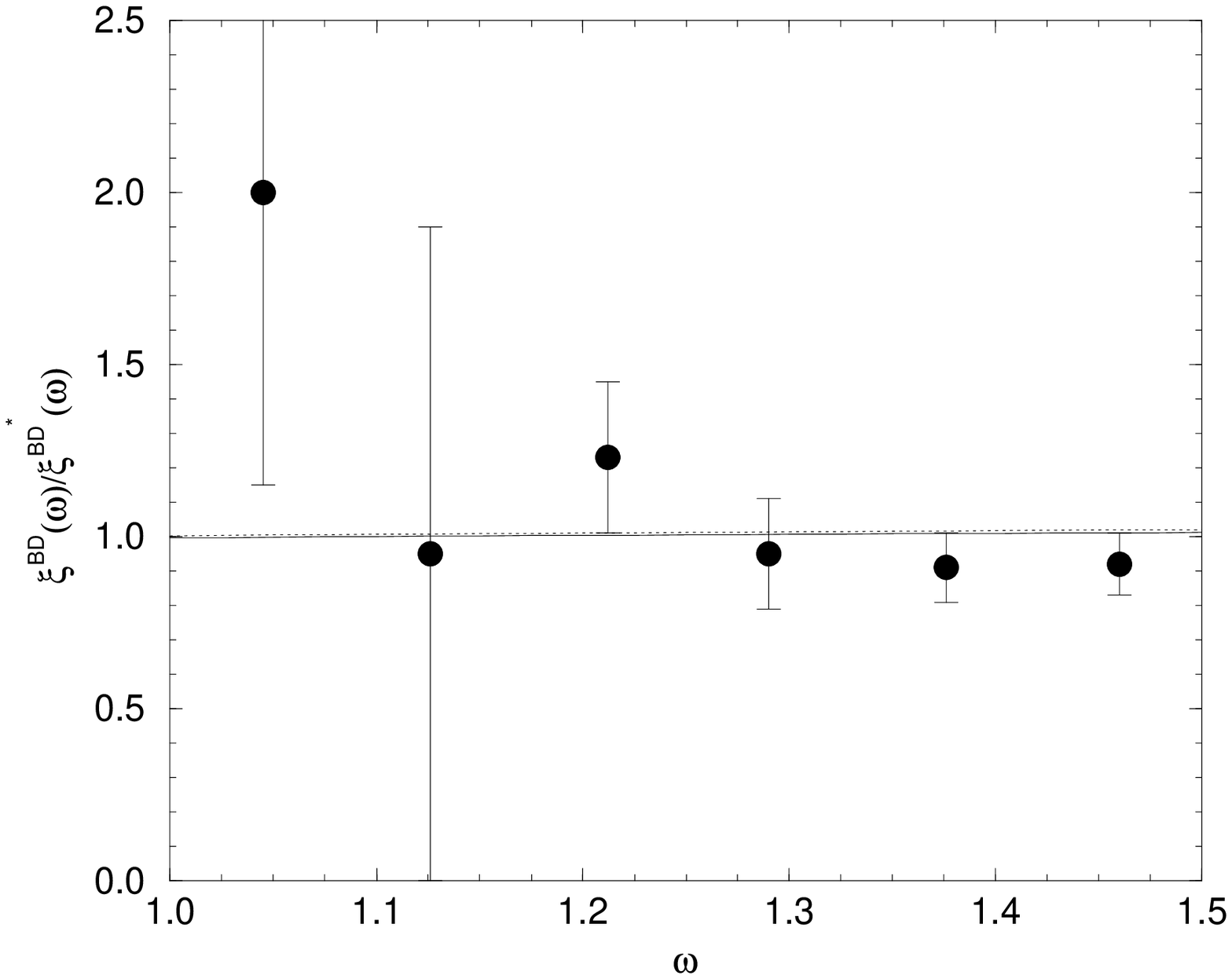,width=3.5in,height=3.5in}} 
\caption{ The two form-factor ratios, $R_1(w)$ (solid line) and
$R_2(w)$ (dotted line) for the HO parameters compared with the
experimental data~\protect\cite{Dubo} of 
$R_1(1)$ (circle) and $R_2(1)$ (square).
7.25: Our predictions of the ratio of $\xi^{BD}(w)/\xi^{BD^*}(w)$ 
for both HO (solid line) and linear (dotted line) parameters 
compared with the experimental data~\protect\cite{data}.}
\end{figure} 
The form factors for $0^-\to1^-$ transitions in the heavy-quark limit
are also related to the IW function via
\bea\label{IW_PV}
\xi(w)&=&\frac{2\sqrt{M_P M_D}}{M_P + M_D}V(q^2)
= \frac{2\sqrt{M_P M_D}}{M_P + M_D}A_2(q^2)\nonumber\\
&=&\frac{2\sqrt{M_P M_D}}{M_P + M_D}
\frac{A_1(q^2)}{[1-q^2/(M_P + M_D)^2]}.
\eea 
The slope of the IW function at zero-recoil for the $B\to D^*$ decays 
is obtained as $0.85\leq\rho^2\leq 0.97$ for all possible combinations 
of the gaussian parameters $\beta_B$ and $\beta_{D^*}$. 
Our results are in good agreement with
the data: $\rho^2$=0.91$\pm$0.15$\pm$0.06 (CLEO)~\cite{Dubo} and
$\rho^2_{\rm avg.}$=0.71$\pm$0.11 (PDG)~\cite{data}.
We present in Fig.~7.25 the ratio of the two IW functions 
for $B\to D$ and $B\to D^*$, i.e., $\xi^{BD}(w)/\xi^{BD^*}(w)$, 
where $\xi^{BD}(w)$ and $\xi^{BD^*}(w)$ is obtained by 
the form factors $f_+(q^2)$ and 
$V(q^2)$, respectively, and compare with the data~\cite{data}. 
Our results for both HO (solid line) and 
linear (dotted line) parameters are almost equal
to 1, the value expected from the heavy quark 
limit ($m_b,m_c\to\infty$) and thus show a good agreement with
the data~\cite{data}. 

Moreover, the decay $\bar{B}^0\to D^{*+}\ell^{-}\bar{\nu}_{\ell}$ 
allows one to measure the chirality of the weak $b\to c$ 
transition~\cite{Neubert,Argus,Dubo,KS}, which is only possible if the
daughter meson has spin $J>0$.
This $b\to c$ chirality can be tested by measuring the forward-backward
asymmetry~\cite{Neubert,Argus,San,KS}: 
\bea\label{Afb}
A_{\rm FB}&=&\frac{3}{4}
\frac{\int K_f(q^2) q^2(|H_{-}|^2 - |H_{-}|^2)dq^2dp_{\ell}}
{\int K_f(q^2) q^2(|H_{+}|^2 + |H_{-}|^2 + |H_0|^2)dq^2dp_{\ell}},
\nonumber\\
&=&\frac{3}{4}\frac{\Gamma_{-}-\Gamma_{+}}{\Gamma}.
\eea  
The helicity alignment of the $W^-$ is given by~\cite{Argus}
\be\label{BD_pole}
A_{\rm pol}=2\frac{\Gamma_{0}}{\Gamma_{+}+\Gamma_{-}} -1,
\ee
which describes the $D^{*+}$ polarization extracted
from the $D^{*+}$ decay angle distribution (see for 
more detail in Ref.~\cite{Neubert,KS}).
From Eqs.~(\ref{Afb}) and (\ref{pol}), we obtain
$A^{BD^*}_{\rm FB}$= 0.195 (0.193) and  
$A^{BD^*}_{\rm pol}$= 1.396 (1.510) 
for the HO (linear) parameters.
Our results for $B\to D^*$ are in excellent agreement with the
available experimental data: 
$A_{\rm FB}$=0.20$\pm$0.08$\pm$0.06, 
$A_{\rm pol}$=1.1$\pm$0.4$\pm$0.2~\cite{Argus} and
$A_{\rm FB}$=0.197$\pm$0.033$\pm$0.016,
$A_{\rm pol}$=1.55$\pm$0.26$\pm$0.13~\cite{Dubo}.  
Similarly, we also obtain $A^{B_sD^*_s}_{\rm FB}$= 0.194 (0.194) and
$A^{B_sD^*_s}_{\rm pol}$= 1.376 (1.422) for the HO (linear) parameters
anticipating future experimental data for $B_s\to D^{*}_s$.
Finally, we summarize in Table~\ref{PPt} and~\ref{PVt} our results of the
decay rates for $0^-\to 0^-$ and $0^-\to 1^-$ transitions, respectively.
\begin{table}
\caption{ Decay rate $\Gamma$ (in unit of ps$^{-1}$) 
and branching ratio ${\rm Br}(0^-\to0^-\ell\nu_{\ell})$
for various semileptonic decays. 
We use $\theta^{\eta-\eta'}_{SU(3)}$=$-19^{\circ}$ to obtain the
branching ratio for $D_{s}$$\to$$\eta(\eta')$ decays.
The used CKM matrix elements are 
$|V_{cs}|$=1.04$\pm$0.16, $|V_{cd}|$=0.224$\pm$0.016, 
$|V_{ub}|$=(3.3$\pm$0.4$\pm$0.7)$\times$10$^{-3}$,
and $|V_{bc}|$=0.0395$\pm$0.003~\protect\cite{data}.}\label{PPt}
\centering
\begin{tabular}{|l|c|c|c|c|}
\hline
Mode & & $\Gamma$ & Br. & Expt.(Br.)\\ \hline
$D\to\pi$ & HO& 0.110$|V_{cd}|^2$ & $(2.30\pm 0.33)\cdot 10^{-3}$&
$(3.7\pm 0.6)\cdot 10^{-3}$ \\
&Linear& 0.113$|V_{cd}|^2$ & $(2.36\pm 0.34)\cdot 10^{-3}$& \\ \hline
$D\to K$ & HO & 8.26$\cdot 10^{-2}|V_{cs}|^2$ 
& $(3.71\pm 1.14)\%[e^{+}\nu_{e}]$ & $(3.66\pm0.18)\%$ \\
& & 6.40$\cdot 10^{-2}|V_{cs}|^2$& $(2.87\pm 0.88)\%[\mu^{+}\nu_{\mu}]$&
\mbox{for} $D^{0}\to K^{-}e^{+}\nu_{e}$ \\
&Linear &  8.36$\cdot 10^{-2}|V_{cs}|^2$ 
& $(3.75\pm 1.16)\%[e^{+}\nu_{e}]$ & $(3.23\pm0.17)\%$\\
& & 6.43$\cdot 10^{-2}|V_{cs}|^2$& $(2.89\pm 0.90)\%[\mu^{+}\nu_{\mu}]$&
\mbox{for} $D^{0}\to K^{-}\mu^{+}\nu_{\mu}$\\ \hline
$D_{s}\to\eta$ &HO& 0.100$\cos^2\delta|V_{cs}|^2$ 
& $(1.7\pm 0.5)\%$ & $(2.5\pm0.7)\%$\\
&Linear& 0.104$\cos^2\delta|V_{cs}|^2$ 
& $(1.8\pm 0.6)\%$ &\\ \hline
$D_{s}\to\eta'$ &HO& 0.026$\sin^2\delta|V_{cs}|^2$  
& $(8.7\pm 2.7)\cdot 10^{-3}$ &
$(8.8\pm3.4)\cdot 10^{-3}$\\
&Linear& 0.028$\sin^2\delta|V_{cs}|^2$ 
&$(9.3\pm 2.9)\cdot 10^{-3}$ & \\ \hline
$B\to\pi$ &HO & 7.06$|V_{ub}|^2$ & $(1.20\pm 0.29)\cdot 10^{-4}$ &
$(1.8\pm0.6)\cdot 10^{-4}$\\
& Linear & 8.16$|V_{ub}|^2$ &
$(1.40\pm 0.34)\cdot 10^{-4}$ & \\ \hline
$B\to D$ &HO& 8.93$|V_{cb}|^2$ & $(2.17\pm 0.19)\%$ &
$(2.00\pm0.25)\% $\\
&Linear& 9.39$|V_{cb}|^2$ & $(2.28\pm 0.20)\%$ & \\ \hline
$B_{s}\to D_{s}$ & HO& 8.80$|V_{cb}|^2$ & $(2.11\pm 0.18)\%$ &
--\\
&Linear& 9.30$|V_{cb}|^2$ & $(2.23\pm 0.20)\%$ & \\
\hline
\end{tabular}
\end{table}
\begin{table}
\caption{Decay rate $\Gamma$ (in unit of $|V_{q_1q_2}|^2$ ps$^{-1}$)
and branching ratio ${\rm Br}(0^-\to1^-\ell\nu_{\ell})$
for various semileptonic decays.
We use $|\delta_V|$=3.3$^\circ$ to obtain the branching ratio
for $D_s\to\phi$ decays.
The $\Gamma_L$ and $\Gamma_T$ are the decay rates for the longitudinal
and transverse helicity contributions, respectively.}\label{PVt}
\begin{tabular}{|c|c|c|c|c|c|}
\hline
Mode &  & $\Gamma$ & $\Gamma_{L}/\Gamma_{T}$ & Br. & Expt.(Br.) \\ \hline
$D^{0}\to K^{*}$
 & HO     & 5.856$\cdot10^{-2}$ 
          & 1.37 & $(2.63\pm 0.81)\%$ & $(2.02\pm0.33)\%$ \\
 & Linear & 5.846$\cdot10^{-2}$ 
          & 1.42 & $(2.62\pm 0.81)\%$ &  \\ \hline
$D^{0}\to\rho^{-}$
 & HO     & 6.382$\cdot10^{-2}$  
          & 1.39 & $(1.33\pm 0.19)\cdot 10^{-3}$
          & $(1.87\pm 0.9)\cdot 10^{-3}$ \\
 & Linear & 6.610$\cdot10^{-2}$  
          & 1.47 & $(1.38\pm 0.19)\cdot 10^{-3}$
          & \\ \hline
$D_{s}\to\phi$
 & HO     & 4.994$\cdot10^{-2}$  
          & 1.33 & $(2.51\pm 0.77)\%$ & $(2.0\pm 0.5)\%$ \\
 & Linear & 5.137$\cdot10^{-2}$  
          & 1.37 & $(2.59\pm 0.80)\%$ &  \\ \hline
$B^{0}\to\rho^{-}$
 & HO     & 11.445 
          & 1.07 & $(1.94^{+1.51}_{-1.08})\cdot 10^{-4}$ &
$(2.5^{+0.8}_{-1.0})\cdot 10^{-4}$ \\
 & Linear & 14.254 
          & 1.19 & $(2.36^{+1.84}_{-0.52})\cdot 10^{-4}$ & \\ \hline
$B^{0}\to D^{*}$
 & HO     & 22.133 
          & 1.20 & $(5.39\pm 0.82)\%$ & $(4.60\pm0.27)\% $ \\
 & Linear & 22.904 
          & 1.23 & $(5.58\pm 0.85)\%$ & \\
\hline
$B^{0}_s\to D^{*}_s$
 & HO     & 21.341 
          & 1.19 & $(5.13\pm 0.75)\%$ & -- \\
 & Linear & 22.253 
          & 1.21 & $(5.35\pm 0.78)\%$ & \\
\hline
\end{tabular}
\end{table}
\section{Summary and Discussion}
In conclusion, in this Chapter, we analyzed the exclusive
$0^{-}$$\to$$0^{-}$ and $0^{-}$$\to$$1^{-}$ semileptonic heavy meson 
decays extending our LFQM constrained by the variational principle for the 
QCD-motivated effective Hamiltonian disscussed in Chapter 4.
Our model not only provided overall a good agreement with the available
experimental data and the lattice QCD results for the weak transition
form factors and branching ratios of the light-to-light ($K_{\ell3}$),
heavy-to-light and heavy-to-heavy meson decays 
but also rendered a large number of predictions to the heavy
meson mass spectra and decay constants. Our predicted meson mass spectra 
given by Table~\ref{t73} were overall in a good agreement with the
data~\cite{data}. Our values of the decay constants given by
Table~\ref{dcon2} were also in a good agreement with the results of lattice 
QCD~\cite{Flynn,Bern2} anticipating future accurate experimental data. 

As an application of our model, we first calculated the spacelike EM form 
factors of $D$ and $B$ mesons as well as $\pi$ and $K$ mesons in $q^+$=0 
frame and estimated the nonvalence contributions of the EM form factors
from the $q^{+}$$\neq$0 frame by comparing the valence contributions
obtained from the $q^{+}$$\neq$0 frame with those obtained from the
$q^{+}$=0 frame.  
The nonvalence contributions from the $q^{+}$$\neq$0 frame are highly
suppressed as the quark mass increases, i.e., in the order of
$\pi>K>D>B$, as one can see from Figs.~7.2-7.5.

We have overcome the difficulty associated with the nonvalence Z-graph
contribution in timelike region by the analytic continuation of weak form
factors from the spacelike region.
In other words, the form factors $f_{\pm}$ for $0^-\to 0^-$ decays and
the form factors $V$, $A_1$ and $A_2$ 
were obtained in the $q^{+}$=0 frame and then
analytically continued to the timelike region by changing $q_{\perp}$
to $iq_{\perp}$ in the form factors. The matrix element of the
$``\perp"$ component of the current $J^{\mu}$ was used to obtain
the form factor $f_{-}$, since the $J^{-}$ is not immune to the
zero mode contributions even in the $q^{+}$=0 frame~\cite{zm}.
Our numerical computation confirmed the equivalence of our analytic
continuation method and the dispersion relation
method~\cite{Melikhov}\footnote{If we were to use the model parameters given 
in \cite{Mel}, we obtain, for example,  the values of 
$f_{+}(0)-f_{+}(q^{2}_{\rm max})$ as follows: 
0.783 (0.781) - 1.2 (1.2) for $D\to K$, 0.682 (0.681) - 1.61 (1.63)
for $D\to\pi$, 0.682 (0.684) - 1.12 (1.12) for $B\to D$, and
0.293 (0.293) - 2.7 (2.3) for $B\to\pi$, where the values in parentheses 
are the results obtained by the auther in~\cite{Melikhov}.}.
The nonvalence contributions from the $q^{+}$$\neq$0
were also quantified for the $0^-\to 0^-$ decays by comparing the valence 
contributions between the $q^{+}$=0 and $q^{+}$$\neq$0 frames. 
Another interesting thing is that our predicted value of the
$\eta$-$\eta'$ mixing angle, i.e., $\theta_{SU(3)}\approx -19^{\circ}$,
works fairly well for the predicitions of the decay rates for
$D_{s}\to\eta(\eta')$ processes.

Furthermore, our analytic solutions of the form factors for both
$0^-\to 0^-$ and $0^-\to 1^-$ decays are well approximated by 
monopole-type form factors except heavy-to-light processes such
as $B\to\pi$ and $B\to\rho$. For these two processes, our analytic
solutions of the form factors are neither approximated by monopole-type
form factors nor the Eq.~(\ref{ansatz}). In order to explain the
discrepancies between our analytic solutions and the approximations
of the pole dominace model, more accurate experimental data or the
lattice results should be accomodated. Nonetheless, if we assume the pole
dominance model is a good approximation for the heavy-to-light decays,
this may indicate that our LQFM with a simple picture of $q\bar{q}$ contents
for the description of a meson is not valid for really high $q^2$ region. 
Anyway, the comparison between our analytic solutions and the results
of the pole dominance model provides a good testing ground for the
validity of our model calculations.

%% file: Conclusion.tex
\newpage
\setcounter{equation}{0}
\setcounter{figure}{0}
\renewcommand{\theequation}{\mbox{6.\arabic{equation}}}
\chapter{Conclusion}
In this thesis, we have investigated the electroweak form factors and
the semileptonic decays of pseudoscalar and vector mesons within
the framework of light-front constituent quark model. 
In Chapters 2 and 3 of this thesis, we have discussed the two 
different schemes in treating the meson masses, e.g., SM and  
IM schemes. Regardless of the difference between
the two schemes, once the best fit parameters were used, both schemes
provided the predictions that were not only pretty similar with each
other but also remarkably good compared to the available experimental
data for form factors, decay constants, chargi radii, etc., of various
light pseudoscalar and vector mesons as well as their radiative decay 
widths. Similarly, once the best fit parameters were chosen,  
the difference from the Jacobi factor in the IM model wave
functions is substantially reduced in the numerical predictions for the 
physical observables. 
However, the difference in the choice of the radial wave function, 
e.g., harmonic-oscillator (HO) wave function versus power-law (PL)
wave function, was appreciable no matter what parameters
were used. For example, in the phenomenology of various meson radiative
decays at low $Q^2$, we observed that the Gaussian type wave function
was clearly better than the PL wave function in comparison with the
available experimental data. 

In order to justify the model wave function as a solution of the 
QCD-inspired dynamic equation, we applied the variational principle in Chapter 4 
of this thesis to the QCD-motivated potential, which includes not only the Coulomb plus
confining potential but also the hyperfine interaction, to obtain
the correct $\rho$-$\pi$ splitting. For the confining potential, we
took both HO ($\sim r^2$) and linear ($\sim r$) type potentials and 
compared the numerical results for these two cases. 
The variational principle for the effective Hamiltonian turns out to be crucial
to find the optimum values of our model parameters as shown in Chapters 4 and 7. 
We adopted the IM scheme to assure the orthogonality of model wave functions. 
As shown in Figs.~4.1(a) and 4.1(b), our central potentials for both
HO and linear potentials in Eq.~(\ref{potent}) are
not only very similar to each other but also quite close to the 
Isgur-Scora-Grinstein-Wise model 2(ISGW2)~\cite{isgw2} potentials. 
Using our quark potential model discussed in Chapter 4, we predicted 
the ground state meson spectra (see Table~\ref{t73}).
Our predictions of ground state meson mass spectra agreed quite well with
the experimental data~\cite{data} (within 6$\%$ error).
Furthermore, our model predicted the two unmeasured mass
spectra of $^{1}S_{0}(b\bar{b})$ and $^{3}S_{1}(b\bar{s})$ systems as
$M_{b\bar{b}}$=9295 (9657) MeV and $M_{b\bar{s}}$=5471 (5424) MeV
for the HO (linear) potential, respectively.

We have also predicted the $\omega$-$\phi$ and 
$\eta$-$\eta'$ mixing angles using the parametrization to 
incorporate the quark annihilation diagrams~\cite{isgur1,georgi,scadron} 
mediated by gluon exchanges and the SU(3) symmetry breaking, i.e., 
$m_u(d)\neq m_s$. Our results of the $\omega$-$\phi$ and
$\eta$-$\eta'$ mixing angles are $|\delta_{V}|\approx
4.2^{\circ} (7.8^{\circ})$ and 
$\theta_{SU(3)}\approx -19.3^{\circ} (-19.6^{\circ})$ for the
HO (linear) potential model, respectively. 
The sensitivity of the ($\omega,\phi$) mass spectra for 
$\sim 1^{\circ} (5^{\circ})$ variation of
$\delta_{V}$, i.e., from $\delta_{V}=4.2^{\circ} (7.8^{\circ})$ to
$3.3^{\circ}$ for the HO (linear) potential case, is within the $1\% (5\%)$ level.
As we discussed in the appendix E, the sign of mixing
angle depends on the sign of SU(3) breaking parameter $X$ 
(see Eqs. (E.3)-(E.5)). While $X_{P}>0$ is
well supported by the Particle Data Group~\cite{data}
($-23^{\circ}\leq\theta^{\eta-\eta'}_{SU(3)}\leq -10^{\circ}$), 
the sign of $X_{V}$ is not yet definite at the present stage
of phenomenology. Regarding the sign of $X_{V}$, it is interesting
to note that $\delta_{V}\approx -3.3^{\circ} 
(=\theta_{SU(3)}-35.26^{\circ})$ (i.e., $X_{V}<0$)
is favored in Refs.~\cite{jaus,Das,Sakurai,Coleman}, while
the conventional Gell-Mann-Okubo mass formula for the exact
SU(3) limit ($X\to 1$) predicts $\delta_{V}\approx 0^{\circ}$
in the linear mass scheme and $\delta_{V}\approx +3.3^{\circ}$ 
(i.e., $X_{V}>0$) in the quadratic mass scheme~\cite{data}. 
Even though it is not yet clear which sign of $\omega$-$\phi$ mixing
angle should be taken, the overall agreement between our
HO potential model with the positive sign, i.e.,
$\delta_{V}\sim +3.3^{\circ}$, and the
available experimental data seem to be quite good.
If we were to choose the sign of $X$ as $X>0$ in Eq. (E.4), then
the fact that the mass difference $m_{\omega}-m_{\rho}$ is
positive is correlated with the sign of the $\omega$-$\phi$
mixing angle~\cite{private}. In other words, $m_{\omega}>m_{\rho}$
implies $\delta_{V}>0$ from Eqs. (E.3)-(E.5).
Perhaps, the precision measurement of $\phi\to\eta'\gamma$
envisioned in the future at TJNAF experiments might be helpful to
give a more stringent test of $\delta_{V}$.
We also predicted the decay constants for various pseudoscalar and
vector mesons as summarized in Tables~\ref{dcon1} and \ref{dcon2}. 
Our predictions for the decay constants are quite
consistent with the available experimental data~\cite{data} as well as
the lattice QCD~\cite{Flynn,Bern2}. For the heavy meson decay 
constants, however, the current experimental data
have large error-bars and the more accurate data are needed. 

We further applied our quark potential model to compute the 
radiative (see Chapter 4) and the semileptonic (see Chapter7) decays 
of pseudoscalar and vector mesons.
One of the distinguished features in our LF approach is the ability to compute
the timelike form factors as shown in $0^-\to0^-$ and $0^-\to 1^-$ semileptonic 
decays. While we calculated the exclusive processes 
in the Drell-Yan-West ($q^+$=0) frame, we 
analytically continued the amplitudes to the timelike region.
Our analytic continuation method (i.e. changing $q_\perp$ to $iq_\perp$ in
the form factors) avoided encountering the nonvalence
diagram (black blob in Fig. 1.1(b)) and yielded the outcome  
of timelike form factors. 
As summarized in Tables~\ref{PPt} and \ref{PVt} for the $0^-\to0^-$ and $0^-\to 1^-$
semileptonic decays, respectively, our numerical results for both HO and linear 
parameters were overall not only quite similar to each other but also in
good agreement with the available experimental data~\cite{data}.
To assure our method of analytic continuation, we used the exactly solvable model 
of scalar field theory interacting with gauge fields and analyzed the 
full information for the nonvalence diagram. As shown in Chapter 5, we were able 
to check if the analytic continuation of the results in the $q^+$=0 frame
(without the black blob) indeed reproduces the exact results 
in timelike region. We established that this method can actually broaden the 
applicability of the standard
LF frame \`{a} la Drell-Yan-West to the timelike form factor calculation.
Exploring the timelike region, we have also found the importance of zero-mode
contributions.  The estimation of the zero-mode contribution was 
presented in Chapter 6. To the extent that the zero-modes
have a significant contribution to some physical observables~\cite{Kaon},
it seems conceivable that the condensation of zero-modes could lead to
the nontrivial realization of chiral symmetry breaking in the LF 
quantization approach.

Throughout the thesis, we attempted to 
fill the gap between the model wave function and the QCD-motivated
potential and explore covering as many observables as possible.
We think that the success of our model hinges on
the advantage of LF quantization realized by the rational energy-momentum
dispersion relation.
It is crucial to calculate the ``good" components of the current in the
reference frame which deletes the complication from the nonvalence Z-graph
contribution.
We anticipate further stringent tests of our model with more accurate data
from future experiments and lattice QCD calculations.


%% file: appendix1.tex
\setcounter{chapter}{0}
\setcounter{equation}{0}
\setcounter{figure}{0}
\renewcommand{\theequation}{\mbox{A.\arabic{equation}}}
\chapter{Spin-Orbit Wave Functions \protect
${\cal R}_{\lam\bar{\lam}}^{JJ_3}(x,{\bf k}_{\perp})$ }
The $4\times4$ Dirac matrices $\gamma_\mu$ are given by
\be
\gamma^\mu\gamma^\nu+\gamma^\nu\gamma^\mu=2g^{\mu\nu}
\ee
with $\gamma^0$ hermitian and $\gamma^i$ antihermitian.
In the chiral representation\cite{jaus} 
\be 
\gamma^0\: = 
\: \left(
\begin{array}{cc}
0 & I \\
I & 0
\end{array} \; \right), 
\;\gamma^i = \left(
\begin{array}{cc}
0 & \sigma^i \\
-\sigma^i & 0
\end{array} \; \; \right),
\ee 
where $I$ is the $2\times 2$ unit matrix and $\sigma^{i}$ are 
Pauli matrices defined as 
\be
\sigma^1\:= \left(
\begin{array}{cc}
0 & 1 \\ 1 & 0
\end{array}\; \right), 
\;\sigma^2= \left(
\begin{array}{cc}
0 & -i \\ i & 0
\end{array}\; \right),
\; \sigma^3= \left(
\begin{array}{cc}
1 & 0 \\ 0 & -1
\end{array}\; \right). 
\ee
We frequently need $\gamma^{+}\equiv\gamma^0+\gamma^3$
and
$\gamma^5\equiv i\gamma^0\gamma^1\gamma^2\gamma^3$,
which are given as
\be
\gamma^{+}\: =\left(
\begin{array}{cccc} 
             0 & 0 & 2 & 0 \\
             0 & 0 & 0 & 0 \\
             0 & 0 & 0 & 0 \\
             0 & 2 & 0 & 0
\end{array}\; \right), 
\;\gamma^{5}=\left(
\begin{array}{cccc}
             -1 & 0 & 0 & 0 \\
             0 & -1 & 0 & 0 \\
             0 & 0 & 1 & 0 \\
             0 & 0 & 0 & 1
\end{array}\; \right). 
\ee
The constituent quarks are spin $\frac{1}{2}$
particles and can be described by Dirac spinors
$u(k,\lam)$ and $v(k,\lam)$ satisfying the
Dirac equation
\be 
  (/\!\!\!k-m)u(k,\lam)=0,\;\;\;
  (/\!\!\!k+m)v(k,\lam)=0, 
\ee
where $/\!\!\!k=k_\mu \gamma^\mu$.
Using the $\gamma$-matrices given by Eqs.~(A.1)-(A.4),  
we now introduce the spinor basis: 
\be
u(k,\lam) 
    = \frac{1}{\sqrt{k^+}}(/\!\!\!k+m)u(\lam),\; \;
    v(k,\lam)
    = \frac{1}{\sqrt{k^+}}(/\!\!\!k-m)v(\lam), 
\ee
\be
u\left(\frac{1}{2}\right)\:
  =\left(
\begin{array}{c}
       1 \\
       0 \\
       0 \\
       0
\end{array}\; \right),
\; \; u\left(-\frac{1}{2}\right)
  =\left(
     \begin{array}{c}
       0 \\
       0 \\
       0 \\
       1
     \end{array} \right), 
\ee 
where $v(\lam)=u(-\lam)$.
We then can derive the spinors $u(k,\lam)$ and $v(k,\lam)$
as well as the antispinors $\bar{u}(k,\lam)$ and
$\bar{v}(k,\lam)$, where we use $\bar{u}=u^+\gamma^0$ and
$\bar{v}=v^+\gamma^0$, as
\be
u\left(k,\frac{1}{2}\right)\:=
\frac{1}{\sqrt{k^+}}
\left(
\begin{array}{c}
m \\ 0 \\ k^+ \\ k^R
\end{array}\; 
\right), \; 
\;u\left(k,-\frac{1}{2}\right)=
\frac{1}{\sqrt{k^+}}
\left(
\begin{array}{c}
-k^L \\ k^+ \\ 0 \\ m
\end{array}\; \right),
\ee
\be
v\left(k,\frac{1}{2}\right)\:=
\frac{1}{\sqrt{k^+}}
\left(
\begin{array}{c}
        -k^L \\ k^+ \\ 0 \\ -m
      \end{array}
    \right),\; 
\; v\left(k,-\frac{1}{2}\right)=
    \frac{1}{\sqrt{k^+}}
    \left(
      \begin{array}{c}
        -m \\ 0 \\ k^+ \\ k^R
      \end{array}\;
    \right),
\ee
\be
  \bar{u}\left(k,\frac{1}{2}\right)
  =
  \frac{1}{\sqrt{k^+}}\left(k^+,k^L,m,0\right),
  \; 
  \bar{u}\left(k,-\frac{1}{2}\right)
  =
  \frac{1}{\sqrt{k^+}}\left(0,m,-k^R,k^+\right),
\ee 
\be
  \bar{v}\left(k,\frac{1}{2}\right)
  =
  \frac{1}{\sqrt{k^+}}\left(0,-m,-k^R,k^+\right),
  \; 
  \bar{v}\left(k,-\frac{1}{2}\right)
  =
  \frac{1}{\sqrt{k^+}}\left(k^+,k^L,-m,0\right).
\ee
Here, $k^R$ and $k^L$ are defined as
$k^R\equiv k^1+ik^2$ and $k^L\equiv k^1-ik^2$,
respectivly.
The Melosh transformation ${\cal R}_M(x,{\bf k}_\perp, m)$
transforms the usual instant frame spin basis into
the LF frame spin basis and thus enables one to
assign proper total angular momentum quantum numbers
to hadronic states.
In the basis given Eq. (A.8)-(A.11), the spin-orbit wave functions 
of pseudoscalar and vector mesons can be represented in the 
following covariant way:
\be\label{R00_A} 
{\cal R}_{\lam\bar{\lam}}^{00}(x,{\bf k}_{\perp})
=\frac{-1}{\sqrt{2}[M^{2}_{0}-(m_1-m_2)^{2}]^{1/2}}
\bar{u}(p_1,\lam)\gamma_5 v(p_2,\bar{\lam}),
\ee
and
\bea\label{R1J_A}
{\cal R}_{\lam\bar{\lam}}^{1J_3}(x,{\bf k}_{\perp}) 
&=&\frac{-1}{\sqrt{2}[M^{2}_{0}-(m_1-m_2)^{2}]^{1/2}}\nonumber \\
&\times&
\bar{u}(p_1,\lam)\biggl[
/\!\!\!\vep(J_3) -\frac{\vep\cdot(p_1-p_2)}{M_0 + m_1 + m_2}\biggr]
v(p_2,\bar{\lam}).  
\eea 
The four-vectors $p_1,p_2$
are given in terms of the LF relative momentum variables 
$(x,{\bf k}_{\perp})$ as follows
\bea 
&&p^+_1= xP^+,\;\; p^+_2=(1-x)P^+, \nonumber\\
&&{\bf p}_{1\perp} = x{\bf P}_\perp + {\bf k}_{\perp},\;
\; {\bf p}_{2\perp}=(1-x){\bf P}_{\perp} - {\bf k}_{\perp},
\eea 
and satisfied by $p^2_i=m^2_i(i=1,2)$. 
The polarization vectors are given by 
\bea 
&&\vep^{\mu}(\pm)=\biggl[
0,\frac{2}{P^{+}}{\bf\vep}_{\perp}(\pm)\cdot{\bf P}_{\perp},
{\bf \vep}_{\perp}(\pm)\biggr],\nonumber\\
&&{\bf \vep}_{\perp}(\pm)=\mp(1,\pm i)/\sqrt{2},\nonumber\\ 
&&\vep^{\mu}(0)=\frac{1}{M_{0}}\biggl[
P^+,\frac{-M^{2}_{0}+P^{2}_{\perp}}{P^{+}},{\bf P}_{\perp}
\biggr].
\eea 
The explicit form of Eqs. (A.12) and (A.13) are given by
\be
{\cal R}^{00}_{\lam\bar{\lam}}\:=
\frac{{\cal R}_0}{\sqrt{2}}
\left(
\begin{array}{cc}
        -k^L & xm_2 + (1-x)m_1\\ 
        -xm_2 - (1-x)m_1 & -k^R 
      \end{array}
    \right),\;
\ee
\be
{\cal R}^{11}_{\lam\bar{\lam}}\:=
{\cal R}_0
\left(
\begin{array}{cc}
xm_2 + (1-x)m_1 +\frac{k^2_\perp}{M_0+m_1+m_2} &
k^R\frac{xM_{0} + m_1}{M_0 + m_1 + m_2}\\
-k^R\frac{(1-x)M_0 + m_2}{M_0 + m_1 + m_2} &
-\frac{(k^R)^2}{M_0 + m_1 + m_2}
      \end{array}
    \right),\;
\ee
\be
{\cal R}^{10}_{\lam\bar{\lam}}\:=
\frac{{\cal R}_0}{\sqrt{2}}
\left(
\begin{array}{cc}
k^L\frac{(1-2x)M_0 + m_2 - m_1}{M_0+m_1+m_2} &
xm_2 + (1-x)m_1 + \frac{2k^2_\perp}{M_0 + m_1 + m_2}\\
xm_2 + (1-x)m_1 + \frac{2k^2_\perp}{M_0 + m_1 + m_2} &
-k^R\frac{(1-2x)M_0 + m_2 - m_1}{M_0+m_1+m_2} 
      \end{array}
    \right),\;
\ee
\be
{\cal R}^{1-1}_{\lam\bar{\lam}}\:=
{\cal R}_0
\left(
\begin{array}{cc}
-\frac{(k^L)^2}{M_0 + m_1 + m_2} &
k^L\frac{(1-x)M_0 + m_2}{M_0 + m_1 + m_2} \\
-k^L\frac{xM_{0} + m_1}{M_0 + m_1 + m_2} &
xm_2 + (1-x)m_1 +\frac{k^2_\perp}{M_0+m_1+m_2}
      \end{array}
    \right),\;
\ee
where
\be
{\cal R}_0= \frac{1}{\sqrt{x(1-x)}[M^2_0-(m_1-m_2)^2]^{1/2}},
\ee
and
\be
M^2_0=\frac{k^2_\perp + m^2_1}{x} +
\frac{k^2_\perp + m^2_2}{1-x}.
\ee

%% file: appendix2.tex
\newpage
\setcounter{equation}{0}
\renewcommand{\theequation}{\mbox{B.\arabic{equation}}}
\chapter{Spin-Averaged Meson Masses of $\eta,\eta',\omega$ and $\phi$}
We list our flavor wave functions of the neutral
pseudoscalar $(\eta,\eta')$ and vector $(\omega,\phi)$ nonet states
to show explicitly how we obtained the values of spin-averaged mass
used in Chapter 2:
\begin{eqnarray}
&&\eta = \frac{1}{\sqrt{6}}(u\bar{u} + d\bar{d} - 2s\bar{s}),\\
&&\eta' = \frac{1}{\sqrt{3}}(u\bar{u} + d\bar{d} + s\bar{s}),\\
&&\omega =\frac{1}{\sqrt{2}}(u\bar{u} + d\bar{d}),\\
&&\phi = s\bar{s}.
\end{eqnarray}
For ideally mixed isocalar and isovector mesons, we take
the flavor wave function as
\begin{eqnarray}
\chi_{ns}^{P(V)}=\frac{1}{\sqrt{2}}(u\bar{u} +
d\bar{d}),\hspace{0.5cm} \chi_{s}^{P(V)}= s\bar{s},
\end{eqnarray}
where $P(V)$ denotes pseudoscalar(vector) meson states.
In terms of this basis, the neutral meson nonet states are given by
\begin{eqnarray}
&&\eta_{1} = \sqrt{\frac{2}{3}}\chi_{ns}^{P} +
\sqrt{\frac{1}{3}}\chi_{s}^{P},
\hspace{.5cm}
\omega_{1} = \sqrt{\frac{2}{3}}\omega +
\sqrt{\frac{1}{3}}\phi,\nonumber\\
&&\eta_{8}= \sqrt{\frac{1}{3}}\chi_{ns}^{P} -
\sqrt{\frac{2}{3}}\chi_{s}^{P},
\hspace{.5cm}
\omega_{8} =  \sqrt{\frac{1}{3}}\omega - \sqrt{\frac{2}{3}}\phi
\end{eqnarray}
where $\eta = \eta_{8}$, $\eta'= \eta_{1}$, $\omega = \chi_{ns}^{V}$
and $\phi = \chi_{s}^{V}$.
We define the spin-averaged masses $\bar{M}$ of $\chi_{ns}$
and $\chi_{s}$ as follows:
\begin{eqnarray}
&&\bar{M}_{ns} = \frac{1}{4}M_{ns}^{P} + \frac{3}{4}M_{ns}^{V},
\hspace{0.5cm}
\bar{M}_{s} = \frac{1}{4}M_{s}^{P} + \frac{3}{4}M_{s}^{V}.
\end{eqnarray}
Using Eqs. (B.3)-(B.6), we assign
\begin{eqnarray}
&&M_{\eta}^{\rm exp} = \frac{1}{3}M_{ns}^{P}
+ \frac{2}{3}M_{s}^{P},\hspace{0.3cm}
M_{\eta'}^{\rm exp} = \frac{2}{3}M_{ns}^{P} +
\frac{1}{3}M_{s}^{P}, \hspace{0.3cm}
M_{\omega}^{\rm exp} = M^{V}_{ns},\nonumber\\
&& M_{\phi}^{\rm exp} = M^{V}_{s}.
\end{eqnarray}
Then, we obtain the following spin averaged masses of $\eta$ and
$\eta'$:
\begin{eqnarray}
&&\bar{M}_{\eta} = \frac{1}{3}\bar{M}_{ns} + \frac{2}{3}\bar{M}_{s}
                = 842 \;{\rm MeV},\\
&&\bar{M}_{\eta'} = \frac{2}{3}\bar{M}_{ns} + \frac{1}{3}\bar{M}_{s}
                = 885 \;{\rm MeV}.
\end{eqnarray}
Likewise, from Eq. (B.6), we obtain
\begin{eqnarray}
&&\bar{M}_{\omega_{1}} = \frac{2}{3}\bar{M}_{\omega}
 + \frac{1}{3}\bar{M}_{\phi},\\
&&\bar{M}_{\omega_{8}} = \frac{1}{3}\bar{M}_{\omega}
 + \frac{2}{3}\bar{M}_{\phi}.
\end{eqnarray}
Using $\bar{M}_{\eta} = \bar{M}_{\omega_{8}}$ and
$\bar{M}_{\eta'} = \bar{M}_{\omega_{1}}$,
we can then evaluate the spin-averaged masses of $\omega$ and $\phi$ as
$\bar{M}_{\omega} = 928$ MeV and $\bar{M}_{\phi} = 799$ MeV,
respectively.

To calculate the spin-averaged meson masses depending on the
schemes of flavor mixing, let's consider the
``perfect mixing"($\theta_{SU(3)} = -10^\circ$) $\tilde{\eta}$ and
$\tilde{\eta'}$ states defined by\cite{isgur1,isgur2},
\begin{eqnarray}
\tilde{\eta} = \frac{1}{\sqrt{2}}(\chi_{ns}^{P} -
\chi_{s}^{P}),\hspace{.5cm}
\tilde{\eta'} = \frac{1}{\sqrt{2}}(\chi_{ns}^{P} +
\chi_{s}^{P}).
\end{eqnarray}
Using Eqs. (B.13) and Eq. (B.6), we obtain
\begin{eqnarray}
&&\eta_{1}=\frac{\sqrt{2}-1}{\sqrt{6}}\tilde{\eta} +
\frac{\sqrt{2}+1}{\sqrt{6}}\tilde{\eta'},
\hspace{0.5cm}
\eta_{8}=\frac{1+\sqrt{2}}{\sqrt{6}}\tilde{\eta} +
\frac{1-\sqrt{2}}{\sqrt{6}}\tilde{\eta'}.
\end{eqnarray}
Thus, the spin-averaged masses $\bar{m}$ of the $\tilde{\eta}$ and
$\tilde{\eta'}$ for the ``perfect
mixing" scheme are related to $\bar{M}_{\eta}$ and $\bar{M}_{\eta'}$
calculated in our scheme (see Eqs. (B.9) and (B.10)), respectively,
as follows:
\begin{eqnarray}
&&\bar{M}_{\eta'} = \frac{(\sqrt{2}-1)^{2}}{6}\bar{m}_{\tilde{\eta}}
+ \frac{(\sqrt{2}+1)^{2}}{6}\bar{m}_{\tilde{\eta'}},\nonumber\\
&&\bar{M}_{\eta} = \frac{(\sqrt{2}+1)^{2}}{6}\bar{m}_{\tilde{\eta}}
+ \frac{(\sqrt{2}-1)^{2}}{6}\bar{m}_{\tilde{\eta'}}.
\end{eqnarray}
From Eq. (B.15), the spin-averaged masses for
the ``perfect mixing" states are given by $\bar{m}_{\tilde{\eta}} = 843$
MeV and $\bar{m}_{\tilde{\eta'}} = 884$ MeV, respectively.
Using the same method as above, for the $\theta_{SU(3)}= -23^\circ$
mixing scheme, we obtain the following spin averaged meson masses:
$m_{\eta} = 838$ MeV and $m_{\eta'} = 873$ MeV, respectively.

\newpage
\setcounter{equation}{0}
\renewcommand{\theequation}{\mbox{C.\arabic{equation}}}
\chapter{The Electromagnetic Decay Width $\Gamma(A\to B + \gamma)$}
In this appendix, we derive the decay widths of $A\to B+\gamma$
($A = \rho,K^{*},\omega,\phi,$ and $B = \pi, K,\eta,\eta'$) and
$A_{1}\to \pi+\gamma$.

Consider the electromagnetic decay process $A\to B + \gamma$.
If the masses of A and B mesons are given by $m_{A}$ and $m_{B}$,
respectively, then the decay rate
$\Gamma(A\to B + \gamma)$ in the rest frame of $A$ is given by
$(\hbar=c=1)$
\begin{eqnarray}
\Gamma(A\to B + \gamma) &=& \frac{1}{(2\pi)^{6}}\frac{S}{2m_{A}}
\int\frac{d^{3}{\bf p}_{B}}{2E_{B}}\frac{d^{3}{\bf p}_{\gamma}}{2E_{\gamma}}
|{\cal M}|^{2}(2\pi)^{4}\delta^{3}({\bf p}_{B} + {\bf p}_{\gamma})
\nonumber\\
&\times& \delta(m_{A} - E_{B} - E_{\gamma}),
\end{eqnarray}
where $S = 1/j!$ for each group of $j$ identical particles in the final
states.  Here, ${\cal M}$ is the transition matrix element defined by
\begin{eqnarray}
{\cal M} =
\vep_{\mu}^{*}(\lambda_{\gamma})\la B(P')|J^{\mu}|A(P,\lambda)\ra.
\end{eqnarray}
To allow decays to all possible spin configurations, we consider
the replacement
\begin{eqnarray}
|{\cal M}|^{2}\to {\overline{|{\cal M}|}}^{2}\equiv
\frac{1}{2S_{A} + 1}\sum_{\lambda_{\gamma}=\pm 1}|{\cal M}|^{2},
\end{eqnarray}
where $S_{A}$ is the spin of particle A and $\lambda_{\gamma}$
corresponds to the state of transverse polarization of the emitted
photon. The above replacement means the average over the initial spin
and the sum over the helicities of the emitted photon.
Then, from Eq. (C.1), we get
\begin{eqnarray}
\Gamma(A\to B + C) &=& \frac{S}{2(4\pi)^{2}m_{A}}\frac{1}{2S_{A} + 1}
\int |{\bf p}_{B}|^{2}d|{\bf p}_{B}|d\Omega \nonumber\\
&\times&\frac{\delta(m_{A} - \sqrt{m_{B}^{2} + {\bf p}_{B}^{2}} - |{\bf
p}_{B}|)} {|{\bf p}_{B}|\sqrt{m_{B}^{2} + {\bf p}_{B}^{2}}}
\sum_{\lambda_{\gamma = \pm 1}}|{\cal M}|^{2}.
\end{eqnarray}
If we let $|{\bf p}_{B}| = \rho$ and
$E\equiv \sqrt{m_{B}^{2} + \rho^{2}} + \rho$, then
\begin{eqnarray}
\Gamma(A\to B + \gamma)=\frac{S}{8\pi m_{A}}\frac{1}{2S_{A} + 1}
\int^{\infty}_{m_{B}}dE\frac{\rho}{E}\delta(m_{A} - E)
\sum_{\lambda_{\gamma = \pm 1}}|{\cal M}|^{2}.
\end{eqnarray}
Therefore, we get
\begin{eqnarray}
\Gamma(A\to B + \gamma) &=& \frac{S}{8\pi
m_{A}^{2}}\frac{\rho_{0}}{2S_{A} + 1}
\sum_{\lambda_{\gamma = \pm 1}}|{\cal M}|^{2},
\end{eqnarray}
where
\begin{eqnarray}
S = 1, \hspace{.2in}{\rm if \hspace {.2in}B \neq \gamma},
\nonumber\\
S = \frac{1}{2}, \hspace{.2in}{\rm if \hspace {.2in}B = \gamma}.
\end{eqnarray}
Here, $\rho_{0}$ is the value of $\rho$ when $E = m_{A}$, i.e.,
$\rho_{0} = (m_{A}^{2} - m_{B}^{2})/2m_{A}$.
In the rest frame of $A$, i.e., $P_{A} = (m_{A}, {\vec 0})$
,$P'_{B} = (m_{B}, {\bf p}_{B})$ and
$P_{\gamma} = (|{\bf p}_{B}|, - {\bf p}_{B})$, the invariant amplitude
square is given by
\begin{eqnarray}
\sum_{\lambda_{\gamma = \pm 1}}|{\cal M}|^{2}
&=& \sum_{\lambda_{\gamma = \pm 1}}|eG_{AB}(Q^{2})
\epsilon^{\mu\nu\alpha\beta}\vep^{*}_{\mu}(\lambda_{\gamma})
\vep_{\nu}(P,\lambda)P'_{\alpha}
P_{\beta}|^{2}\nonumber\\
&=& e^{2}|G_{AB}(Q^{2})|^{2}
2|\epsilon^{12\alpha\beta}P'_{\alpha}P_{\beta}|^{2}
= e^{2}|G_{AB}(Q^{2})|^{2}2m_{A}^{2} {\bf p}_{B}^{2}\nonumber\\
&=& \frac{(m_{A}^{2} - m_{B}^{2})^{2}}{2}e^{2}|G_{AB}(Q^{2})|^{2}.
\end{eqnarray}
Therefore, we get the following decay width $\Gamma(A\to B + \gamma)$:
\begin{eqnarray}
\Gamma(A\to B + \gamma) = \frac{\alpha}{2S_{A} + 1}|G_{AB}(0)|^{2}
\biggl(\frac{m_{A}^{2} - m_{B}^{2}}{2m_{A}}\biggr)^{3},
\end{eqnarray}
where $\alpha(= e^{2}/4\pi)$ is the fine-structure constant.

The decay width of $A_{1}\to\pi\gamma$ can be calculated
in the same manner using our definition of the transition matrix
element given by Eq. (2.26) as follows:
\begin{eqnarray}
&&\sum_{\lambda_{\gamma = \pm 1}}|{\cal M}|^{2}
=|\epsilon_{\mu}^{*}(\lambda_{\gamma})
\la\pi(P')|J^{\mu}|A_{1}(P,\lambda)\ra|^{2}
= 2\biggl |e\frac{G_{1}(Q^{2})}{m_{A_{1}}}(P_{A_{1}} + P'_{\pi})\cdot
P_{\gamma}\biggr|^{2} \nonumber\\
&&\;\;\;= 2\biggl | e\frac{G_{1}(Q^{2})}{m_{A_{1}}}\biggr |
{\bf p}_{\pi}| (m_{A_{1}} + m_{\pi} + |{\bf p}_{\pi}|)|^{2}
= 2e^{2}\biggl |\frac{G_{1}(Q^{2})}{m_{A_{1}}}\biggr |^{2}
(2m_{A_{1}}{\bf p}_{\pi})^{2},
\nonumber\\
\end{eqnarray}
and the result is given by
\begin{eqnarray}
\Gamma(A_{1}\to\pi\gamma)
= \frac{4\alpha}{3}\biggl |\frac{G_{1}(0)}{m_{A_{1}}}\biggr |^{2}
\biggl(\frac{m_{A_{1}}^{2} - m_{\pi}^{2}}{2m_{A_{1}}}\biggr)^{3}.
\end{eqnarray}


%% file: appendix3.tex
\newpage
\setcounter{equation}{0}
\renewcommand{\theequation}{\mbox{D.\arabic{equation}}}
\chapter{Fixation of the Model Parameters Using the Variational
Principle}
In this appendix, we discuss how to fix the parameters of our model,
i.e., quark masses ($m_{u},m_{s}$), Gaussian parameters (
$\beta_{u\bar{u}}=\beta_{u\bar{d}}, \beta_{u\bar{s}},\beta_{s\bar{s}}$),
and potential parameters $(a,b,\kappa)$ in $V_{q\bar{q}}$ given by 
Eq.~(\ref{potent}).
In our potential model, the $\rho$-$\pi$ mass splitting is obtained by
the hyperfine interaction $V_{\rm hyp}$.

Our variational method first evaluates
$\langle\Psi|[H_{0}+V_{0}]|\Psi\rangle$ with a trial function
$\phi_{10}(k^{2})$
that depends on the parameters $(m,\beta)$ and varies these parameters
until the expectation value of $H_{0}+V_{0}$ is a minimum.
Once these model parameters are fixed, then,
the mass eigenvalue of each meson is obtained by
$M_{q\bar{q}}=\langle\Psi|[H_{0}+V_{0}]|\Psi\rangle +
\langle\Psi|H_{\rm hyp}|\Psi\rangle$
\footnote{As we will see later, in our fitting of the $\rho$-$\pi$
splitting, the rather big mass shift due to the hyperfine interaction
is attributed to the large QCD coupling constant,
$\kappa=0.3-0.6$.}.
In this approach, we do not discriminate the Gaussian parameter set
${\bf\beta}=(\beta_{u\bar{u}},\beta_{u\bar{s}},\beta_{s\bar{s}})$
by the spin structure of mesons.

Let us now illustrate our detailed procedures of finding the optimized
values of the model parameters using the variational principle:
\be\label{variation}
\frac{\partial\langle\Psi|[ H_{0}+ V_{0}]|\Psi\rangle}{\partial\beta}= 0.
\ee
From Eqs.~(\ref{hamil})-(\ref{potent}) and 
Eq.~(\ref{variation}), we obtain the following equations
for the HO and linear potentials:
\begin{eqnarray}
{\rm HO}&:&\;b_{h}= \frac{\beta^{3}}{3}\biggl\{
\frac{\partial\langle\Psi|H_{0}|\Psi\rangle}{\partial\beta}
- \frac{8\kappa_{h}}{3\sqrt{\pi}} \biggr\},\;\;\;\;\;\\
{\rm Linear}&:&\;b_{l}= \frac{\sqrt{\pi}\beta^{2}}{2}\biggl\{
\frac{\partial\langle\Psi|H_{0}|\Psi\rangle}{\partial\beta}
- \frac{8\kappa_{l}}{3\sqrt{\pi}} \biggr\},\;\;\;\;\;
\end{eqnarray}
where the subscript $h$ $(l)$ represents the HO (linear) potential
parameters. Equations (D.2) and (D.3) imply that the variational principle
reduces a degree of freedom in the parameter space.
Thus, we have now four parameters, i.e.,
$\{m_{u},\beta_{u\bar{d}},a,\kappa$ (or $b)\}$. However, in order to
determine these four parameters from the two experimental values of
$\rho$ and $\pi$ masses, we need to choose two input parameters.
These two parameters should be carefully chosen.
Otherwise, even though the other two parameters
are fixed by fitting the $\rho$ and $\pi$ masses, our
predictions would be poor for other observables such as the ones
in Section 4.2 as well as other mass spectra. From our trial and error
type of analyses, we find that $m_{u}$=0.25 (0.22) GeV is the best
input quark mass parameter for the HO (linear) potential
among the widely used $u(d)$ quark mass, $m_{u}$=0.22 GeV~\cite{isgur2},
0.25 GeV~\cite{jaus}, and 0.33 GeV~\cite{dziem1,isgw,isgw2}.
For the linear potential, the string tension $b_{l}=0.18$ GeV$^{2}$ is
well known from other quark model analyses~\cite{isgur2,isgw,isgw2}
commensurate with Regge phenomenology.
Thus, we take $m_{u}$=0.22 GeV and
$b_{l}$=0.18 GeV$^{2}$ as our input parameters for the linear potential
case. However, for the HO potential, there is no
well-known quantity corresponding to the string tension and thus we
use the parameters of $m_{u(d)}$=0.25 GeV and
$\beta_{u\bar{d}}$=0.3194 GeV as our input
parameters which turn out to be good values to
describe various observables of both the $\pi$ and
$\rho$ mesons for the Gaussian radial wave function~\cite{jaus}.

Using Eqs. (\ref{hamil}), (D.2) and (D.3) with the input value sets of
(1) ($m_{u}$=0.25 GeV, $\beta_{u\bar{d}}$=0.3194 GeV)
for the HO potential and (2) ($m_{u}$=0.22 GeV,
$b_{l}$=0.18 GeV$^{2}$) for the linear potential, we obtain the
following parameters from the $\rho$ and $\pi$ masses, viz.,
$\langle\Psi|H^{V(P)}_{u\bar{d}}|\Psi\rangle=M^{V(P)}_{u\bar{d}}=
m_{\rho(\pi)}$ ($P$= pseudoscalar and $V$= vector):
\begin{eqnarray}
{\rm HO}&:&\;a_{h}=-0.144\;\mbox{GeV},\;
b_{h}= 0.010\;\mbox{GeV}^{3},\;
\kappa_{h} = 0.607,\;\;\;\;\;\;\\
{\rm Linear}&:&\;a_{l}= -0.724\;\mbox{GeV},\;
\beta_{u\bar{d}}= 0.3659\;\mbox{GeV},\;
\kappa_{l} = 0.313.\;\;\;\;\;\;
\end{eqnarray}
As shown in Fig. 4.1(a), it is interesting to note that our two
central potentials,
Coulomb plus HO (solid line) and Coulomb plus linear
(dotted line) potentials, are not much different from each other and
furthermore quite comparable to the Coulomb plus linear quark potential
model suggested by Scora and Isgur (ISGW2)~\cite{isgw2} (long-dashed line for
$\kappa$=0.3 and dot-dashed line for $\kappa$=0.6) up to the range of
$r \leq 2$ fm. Those four potentials (HO, linear, and ISGW2) are also
compared with the GI potential
model~\cite{isgur2} (short-dashed line) in Fig. 4.1(a).
The corresponding string tensions, i.e.,
$f_{0}(r)=-dV_{0}(r)/dr$, are also shown in Fig. 4.1(b).

Next, among various sets of $\{m_{s},\beta_{u\bar{s}}\}$
satisfying Eqs. (D.2) and (D.3), we find $m_{s}$=0.48 [0.45] GeV and
$\beta_{u\bar{s}}$=0.3419 [0.3886] GeV for HO [linear]
potential by fitting optimally the
masses of $K^{*}$ and $K$, i.e., $M^{V(P)}_{u\bar{s}}=m_{K^{*}(K)}$.
Once the set of $\{m_{s},\beta_{u\bar{s}}\}$ is fixed, then
the parameters $\beta_{s\bar{s}}$=0.3681 [0.4128] GeV for
the HO [linear] potential can be obtained from Eq. (D.2)[(D.3)].
Subsequently, $M^{V}_{s\bar{s}}$ and
$M^{P}_{s\bar{s}}$ are predicted as 996 [952] MeV and 732 [734] MeV
for the HO [linear] potential, respectively.
As shown in Fig. 4.2(a) [4.2(b)], the solid, dotted, and dot-dashed lines 
are fixed by the HO[linear] potential
parameter sets of $\{m_{u},\beta_{u\bar{d}}\}$,
$\{m_{s},\beta_{u\bar{s}}\}$, and $\beta_{s\bar{s}}$, respectively,
and these three lines cross the same point in the space of $b$ and
$\kappa$ if the parameters in Table 4.1 are used.

We have also examined the sensitivity of our variational
parameters and the corresponding mass spectra
using a Gaussian smearing function to weaken the singularity of
$\delta^{3}(r)$ in hyperfine interaction, viz.,
\begin{eqnarray}\label{smear}
\delta^{3}(r)\to\frac{\sigma^{3}}{\pi^{3/2}}\exp(-\sigma^{2}r^{2}).
\end{eqnarray}
By adopting the well-known cutoff value of 
$\sigma$=1.8~\cite{isgur2,Capstick} and repeating the same
optimization procedure as the contact term [i.e., $\delta^{3}(r)$] case,
we obtain the following parameters\footnote{For the sensitivity check of
smearing out $\delta^{3}(r)$ [Eq.~(\ref{smear})], we kept
$\beta_{u\bar{d}}=0.3659$ GeV for the linear potential case given
by Eq. (D.5) as an input value and checked how much $b_{l}$
changed.} for each potential:
\begin{eqnarray}
{\rm HO}&:&\; a_{h}= -0.123\;\mbox{GeV},\;
b_{h}= 9.89\times 10^{-3}\;\mbox{GeV}^{3},\;
\kappa_{h} = 0.636,\;\;\;\;\;\\
{\rm Linear}&:&\; a_{l}= -0.7\;\mbox{GeV},\;
b_{l}= 0.176\;\mbox{GeV}^{2},\;
\kappa_{l} = 0.332.\;\;\;\;\;
\end{eqnarray}
The changes of other model parameters and mass spectra are given in
Tables 4.1 and 4.2. As one can see in Eqs. (D.7), (D.8) and 
Tables 4.1, 4.2, the effects of smearing out $\delta^{3}(r)$ are quite 
small and the smearing effects are in fact negligible for our numerical
analysis in Section 4.2.

\newpage
\setcounter{equation}{0}
\renewcommand{\theequation}{\mbox{E.\arabic{equation}}}
\chapter{Mixing Angles of ($\eta$,$\eta'$) and ($\omega$,$\phi$)}
In this appendix, we illustrate the mixing angles of
($\eta$,$\eta'$) and ($\omega$,$\phi$) by adopting the formulation
to incorporate the quark-annihilation diagrams and the effect of
SU(3) symmetry breaking in the meson mixing angles.

Equations~(\ref{f1f2}) satisfy the (mass)$^{2}$ eigenvalue equation
\be
{\cal M}^{2}|f_{i}\ra= M^{2}_{f_{i}}|f_{i}\ra\hspace{.3cm}(i=1,2).
\ee
Taking into account SU(3) symmetry breaking, we use the following
parametrization for ${\cal M}^{2}$ suggested by Scadron\cite{scadron}:
\begin{eqnarray}
{\cal M}^{2}=\left(\begin{array}{cc}
M^{2}_{n\bar{n}} + 2\lambda & \sqrt{2}\lambda X \\
\sqrt{2}\lambda X & M^{2}_{s\bar{s}} + \lambda X^{2}
\end{array} \right).
\end{eqnarray}
The parameter $\lambda$ characterizes the strength of the
quark-annihilation graph which couples the $I$=0 $u\bar{u}$ state
to $I$=0 $u\bar{u},d\bar{d},s\bar{s}$ states with equal strength
in the exact SU(3) limit. The parameter $X$, however, pertains to
SU(3) symmetry breaking such that the quark-annihilation graph
factors into its flavor parts, with $\lambda$, $\lambda X$,
and $\lambda X^{2}$ for the $u\bar{u}\to u\bar{u}(d\bar{d})$,
$u\bar{u}\to s\bar{s}$(or $s\bar{s}\to u\bar{u})$, and
$s\bar{s}\to s\bar{s}$ processes, respectively.
Of course, $X\to 1$ in the SU(3) exact limit.
Also, in Eq. (E.2), $M^{2}_{n\bar{n}}$ and
$M^{2}_{s\bar{s}}$ describe the masses of the corresponding mesons
in the absence of mixing.

Solving Eqs.~(\ref{f1f2}), (E.1), and (E.2), 
we obtain Eq.~(\ref{tan2}) and
\begin{eqnarray}
\lambda&=& \frac{(M^{2}_{f_{1}} - M^{2}_{n\bar{n}})(M^{2}_{f_{2}}
- M^{2}_{n\bar{n}})}{2(M^{2}_{s\bar{s}}- M^{2}_{n\bar{n}})},\\
X^{2}&=& \frac{2(M^{2}_{f_{2}} -M^{2}_{s\bar{s}})(M^{2}_{s\bar{s}}
- M^{2}_{f_{1}})}{(M^{2}_{f_{2}} - M^{2}_{n\bar{n}})(M^{2}_{f_{1}}
- M^{2}_{n\bar{n}})},\\
\tan2\delta&=& \frac{2\sqrt{2}\lambda X}{(M^{2}_{s\bar{s}}
- M^{2}_{n\bar{n}} + \lambda X^{2} - 2\lambda)}.
\end{eqnarray}
Equations (E.3) and (E.4) are identical
to the two constraints Tr(${\cal M}^{2}$)= Tr($M^{2}_{f_{i}}$) and
det(${\cal M}^{2}$)= det($M^{2}_{f_{i}}$).
The sign of $\delta$ is fixed by the signs of the $\lambda$ and $X$
from Eq. (E.5). Also, since Eq. (E.2) is decoupled from the subspace of
$(u\bar{u}-d\bar{d})/\sqrt{2}$, the physical masses of $m_{\pi}$ and
$m_{\rho}$ are confirmed to be the masses of $M^{P}_{n\bar{n}}$ and
$M^{V}_{n\bar{n}}$, respectively, as we used in 
Section 4.1 to fix the parameters ($a,b,\kappa$).

Given the fixed physical masses of $M^{P}_{n\bar{n}}$=$m_{\pi}$
and $M^{P}_{n\bar{n}}$=$m_{\rho}$ together with $M_{f_{i}} (i=1,2)$,
the magnitudes of mixing angles for $\eta$-$\eta'$ and $\omega$-$\phi$
now depend only on the masses of $M^{P}_{s\bar{s}}$
and $M^{V}_{s\bar{s}}$, respectively, from Eqs.~(\ref{f1f2}).
However, from Eqs. (E.3)-(E.5), one can see that the sign of mixing
angle depends on the sign of parameter $X$. While $X_{P}>0$ is
well supported by the Particle Data Group~\cite{data} 
($-23^{\circ}\leq\theta^{\eta-\eta'}_{SU(3)}\leq
-10^{\circ}$), the sign of $X_{V}$ is not yet definite at the present stage 
of phenomenology. Regarding the sign of $X_{V}$, it is interesting
to note that $\delta_{V}\approx
-3.3^{\circ} (=\theta_{SU(3)}-35.26^{\circ})$ (i.e., $X_{V}<0$)
is favored in Refs.~\cite{jaus,Das,Sakurai,Coleman}, while
the conventional Gell-Mann-Okubo mass formula for the exact
SU(3) limit ($X\to 1$) predicts $\delta_{V}\approx 0^{\circ}$
in the linear mass scheme and
$\delta_{V}\approx +3.3^{\circ}$ (i.e., $X_{V}>0$) in the
quadratic mass scheme~\cite{data}. Our predictions
for the $\omega$-$\phi$ and $\eta$-$\eta'$ mixing angles
are given in Section 4.1.

The corresponding results of the mixing parameters $\lambda_{V(P)}$
and $X_{V(P)}$ in Eqs.~(E.3) and (E.4) are obtained for the
HO (linear) potential as follows:
\begin{eqnarray}
&&\lambda_{V}= 0.57\; (0.73)m^{2}_{\pi}\;\mbox{GeV}^{2},
\; X_{V}= \pm2.10\; (\pm3.08),\nonumber\\
&&\lambda_{P}= 13.5 \;(13.3)m^{2}_{\pi}\;\mbox{GeV}^{2},
\; X_{P}= 0.84\; (0.85).
\end{eqnarray}
Our values of $\lambda_{V}$ and $\lambda_{P}$ for both HO
and linear potential cases are not much different from the predictions
of Ref.~\cite{scadron}. The reason why $\lambda_{V}$
is much smaller than $\lambda_{P}$, i.e.,
$\lambda_{P}\approx 23\lambda_{V} (18\lambda_{V})$ in our HO (linear)
case and $\lambda_{P}\approx 18\lambda_{V}$ in Ref.~\cite{scadron},
may be attributed to the fact that in the quark-annihilation graph,
the $1^{--}$ annihilation graph involves one more gluon
compared to the $0^{-+}$ annihilation graph.
This also indicates the strong departure of $\eta$-$\eta'$ from 
the ideal mixing.


%% file: appendix4.tex
\newpage
\setcounter{equation}{0}
\renewcommand{\theequation}{\mbox{F.\arabic{equation}}}
\chapter{Derivation of the Matrix Element of the Weak
Vector Current }
\section{Drell-Yan-West ($q^+$=0) Frame}
In this appendix, we show the derivation of the matrix element of the
weak vector current $<P_{2}|\bar{q}_{2}\gamma^{\mu}Q_{1}|P_{1}>$ for
$0^-\to0^-$ decays given in Eq.~(\ref{PP_form}).

In the LFQM, the matrix element of the weak vector
current can be calculated by the convolution of initial and final
LF wave function of a meson as follows:
\begin{eqnarray}
<P_{2}|\bar{q}_{2}\gamma^{\mu}Q_{1}|P_{1}>&=&
\sum_{\lambda_{1},\lambda_{2},\bar{\lambda}}
\int dp^{+}_{\bar{q}}d^{2}{\bf k}_{\perp}
\phi^{\dagger}_{2}(x,{\bf k'}_{\perp})\phi_{1}(x,{\bf k}_{\perp})
\nonumber\\
&\times&
{\cal R}^{00\dagger}_{\lambda_{2}\bar{\lambda}}(x,{\bf k'}_{\perp})
\frac{\bar{u}(p_{2},\lambda_{2})}{\sqrt{p^{+}_{2}}}\gamma^{\mu}
\frac{u(p_{1},\lambda_{1})}{\sqrt{p^{+}_{1}}}
{\cal R}^{00}_{\lambda_{1}\bar{\lambda}}(x,{\bf k}_{\perp}),
\end{eqnarray}
where the spin-orbit wave function ${\cal R}^{JJ_{3}}(x,{\bf k}_{\perp})$
for pseudoscalar meson($J^{PC}=0^{-+}$) obtained from Melosh transformation
is given by
\be
{\cal R}^{00}_{\lambda_{i}\bar{\lambda}}=
\frac{1}{\sqrt{2}\sqrt{ M^{2}_{i0} -(m_{i}-m_{\bar{q}})^{2} }}
\bar{u}(p_{i},\lambda_{i})\gamma^{5}v(p_{\bar{q}},\bar{\lambda}),
\ee
and
\be
M^{2}_{i0}=\frac{k_{\perp}^{2}+m^{2}_{i}}{1-x}
+ \frac{k_{\perp}^{2}+m^{2}_{\bar{q}}}{x}.
\ee
Subsituting Eq.~(F.2) into Eq.~(F.1) and using the quark momentum variables
given in Eq.~(\ref{qq_var}), one can easily obtain
\begin{eqnarray}
<P_{2}|\bar{q}_{2}\gamma^{\mu}Q_{1}|P_{1}>
&=& -\int dxd^{2}{\bf k}_{\perp}
\frac{\phi^{\dagger}_{2}(x,{\bf k'}_{\perp})\phi_{1}(x,{\bf k}_{\perp})}
{2(1-x)\prod^{2}_{i}\sqrt{ M^{2}_{i0} -(m_{i}-m_{\bar{q}})^{2} }}
\nonumber\\
&\times&{\rm Tr} \biggl[\gamma_{5}({\not\! p}_{2}+m_{2})
\gamma^{\mu}({\not\!p}_{1}+m_{1})\gamma_{5}
({\not\! p}_{\bar{q}}-m_{\bar{q}})\biggr],
\end{eqnarray}
where we used the following completeness relations of the Dirac spinors
\be
\sum_{\lambda_{1,2}}u(p,\lambda)\bar{u}(p,\lambda)={\not\!p}+m,
\;\;
\sum_{\lambda_{1,2}}v(p,\lambda)\bar{v}(p,\lambda)={\not\!p}- m.
\ee
In the standard $q^{+}$=0 frame where the decaying hadron is at
rest, the trace terms in Eq.~(F.4)
for the  ``+" and ``$\perp$" components of the vector current
$J^{\mu}$=$\bar{q}_{2}\gamma^{\mu}q_{1}$, respectively,
are obtaind as follows
\begin{eqnarray}
& &{\rm Tr} \biggl[\gamma_{5}({\not\! p}_{2}+m_{2})
\gamma^{\mu}({\not\!p}_{1}+m_{1})\gamma_{5}
({\not\! p}_{\bar{q}}-m_{\bar{q}})\biggr]\nonumber\\
& &= -4\biggl[ p^{\mu}_{1}(p_{2}\cdot p_{\bar{q}} + m_{2}m_{\bar{q}})
+ p^{\mu}_{2}(p_{1}\cdot p_{\bar{q}}+ m_{1}m_{\bar{q}})
+ p^{\mu}_{\bar{q}}(-p_{1}\cdot p_{2}+ m_{1}m_{2})\biggr]\nonumber\\
& & = -\frac{4P^{+}}{x}\biggl[ {\cal A}_{1}{\cal A}_{2}
+ {\bf k}_{\perp}\cdot{\bf k'}_{\perp}\biggr],\;\; 
\mbox{for}\,\mu= +  \\
& & = -2\biggl[ \frac{({\cal A}^{2}_{1}+ k_{\perp}^{2})}{x(1-x)}
({\bf k}_{\perp}-{\bf q}_{\perp}) + \frac{({\cal A}^{2}_{2}
+ k_{\perp}^{'2})}{x(1-x)}{\bf k}_{\perp} + [ (m_{1}-m_{2})^{2}
+ q_{\perp}^{2}]{\bf k}_{\perp}\biggr], \;\; \mbox{for}\,\mu=\perp 
\nonumber\\
\end{eqnarray}
where ${\cal A}_{i}$=$m_{i}x + m_{\bar{q}}(1-x)$
and ${\bf k'}_{\perp}$=${\bf k}_{\perp}-x{\bf q}_{\perp}$. Our convention
of the scalar product, $p_{1}\cdot p_{2}$=$(p^{+}_{1}p^{-}_{2}
+ p^{-}_{1}p^{+}_{2})/2 - {\bf p}_{1\perp}\cdot{\bf p}_{2\perp}$ were
used to derive Eqs.~(F.6) and (F.7) from the second line of the above
equation. Substituting Eqs.~(F.6) and (F.7) into Eq.~(F.4), we now 
obtain the matrix element of the weak vector current
$<P_{2}|\bar{q}_{2}\gamma^{\mu}Q_{1}|P_{1}>$ for $\mu=+$ 
[see Eq.~(\ref{fp})]
and $\perp$ [see Eq.~(\ref{Jperp2})] in $q^{+}$=0 frame, respectively.

\section{Valence Contributions in $q^{+}$$\neq$0 Frame}
\subsection{ Electromagnetic Form Factors}
The analysis in $q^{+}$$\neq$0 frame is carried out
in a reference frame where the momentum transfer is purely
longitudinal, $i.e.$, ${\bf q}_{\perp}$=0 and $q^{2}$=$q^{+}q^{-}$.
In this case, we have
for the meson rest frame ${\bf P}_{1\perp}$=${\bf P}_{2\perp}$=0;
\begin{eqnarray}
&&P_{1}=(P^{+}_{1},\frac{M_{1}^{2}}{P^{+}_{1}},{\bf 0}_{\perp}),
\hspace{0.3cm}
P_{2}=(P^{+}_{1}-q^{+},\frac{M_{2}^{2}}{P^{+}_{1}-q^{+}},{\bf 0}_{\perp}),
\nonumber\\
&&q=(q^{+}, \frac{M_{1}^{2}}{P^{+}_{1}}-
\frac{M_{2}^{2}}{P^{+}_{1}-q^{+}},{\bf 0}_{\perp}),
\end{eqnarray}
where $M_{1}(M_{2})$ is the mass of the initial(final) meson.
The momentum transfer square $q^{2}$ can be rewritten in terms of the
fraction $r$=$P^{+}_{2}/P^{+}_{1}$=$1-q^{+}/P^{+}_{1}$ as follows
\be
q^{2}= -\frac{M^{2}}{r}(1-r)^{2},
\ee
where $M$=$M_{1}$=$M_{2}$.
Note from the constraint $0<r<1$(or equivalently, $0<q^{+}<P^{+}_{1}$)
that the square of the momentum transfer is always spacelike, $i.e.$,
$q^{2}<0$. Constraining $r$ as $0<r<1$, we obtain the solution for
$r$ from Eq.~(F.9) as follows
\be
r=\biggl(1+ \frac{Q^{2}}{2M^{2}}\biggr)-\biggl[
\biggl(1+ \frac{Q^{2}}{2M^{2}}\biggr)^{2}-1\biggr]^{1/2},
\ee
where $Q^{2}=-q^{2}>0$. For given $P_{1}$ and $P_{2}$,
the relevant quark momentum variables are
\begin{eqnarray}
p^{+}_{1}&=&(1-x)P^{+}_{1},\hspace{.5cm}
p^{+}_{\bar{q}}=xP^{+}_{1},\nonumber\\
{\bf p}_{1\perp}&=&(1-x){\bf P}_{1\perp} + {\bf k}_{\perp},
\hspace{.5cm}
{\bf p}_{\bar{q}\perp}= x{\bf P}_{1\perp} - {\bf k}_{\perp},
\nonumber\\
p^{+}_{2}&=&(1-x')P^{+}_{2},\hspace{.5cm}
p'^{+}_{\bar{q}}=x'P^{+}_{2},\nonumber\\
{\bf p}_{2\perp}&=&(1-x'){\bf P}_{2\perp} +
{\bf k'}_{\perp},\hspace{.5cm}
{\bf p'}_{\bar{q}\perp}= x'{\bf P}_{2\perp} - {\bf k'}_{\perp},
\end{eqnarray}
where $x(x'=x/r)$ is the momentum fraction carried by the spectator
$\bar{q}$ in the initial(final) state. Note that the spectator model
satisfies $p^{+}_{\bar{q}}$=$p'^{+}_{\bar{q}}$
and ${\bf p}_{\bar{q}\perp}$=${\bf p'}_{\bar{q}\perp}$.

Taking a Lorentz frame given in Eq.~(F.8), the matrix element
of the ``good" current $J^{+}$ in LHS of Eq.~(\ref{eq71}) has the 
form 
\begin{eqnarray}
\la P_{2}|J^{+}_{\rm em}|P_{1}\ra&=&
\sum_{\lambda_{1},\lambda_{2},\bar{\lambda}}
\int P^{+}_{1}dxd^{2}{\bf k}_{\perp}
\phi_{2}(x',{\bf k}_{\perp})\phi_{1}(x,{\bf k}_{\perp})\nonumber\\
&\times&R^{00\dagger}_{\lambda_{2}\bar{\lambda}}(x',{\bf k}_{\perp})
\frac{\bar{u}(p_{2},\lambda_{2})}{\sqrt{p^{+}_{2}}}\gamma^{+}
\frac{u(p_{1},\lambda_{1})}{\sqrt{p^{+}_{1}}}
R^{00}_{\lambda_{1}\bar{\lambda}}(x,{\bf k}_{\perp}).
\end{eqnarray}
Using the covariant form of the spin-orbit wave function given in
Eq.~(\ref{R00_A}), Eq.~(F.12) is transformed into
\begin{eqnarray}
\la P_{2}|J^{+}_{\rm em}|P_{1}\ra&=&
-\sqrt{1/r}\int dxd^{2}{\bf k}_{\perp}
\phi_{2}(x',{\bf k}_{\perp})\phi_{1}(x,{\bf k}_{\perp})
\nonumber\\
&\times&\frac{{\rm Tr} \biggl[\gamma_{5}({\not\! p}_{2}+m_{2})
\gamma^{+}({\not\! p}_{1}+m_{1})\gamma_{5}
({\not\! p}_{\bar{q}}-m_{\bar{q}})\biggr]}
{2\tilde{M}_{10}\tilde{M'}_{20}\sqrt{(1-x)(1-x')}},
\end{eqnarray}
where $\tilde{M}_{i0}$=$\sqrt{ M^{2}_{i0}-(m_{i}-m_{\bar{q}})^{2}}$.
After some manipulations, the trace term in Eq.~(F.13) can be
reduced to
\be
{\rm Tr} \biggl[\gamma_{5}({\not\! p}_{2}+m_{2})
\gamma^{+}({\not\! p}_{1}+m_{1})\gamma_{5}
({\not\! p}_{\bar{q}}-m_{\bar{q}})\biggr]=
-\frac{4P^{+}_{1}}{x'}({\cal A}_{1}{\cal A}_{2} + k_{\perp}^{2}),
\ee
where
\be
{\cal A}_{1}= m_{q}x + m_{\bar{q}}(1-x),\;\;
{\cal A}_{2}= m_{q}x' + m_{\bar{q}}(1-x').
\ee
One can also easily show the following identity
\begin{eqnarray}
\tilde{M}_{0}\sqrt{1-x}\tilde{M'}_{0}\sqrt{1-x'}&=&
\sqrt{{\cal A}_{1}^{2}+ k^{2}_{\perp}}
\sqrt{{\cal A}_{2}^{2}+ k^{2}_{\perp}}.
\end{eqnarray}
Finally, using Eqs.~(\ref{eq71}) and (F.13)-(F.16),
we obtain the EM form factor resulted from the valence
contribution in $q^{+}$$\neq$0 frame as follows
\begin{eqnarray}
F(Q^{2}) &=& e_{q}\frac{2}{1+r}\int^{r}_{0}dx\int d^{2}{\bf k}_{\perp}
\phi_{2}(x',{\bf k}_{\perp})\phi_{1}(x,{\bf k}_{\perp})\nonumber\\
&\times&\frac{{\cal A}_{1}{\cal A}_{2}+ k^{2}_{\perp}}
{\sqrt{{\cal A}_{1}^{2}+ k^{2}_{\perp}}
\sqrt{{\cal A}_{2}^{2}+ k_{\perp}^{2}}}
+ e_{\bar{q}}(q\leftrightarrow\bar{q}\hspace{.2cm}
{\rm of\hspace{.1cm}the\hspace{.1cm}first\hspace{.1cm}term}),
\end{eqnarray}
which is normalized to unity at $Q^{2}$=0.
\subsection{ $0^{-}\to 0^{-}$ Semileptonic Decays}
While the momentum transfer square $q^{2}$ for the EM transitions of
a pseudoscalar meson in $q^{+}$$\neq$0 frame is always negative [see
Eq.~(F.9)],  the $q^{2}$ for the $P\to P$ semileptonic decays
is positive, i.e., $m^{2}_{l}\leq q^{2}\leq (M_{1}-M_{2})^{2}$.
One can easily obtain $q^{2}$ in terms of the fraction $r$ as follows
\be
q^{2}=(1-r)(M^{2}_{1}-\frac{M^{2}_{2}}{r}).
\ee
Accordingly, one gets the two solutions for r:
\be
r_{\pm}=\frac{M_{2}}{M_{1}}\biggl[ v_{1}\cdot v_{2}
\pm \sqrt{(v_{1}\cdot v_{2})^{2}-1} \biggr] ,
\ee
where $v_{1}(v_{2})$ being the four velocity of the parent(daughter)
meson and
\be
v_{1}\cdot v_{2}=\frac{ M^{2}_{1}+M^{2}_{2}-q^{2}}{2M_{1}M_{2}}.
\ee
The $+(-)$ signs in Eq.~(F.19) correspond to the daughter meson
recoiling in the positive(negative) $z$-direction relative to
the parent meson. At zero recoil($q^{2}$=$q^{2}_{\rm max}$) and
maximum recoil($q^{2}$=0), $r_{\pm}$ are given by
\begin{eqnarray}
&&r_{+}(q^{2}_{\rm max})=r_{-}(q^{2}_{\rm max})=\frac{M_{2}}{M_{1}},
\nonumber\\
&&r_{+}(0)= 1,\hspace{0.5cm}
r_{-}(0)=\biggl(\frac{M_{2}}{M_{1}}\biggr)^{2}.
\end{eqnarray}
In order to obtain the form factors $f_{\pm}(q^{2})$ which are
independent of $r_{\pm}$, by putting\cite{Dem,Cheng}
\be
\la P_{2}|\bar{Q_{2}}\gamma^{+}Q_{1}|P_{1}\ra|_{r=r_{\pm}}\equiv
2P_{1}^{+}H(r_{\pm}),
\ee
one obtains from Eq.~(\ref{PP_form})
\be
f_{\pm}(q^{2})=\pm \frac{(1\mp r_{-})H(r_{+}) - (1\mp r_{+})H(r_{-})}
{r_{+}-r_{-}},
\ee
Using a Lorentz frame given in Eq.~(F.8) and the quark momentum
variables in Eq.~(F.11), we obtain $H(r)$ in Eq.~(F.22) as
follows
\be
H(r)=\int^{r}_{0}dx\int d^{2}{\bf k}_{\perp}
\phi_{2}(x',{\bf k}_{\perp})\phi_{1}(x,{\bf k}_{\perp})
\frac{{\cal B}_{1}{\cal B'}_{2}+ k^{2}_{\perp}}
{\sqrt{{\cal B}_{1}^{2}+ k^{2}_{\perp}}
\sqrt{{\cal B'}_{2}^{2}+ k_{\perp}^{2}}},
\ee
where ${\cal B}_{i}$=$m_{i}x + m_{\bar{q}}(1-x)$ and
${\cal B'}_{i}$=$m_{i}x' + m_{\bar{q}}(1-x')$.

%% file: My_resume.tex
\newpage
\chapter*{ Publications}
{\bf Refereed Journals:}

1. Light-Front Quark Model Analysis of Exclusive $0^{-}\to0^{-}$
Semileptonic Heavy Meson Decays,
Phys. Lett. B {\bf460}, 461 (1999) (C.-R. Ji). 

2. Mixing Angles and Electromagnetic Properties
of Ground State Pseudoscalar 
and Vector Meson Nonets in the Light-Cone Quark Model,
Phys. Rev. D {\bf 59}, 074015 (1999) (C.-R. Ji). 

3. Kaon Electroweak Form Factors in the
Light-Front Quark Model,
Phys. Rev. D {\bf 59}, 034001 (1999) (C.-R. Ji).

4. Nonvanishing Zero-Modes in the Light-Front Current, 
Phys. Rev. D {\bf 58}, 071901 (1998) (C.-R. Ji).

5. Relations among the Light-Cone Quark
Models with the Invariant Meson 
Mass Scheme and the Model Predictions of $\eta-\eta'$ Mixing Angle, 
Phys. Rev. D {\bf 56}, 6010 (1997) (C.-R. Ji).

6. Light-Cone Quark Model Predictions for Radiative Meson Decays,
Nucl. Phys. A {\bf 618}, 291 (1997) (C.-R. Ji). 

7. Non-Abelian Berry's Phase in a Slowly Rotating Electric Field, 
J. Math. Phys. {\bf 38}, 1281 (1997) (C.-R. Ji and H.-K. Lee). 

{\bf Preprints and Work in Progress:}

Exploring the timelike region for the elastic form factors in
a scalar field theory, hep-ph/9906225 (C.-R. Ji).

Light-Front Quark Model Analysis of Exclusive $0^{-}\to1^{-}$
Semileptonic Heavy Meson Decays, in preparation (C.-R. Ji). 

{\bf Other Publications}

1. Exploring Timelike Region of QCD Exclusive Processes in
Relativistic Quark Model, in {\em 11th International Light-Cone Workshop 
"New Directions in QCD"}, to be published in AIP proceedings (C.-R. Ji),
{\bf Kyungju}, Korea, June 21-25, 1999.

2. Light-front quark model predictions of meson elastic and transition
form factors,
in {\em Workshop on Exclusive and Semi-exclusive Processes at High
Momentum Transfer}, to be published in the proceedings (C.-R. Ji),
{\bf Jefferson Lab.}, May 20-22, 1999.

3. Meson Electroweak Form Factors in the Light-Cone Approach, 
in {\em Nuclear and Particle Physics with CEBAF at JLAB}, 
Fizika B {\bf 8}, 321 (1999) (C.-R. Ji), 
{\bf Dubrovnik}, November 4-9, 1998.

4. Light-Cone Approach for Exclusive Reactions and Decays in QCD,
in {\em 1998 YITP-Workshop on QCD and Hadron Physics},
to be published in the Proceedings by World Scientific (C.-R. Ji),
{\bf Kyoto}, October 14-16, 1998.

5. Semileptonic Decays of Pseudoscalar Mesons in the Light-Front
Quark Model, in {\it 16th European Conference on Few-Body problems
in Physics}, Few-Body System Suppl. {\bf 10}, 131 (1999) (C.-R. Ji), 
{\bf Autran}, June 1-6, 1998.

6. Mixing Angles and Electromagnetic Properties of Ground State
Pseudoscalar and Vector Meson Nonets in the Light-Front Quark Model, 
in {\it The structure of mesons, baryons and nuclei and
MESON98 Workshop}, Acta Physica Polonica {\bf B 29}, 3363 (1998) 
(C.-R. Ji), {\bf Krakow}, May 27-June 2, 1998.

7. Lightcone Vacuum and Radiative Meson Decays,
in {\it International School-Seminar ``Structure of Particles and Nuclei
and their} {\it Interactions"}, to be published in the Proceedings (C.-R. Ji),
{\bf Tashkent}, October 6-13, 1997.

8. Radiative Meson Decays in the Light-cone quark model,
in {\it 10th Summer School $\&$ Symposium on Nuclear Physics
`` QCD, Lightcone} {\it Physics and Hadron Phenomenology"},
pp. 255-259, published by World Scientific Publishing Co. in 1998
and edited by C.-R. Ji and D.-P. Min, 
{\bf Seoul}, June 23-28, 1997.

9. Radiative Meson Decays in a Relativistic Quark Model,
in {\it the Conference on Perspectives in Hadronic Physics},
pp. 539-548, published by World Scientific Publishing Co. in 1998
and edited by S. Boffi, C. Atti and M. Giannini (C.-R. Ji),
{\bf Trieste,} May 12-16, 1997.
 
10. Radiative Decays and Transition Form Factors of Strange Mesons in the 
Relativistic Costituent Quark Model, in {\it Strangeness '96},  
Heavy Ion Physics {\bf Vol. 4}, 369-379(1996) (C.-R. Ji), 
{\bf Budapest}, May 15-17,1996.

{\bf  Talks at Conferences and Workshops}

1. Form factors of two-body bound states and the analytic
continuation method in (3+1) dimensional scalar theory,
in {\em Centennial meeting of the American Physical Society},
Bull. Am. Phys. Soc. {\bf 44}, No. 1, 169 (1999),
{\bf Atlanta}, March 20-26, 1999.

2. Nonvanishing zero-modes in the light-front current,
in {\em 65th Annual Southeastern Section Meeting of the American
Physical Society}, Bull.Am.Phys.Soc. {\bf Vol. 43}, No. 7, 1607 (1998),
{\bf Miami}, November 13-15, 1998.
 
3. $P\to P$ Semileptonic decays in the light-cone quark model,
in {\em Joint APS/AAPT Meeting of the American Physical Society}, 
{\bf Columbus}, April 18-21, 1998.

4. Effective potential and mixing angles of
$(\omega,\phi)$ and $(\eta,\eta')$ in the light-cone quark model,
in {\em 64th Annual Southeastern Section Meeting of the American
Physical Society}, Bull.Am.Phys.Soc. {\bf Vol. 42}, No. 9, 1805 (1997),
{\bf Nashville}, November 6-8, 1997.

5. Light-cone quark model studies of radiative meson decays,
in {\it 8th International Workshop on Light-Cone QCD and Nonperturbative}
{\it Hadronic Physics}, {\bf Lutsen}, August 11-22, 1997.

6. Relations among the light-cone quark models with the invariant meson
mass scheme and the model prediction of $\eta-\eta'$ mixing,
in {\em Joint APS/AAPT Meeting of  the American Physical Society}, 
{\bf Washington}, April 18-21, 1997.

7. Light-cone quark model predictions for radiative meson decays,
in {\em  63rd Annual Southeastern Section Meeting of the American Physical
Society}, {\bf Decatur}, November 14-16, 1996.

8. Light-cone quark model predictions for radiative meson decays,
in {\em Joint APS/AAPT Meeting of the American Physical Society}, 
{\bf Indianapolis}, May 2-5, 1996.


%% file: diss_main.bbl
\begin{thebibliography}{11a}
\bibitem{QCD} D. J. Gross and F. Wilczek, 
\Journal{\PRL}{30}{1343}{1973}; H. D. Politzer, 
{\em ibid.} {\bf 30}, 1346 (1973).  

\bibitem{chernyak} V. L. Chernyak and I. R. Zhitnitsky, 
Phys. Rep. {\bf 112}, 1738 (1984).

\bibitem{narison} S. Narison, \Journal{\PLB}{224}{184}{1989}. 

\bibitem{rad} A. V. Radyushkin, presented at 2nd European Workshop on
Hadronic Physics with Electrons Beyond 10-Gev, Dourdan, France, 1990.

\bibitem{gott} S. Gottlieb and A. S. Kronfeld, 
\Journal{\PRL}{55}{2531}{1985}.

\bibitem{loft} T. A. DeGrand and R. D. Loft, 
\Journal{\PRD}{38}{954}{1988}. 

\bibitem{gupta} D.  Daniel, R. Gupta, and D. G. Richards, 
\Journal{\PRD}{43}{3751}{1991}.

\bibitem{teren} M. V. Terent'ev, Yad. Fiz. {\bf 24}, 207 (1976) 
[ Sov.J. Nucl.  Phys. {\bf 24}, 106 (1976)]; V. B. Berestetsky and 
M. V. Terent'ev, $ibid$.  {\bf 24}, 1044 (1976) [ {\bf 24}, 547 (1976)]; 
{\bf 25}, 653 (1977)[ {\bf 25}, 347 (1977)].

\bibitem{jaus1} W. Jaus, Phys. Rev. D {\bf 41}, 3394 (1990).

\bibitem{jaus} W. Jaus, \Journal{\PRD}{44}{2851}{1991}.  

\bibitem{chung2} P. L. Chung, F. Coester, and W. N. Polyzou, 
Phys. Lett. B {\bf 205}, 545 (1988).

\bibitem{IM} H.-M. Choi and C.-R.Ji, Phys. Rev. D {\bf 56}, 6010 (1997).

\bibitem{Mix} H.-M. Choi and C.-R. Ji, \Journal{\PRD}{59}{074015}{1999}. 

\bibitem{Bag} A. S. Bagdasaryan and S. V. Esaybegyan,
Sov. J. Nucl. Phys. {\bf 42}, 278 (1985).

\bibitem{azn} I. G. Aznauryan and K. A. Oganessyan, 
\Journal{\PLB}{249}{309}{1990}.

\bibitem{Kar} V. A. Karmanov, Nucl. Phys. B {\bf 166}, 378 (1980);
Nucl. Phys. A {\bf 362}, 331(1981).

\bibitem{card} F. Cardarelli et al., Phys. Lett. B {\bf 349},
393 (1995); {\bf 359}, 1 (1995); {\bf 332}, 1 (1994).

\bibitem{tao} T. Huang, B.-Q. Ma, and Q.-X. Shen, 
\Journal{\PRD}{49}{1490}{1994}. 

\bibitem{sch} F. Schlumpf, \Journal{\PRD}{50}{6895}{1994}.

\bibitem{dziem1} Z. Dziembowski and L. Mankiewicz,
\Journal{\PRL}{58}{2175}{1987}; Z. Dziembowski,
{\em ibid.} {\bf 37}, 2030 (1988); {\bf 37}, 768 (1988);
{\bf 37}, 778 (1988).

\bibitem{ji} C.-R. Ji and S. R. Cotanch, 
\Journal{\PRD}{41}{2319}{1990}; 
C.-R. Ji, P. L. Chung, and S. R. Cotanch, 
{\em ibid.} {\bf 45}, 4214 (1992). 

\bibitem{spin} H.-M. Choi and C.-R. Ji, 
\Journal{\NPA}{618}{291}{1997}.

\bibitem{BPP} S. J. Brodsky, H.-C. Pauli, and S. S. Pinsky, 
Phys. Rept. {\bf 301}, 299 (1998).

\bibitem{Dirac} P. A. M. Dirac, Rev. Mod. Phys. {\bf 21}, 392 (194).

\bibitem{Lepage} G. P. Lepage, S. J. Brodsky, T. Huang, and P. Mackenzie, 
in {\em Particle and Fields 2, Proceedings of the Banff Summer Institute,}
Banff, Alberta, 1981, edited by A. Z. Capri and A. N. Kamal(Plenum,
New York, 1983), p.83.

\bibitem{zm} H.-M. Choi and C.-R. Ji, \Journal{\PRD}{58}{071901}{1998}.

\bibitem{zero} M. Burkardt, \Journal{\NPA}{504}{762}{1989}; 
\Journal{\PLB}{268}{419}{1991}; \Journal{\PRD}{47}{4628}{1993};
K. Yamawaki, ``QCD, Lightcone Physics and
Hadron Phenomenology'', lectures given at 10th Annual Summer School and
Symposium on Nuclear Physics (NUSS 97), Seoul, Korea, June 23-28,
1997, hep-th/9802037

\bibitem{BH} S. J. Brodsky and D. S. Hwang, 
\Journal{\NPB}{543}{239}{1998}.

\bibitem{Fred} J. P. B. C. de Melo, J. H. O. Sales, T. Frederico and
P. U. Sauer, \Journal{\NPA}{631}{574}{1998}; J. P. B. C. de Melo,
H. W. L. Naus and T. Frederico, 
\Journal{\PRC}{59}{2278}{1999}.

\bibitem{JS} C.-R. Ji and Y. Surya, \Journal{\PRD}{46}{3565}{1992}.

\bibitem{Sop} D. E. Soper, Ph.D. Thesis, SLAC Report No. {\bf 137}
(1971).

\bibitem{Mel} H. J. Melosh, \Journal{\PRD}{9}{1095}{1974};
L. A. Kondratyuk and M. V. Terentyev, Sov. J. Nucl. Phys. {\bf 31}, 561
(1983).

\bibitem{JKM}
C.-R. Ji, G.-H. Kim and D.-P. Min, \Journal{\PRD}{51}{879}{1995}.

\bibitem{JCon}
C.-R. Ji, in: DIRKFEST'92, eds. W. Buck, K. Maung and B. Serot (World
Scientific Publishing, Singapore,1992) pp. 147-163;
C.-R. Ji, {\em Challenges in Lightcone Field Quantization},
in Second Pacific Winter School for Theoretical Physics, Sorak, 1995.

\bibitem{BMPP} 
S. J. Brodsky, G. McCartor, H.-C. Pauli and S. Pinsky, {\em The challenge
of light-cone quantization of gauge field theory}, Tech. Report
SLAC-PUB-5811, Stanford Linear Accelerator Center, Stanford
University, Stanford, USA, 1992.

\bibitem{Horn}
K.Hornbostel, {\em Nontrivial vacua from equal time to the light cone}, 
\Journal{\PRD}{45}{3781}{1992}; C.-R. Ji,
{\em The Lightcone Vacuum}, Proceedings of the Cornelius Lanczos
International Centenary Conference, published by siam, 1994, pp.~615--617.

\bibitem{Rey} C.-R. Ji and S.J. Rey, \Journal{\PRD}{53}{5815}{1996}.

\bibitem{Ku} Y. Kuramashi, M. Fukugita et al., 
\Journal{\PRL}{72}{3448}{1994}.

\bibitem{Drell} S. D. Drell and T. M. Yan,
\Journal{\PRL}{24}{181}{1970}; G. West, {\em ibid.} {\bf 24}, 1206 (1970).

\bibitem{LB} G. P. Lepage and S. J. Brodsky,
\Journal{\PRD}{22}{2157}{1980}.

\bibitem{Kaon} H.-M. Choi and C.-R. Ji, \Journal{\PRD}{59}{034001}{1999}.

\bibitem{HI} C. Hayne and N. Isgur, \Journal{\PRD}{25}{1944}{1982}.

\bibitem{anal} H.-M. Choi and C.-R. Ji, ``Exploring the timelike region
for the elastic form factors in a scalar field theory", hep-ph/9906225.

\bibitem{Semi} H.-M. Choi and C.-R. Ji, \Journal{\PLB}{460}{461}{1999}.

\bibitem{BHL1} S. J. Brodsky, T. Huang, and G. P. Lepage, in
{\em Quarks and Nuclear Forces,} edited by D. Fries and B. Zeitnitz,
Springer Tracts in Modern Physics, Vol. 100 (Springer-Verlag,
New York, 1982).

\bibitem{nisgur} N. Isgur, in {\em The New Aspect of Subnuclear Physics},
editted by A. Zichichi (Plenum, New York, 1980) p. 107.

\bibitem{isgur1} N. Isgur, \Journal{\PRD}{12}{3770}{1975};
{\bf 13}, 122 (1976).

\bibitem{isgur2} S. Godfrey and N. Isgur, \Journal{\PRD}{32}{189}{1985}.

\bibitem{data} Particle Data Group, C. Caso et al., 
Eur. Phys. J. C {\bf 3}, 1 (1998). 

\bibitem{smilga} B. L. Ioffe and A. V. Smilga,
\Journal{\NPB}{216}{373}{1983}; \Journal{\PLB}{114}{353}{1982}.

\bibitem{cello1} CELLO coll., H.-J. Behrend et al., 
\Journal{\ZPC}{49}{401}{1991}.

\bibitem{cello2} CELLO coll., V. Savinov et al.,
PHOTON 95 Conference, Sheffield (1995).

\bibitem{tpc} TPC/2$\gamma$ coll., H. Aihara et al., 
\Journal{\PRL}{64}{172}{1990}.

\bibitem{arnold} R. G. Arnold, C. E. Carlson, and F. Gross,
\Journal{\PRC}{21}{1426}{1980}.

\bibitem{chung} P. L. Chung, F. Coester, B. D. Keister and
W. N. Polyzou, \Journal{\PRC}{37}{2000}{1988}.

\bibitem{gross} C. E. Carlson and F. Gross,
\Journal{\PRL}{53}{127}{1984}.

\bibitem{hiller} S. J. Brodsky and J. R. Hiller,
\Journal{\PRD}{46}{2141}{1992}. 

\bibitem{grach} I. L. Grach and L. A. Kondratyuk,
Sov. J. Nucl. Phys. {\bf 39}, 198 (1984).

\bibitem{keister} B. D. Keister, \Journal{\PRC}{49}{1500}{1994}.

\bibitem{Ito} H. Ito and F. Gross, \Journal{\PRL}{71}{2555}{1993}.

\bibitem{dzh} R. I. Dzhelyadin et al.,
\Journal{\PLB}{102}{296}{1981}.

\bibitem{munz} C. R. M$\ddot{u}$nz et al.,
\Journal{\PRC}{52}{2110}{1995}.

\bibitem{braun} V. M. Braun and I. E. Filyanov,
\Journal{\ZPC}{44}{157}{1989}; {\em ibid}. C {\bf 48}, 239 (1990).

\bibitem{bel} V. M. Belyaev, \Journal{\ZPC}{65}{93}{1995}.

\bibitem{rose1} J. Babcock and J. L. Rosner, \Journal{\PRD}{14}{1286}{1976}.

\bibitem{rose2} J. L. Rosner, Phys. Rev. D {\bf 23}, 1127 (1981).

\bibitem{ziel1} M. Zielinski et al., Phys. Rev. Lett. {\bf 52}, 1195 (1984).

\bibitem{ziel2} M. Zielinski, Phys. Rev. Lett. {\bf 58}, 2002 (1987).

\bibitem{coester} F. Coester, in {\em Constraint's Theory and
Relativistic Dynamics}, edited by G. Longhi and L. Lusanna(World 
Scientific, Singapore, 1987), p. 159; in {\em The Three-Body Force in 
the Three-Nucleon System}, edited by B. L. Berman and 
B. F. Gibson(Springer, New York,1986), p. 472.

\bibitem{amendolia} 
S. R. Amendolia et al., Phys. Lett. B {\bf 146}, 116 (1984).


\bibitem{adler} S. L. Adler, Phys. Rev. {\bf 117}, 2426 (1969);
J. S. Bell and R. Jackiw, Nuovo Cimento A {\bf 60}, 47 (1969).

\bibitem{gasser} 
J. Gasser and H. Leutwyler, Nucl. Phys. B {\bf 250}, 465 (1985).

\bibitem{qft} C. Itzykson and J. -B. Zuber, {\em Quantum Field Theory}
(McGRAW-HILL International editions, 1985), p. 550.

\bibitem{john} J. F. Donoghue, B. R. Holstein, and Y. C. R. Lin,
Phys. Rev. Lett. {\bf 55}, 2766 (1985).

\bibitem{georgi} A. De R\'{u}jula, H. Georgi, and S. Glashow,
Phys. Rev. D {\bf 12}, 147 (1975).

\bibitem{scadron} M. D. Scadron, Phys. Rev. D {\bf 29}, 2076 (1984).

\bibitem{lucha} W. Lucha, F. F. Sch\"{o}berl, and D. Gromes,
Phys. Rep. {\bf 200}, 127 (1991).

\bibitem{Karl} N. Isgur and G. Karl, Phys. Lett. B {\bf 72}, 109 (1977).

\bibitem{gromes} D. Gromes and I. O. Stamatescu, Nucl. Phys. B {\bf 112},
213 (1976).

\bibitem{isgw} N. Isgur, D. Scora, B. Grinstein, and M. B. Wise,
Phys. Rev. D {\bf 39}, 799 (1989).

\bibitem{isgw2} D. Scora and N. Isgur, Phys. Rev. D {\bf 52}, 2783 (1995).

\bibitem{Capstick} S. Capstick and N. Isgur, Phys. Rev. D{\bf 34},
2809 (1986).

\bibitem{Das} T. Das, V. S. Mathur, and S. Okubo, Phys. Rev. Lett. {\bf 19},
470 (1967); J. J. Sakurai, $ibid$. {\bf 19}, 803 (1967).

\bibitem{Sakurai} R. J. Oakes and J. J. Sakurai, Phys. Rev. Lett. {\bf 19},
1266 (1967).

\bibitem{Coleman} S. Coleman and H. J. Schnitzer, Phys. Rev. {\bf 134},
B863 (1964); N. M. Kroll, T. D. Lee, and B. Zumino, $ibid$. {\bf 157},
1376 (1967).

\bibitem{Bebek} C. J. Bebek et al., Phys. Rev. D {\bf 17}, 1693 (1978).

\bibitem{bell} J. J. Sakurai, K. Schilcher and  M. D. Tran, Phys. Lett.
B {\bf 102}, 55 (1981); J. S. Bell and J. Pasupathy, {\em ibid.} B {\bf 83},
389 (1970).

\bibitem{amen2} R. A. Amendolia et al., Phys. Lett. B {\bf 178},
435 (1986).

\bibitem{dono} J. F. Donoghue, B. R. Holstein, and Y. C. R. Lin,
Phys. Rev. Lett. {\bf 55}, 2766 (1985).

\bibitem{r1} {\em The second DA$\Phi$NE physics handbook}, edited by
L. Maiani, G. Pancheri, and N. Paver, INFN-LNF-Divisione Ricerca, 
Report No. ISBN 88-86409-02-8.

\bibitem{r2} F. Anulli et al., {\em Measurement of two photon
interactions with the KLOE small angle tagging system}, p. 607
in Vol. II of Ref.~\cite{r1}.

\bibitem{private} private communication with Prof. Nathan Isgur.

\bibitem{SM}  M. Sawicki and L. Mankiewicz,
\Journal{\PRD}{37}{421}{1988}; L. Mankiewicz and M. Sawicki,
{\it ibid}. {\bf 40}, 3415 (1989).

\bibitem{GS} S. Glazek and M. Sawicki,
\Journal{\PRD}{41}{2563}{1990}.

\bibitem{Sa} M. Sawicki, \Journal{\PRD}{44}{433}{1991};
{\it ibid}. {\bf 46}, 474 (1992).

\bibitem{Ba} N.C.J. Schoonderwoerd and B.L.G. Bakker,
\Journal{\PRD}{57}{4965}{1998}; {\it ibid}. {\bf 58}, 025013 (1998).

\bibitem{SDrell} J. D. Bjorken and S. D. Drell,
{\em Relativistic Quantum Fields} (McGraw-Hill, New York, 1965), pp. 209-282.

\bibitem{Gasio} S. Gasiorowicz, {\em Elementary Particle Physics}
(Wiley, New York, 1966), pp. 348-362.

\bibitem{Man} S. Mandelstam, \Journal{\PRL}{4}{84}{1960}.

\bibitem{Gell} M. Gell-Mann, M. L. Goldberger, and W. Thirring,
Phys. Rev. {\bf 95}, 1612 (1954); M. Goldberger, Phys. Rev. {\bf 99},
979 (1955).

\bibitem{Brodsky} S. J. Brodsky, R. Roskies, and R. Suaya, Phys. Rev. D
{\bf 8}, 4574(1973).

\bibitem{Flynn} J. M. Flynn and C. T. Sachrajda, ``Heavy Quark
Physics From Lattice QCD", to appear in Heavy Flavor (2nd edition) edited 
by A. J. Buras and M. Lindner (World Scientific, Singapore),
hep-lat/9710057.

\bibitem{Flynn2} UKQCD Collaboration, J. M. Flynn et al.,
\Journal{\NPB}{461}{327}{1996}.

\bibitem{UK2} UKQCD Collaboration, L. Del Debbio et al., 
\Journal{\PLB}{416}{392}{1998}.

\bibitem{ALL} APE Collaboration, C. R. Allton et al., 
\Journal{\PLB}{345}{513}{1995}.

\bibitem{LMMS} V. Lubicz, G. Martinelli, M. McCarthy and
C. T. Sachrajda, \Journal{\PLB}{274}{415}{1992}.
 
\bibitem{Bern2} MILC Collaboration, C. Bernard et al., 
\Journal{\PRL}{81}{4812}{1998}. 

\bibitem{UKQCD} UKQCD Collaboration, D. R. Burford et al.,
\Journal{\NPB}{447}{425}{1995}.

\bibitem{Bernard} C. W. Bernard, A. X. El-Khadra, and A. Soni,
\Journal{\PRD}{43}{2140}{1991}; $ibid$ {\bf 45}, 869 (1992).

\bibitem{Bowler} UKQCD Collaboration, K. C. Bowler et al., 
\Journal{\PRD}{51}{4905}{1995}.

\bibitem{Abada} ELC Collaboration, A. Abada et al., 
\Journal{\NPB}{416}{675}{1994}.

\bibitem{BB1} P. Ball and V. M. Braun, \Journal{\PRD}{58}{094016}{1998}.

\bibitem{BB2} P. Ball and V. M. Braun, \Journal{\PRD}{55}{5561}{1997}.

\bibitem{Ball} P. Ball, \Journal{\PRD}{48}{3190}{1993}.

\bibitem{Nar2}
S. Narison, \Journal{\PLB}{283}{384}{1992}; $ibid$ {\bf 322}, 247 (1994).

\bibitem{BBD} P. Ball, V. M. Braun and H. G. Dosch,
\Journal{\PRD}{44}{3567}{1991}.

\bibitem{IW} N. Isgur and M. B. Wise, \Journal{\PLB}{232}{113}{1989};
$ibid$ {\bf 237}, 527 (1990).

\bibitem{Wirbel} M. Wirbel, B. Stech, and M. Bauer,
\Journal{\ZPC}{29}{637}{1985}; M. Bauer, B. Stech, and M. Wirbel, $ibid$.
{\bf 42}, 671 (1989).

\bibitem{Dem} N.B. Demchuk, I. L. Grach, I. M. Narodetskii, and
S. Simula, \Journal{\PAN}{59}{2152}{1996}.

\bibitem{Cheng} H.-Y. Cheng, C.-Y. Cheng, and C.-W. Hwang,
\Journal{\PRD}{55}{1559}{1997}.

\bibitem{OD1} P. J. O'Donnell and Q. P. Xu,
\Journal{\PLB}{325}{219}{1994}; {\bf 336}, 113(1994).

\bibitem{OD2} P. J. O'Donnell, Q. P. Xu, and H. K. K. Tung,
\Journal{\PRD}{52}{3966}{1995}.

\bibitem{Melikhov} D. Melikhov, \Journal{\PLB}{394}{385}{1997};
\Journal{\PRD}{53}{2460}{1996}.

\bibitem{Mel2} M. Beyer and Dmitri Melikhov,
\Journal{\PLB}{436}{344}{1998}.

\bibitem{Jaus3} W. Jaus, \Journal{\PRD}{53}{1349}{1996}.

\bibitem{Roos} H. Leutwyler and M. Roos, Z. Phys. C {\bf 25}, 91 (1984).

\bibitem{Gasser} J. Bijnens, G. Colangelo, G. Ecker, and J. Gasser,
in {\em The Second DA$\Phi$NE Physics Handbook}, edited by L. Maiani,
G. Pancheri, and N. Paver (INFN, Rome, Italy, 1995), p. 315.

\bibitem{CL} E. P. Shabalin, Sov. J. Nucl. Phys. {\bf 51}, 296 (1990).

\bibitem{VMD} L.-M. Chounet, J.-M. Gailard, and M. K. Gailard,
Phys. Rep. {\bf 4}, 199 (1974).

\bibitem{Andrei} A. Afanasev and W. W. Buck, Phys. Rev. D {\bf 55},
4380 (1997).

\bibitem{Yuri} Yu. Kalinovsky, K. L. Mitchell, and C. D. Roberts,
Phys. Lett. B {\bf 399}, 22 (1997).

\bibitem{Neubert} M. Neubert, Phys. Rep. {\bf 245}, 261(1994).

\bibitem{Gatto} M. Ademollo and R. Gatto, Phys. Rev. Lett. {\bf 13},
264 (1964).

\bibitem{IS} N. Isgur, \Journal{\PRD}{12}{3666}{1975}.

\bibitem{data2} A. Ryd, talk presented at Seventh Int. Symp. on 
Heavy Flavors, Santa Barbara, California, July 1997.

\bibitem{E653} Fermilab E653 Collaboration, K. Kondama et al.,
\Journal{\PLB}{286}{187}{1992}.

\bibitem{E687} Fermilab E687 Collaboration, P. L. Frabetti et al.,
\Journal{\PLB}{307}{262}{1993}.

\bibitem{Cleo} J. P. Alexander et al., \Journal{\PRL}{77}{5000}{1996}. 

\bibitem{Ivan} M. A. Ivanove, Yu. L. Kalinovsky, P. Maris, and
C. D. Roberts, \Journal{\PRC}{57}{1991}{1998}.

\bibitem{Argus} H. Albrecht et al., \Journal{\ZPC}{57}{249}{1993}.

\bibitem{Athanas} M. Athanas et al., \Journal{\PRL}{79}{2208}{1997}.

\bibitem{ARLT} ARGUS Collaboration, H. Albrecht et al.,
\Journal{\PLB}{219}{121}{1989}.

\bibitem{CLLT} CLEO Collaboration, D. Bortoletto et al.,
\Journal{\PRL}{63}{1667}{1989}.

\bibitem{OPAL} K. Ackerstaff et al., OPAL Collaboration,
\Journal{\PLB}{395}{128}{1997}.

\bibitem{ALEPH} D. Buskulic et al., ALEPH Collaboration,
\Journal{\PLB}{395}{373}{1997}.

\bibitem{DELPHI} P. Abreu et al., DELPHI Collaboration,
\Journal{\ZPC}{71}{539}{1996}.

\bibitem{Barish} CLEO Collaboration, B. Barish et al.,
\Journal{\PRD}{51}{1014}{1995}.

\bibitem{San} CLEO Collaboration, S. Sanghera et al.,
\Journal{\PRD}{47}{791}{1993}.

\bibitem{Dubo} CLEO Collaboration, J. E. Duboscq et al.,
\Journal{\PRL}{76}{3898}{1996}.

\bibitem{KS} J. G. K$\ddot{o}$rner, G. A. Schuler,
\Journal{\ZPC}{38}{511}{1988}; \Journal{\PLB}{226}{185}{1989}.

\end{thebibliography}
